*To my wife Barbara and*

*in memory to Berra Ferruccio*

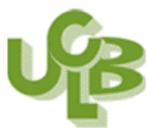

**Université Claude Bernard Lyon I**

**Institut de Physique Nucléaire**

**THESE**

pour l'obtention

du **DIPLOME DE DOCTORAT**

présentée et soutenue publiquement le

**14 Octobre 2005**

par Paolo Berra

# *Conception, construction et essai d'un accélérateur linéaire à protons impulsé à 3 GHz (LIBO) pour la thérapie du cancer*

**Jury :**      **Ugo Amaldi (Rapporteur) (TERA, Genève et Milan)**

**Jean-Marie De Conto (Rapporteur) (LPSC, Grenoble)**

**Albert Demeyer (directeur de thèse) (UCBL, Lyon)**

**Jean-Pierre Gerard (CAL, Nice)**

**Hans Falk Hoffmann (CERN, Genève)**

**Joseph Remillieux (UCBL, Lyon)**

**membres invités :**      **Marcel Bajard (UCBL, Lyon)**

**Ettore Rosso (CERN, Genève)**

**Balazs Szeless (CERN, Genève)**

**Mario Weiss (TERA, Genève)**

*Main index*

---

*    The sections marked with an asterisk are presented as optional material, which can be omitted for a first reading without disrupting the overall continuity of the presentation.

***Conception, construction et essai d'un accélérateur linéaire à protons***
***impulsé à 3 GHz (LIBO) pour la thérapie du cancer***

## Préface

Dans le courant des dix dernières années, l'utilisation de faisceaux de protons est devenue un outil clinique pour le traitement des tumeurs profondes. LIBO (acronyme anglais pour LInac BOoster) est un accélérateur linéaire à protons (SCL) de prix modéré pour la thérapie du cancer. Il a été conçu par la Fondation TERA (Italie) pour être installé en aval d'un cyclotron à protons existant et, impulsé à une radiofréquence de 3 GHz, incrémenter l'énergie de son faisceau jusqu'aux 200 MeV nécessaires à une pénétration de ~25 cm dans le corps humain. Cette solution permet de transformer des cyclotrons commerciaux, normalement utilisés dans la thérapie du mélanome oculaire, dans la production d'isotopes et en physique nucléaire, en un accélérateur pour la cure des tumeurs profondes. Un module prototype du LIBO a été construit et testé avec succès à puissance RF maximale au CERN puis avec un faisceau de protons à Catane (Italie) auprès du Laboratorio Nazionale del Sud (LNS) de l'INFN, par une collaboration internationale formée par la Fondation TERA, le CERN et les Universités et les Sections INFN de Milan et Naples. Le but du projet à moyen terme concerne le transfert technologique des connaissances acquises dans ce domaine à un consortium de sociétés industrielles en vue d'installer dans des hôpitaux ce nouvel outil médical. La conception, la construction et les essais du module prototype du LIBO sont décrits en détail.

***Design, construction and tests of a 3 GHz proton linac booster (LIBO)***
***for cancer therapy***

## Preface

In the last ten years the use of proton beams in radiation therapy has become a clinical tool for treatment of deep-seated tumours. LIBO is a RF compact and low cost proton linear accelerator (SCL type) for hadrontherapy. It is conceived by TERA Foundation as a 3 GHz Linac Booster, to be mounted downstream of an existing cyclotron in order to boost the energy of the proton beam up to 200 MeV, needed for deep treatment (~25 cm) in the human body. With this solution it is possible to transform a low energy commercial cyclotron, normally used for eye melanoma therapy, isotope production and nuclear physics research, into an accelerator for deep-seated tumours. A prototype module of LIBO has been built and successfully tested with full RF power at CERN and with proton beam at INFN Laboratori Nazionali del Sud (LNS) in Catania, within an international collaboration between TERA Foundation, CERN, the Universities and INFN groups of Milan and Naples. The mid-term aim of the project is the technology transfer of the accumulated know-how to a consortium of companies and to bring this novel medical tool to hospitals. The design, construction and tests of the LIBO prototype are described in detail.

## *Introduction*

After my graduation in nuclear engineering at Politecnico of Milan and with a one-year experience on the design of a synchrotron for hadrontherapy, I started to work with CERN and TERA Foundation on LIBO project in 1998 in the frame of an international collaboration. The main activities presented in this thesis have been performed in the period 1998-2002, most of which I spent at CERN in co-operation with senior specialists in mechanics, Radio Frequency (RF) and particle accelerator technology.

The objective of the thesis is to present the results of these activities that have generated the first LIBO prototype module. In particular chapters 1 and 3 are used as a general introduction to cancer therapy with hadrons and to the physics of linear accelerators, giving some background information to be used later. A brief description of LIBO studies is shown in chapters 2 and 4 while chapters 5 to 7 are devoted to the design, construction and tests of the prototype, with particular emphasis on the engineering subjects. In this part I was directly involved and then it represents the main core of the present thesis. Finally chapter 8 shows some ideas about future possibilities generated by the success of the prototype, such as the IDRA project, the technology transfer to industry for LIBO production and the new linac for cancer therapy with carbon ions of TERA Foundation.

# Chapter 1

# Conventional radiotherapy and hadrontherapy:

# the physics for medicine

*"... the specific ionisation or dose is many times less where the proton enters the tissue at high energy than it is in the last centimetre of the path where the ion is brought to rest; ... these properties make it possible to irradiate intensely a strictly localised region within the body, with but little skin dose; ... since the range of the beam is easily controllable, precision exposure of well defined small volumes within the body will soon be feasible..."*

*[R. R. Wilson, 1946]*

## 1.1     Introduction

Cancer is the second major cause of death (after cardiovascular diseases) in the developed countries. Today the three available therapeutic approaches are: surgery, radiation therapy and chemotherapy. With these techniques successful treatments can be achieved for about 45% of all cancer patients. For about 2/3 of the patients the disease is still well localised within a specific region of the body at the time of the diagnosis of cancer. For these patients the chances of cure using a local therapy (surgery or radiation therapy) are reasonably good. Reduced sizes of tumours and early diagnosis allow big probability for good therapeutic results. In this context screening plays a crucial role in the early detection of the disease. Surgery is the most successful modality. Radiation therapy, used also in combination with surgery, is the second most effective treatment and it is used when the tumour is still well localised but it is inoperable. Chemotherapy is used when the disease has already spread in the whole body with distant metastases and with the intent to eliminate the diffused cancer cells.

Today big pharmaceutical companies on new developments based on genetic technologies undertake the largest effort in cancer research. However in the near future, while waiting for an eventual breakthrough from these new methods, surgery and radiation therapy will continue to play a major role in the control of primary solid tumours. It is also important to note that the different types of therapy are not necessarily exclusive and are often used in a complementary way. Many recent successes in cancer management are in fact based on the combined use of different modalities. About 2/3 of all cancer patients receives radiation therapy alone or in combination with other modalities. A reduction of the toxicity of a therapy modality automatically improves the tolerance of the others in a combined treatment. Improvements in radiation therapy were achieved in the past by using advanced treatment techniques (for example conformal radiation therapy) and/or by using new types of radiation where beams of protons or light ions are used, the so-called hadrontherapy [B.1]. Protons have a well-defined range of penetration in materials (the range depends on the selected initial energy of the beam), and they show a peak dose in the region where the beam stops.

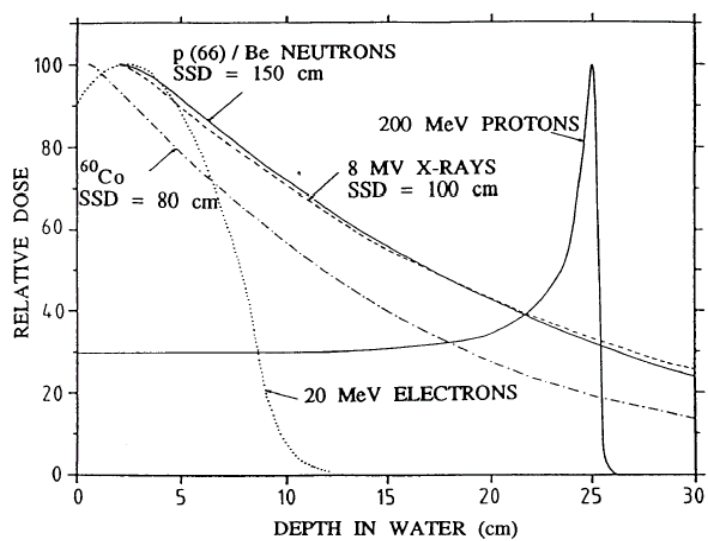

**_Figure 1.1_** _Dose distributions for photon, electron, neutron and proton beams [B.2]._

This region is called the Bragg peak [B.2]. Figure 1.1 shows the dose deposition of a mono-energetic proton beam as a function of the depth in water compared with the corresponding dose profile of clinically used photons, which show a characteristic exponential fall-off of the dose with increasing depth. With proton energy modulation one could generate the so-called Spread Out Bragg Peak (SOBP). The possibility of using the Bragg peak as a practical tool for the localisation of the dose in depth has been known for more than 50 years, but unfortunately at that time modern computer technology for treatment planning and the precise screening of the patient were not available. Technological advances in these areas provided by the Computerised Tomography (CT), the Magnetic Resonance Imaging (MRI), and the Positron Emission Tomography (PET) allow today to take full advantage of the inherent precision of this method (figure 1.2, left). The main advantage of proton therapy is expected from the superior capability to confine the dose on the target volume and so to reduce the dose burden to the surrounding healthy tissues. The major source of comparison for proton (and ion) therapy is clearly conventional therapy with photons. Protons are expected to produce superior results for the treatment of large tumours of complex shape, where a significant reduction of the dose outside of the target volume is clinically desirable. This dose sparing, a factor two or more, could be relevant in many situations, for example for radiation therapy of pediatric tumours. Children surviving cancer often present severe consequences of the treatment in adulthood, like reduced intelligence after brain irradiation or an abnormal growth after treatments covering parts of the skeleton. By using protons one can expect to reduce these negative effects. It is very important to note that the clinical goal of proton therapy is not only to increase patient survival but also to achieve a better quality of life after the treatment. From the radiobiological point of view protons behave like photons in conventional therapy. The experience acquired in the hospitals on the tolerances of the different organs and the response to radiation of the various types of cancers can be directly applied to protons.

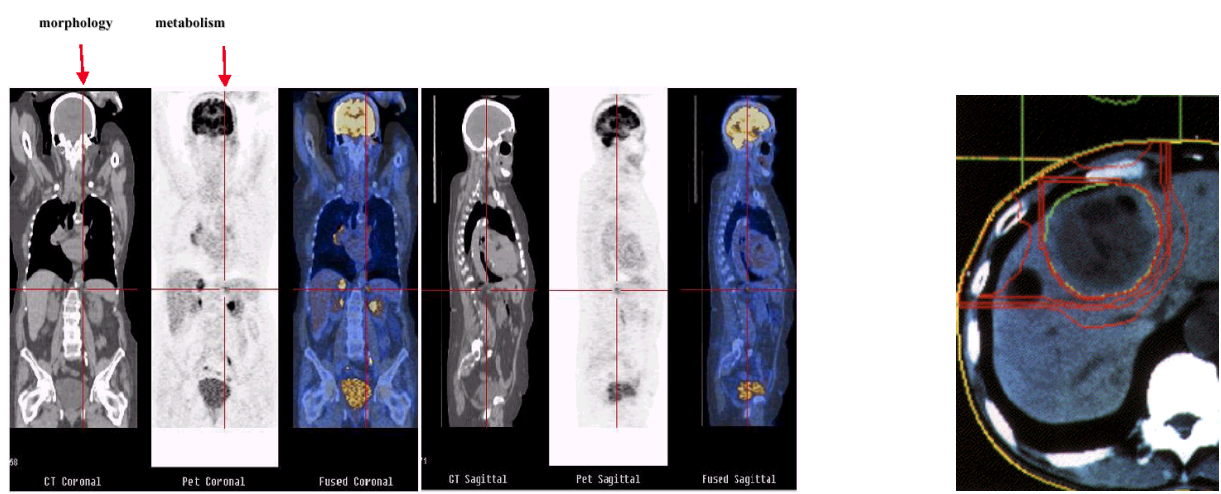

**Figure 1.2** *On the left a series of images of human body obtained with Computer Tomography (CT) and Positron Emission Tomography (PET) is shown. The first gives the morphology while the second shows the metabolism [B.2, B.7]. On the right proton iso-dose lines for bladder cancer show that radiation dose deposited in the normal tissue of the liver surrounding the tumour is minimal (Tsukuba, Japan).*

The major disadvantage of proton therapy is the large size of the accelerator and of the beam lines needed to transport the beam. For eye melanomas, as for the treatment of macular degeneration, protons of energies in the range of 60–70 MeV are enough, while the maximum energy needed for protons on deep-seated tumours therapy is of the order of 200-250 MeV. Because of the resultant magnetic rigidity, the beam lines are heavy and the accelerators are rather large compared to electron linacs. It is therefore simply a question of size and costs, which hinders proton therapy to be more widely spread in the hospitals. Concerning the improved localisation of the dose heavier ions behave similarly (or better) to protons, but the magnetic rigidity needed for beam transport is by a factor of 3 higher than with protons. The accelerator and beam lines are correspondingly larger and therefore even more expensive. The most important issue is however the difference in the radiobiological behaviour of these high-LET beams, with the consequent inhibition of spontaneous repair of radiation damages affecting both the cancer and healthy tissue cells. The high LET could bring advantages for the treatment of certain types of radio-resistant tumours but could also be a disadvantage for many other treatments with respect to a possibly higher rate of late complications of normal tissues. Ion therapy represents a very interesting addition to photon and proton therapy, but at this time more clinical results are needed. For this reason ion therapy is still a matter of research or at least of few centres of excellence, as opposed to proton therapy, which could become soon a business reality.

## 1.2    European strategies for cancer therapy*

Figure 1.3 shows the European situation regarding cancer therapy. At the moment 45% of all the patients are "cured", which means that these patients have a symptom-free survival period exceeding five years. About 90% of the cured patients (i.e. 40% of the total) are cured by loco-regional control of the primary tumour, with surgery and radiotherapy. Actually surgery and radiotherapy alone are successful in 22% and 12% of the cases respectively. When combined, they account for another 6% of the cases so that radiotherapy is involved in almost half of the curative treatments of loco-regional type. The above figure also shows that, contrary to a widespread belief, all the other systemic treatments account for 5% only of the cured patients. 37 % of the tumours are metastasised at the moment of diagnosis and cannot be cured with loco-regional treatments alone. Moreover 18% of all the patients die because of a primary tumour without metastases. There is big space for improvements here and early detection with improving loco-regional treatments is then a must.  The percentage of the cured patients could pass then from 45% to 63 %, or even better, if all primary tumours could be locally controlled.

In connection with the above considerations the _Cancer Research Working Part_y [A.1] has proposed the following strategic approaches:

❑   _Early detection and improved diagnosis_ based on widespread screening with the aim of reducing the number of late diagnoses.

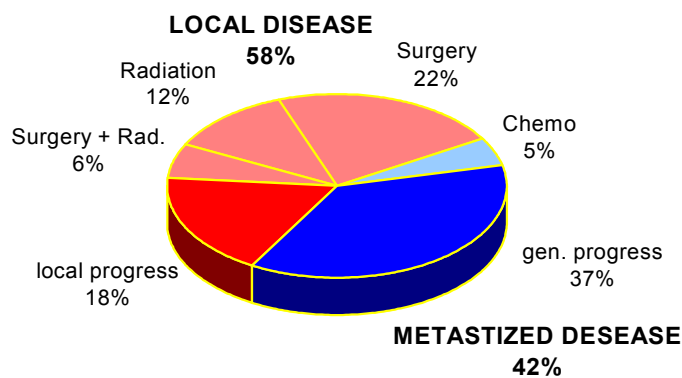

**Figure 1.3** _European situation for cancer therapy._

- ❑ *Improved local treatmen*t, to avoid poor treatments, to treat tumours with difficult localisation and tumours that are radio-resistant to conventional radiotherapy.

- ❑ *Improved systemic treatments* combined with local treatments, which are able to reduce the tumour mass significantly.

Radiotherapy has the main task to improve the local treatments, following two main different ways:

- *Improvements in the results of existing local treatments.*
- *Development of new local treatment*s.

As mentioned in the European working party recommendation*, "…local control of the primary tumour is a basic prerequisite condition for cancer cure. An unsuccessful treatment of the primary tumour will nearly always lead to death, due to local tumour progression or development of distant metastasis. An improved local treatment will lead to a higher cure rate, by avoiding uncontrollable local tumour progression. A more effective local treatment with irradiation will also result in organ preservation in a larger group of patients and, therefore, improve the quality of life. Tumour sites where an improved local control rate has led to a higher survival rate with radiotherapy are: e.g. head and neck, cervix, breast, bladder and prostate. It has been estimated that 30 to 50% of the patients who failed after radiotherapy, developed their first relapse at the primary site…".*

After stating that advanced X-ray therapy and protontherapy are by now reasonably developed modalities, the European Working Party paper concentrates also on the developments of new local treatments.

*"…Although in radiotherapy one generally aims at minimising the effect of the dose to normal tissue, that effect is still a limiting factor for the total dose that can be given to the tumour. The ideal situation in radiotherapy is for a large amount of energy to be deposited in the tumour volume and none outside, in the surrounding healthy tissue. This means, in principle, the avoidance of any sideways spreading out of the beam and the achievement of a very well-defined distribution of the energy in depth.*

*Another crucial factor is that about 20% of tumours are radio-resistant, i.e. they do not respond to treatment with photons or electrons; they are, however, sensitive to beams that deposit their energy in microscopically*

*localised packets (high energy deposition), a process that increases the probability of giving a lethal dose to every cell (e.g. beam of light ions)…".*

In conclusion protons should be considered as a new standard medical technique, while ion therapy is now under study using carbon, oxygen and neon beams, although the first is producing the most promising clinical results. It is the experts' opinion that light ion treatments need a direct intervention form the European Commission, while the proton therapy, in view of technical and medical progresses performed in the last years, should be a task of the single Member States.

## 1.3    Fundamentals of radiation therapy*

Oncology surgeons and radiation oncologists believe that the key to curing patients with solid tumours is controlling those tumours at their site or region of origin: removing them totally or inactivating the cells completely. Studies show that local control can influence inactivation or survival [B.2, B.13, B.14]. Surgical and radiation oncologists try to minimise damage to uninvolved tissues: the surgeon by minimising removal of normal tissue; the radiation oncologist by controlling the radiation distribution so as to concentrate the radiation in the tumour mass and surrounding volume at risk for microextensions of tumour, while minimising the radiation received by deep tissues.

In radiobiology one of the most important parameter is the energy deposition in the target region of the particles and represents the dosimetric properties of the particles at the macroscopic scale. The absorbed dose, measured in Gray (Gy or J/kg), is defined as the ratio between the absorbed energy (due to the radiation) in a small target volume and the mass of the volume.

The main task of radiotherapy is then the local control of the dose distribution in the tumour and, in particular situations of the surrounding diffusion paths. For this reason it is necessary to depose only in the tumour a sufficiently high dose.

Supposing to have a fairly accurate identification of the cancer in the human body, it is possible to evaluate the probability to obtain local control of the cancer through the analysis of the dose-effect curves. In these curves it is possible to extrapolate, for the tumour tissues, the desired effect (control), and for the healthy tissues, the probability to generate serious and irreversible damage (complications) both as function of the dose delivered. The figure 1.4 shows dose-effect curves for a certain tumour tissue and for the surrounding healthy tissue. It is evident that to the absorbed dose necessary to achieve a probability of 100% of obtaining local control of the tumour corresponds also a very high probability of producing serious complications in the healthy tissue. With these considerations it is evident that the probability to kill the cancer without unwanted side effects increases proportionally to the conformity of the irradiation delivered. Due to the dose deposition of the hadrons in the tissues, as mentioned previously, one can increase the probability to cure the tumour because the absorbed dose is more concentrated with respect to electrons and photons.

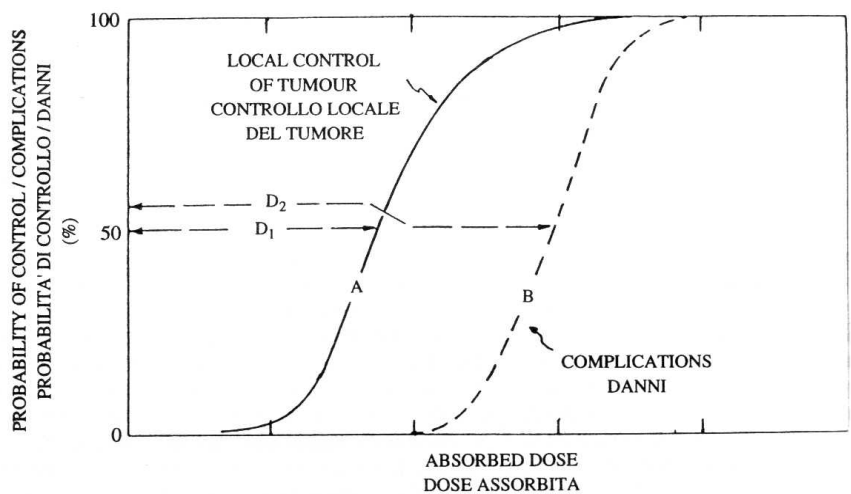

**_Figure 1.4_** _Dose-effect curves for neoplastic (A) and normal (B) tissues. The therapeutic ratio is defined as the ratio $D_2/D_1$ between the dose $D_2$ having a probability of 50% to give serious complications to healthy tissues and the dose $D_1$ having the same probability to produce the local control of the tumour._

It is therefore possible, maintaining the same dose delivered to the target volume, to reduce the dose absorbed by the surrounding normal tissues and, as a consequence, the probability of serious and irreversible complications. At the same time one can also increase the dose to the tumour and therefore the probability of local control of the tumour, maintaining at low levels the dose delivered to healthy tissues.

In order to understand the behaviour of the tumour under irradiation, one may consider, in addition to the dose, which expresses, at the macroscopic level the local deposition of energy, another parameters: the LET (Linear Energy Transfer or $L_\infty$). The LET expresses, at the microscopic scale, the modalities of energy transfer from the radiation to the tissues and it can be identified as the density of energy deposition along the track of the particles in the tissue. In formula $L_\infty = \Delta E/\Delta x$ (keV/µm in water), where $\Delta E$ represents the energy deposited by the charged particles in the tissue in a short path $\Delta x$. In general charged hadrons are defined as densely ionising radiation in contrast to the sparsely ionising radiation, such as X, $\gamma$ rays and electrons. Ions have LET considerably higher (15-200 keV/µm) than the conventional photon beams (0.2-2 keV/µm). In particular it is evident that carbon ions have a LET between hundred and thousand times larger than the LET of photons. Protons are considered as sparsely ionising radiation at high energy entrance point (0.4 keV/µm at 200 MeV), in contrast to the slow protons around the Bragg peak, that have traversed a substantial amount of tissue and are considered highly densely ionising radiation [B.4].

In general the absorbed dose is not a good parameter of biological effects. For this reason the Relative Biological Effectiveness (RBE) is used to indicate the enhancement in biological effects produced by densely ionising radiation with respect to sparsely ionising radiation at the same dose.

RBE of a certain radiation is defined as the ratio between the absorbed dose of a reference radiation and the test radiation, required generating the same biological effect (figure 1.6). In formula RBE=$D_\gamma/D$, where D is the absorbed dose needed to produce the studied effect in the irradiated body with the chosen beam and $D_\gamma$ is the dose generated by photons that produces the same effects.

RBE for cell inactivation is one of the most important aspects in view of the use of hadrons for tumour sterilisation. It depends on the LET value (figure 1.7). Experimental results show that high LET particles are more effective (in terms of damaging the hit cells) than electrons and photons by a factor three at 10% survival level, and smaller then one at very high values of LET. Studies show that the RBE tends to increase as the energy decreases: one of the most interesting conclusions is that protons in the plateau are as lethal as photons (or less effective), while protons in the Bragg peak region show a higher effectiveness. These aspects give the evidence of a different quality of lesions produced by densely ionising radiation and provide further support to the thesis that increasing ionisation density leads to an increase in the complexity of DNA lesions, with a consequent decrease in their reparability. When a beam of high-LET particles (like ions) traverses a cell, it leaves a very dense pathway of ionisation that causes much disruption in both normal cells and tumour cells, and overwhelms the cells' natural capability to repair the damage. Low-LET particles (photons, electrons and protons), on the other hand, leave a sparsely ionising pathway, which results in sparse ionisation; the cell is left sufficiently intact to repair itself. As the damage to cells increases, the cells' repair ability decreases. In the same way it is known that cancer cells have diminished repair capability. This provides then the mechanism for selective destruction of cancer cells.

_Radiosensitive tumours are generally treated with low-LET radiation, to take advantage of this selective effect. High-LET particles may have an advantage over low-LET particles in situations where the cancer cells are more like normal tissue in terms of radiation tolerance._

Another important parameter is the OER (Oxygen Enhancement Ratio) that represents the ratio between the doses required producing a given effect in the absence and presence of oxygen. The oxygen content is generally low in the scarcely vascularised tumours and the biological effects usually decrease when the oxygen content is reduced.

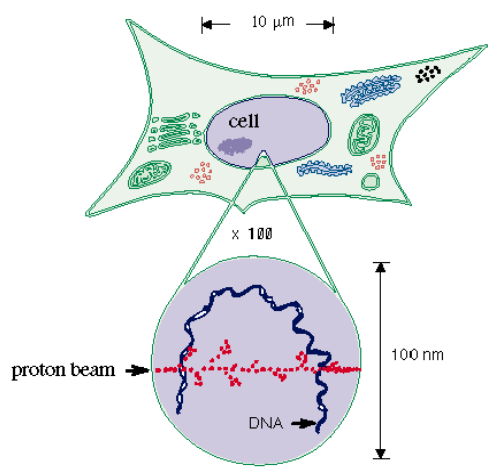

**Figure 1.5** _Ionisation in a cell produces chemically active radicals that damage DNA in a chromosome. Cells have the capability to recovery from such damage to some extent. However, when the damage exceeds a certain level, cells no longer proliferate. This is the reason why hadron beams are used for cancer therapy. If recovery of normal tissue is greater than that of the tumour, fractionating irradiation can eliminate tumour cells with minimum damage to normal tissue surrounding the tumour._

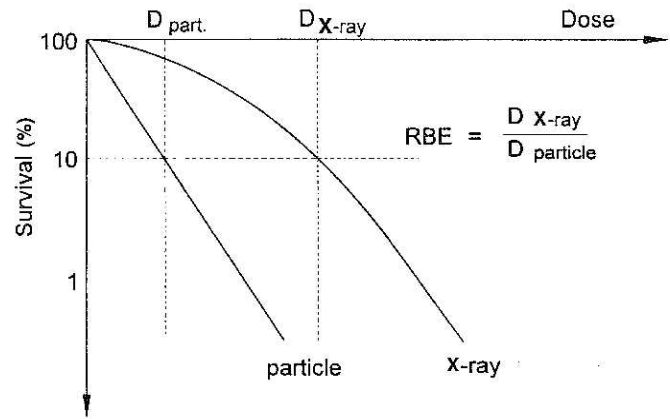

**Figure 1.6** *Definition of the relative biological efficiency (RBE) is shown as the ratio of the X-rays to proton dose producing the same biological effect [B.2].*

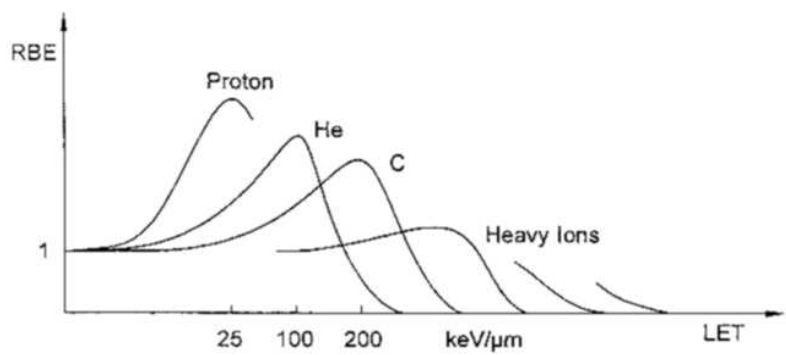

**Figure 1.7** *RBE-LET relationship is given schematically for different particles. For protons the RBE maximum is found at a LET value of 25 keV/μm and for Carbon ions at 250 keV/μm [B.2].*

In formula OER=D/$D_0$, where D is the dose needed to produce the effect in the actual tissue and $D_0$ is the dose which would be needed if the tissue was fully oxygenated in normal air under standard pressure. Summary of experimental data on the LET dependence of OER shows that the negative effect on cancer control due to the lack of oxygenation of irradiated tissues is reduced in high LET radiation with respect to the conventional low LET beam like electrons and photons. This is due to the fact that for high LET radiation, the effects, more than being mediated by free radicals (indirect effect) are due to the local high deposition of energy at the microscopic level with consequent break of the DNA of the cell which is hit (direct effect). OER is in the range 2.5-3.2 for photons, while ions are in the range of 1-3, which implies that the unfavourable therapeutic ratio of non-oxygenated tumours treated with photons and electrons can be modified with ion treatment. Therefore another method to improve the local control of some tumours is the use of high LET beams such as carbon or oxygen ions [B.17, B.26, B.27].

## 1.4      Scientific comparison between conventional radiation therapy and hadrontherapy

Particle accelerators are fully integrated and used in several sectors of our society, from high-energy research to industry, material science and medicine. In particular table 1.1 shows that about half of all running particle accelerators in the world are used in biomedicine, and the majority is implemented for radiotherapy. The state of the art in conventional radiation therapy is based on the use of very compact electron linacs mounted in the head of a rotating gantry [B.2, B.28]. These accelerators, generally standing wave structures, use 3 GHz micro-waves generated by klystrons. They can produce electron beams with energy of 3-25 MeV and cross sections between few $cm^2$ and few tens of $cm^2$ at the treatment distance of about 1 metre, and photon beams, obtained by slowing down the accelerated electrons in a heavy target, with continuous energy spectrum (with maximum energy equal to the electron energy) and transverse cross sections as large as 30 x 30 $cm^2$. Electrons and photons have the same biological effects on the irradiated cells, but the delivered doses have very different spatial distributions. This can be seen from figure 1.1, which shows the relative absorbed dose (in water), for electron and photon beams produced by 8 MeV electrons, together with the curves representing the energy depositions in matter due to the hadron beams. Electron beams are characterised by a maximum range in the tissue (depending on the initial energy of the beam) beyond which one can see a low intensity tail due to bremsstrahlung photons. Due to these characteristics, electron beams are suitable for the treatment of superficial or semi-deep tumours and are used in about 10% of all conventional treatments.

On the other hand, photon beams (or X-beams) are characterised by an exponential absorption, after a maximum which is reached at 2 cm for beams of 8 MeV maximum energy. This depth corresponds to the maximum range of the secondary electrons produced by the primary photons in the more superficial layers of the irradiated tissue. As a consequence of this "build-up effect", in a high-energy X-ray irradiation the skin dose is relatively low. In spite of the roughly exponential decrease of the dose with depth, X-ray beams from a linear accelerator are suitable for an efficient treatment of even "deep-seated" tumour targets.

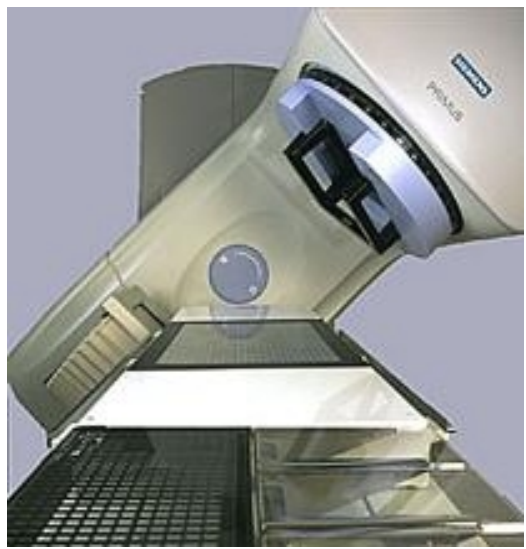

***Figure 1.8*** *Conventional linear accelerator (Simens).*

**Table 1.1** *Particle accelerators in the world. Linacs used in radiotherapy represent about 40% of all running accelerators (France, Germany, Italy: 4-5 units per million inhabitants, Switzerland: 11 units per million inhabitants).*

| Category | Number |
|---|---|
| Particle accelerators for high energy research | 120 |
| Particle accelerators in industry | 1500 |
| Particle accelerators for ion implantation and surface modification | > 7000 |
| Synchrotron radiation sources | > 100 |
| Particle accelerators for conventional radiotherapy | > 7500 |
| Particle accelerators for hadrontherapy | 30 |
| Particle accelerators for biomedical research | 1000 |
| Particle accelerators for medical radioisotope production | 200 |
| **Total** | **> 17450** |

In order to irradiate selectively such targets, radiotherapists use multiple beam entry ports onto a point usually coinciding with the geometrical centre of the target. In order to apply these irradiation techniques, it is necessary for the whole electron linac to rotate around a horizontal axis. Through the use of a rotating gantry the beam can be directed onto the supine patient from several directions. The localisation of the dose in depth is then achieved through the superposition of several converging beams. It has to be underlined that even a small increase of the maximum dose is worthwhile: for a typical tumour which is controlled with a 50% probability, a 10% increase of the dose usually improves this probability by 15- 20%, so that the control rate increases from 50% to 65-70%. This is a sizeable effect since it corresponds to a reduction of the failure rate from the initial 50% to 30-35%. In a typical treatment one delivers in each session 2-2.5 Gy to the tumour, while giving less than 1-1.2 Gy to any of the organ at risk. Since the treatment lasts about 30 session, usually spread over 6 weeks, the target will have eventually received 60-75 Gy.

The use of dynamic computer-controlled multi-leaf collimators offers here new additional possibilities. The most interesting is to apply the dose with a non-uniform distribution of photon fluence for each of the constituent dose fields. The superposition of non-homogeneously shaped dose distributions can produce a resultant dose distribution of superior quality shaped in all 3 dimensions (with a higher degree of conformity). This new approach is called intensity-modulated radiotherapy (IMRT).

As already mentioned the depth-dose curves of hadrons (protons and ions) are completely different from those of photons, generating the so-called Bragg peak. In order to treat deep-seated tumours it is necessary to reach depths of this peak of more than 25 cm: proton and carbon ion beams must then have an initial energy not lower than 200 MeV and 4000-4500 MeV respectively (i.e. 330-375 MeV/u).

The depth of the Bragg peak depends on the initial energy of the protons and its width on the energy spread of the beam. By varying the energy during the irradiation in a controlled way, one can superimpose many narrow Bragg peaks and obtain a Spread-Out Bragg Peak. In hadrontherapy the time needed for the preparation of the patient before the treatment is usually a significant fraction of the total treatment time.

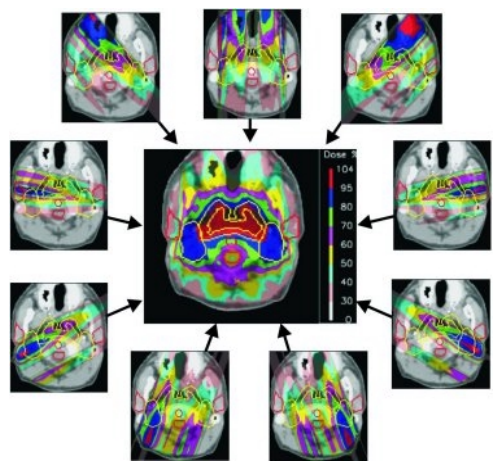

**Figure 1.9** _Intensity modulated treatment planning with photons (IMRT). With 9 fields it is possible to construct a highly conformal dose distribution with good dose sparing in the region of the brain stem (PSI, Switzerland)._

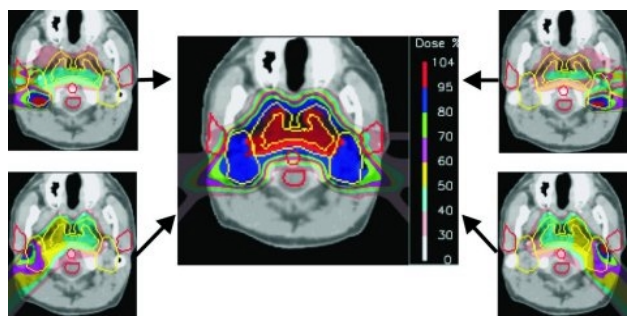

**Figure 1.10** _Intensity modulated therapy with protons (IMPT): high conformity is achieved by a low number of dose fields. The advantage is the reduction of dose burden outside of the target volume (PSI, Switzerland)._

This is why all new dedicated hadron facilities are designed with one accelerator for the delivery of the beam sequentially into several treatment rooms. Radiotherapists use rotating linacs to treat patients with X-ray beams and it is important to have the same possibility when using hadron beams.

The magnetic rigidity of 200 MeV protons is such that the magnetic channel capable of doing this has a typical radius of 5 m. For this reason fixed (mainly horizontal) proton beams have been used world-wide till the first hospital based center became operational. Since then the new facilities have usually one or more gantries, which are large mechanical structures which rotate around a horizontal axis and rigidly support the needed bending magnets and quadrupoles. Having chosen the irradiation angle, it is necessary to conform the dose to the tumour target both transversally and in depth.

Till 1998 _passive spreading systems_ have been used, where a scatterer diffuses the protons and their energy is adapted to the distal form of the tumour by using appropriate absorbers. The transverse form of the irradiation field is defined by collimators. Unfortunately the passive scattering requires a long throw of the beam on the gantry head. With passive system protons are treated as X-rays, but charged particles can be much better used since, at variance with X-rays, they can be bent by a magnetic field and directed on a target at will. In 1997 at PSI (Villigen) the first rotating gantry with a 250 MeV proton beam came into operation with an _active spreading system_ [B.40, B.42]. The target is subdivided into many thousands of _voxels_ and each one is irradiated at successive steps. This system is capable to deliver the so-called proton intensity modulated proton therapy (IMPT). Figures 1.9 and 1.10 show examples of the dose distribution for a nasopharyngeal tumour for both IMRT and IMPT therapies. For IMRT the dose is delivered using 9 photon fields, each field is applied with modulation of the photon flux. For IMPT only 4 modulated fields can deliver a highly conformal dose to the primary target and a reduced dose to the affected lymph nodes (the secondary target) with a maximal sparing of the organs at risk (brain stem and parotid glands). With beam scanning, IMPT can be calculated and delivered just under computer control without the need for any patient specific hardware. The advantage in terms of sparing of critical organs, for the same dose to the target, is clear. The physical reasons are that in a SOBP the distal dose drops from 80% to 20% in a few millimetres and the dose to the skin is typically equal to 70% of the dose in the SOBP. The proton diffusion in matter is such that the lateral edges of the irradiated target volume have a fall-off thickness of about 5 mm from 80% to 20% of the dose at a depth of 25 cm. Then one can summarised the main difference of proton IMPT compared to photon IMRT as the absence with protons of the "dose bath" to the whole brain.

In order to prove the above explanations, table 1.2 and figure 1.11 show the experience of Cantonal Hospital of Bellinzona (Switzerland). In particular the dose distributions for head and neck carcinomas have been obtained with different techniques [B.2]. The spot scanning dose distribution is considered the best solution if one considers tumours near to critical organs. Concerning the irradiation of the target volume, the conventional therapy with intensity modulated photons are extremely selective and approximately equal to protons, but they become critical when the dose at non-cancerous and critical organs and conformity index are considered.

**Table 1.2** _Mean doses to PTV (DminS); EUD equivalent uniform dose; maximum dose at spinal marrow (Dmax); dose at 2/3 of the parotid glandular volume (D2/3V), percent of the irradiated volume defined as the % of target volume that receives a dose bigger than the 50% of the calculated dose (IV50); conformity index (CI), defined as the ratio between the treated volume and the PTV volume [B.28]._

|  | DminS to PTV (Gy) | EUD (Gy) | Dmax (Gy) | D2/3V (Gy) | IV50 (%) | CI |
|---|---|---|---|---|---|---|
| **X rays + electrons** | 42.0 | 49.6 | 39.7 | 51.5 | 28 | 1.98 |
| **Photons (5 fields not modulated)** | 48.2 | 52.6 | 38.8 | 46.4 | 28 | 1.91 |
| **Photons (5 modulated fields)** | 47.3 | 52.4 | 31.1 | 43.3 | 33 | 1.68 |
| **Photons (9 fields)** | 47.5 | 52.7 | 26.2 | 41.1 | 30 | 1.60 |
| **Protons (passive distribution)** | 51.4 | 53.8 | 20.4 | 28.4 | 18 | 1.62 |
| **Protons (spot scanning)** | 49.8 | 53.3 | 17.6 | 23.2 | 16 | 1.23 |

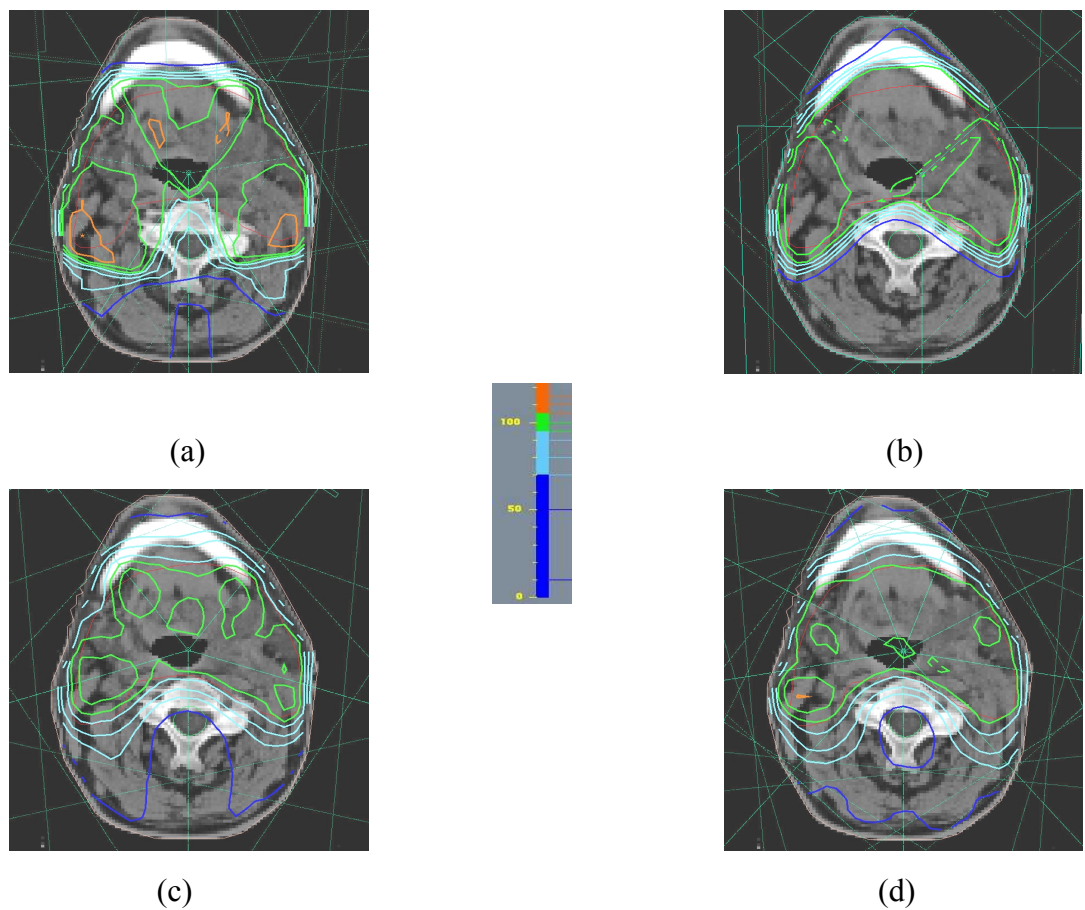

**Figure 1.11** *Oropharynx T4N2 infiltrating base of the skull. (a) Photons with 5 non modulated fields. (b) Protons with passive dose distribution. (c) Photons with 5 intensity modulated fields. (d) Photons with 9 intensity modulated fields (Cantonal Hospital of Bellinzona).*

## 1.5    Brief history and status of hadrontherapy in the world*

The rationale use of protons evolves from radiation therapy's history. Low energy X-ray therapy began in 1896, just months after Roentgen's discovery. Many complications were seen, however, because orthovoltage radiation concentrated energy at the skin surface. In the mid-1950s, megavoltage radiation was added to the radiation oncologist's tools. The high-energy X-rays and cobalt-60 gamma rays helped patients because the beams concentrated their maximum energy one to two cm below the skin surface, closer to many tumour targets. Increased cure rates and decreased complications were soon documented. Supervoltage X-rays became available with betatrons in the 1960s and linear accelerators in the late 1960s and 1970s. Again, improved control of local and regional disease occurred, as did further reductions in complications. This occurred primarily because the maximum energy deposition reached 4 cm in depth. *Conformal radiation therapy* was developed in the late 1980s and early 1990s as a sophisticated tool to confine more precisely the high doses of photon radiation to target tissues. In these investigations protons were seen to possess the properties of energy deposition that had been sought by radiation oncologists since Roentgen's day. The only factor inhibiting the widespread use of protons in the early years was the complex technology and high costs for the treatments.

R. R. Wilson first proposed to use protons for cancer therapy in a scientific article in 1946 [B.1], anticipating the physical characteristics of a proton beam to provide the clinicians with superior controllability of dose distribution when compared to photons or electrons. Two years later, researchers at the Lawrence Berkeley Laboratory (LBL) conducted the first positive treatments on humans consisted of irradiation to destroy the pituitary gland in-patients with metastatic breast cancer that was hormone sensitive. After observing proton therapy in the 1950s at both LBL and Uppsala (Sweden), the Harvard Cyclotron Facility became also interested in using protons for medical treatments. However, this early work was limited due to the inability to perform 3-D imaging and reliance on facilities primarily dedicated to physics research. During the 1970s, the Massachusetts General Hospital conducted the first research on mixed proton/X-ray radiotherapy for the treatment of prostate cancer. Unfortunately, during this decade, the role of protons in radiotherapy was significantly underestimated in favour of neutron therapy. In 1975 the University of California Berkeley started to study helium and heavy-ion irradiation, while the Los Alamos National Laboratory gained experience in using pions. In the 1980s design and construction began on the first dedicated clinical proton facility at Loma Linda University Medical Center [B.30]. To date, the Loma Linda facility has treated over 9500 patients with proton therapy. Loma Linda University selected protons for the new facility because of their apparent relatively low RBE and relatively low cost to produce as compared to other heavy charged particles. Their biologic effects were considered similar to photons. This partially established their potential value, because the data and experience accumulated from using photons and electrons since Roentgen's discovery would apply to protons. The Proton Therapy Co-operative Group (PTCOG) was also created during the 1980s for scientists to exchange ideas on the development of proton therapy, such as the Proton Radiation Oncology Group (PROG) to sponsor the development of clinical trials involving proton therapy.

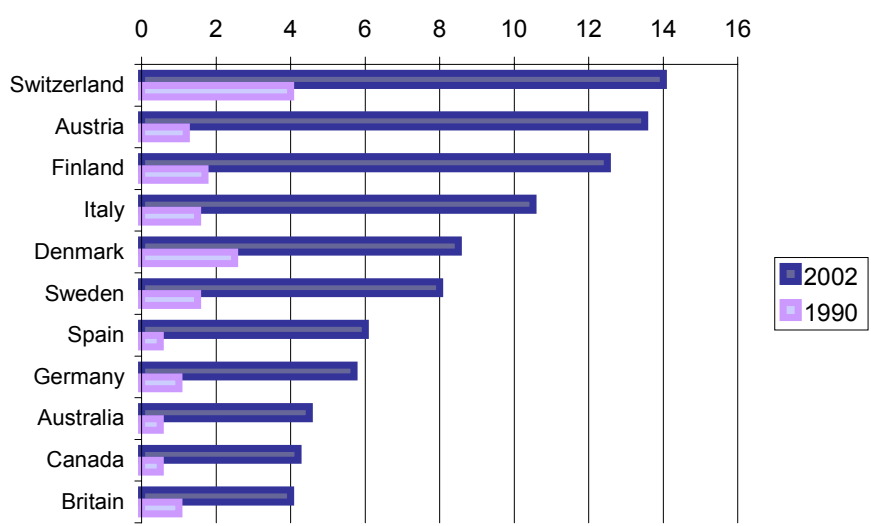

**Figure 1.12** _Magnetic resonance imaging units per million population in the periods 1990 and 2002 (OECD Health Data 2004). The hadrontherapy development is connected with the increasing of the facilities for tumour detection._

Over 40,000 patients have now been treated with proton therapy world-wide (table 1.3). At the end of 2005 five hospital-based centres [B.3, B.6] can treat deep-seated tumours with *protons*, which are equipped with several treatment rooms and rotating gantries: the above mentioned Loma Linda University Centre in California, the Northeastern Proton Therapy Centre in Boston [B.31], and the three centres of Kashiwa, Tsukuba, and Wasaka Bay in Japan. Two hospital-based *ion* centres also exist in Japan: HIMAC in Chiba [B.38] and HARIMAC in Hyogo. All the other centres make use of beams produced by accelerators built for fundamental research in nuclear and particle physics and later adapted to protontherapy. Such centres are running, some since long time, in Europe, Japan, Russia and USA [A.1, B.2, B.3, B.4, B.7].

Eye melanomas need proton energies in the range 60–70 MeV and passive spreading is sufficient: the accelerators are much smaller and there is no need for active dose distribution. Contrary to deep therapy, this type of treatment is well developed in Europe. The most important centres are the Centre Antoine Lacassagne in Nice (France), the PSI-Optis at Villigen (Switzerland), the Clatterbridge Centre for Oncology (UK), the Centre de Protonthérapie d'Orsay (France) and the Louvain-la-Neuve cyclotron (Belgium). In 1998 the Berlin cyclotron was commissioned. Many new protontherapy centres are under construction. Today four companies offer world wide turn-key systems (based on a normal cyclotron, a superconducting cyclotron and two synchrotrons, respectively). At the end of 2005, two ions centres are under construction in Heidelberg-Germany (designed by GSI) and in Pavia-Italy (designed by TERA) [B.7, B.32], while in 2005 the ETOILE project has been approved in Lyon-France [B.35, B.36, B.52] as well as the Med Austron project in Austria. All foresee the use of the PIMMS synchrotron, developed at CERN [B.33].

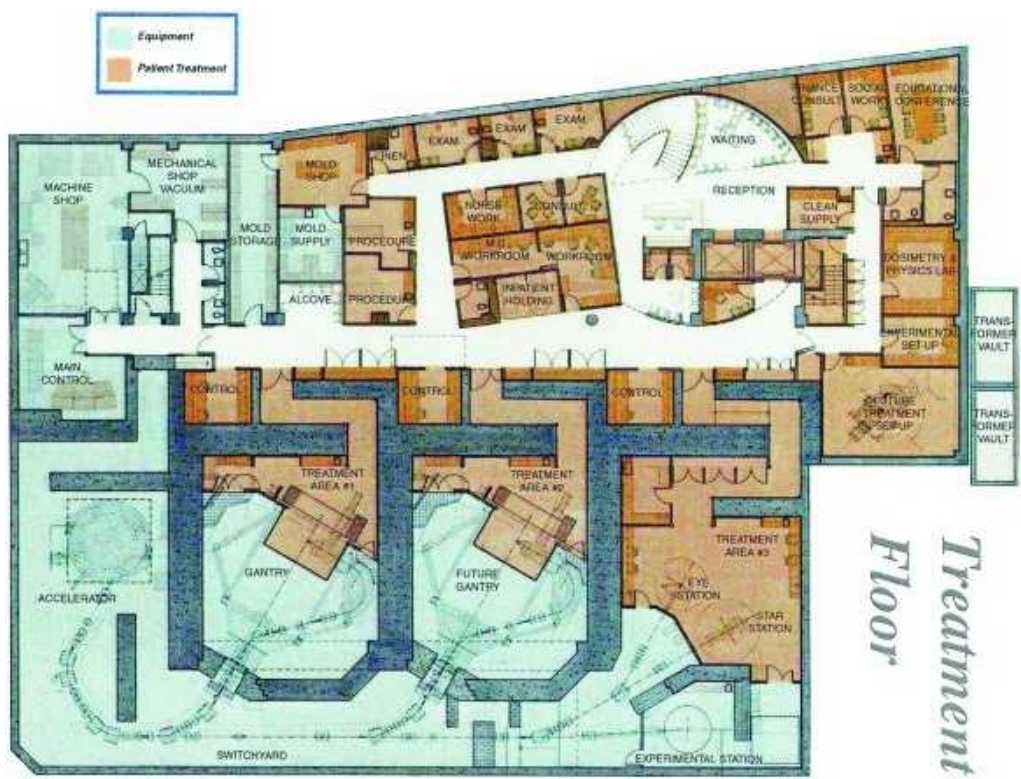

***Figure 1.13*** *Northeast Proton Therapy Center (NPTC) in Boston.*

***Table1.3*** *Patients treated in the world (January 2005)*

| WHO | WHERE | WHAT | DATE | Patients | DATE |
|---|---|---|---|---|---|
| Berkeley 184 | CA. USA | p | 1954 — 1957 | 30 | |
| Berkeley | CA. USA | He | 1957 — 1992 | 2054 | June-91 |
| Uppsala | Sweden | p | 1957 — 1976 | 73 | |
| Harvard | MA. USA | p | 1961 — 2002 | 9116 | |
| Dubna | Russia | p | 1967 — 1996 | 124 | |
| Moscow | Russia | p | 1969 | 3785 | Dec-04 |
| Los Alamos | NM. USA | π - | 1974 — 1982 | 230 | |
| St. Petersburg | Russia | p | 1975 | 1145 | Apr-04 |
| Berkeley | CA. USA | ion | 1975 — 1992 | 433 | June-91 |
| Chiba | Japan | p | 1979 | 145 | Apr-02 |
| TRIUMF | Canada | π - | 1979 — 1994 | 367 | Dec-93 |
| PSI (SIN) | Switzerland | π - | 1980 — 1993 | 503 | |
| PMRC (1), Tsukuba | Japan | p | 1983 — 2000 | 700 | July-00 |
| PSI (72 MeV) | Switzerland | p | 1984 | 4182 | Dec-04 |
| Dubna | Russia | p | 1999 | 296 | Dec-04 |
| Uppsala | Sweden | p | 1989 | 418 | Jan-04 |
| Clatterbridge | England | p | 1989 | 1372 | Dec-04 |
| Loma Linda | CA. USA | p | 1990 | 9585 | Nov-04 |
| Louvain-la-Neuve | Belgium | p | 1991 – 1993 | 21 | |
| Nice | France | p | 1991 | 2555 | Apr-04 |
| Orsay | France | p | 1991 | 2805 | Dec-03 |
| Themba LABS | South Africa | p | 1993 | 468 | Nov-04 |
| MPRI | IN USA | p | 1993 | 34 | Dec-99 |
| UCSF - CNL | CA USA | p | 1994 | 632 | Jun-05 |
| HIMAC, Chiba | Japan | C ion | 1994 | 1796 | Feb-04 |
| TRIUMF | Canada | p | 1995 | 89 | Dec-03 |
| PSI (200 MeV) | Switzerland | p | 1996 | 209 | Dec-04 |
| G.S.I Darmstadt | Germany | C ion | 1997 | 198 | Dec-03 |
| H. M. I, Berlin | Germany | p | 1998 | 546 | Dec-04 |
| NCC, Kashiwa | Japan | p | 1998 | 300 | Oct-04 |
| HIBMC, Hyogo | Japan | p | 2001 | 483 | Dec-04 |
| PMRC (2), Tsukuba | Japan | p | 2001 | 492 | Jul-04 |
| NPTC, MGH | MA USA | p | 2001 | 973 | Dec-04 |
| HIBMC, Hyogo | Japan | C ion | 2002 | 30 | Dec-02 |
| INFN-LNS, Catania | Italy | p | 2002 | 82 | Oct-04 |
| WERC | Japan | p | 2002 | 19 | Oct-04 |
| Shizuoka | Japan | p | 2003 | 100 | Dec-04 |
| MPRI | USA | p | 2004 | 21 | Jul-04 |
| Wanje, Zibo | China | p | 2004 | 1 | Dec-04 |
| | | | | **1100 pions** | |
| | | | | **4511 ions** | |
| | | | | **40801 protons** | |
| | | | **TOTAL 46412 all particles** | | |

## 1.6    Main systems used in hadrontherapy

### 1.6.1    Beam delivery systems

The beam delivery system is used to control the beam during irradiation of the target volume. It can be divided into two main categories: (1) beam widening techniques and (2) conformal techniques.

Beam widening techniques give a homogeneous dose distribution, combined with range modulators and shaped collimators to achieve high dose volumes with flat shapes in the target. For an individual beam it is unavoidable that also adjacent normal tissue is included in the high dose volume. Conformal techniques give 3D dose distributions conformal to the target volume. The treatment volume can be by various combinations of beam widening and depth control techniques. As the depth modulation is uniform over the whole field, some of the volume elements outside the target volume are exposed to the full target volume dose. The relevance of this unwanted exposure is reduced by multy-port irradiation. For conformation therapy the dose distribution is shaped in three dimensions for each port, resulting in an optimal reduction of the integral dose. With help of a multileaf collimator and a bolus to fit the distal shape or the dose distribution, the two dimension shaping or the beam cross can be developed into a three dimensional shaping. The same results can be achieved directly by voxel or spot scanning.

In a typical treatment room the beam, after passing through a vacuum window at the end of the beam transport system, travels through several kind of devices and drift spaces before entering the patient. These devices change the beam range, modulate the range, spread the beam laterally and shape its lateral profile. Dose measuring devices and beam monitoring devices are required for radiation control.

Several systems have been studied at this time [B.42], such as, for example, the nozzle designed by IBA that foresees a passive system with, eventually, an upgrade for active system [B.45, B.46]. The parameters of beam delivery systems lead to competing requirements on the design and placement of the beam line elements. An optimisation process involving several parameters with clinical tradeoffs is often required. In general, the maximum field size, the field uniformity, the treatment time, the beam divergence, the lateral dose penumbra of the field, the background radiation, and the beam fragmentation are important factors to consider. Furthermore, different clinical requirements necessitate different designs, which generally fall into two categories: beam lines for small and large radiation fields. Large radiation field requires more complex systems for lateral spreading of the beam. The small beam spot extracted from the accelerator must be modified to cover areas as great as 20 x 20 cm$^2$ with dose uniformity within ± 2%, and the dose rate on the order of 1-2 Gy/min. The beam lines for small fields are simpler because they can often directly utilise the beam from the accelerator with fewer modifications.

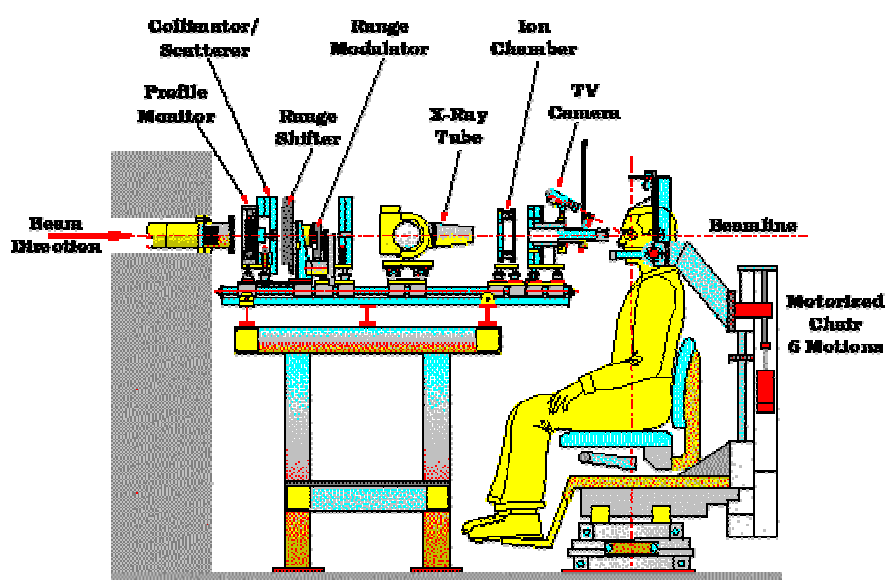

**Figure 1.14** *Example of common equipment of horizontal fixed beam room (TRIUMF, Canada).*

The precision of the dose distribution is connected to several factors that can be divided in two main groups. The first group is generated by the actual physical limits for the precision of the treatment, while the second group includes all possible errors inside the beam delivery device and that can be optimised by the design process. The shape of the distal-off is determined by the energy straggling (a physical limit) and by the momentum band of the beam (a beam delivery parameter). The distal fall-off of the Bragg peak is steepest if the energy of the beam corresponds to the minimal possible range (the depth of the tumour in the patient itself). The distal fall-off, due to range straggling, vary, in first approximation, linearly with the nominal range of the beam. The lateral fall-off of the dose is dictated by multiple Coulomb scattering and by the phase space of the beam. Also here the contribution to lateral fall-off of the dose due to the multiple Coulomb scattering in the patient (a physical limit) vary linearly with the proton range. The dose deposited by the proton beam from the entrance side to the target volume can be reduced by limiting the modulation of the range in depth at least to cover the tumour homogeneously. The deposition in depth is produced by changing the energy of the beam independently of its lateral spread and the modulation is chosen to correspond to the maximal thickness of the target. This approach is called fixed range modulation. If the modulation is done as a function of the transverse position of the protons, taking into account the variable thickness of the target volume, one can have the so-called variable range modulation. With this method it is then possible to produce a complete 3D conformation of the dose.

As a general rule, the best dose distribution is dictated by:

- a low beam energy (minimal range);
- a small beam phase space (smaller than the physical limits given by multi Coulomb scattering and range straggling);
- a reduction of amount of material in the beam (transport of the beam in vacuum up to the patient);
- the distance of the beam port to the patient (since the broadening due to the multi Coulomb scattering depends on the material seen by the protons on the way to the target and on its propagation distance). For the active method the material is placed very close and for the passive method very far from the patient.

On the basis of such very general criteria, the dose deposition technique can be classified in three main groups: the passive systems, the fixed dynamic scanning, and the computer-controlled dynamic scanning.

❑ _Passive scattering method_

The passive scattering method is known and reliable, and represents today the "state of the art" of the beam delivery devices in hadrontherapy. Figure 1.15 summarises the principle of the method. The beam is scattered about 4 m ahead of the patient by a double scattering system, which is designed to generate uniform beams at the patient location. A collimator is used to clean the beam in the lateral directions. A range shifter wheel permits to modulate the dose uniformly in depth, generating the so-called Spread-Out Bragg Peak (SOBP). A compensatory bolus is used to adjust the distal edge of the dose distribution to correspond exactly to the distal edge of the target volume. The compensatory material is shaped with variable thickness transversally to the beam. By this method the dose can be conformed to the distal part of the target surface. However unwanted dose is deposited partially outside of the target volume, and this aspect can be avoided by using a beam delivery system with variable range modulation. Using beams from different directions is possible to distribute the distal dose shaping capability of compensators over the whole surface of the tumour. At the end a good three-dimensional overall conformation of the dose can be obtained, at least when considering the higher dose levels. Improvements foresee multileaf collimators instead of individually shaped collimators and the development of computer controlled compensators [B.7, B.42].

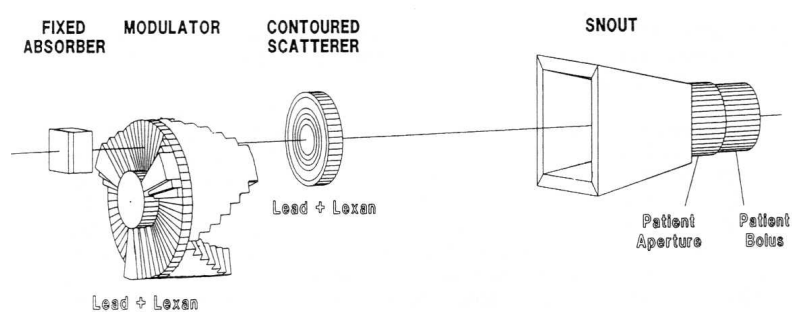

*Figure 1.15* Passive scattering.

❑  *Magnetic beam spreading*

Here a double system of magnets is used to deflect the beam over a distance of about 4 m in order to obtain a good uniformity of the beam at the patient location. Different patterns of scanning were investigated (called wobbling for scanned circles and raster scanning for multiple lines). The shaping of the dose in depth and the adjustment of the distal edge were performed with collimators and compensators as used in the scattering method. The main advantages of this technique, compared with the previous one, can be summarised as: a better use of the beam (the size of the scanned field can be quickly adapted to cover the target volume by changing the reference current in the power supplies of the scanning system); less material in the beam line with less activation problems; less fragmentation of the beam; a simple scanning system without computer control. The disadvantages are: compensators and collimators are still needed as well as a stable beam during the beam scanning, a long throw on the gantry. With this system the beam delivery is still not fully conformal. For this reason the use of multileaf collimators, as for the scattering method, can be done to provide a complete conformation of the dose.

❑  *Computer controlled beam scanning*

The fast dynamic scanning of a proton pencil beam is designed to provide three-dimensional dose conformation as a completely automated routine therapy. In this method a focused proton pencil beam is applied directly inside the patient and scanned with a computer control using two magnets upstream in the beam line. The variation in range can be obtained by either changing the energy of the beam in the accelerator or by using a range shifter system placed closed to the patient.

The total dose is built dynamically as a sum of many individual pencil beam spots. For this reason this method is often called spot scanning method. The individual control of the exposure time for each elementary beam can be achieved either by changing the speed of scanning (continuous scanning) or by switching off the beam at the end of each spot (discrete spot scanning) (annex 2.1). The first approach to spot scanning technique was used in the 1970s at Chiba to 70 MeV proton beams. Facilities have been realised also in Vancouver and at PSI [B.40].

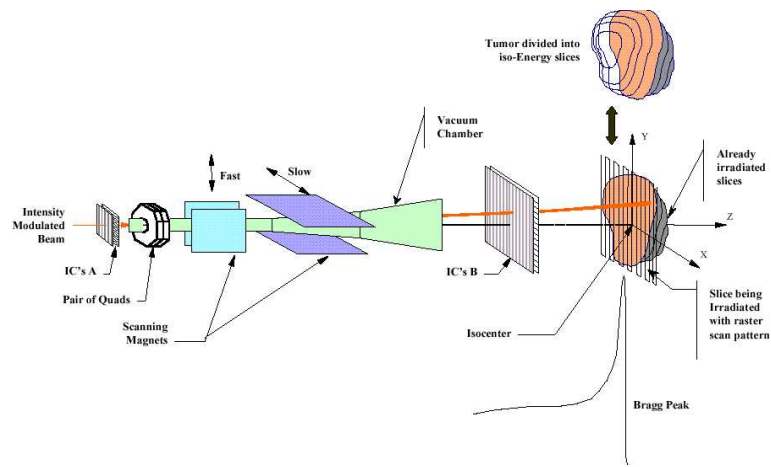

**Figure 1.16** _Active scanning designed by IBA. This method consists in directing lots of small pencil beams into the target to cover the 3D volume. The beam position is adjusted in X & Y by 2 scanning magnets while the depth depends on the beam energy. The 3D volume is divided in several slices parallel to the body surface. The dose in each slice is delivered by controlling simultaneously the beam intensity and the speed of the beam spot in the X & Y direction._

**1.6.2    Gantry solutions**

In conventional radiation therapy the dose is sheared into the surrounding healthy tissue between several entrance ports using different incidences. Since most patients are lying during the treatment, this implies the need of a device providing variable incidence angle between the horizontal plane and the beam: this device is called rotating gantry. It is a rotating mechanical structure which supports and positions the final part of the transfer line and can guide the beam away from the center and bent back towards the axis to cross it orthogonally. With this device the proton beam can be focused and concentrated on the patient's tumour from any angle in the rotation plane, even for tumours with complex shape and close to critical organs. If the beam line ends at the axis (so the patient is located on the axis) the gantry is called isocentric. The gantry is considered as a major component of the hadron facility, due to its complexity and because it takes a significant fraction of the total costs of the installation. Due to the large momentum of proton and ion beams, hadrontherapy needs huge bending and focusing magnets, mounted on the rigid and stable mechanical structure; moreover to minimise treatment preparation time the gantry must rotate rapidly (about 1 rpm). All these aspects must be compatible at the same time with very tight tolerances on the beam positions (typically ±0.5 mm in three dimensions). Generally speaking during the gantry design few constraints of the beam must be considered. At the coupling point to the gantry the beam should be rotationally symmetrical and achromatic in order to maintain the beam independent to the gantry angle. In parallel the optics of the gantry must cancel dispersion in position and angle (due to Δp/p) at the isocentre in order to minimise the sensitivity of the beam position to tuning errors and momentum offsets.

A general classification of gantry geometry is done by Pedroni [B.42]. The main lines of this classification can be summarised into the following categories.

❑   *Long throw designs*

In this category are included the usual gantry design (where the beam is bent two times (90° and 180°)) and the "corsrew gantry" used at Loma Linda General Hospital. Common to these two solutions is the long throw with a radius of about 5-6 m, necessary for the use of passive scattering technique. The main advantage of long throw solution is the possibility to accommodate all different beam delivery systems. Since the spot scanning technique is under development, most of the proposed proton facilities are designed with a "corkscrew gantry" allowing to start the project with a passive system and to upgrade it with an active system at a later stage. In the "corksrew" gantry the bending of the beam is performed in perpendicular planes, reducing the space in the longitudinal direction. In the other designs the bend is all in one plane.

❑   *Short throw designs*

This design, used for the spot scanning technique, foresees an "usual" gantry with a very short distance between the last bending magnet and the patient, with a radius of about 3 m. A disadvantage is the large divergence of the beam and the resulting increased skin dose (the source-to-skin-distance is less than 1 m).

❑   *Insertion of the beam modifiers in the beam optics of the gantry*

Here the design foresees to place the devices used for the broadening of the beam upstream of the patient in the beam transport system in order to reduce the radius (about 3 or 4 m) of the gantry by eliminating the distance between the last magnet and the isocenter. Several solutions have been studied by Enge, Jongen and Pedroni [B.4, B.42] for passive or active systems.

❑   *Eccentric and dynamic patient support*

The use of an eccentric patient support reduces the gantry diameter. Studies have been performed in this sense by the group of Tsukuba [B.37]. Movable support has also been implemented in order to reduce the weight of the rotating structure.

Combinations of different solutions mentioned above have also been adopted. One example is the system used at PSI, which combines the use of insertion of beam modifiers in the gantry beam optics with the eccentric support. Here the system foresees the spot scanning technique.

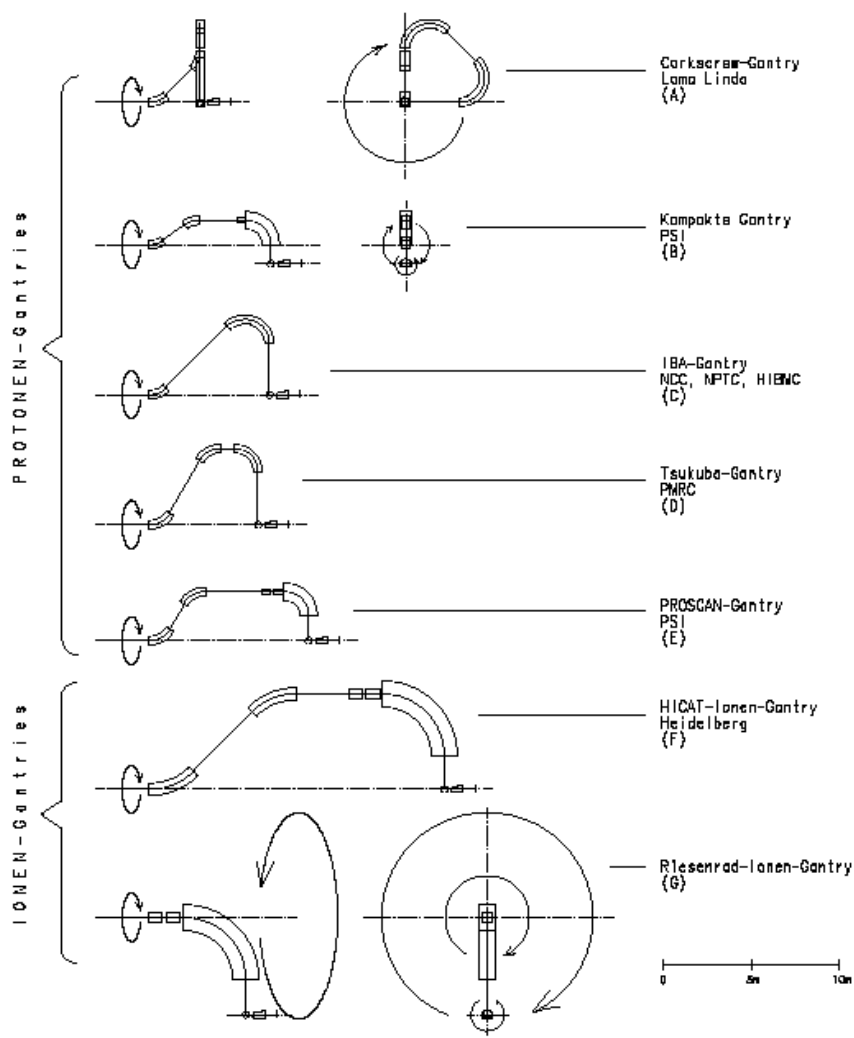

_**Figure 1.17** Different gantry configurations for protons and ions used world-wide in hadrontherapy facilities._

## 1.7    Costs comparison between proton and conventional radiation therapy

The introduction of a new treatment facility into the clinical routine is difficult, especially if it is more expensive in setting up and running than conventional ones, even if it could create potential clinical benefits. In this context these benefits have to be proven in relation to the increased costs of treatment. For this reason, in the past years many scientific projects have been stopped and never come to realisation. However today proton therapy is booming world-wide. In fact hadrontherapy is now in a transition characterised by the imminent availability of many new dedicated hospital facilities.

Generally, the discussion of costs and prices depends on the individual social system and the overall cost level in different countries and a reasonable comparison is difficult because there are no precise data available on the industrial cost of the new facilities as well as the relative financial running costs. The equipment of a proton therapy center is more expensive, in the region of 25-40 M$ (depending on the number of treatment rooms used in the facility), than the 3-5 million needed for the corresponding units in modern radiotherapy. The energy consumption depends on the technique of the accelerator, the running costs are higher, while the clinical personnel is almost equivalent to conventional systems.

Estimations have been performed for the Italian case [B.7] using, as a reference, the data of the Italian National Health Service. Here the average cost of a full proton treatment in dedicated centre is between 9 kEuro and 13 kEuro (for the fees used to cover only the running costs or to cover totally the capital invested respectively). These values should be compared with the cost of traditional radiotherapy (3 kEuro per treatment) and the costs of conformal photon therapy (about 6 kEuro per treatment), where at least 5-8 multi-ports are needed. It can be concluded that the costs of proton therapy are about between 3 and 4 times more expensive than standard radiotherapy, but only 1.5 and 2 times larger than conformal radiotherapy.

This figure is well compatible with estimation published for the German case [B.2], where the medium cost of conventional radiotherapy is about 3.5 kEuro, while an average proton treatment could cost about 7.5 kEuro. These figures have to be compared with the average costs of oncological surgery (7.5 kEuro) and of chemotherapy (32 kEuro). In this context the median oncological treatment for curable cancer diseases costs about 10 kEuro. Moreover if one takes into account not only the cost of the initial treatment but also the costs of failures and their frequency, important initial investments of proton accelerators may be good in comparison with treatments needing small investments (as standard radiotherapy) but giving poorer results.

The possibility to build a proton facility based on free financing (including only the running costs) has to be proven by experts, while positive clinical results of hadrontherapy compared to most advanced conventional radiotherapy are crucial to convince the social partners. At this stage, if one should consider to cover the capital investment and the running costs totally, only clinical data for chordomas and chondrosarcomas are significant enough to see cost benefit in favour of hadrontherapy [B.50]. So despite their indisputable advantages in terms of physical selectivity, protons will have to demonstrate their competitiveness through clinical trials and costs reduction. In these cases hadrontherapy could become more accessible to patients and be considered as an attractive treatment. The construction of compact proton and ion accelerators is certainly going in this direction.

## 1.8     The Italian case: the national network of hadrontherapy*

Starting with the number of patients to be treated by protons and ions, one could extrapolate important considerations about the potential necessity of new hadron facilities. In Italy, the co-ordination and the guaranteeing of a rational distribution of these centres at a national level are covered by AIRO (Associazione Italiana di Radioterapia Oncologica), and in this frame several studies are in process [B.7, B.23]. One should start with the assumption that a treatment room (with or without gantry) could cover about 4500 fractions per year (considering a single treatment of 20 minutes per patient, working 7 hours per day). For proton therapy a full treatment is composed by 11-12 fractions per patient, so that a single room could cover about 400 patients per year. From tables B and C annex 1.1, it is evident that to cover the elective tumour cases, where protons are mandatory, one should need two rooms with proton beams, while for the cases where the proton could generate an advantage in terms of clinical treatment, one should have about 26 rooms at the national level. With these considerations, one should consider at least 4-5 proton therapy centres per 10 million people, uniformly distributed on the national territory.

Similarly one should have for ions about 3000 patients per year, with 6-8 fractions per full treatment. In this case 3-4 rooms with ion beams concentrated in a single centre could be enough.

According to the Italian Hadrontherapy Program, TERA Foundation has promoted the study of the National Centre for Hadrontherapy [B.7] since 1992, a sanitary structure of excellence which will be the Italian focal point of all hadrontherapy activities. This facility will use protons and carbon ions and is now under construction in Pavia. In addition to this, TERA promoted to equip a number of hospitals with compact accelerators, which would ultimately make a national network of hadrontherapy facilities connected between themselves and the National Centre. This will permit coverage of the other needs of radiotherapists at a national level. The aim of these compact accelerators is to reduce the cost of a protontherapy treatment to about 8-10 kEuro [A.1], satisfying the radiotherapists' requests at the same time.

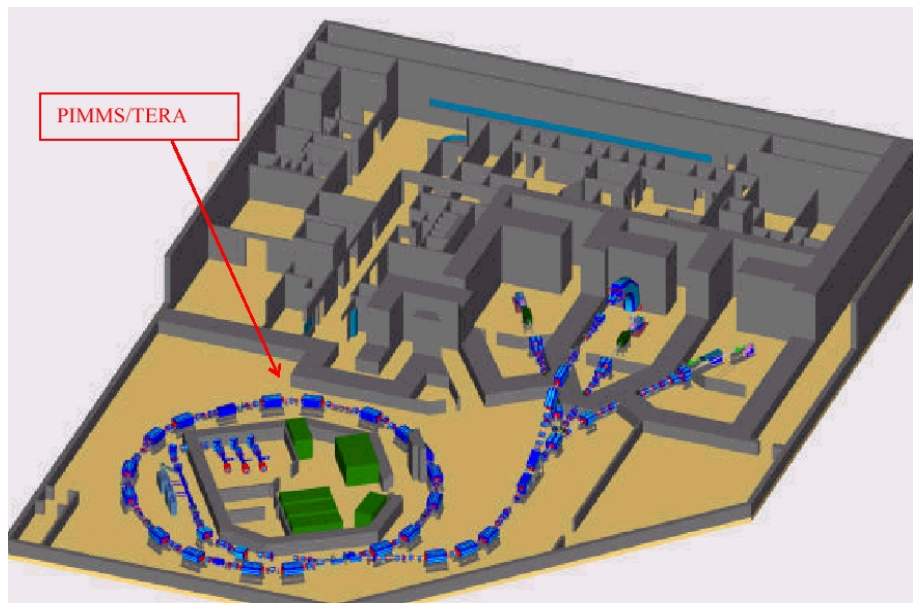

**Figure 1.18** *CNA project: dedicate proton and ion facility designed by PIMMS-TERA group [B.33].*

*Chapter 2*

*LIBO project:*

*a compact linear accelerator for protontherapy*

## 2.1    General overview

Today the majority of the centres involved in radiation therapy in the world use RF electron linear accelerators to treat patients with electron and photon beams. In these cases to reduce the longitudinal and transverse dimensions of the accelerating structures and consequently the relative costs, a typical RF frequency of 3 GHz is adopted. As a consequence many commercial power supplies are now available on the market. The electrons are light particles and they are quickly accelerated approaching the velocity of light ($\beta \cong 1$), so the electron linacs for radiation therapy (with an input energies of few keV and an output energy up to about 25 MeV) operate well at 3 GHz. Protons, on contrary, are about 2000 times heavier and are accelerated much more slowly, so that $\beta$ stays below 0.6. A linac is composed of many resonating cells, which are $\beta\lambda$ or $\beta\lambda/2$ long (see chapter 3), and for high frequencies and low $\beta$, the cell length can become prohibitively short. This generates difficulties in the use of the 3 GHz frequency at low energies, with several complications for the construction. In particular high frequency implies very stringent constraints on the mechanical accuracy of the cavities and a very careful checking of their RF properties.

In 1993 Prof. Ugo Amaldi and the TERA Foundation launched studies of compact accelerators (the PACO project, Italian acronym of COmpact Accelerators Project for protontherapy) in the frame of what was called the "Hadrontherapy Programme" (see chapter 1.8). The basic idea of this work was the design and construction of new accelerators for protontherapy, trying to reduce the total cost, without compromising the health care possibilities.

A compact proton accelerator for cancer therapy was required to have the following features:

❑   acceleration of a minimum of 2 x $10^{10}$ protons/sec to an energy of at least 200 MeV with a running efficiency comparable with the conventional electron linacs (98%);

❑   possibility to install the entire facilities in a bunker covering a total area of 300 $m^2$ or less;

❑   maximum power consumption of 250 kW;

❑   costs, without buildings, not more than the equivalent of 11 Meuros in 1993 money (including a gantry with its controls and dose distribution system, which should be able to run with both passive and active spreading system);

❑   a 60 MeV beam for eye melanoma therapy must be foreseen at low cost;

❑   the beam characteristics required for protontherapy as presented in paragraph 2.3.

Not all these goals were attained in the study that lasted four years (from 1993 to 1996) and concentrated on four projects: (1) a linac, (2) a superconducting cyclotron, (3) a high fields synchrotron and a (4) weak focusing synchrotron. For the linac a high frequency was chosen from the beginning because the allowed electric field gradient in a copper structure is roughly proportional to $f^{1/2}$ (f being the RF frequency), so that a higher frequency corresponds to a shorter accelerator. The final idea was then to use the 3 GHz technologies available on the market also for hadrontherapy.

TERA choose Dr. Luigi Picardi (ENEA) as co-ordinator of the overall PACO study and Ing. Mario Weiss as linac project leader. In this context U. Amaldi proposed to use as linac injector a 30 MeV cyclotron that could be used also to produce radioisotopes [B.3]. The standard sequence of a RFQ and various linacs was studied first. In 1994 Ken Crandall and M. Weiss published the TERA note "Preliminary Design of Compact Linacs for TERA" [A.12]. In this note the dynamics of the proton acceleration was described in the all linac solution, which included a novel structure called SCDTL, operating at 3 GHz and low energies (< 10 MeV) and patented by ENEA, Frascati [B.12].

Thus the study group chaired by M. Weiss concentrated on the sequence of three accelerators (figure 2.1 and tables 2.1.a and b), all in standing wave regime: a radio frequency quadrupole (RFQ) operating at a sub-harmonic of 3 GHz, i.e. at 750 MHz, followed by a novel 3 GHz structure called side coupled drift tube linac (SCDTL) [L. Picardi], and terminated with a 3 GHz side coupled linac (SCL). The beam is extracted from the ion source at 80 keV and transported to the RFQ via the low energy beam transport system. The RFQ output of 5 MeV is sufficient for the SCDTL that accelerates the beam up to 70 MeV. The last accelerator in the chain, the SCL, finally brings the beam to the energy of 200 MeV.

At the end of the PACO project the results of the detailed studies of the four different accelerators  were described extensively in the "Green Book" [A.1] which among the editors had Dr. Martino Grandolfo, then Director of the Physics Department of Istituto Superiore di Sanità (ISS - Rome). This was natural since in 1992 U. Amaldi had proposed to install in ISS the first of the compact accelerators, object of the PACO study. At the beginning of 1996 the Department decided to choose the linac solution and the TOP Project (_Terapia Oncologica con Protoni_) was launched by ISS in collaboration with ENEA. An agreement was signed between ENEA and ISS and in 1997 L. Picardi and collaborators published a detailed report [B.12].

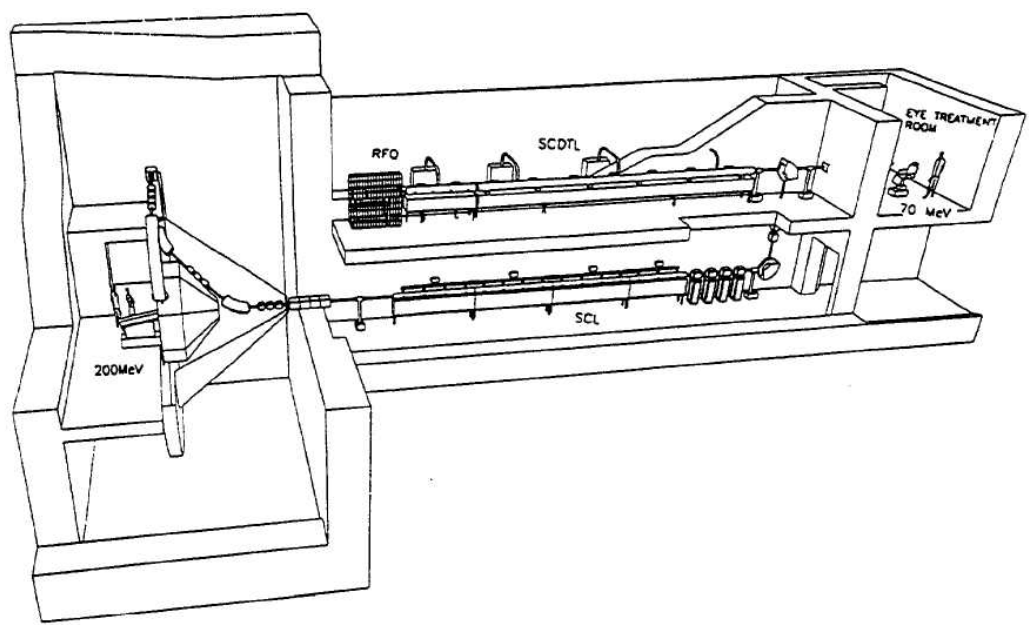

**_Figure 2.1_** _The compact linacs for protontherapy of the TERA Foundation [A.1]._

**Table 2.1.a** *The compact linacs for protontherapy of the TERA Foundation [A.1].*

| |
|---|
| Output energy 70 - 200 MeV |
| Average current 20 nA |
| Beam pulse duration 3 µs |
| Repetition rate 200 Hz |
| Beam duty cycle 0.06 % |
| ***Proton Source and LEBT*** |
| Energy 80 keV |
| Current up to 1 mA |
| Pulse duration 50 µs |
| Repetition rate 200 Hz |
| Beam duty cycle 1 % |
| ***1. RFQ (750 MHz)*** |
| Output energy 5 MeV |
| Beam peak current 33 µA |
| Transverse norm. emittance 0.2 π mm mrad |
| Longitudinal emittance 0.15 π deg MeV |
| Structure length 2.6 m |
| RF peak power 600 kW |
| RF duty cycle 0.1 % |
| ***2. SCDTL (3 GHz)*** |
| Output energy 70 MeV |
| Beam peak current 33 µA |
| Transverse norm. emittance 0.2 π mm mrad |
| Longitudinal emittance 0.6 π deg MeV |
| Energy spread, bunch length (ΔE, Δφ) 0.13 MeV, 4.4° |
| Structure length 11 m |
| Number of tanks 56 |
| Number of permanent magnet quads 56 |
| RF peak power 9 MW |
| RF duty cycle 0.06 % |
| ***3. SCL (3 GHz)*** |
| Output energy 70-200 MeV |
| Output peak current 33 µA |
| Transverse norm. emittance <0.25 π mm mrad |
| Longitudinal emittance 0.6 π deg MeV |
| Synchronous phase angle -13° |
| Energy spread, bunch length (ΔE, Δφ) 0.2 MeV, 3° |
| Structure length 11 m |
| Number of tanks 24 |
| Number of permanent magnet quads 24 |
| RF peak power 30 MW |
| RF duty cycle 0.06% |

**Table 2.1.b** *RF power system of the TERA proton linacs.*

| |
|---|
| Number of klystrons: 1(at 750 MHz), 13 (at 3 GHz) |
| Total klystron RF peak power 40 MW (efficiency 40 %) |
| Modulator average power 60 kW |
| D.C. power for klystron focusing coils 35 kW |
| Mains power for RF ~120 kW |

Generally speaking the applications of linear accelerators to the protontherapy is highly innovative, because, since now, only circular machines, like cyclotrons and synchrotrons, have been used. Linacs at frequencies always smaller than 1 GHz have till now been implemented for two main scopes: either the injection in circular machines or the production of beams with high average current and high power for neutron generation. At the beginning of the 90's some designs for the use of linacs in protontherapy had been investigated on the paper, but they were very preliminary and they have never found the financial support for further developments [B.8, B.9, B.10].

The main advantages in the choice of linacs as medical accelerator can be summarised as:

▪ the output energy is variable as for synchrotrons;

▪ the time structure is very well suited for application to the spot scanning as developed at PSI [B.40]

▪ the transverse beam emittances are smaller than the typical emittances of cyclotrons and synchrotrons, requiring much smaller gaps in the dipole magnets of the gantry, thus reducing considerably their weight and cost.

The linac solution for compact linacs in protontherapy is also convenient for the installation of a light gantry [B.7]. Linear accelerator technology presents also some fundamental advantages, such as simple injection and extraction, modularity of the construction, so that the structure can be built progressively in steps. Moreover the construction procedures (machining and brazing) are well known and a real technological transfer from research centres to industry can be performed.

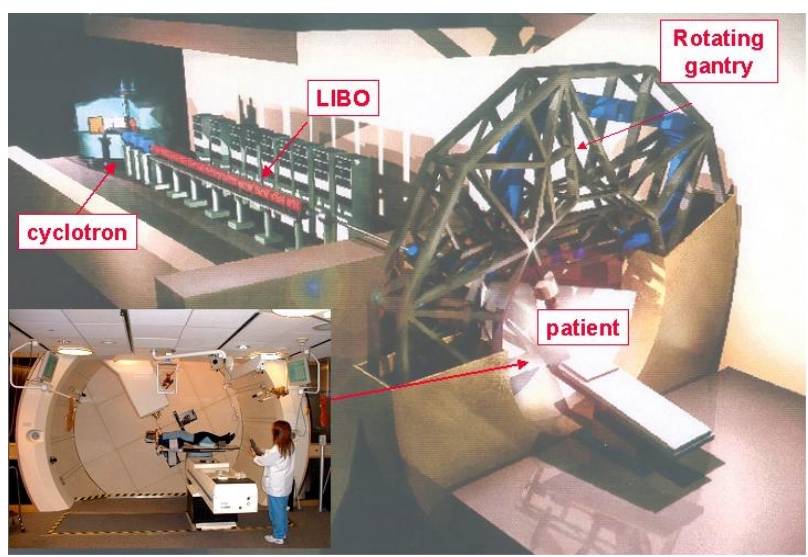

**Figure 2.2** *The booster option: the LIBO with the cyclotron, used as injector, and a rotating light gantry.*

❑ *The booster option: a combination cyclotron-linac for protontherapy*

During the PACO study the proposal of using as injector a high current 30 MeV cyclotron (that can in parallel be used to produce many radiofarmaceuticals for diagnostics and brachitherapy) was modified since in a linac the length of the accelerating cell increases as the proton velocity increases. At the time it was thought that the cells for 30 MeV protons were mechanically too thin, so that it was decided to design a linac booster option starting from about 60 MeV, which requires accelerating cells that are 40% longer than the ones needed for 30 MeV protons.

This design considered the use of the 3 GHz SCL structure, mentioned above, mounted downstream of a low energy cyclotron, in order to boost the energy of the proton beam up to 200 MeV and more. In this way, it is possible to transform a cyclotron, of which many exist in hospitals and physics laboratories in the energy range of 50-60 MeV, in an accelerator for deep-seated tumours. In 1996 it was decided to call this new booster option LIBO, the English acronym of Linac Booster (figure 2.2). As concrete example, in the preparation of the "Green Book" (1996) the 62 MeV cyclotron used in Clatterbridge (UK) was chosen.

In the 3 GHz linacs, shown in figure 2.1, a bunch to bucket matching between the different consecutive accelerating structures has been used. On contrary in the booster option the cyclotron works normally at 25 MHz while the linac at 3 GHz, then a ratio of the operating frequencies of about 120. So the matching technique mentioned above has been abandoned for a new solution where the injected beam is not confined longitudinally, and the particles are trapped only in the stable regions of the linac buckets (figure 2.3).

As previously explained, the beam intensity for protontherapy is relatively small, about 2·10^10 particles per second, i.e. a few nanoamperes, while the cyclotron current is usually much bigger (50-100 μA).

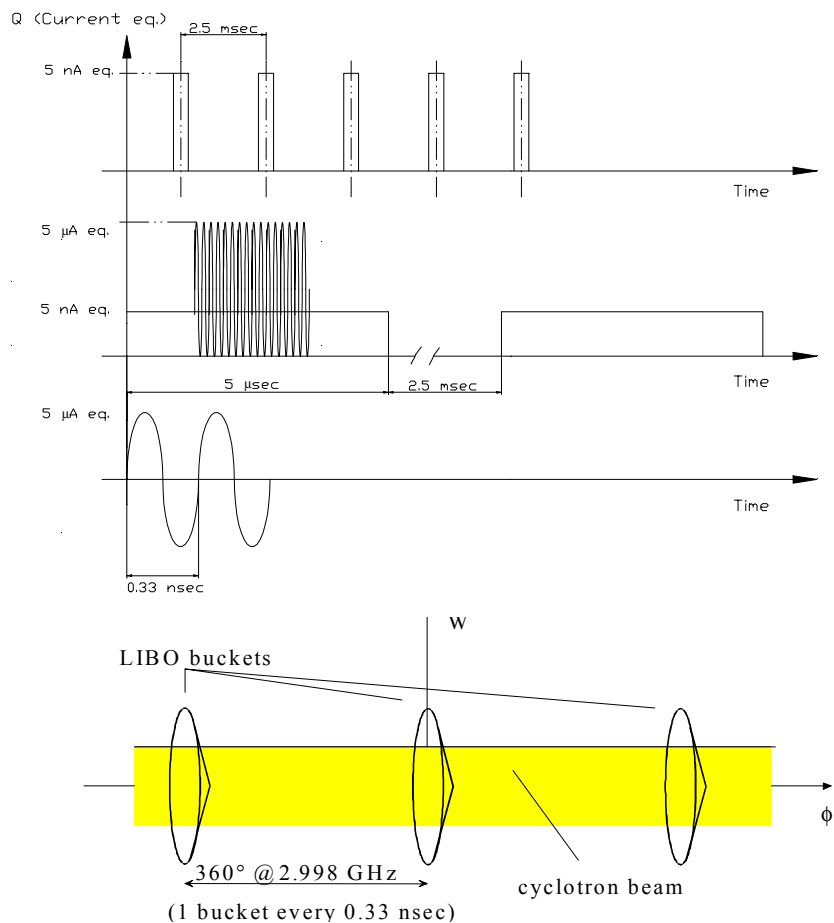

**_Figure 2.3_** _Scheme of principle of LIBO beam characteristics. The typical time-structure of the beam is shown on three different time scales. For illustration the pulse repetition frequency is 400 Hz, macropulse length of 5 μsec, with every bucket filled, so that micropulses or bunches are spaced at the RF period of 0.33 nsec. At the injection the buckets are open and the particles are trapped in the stable regions._

This current can be pulsed each 5 μsec (pulse length) at the repetition rate of 200-400 Hz and even with a small capture efficiency, enough beam passes through the small LIBO aperture to result in an average output current as needed for deep treatment. In fact the beam optics computations have shown that LIBO can have an overall efficiency $\geq 10^{-4}$, so that the 50 μA cyclotron beam would yield at least the 5 nA average current, making then possible the combination cyclotron-linac for cancer therapy [B.4]. For LIBO, the SCL of the compact linac has been modified to accept the bigger beam emittance of the cyclotron.

From the beginning LIBO has been conceived as a modular structure, composed of modules, each of which, fed by RF power supply, can be considered as an RF unit. A module is composed of four tanks for beam acceleration and bridge couplers that connect the tanks together. Each tank is formed by resonant cells (figure 2.4). As injector to the linac, cyclotrons in the range of 50-70 MeV have initially been considered but at present the TERA Foundation is constructing modules for 30 MeV protons, as initially foreseen.

The RF power requirement for the linac is not too large. In fact to produce an average accelerating gradient in each tank of about 15 MV/m (corresponding to an overall average gradient in the module of about 10 MV/m), each tank requires a peak power of about 1.5 MW. For 40 tanks, given the low duty cycle (0.1-0.2%), the average RF power is about 60-120 kW. Another advantage of the linac, shown by detailed dynamics calculations, is that switching off the last klystrons one by one and/or reducing their power, it is possible to obtain a beam of variable energy in the range 100 – 200 MeV. This makes LIBO more similar to a synchrotron than to a cyclotron. On the other hand, as anticipated, the beam structure is very suitable for the _active scanning techniques_ now in use in hadrontherapy, since the 2.5-5 ms that separate two consecutive proton pulses are exactly what is needed to move the beam in the _voxel scanning mode_ used at PSI. Potentially this proton linac has then a wide range of applications, from eye melanoma to deep-seated tumour therapy.

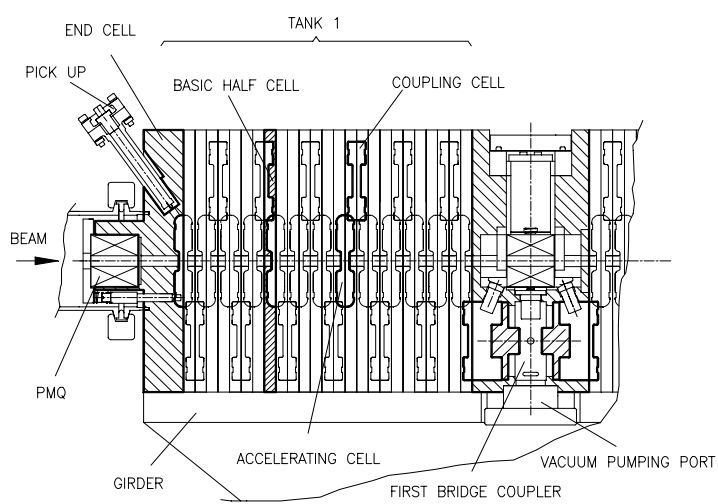

**_Figure 2.4_** _Partial view of a LIBO module: it is a standing wave side-coupled linac (SCL) composed by accelerating (AC) and coupling (CC) cells, bridge couplers for electromagnetic connection between consecutive accelerating tanks, and end half-cell cavities. Permanent magnet quadrupoles (PMQs) are used for beam focusing as well as pick up loops for measurement of the electromagnetic field in the tanks._

❑   *The origin of the LIBO-62 studies and the general frame for the first prototype construction*

In 1997 the Clatterbridge Centre for Oncology (UK) showed a particular interest for the LIBO option. Here a 62 MeV cyclotron, produced by Scanditronix (MC60), exists and it is currently used for eye proton treatment. Moreover, few years before a 1.3 GHz booster linac option had been considered [B.11] to upgrade the facility for deep tumours. For this reasons the LIBO studies have been based on specifications of the Clatterbridge cyclotron (table 4.1). In relation with TERA design, some preparatory work was also performed on this cyclotron in 1997 and 1998 by Clatterbridge physicists, in collaboration with the RF group from Daresbury Laboratory. The measurements made on the Clatterbridge proton beam proved that this cyclotron can be pulsed and that the transverse emittances and the momentum spread can be handled by an adequately designed LIBO [B.11]. For all these reasons the present work will be devoted entirely to this 62 MeV solution (the so-called LIBO-62 facility).

In 1998 an international collaboration, chaired by M. Weiss, was set up among TERA, CERN, the Universities and INFN Sections of Milan and of Naples [A4, A.5, A.6, A.7]. The final goal of the collaboration was the construction and the test of the *first prototype module* of LIBO-62, that accelerates protons from 62 MeV to 74 MeV. The design, construction and tests of this first prototype are the main subject of the present thesis (figure 2.6). The first two years (1998-2000) have been devoted to the construction of this first module, which was done in the CERN workshop. At the end of 2000 the prototype was tested in the LIL tunnel at CERN, using an existing 3 GHz modulator. In October 1999 the Scientific Committee of the INFN *Laboratori Nazionali del Sud* in Catania accepted the proposal to test the prototype with 62 MeV protons produced by the Catania SC cyclotron. In this framework, in the biennium 2000-2002 the LIBO prototype module was installed with its diagnostic equipment on the Catania beam line (summer 2001). The complete acceleration tests were performed between autumn 2001 and spring 2002.

**Table 2.2** *Main milestones of the LIBO history [A.1, A.2, and A.3].*

▪   *1993: Proposal by TERA of 3 GHz proton linear accelerator for cancer therapy and the new linac booster option*

▪   *1993-1996: Design of the 3 GHz protontherapy linac for the "Green Book"*

▪   *1996-1997: General design study of LIBO*

▪   *1998: Final general design of the first LIBO prototype module*

▪   *1998: Setting up of a new collaboration between TERA, CERN, Universities and INFN of Naples and Milan for the prototype construction*

▪   *1999: Final engineering design of LIBO prototype*

▪   *1999-2000: Production at CERN Workshop of LIBO prototype and first installation*

▪   *2000: RF power test of LIBO prototype at CERN*

▪   *2001-2002: Installation and acceleration tests of LIBO prototype at LNS in Catania*

▪   *2001: Proposal by TERA of IDRA project*

▪   *2003-2005: Technology transfer of LIBO construction to industry*

▪   *2003-2004: Proposal by TERA of a new linear accelerator for ions (CLUSTER)*

For the Catania tests, the Collaboration reached an agreement with IBA/Scanditronix and a solid-state modulator and a 3 GHz klystron were lent to TERA and transported to Catania.

The success of the tests, as mentioned in the next chapters, has refuelled interest in the use of a 30 MeV high-current cyclotron as injector: this is the IDRA project proposed in 2001. The machining and brazing techniques developed for the 62 MeV cells could then be adapted to the 40% shorter cells used in a 30 MeV accelerating structure. In this design, with a 15 m long structure, one would accelerate the proton beam supplied by a 30 MeV cyclotron to more than 210 MeV. A machine requiring lower injection energy would be of interest to a much greater number of hospitals, due to the fact that 30 MeV commercial cyclotrons can be used to produce radioactive isotopes in situ for cancer diagnosis and therapy. By acquiring the so-called LIBO-30, such hospitals could then increase the scope of their activities including many new radiofarmaceuticals and proton beam treatment of deep-seated tumours (see chapter 8).

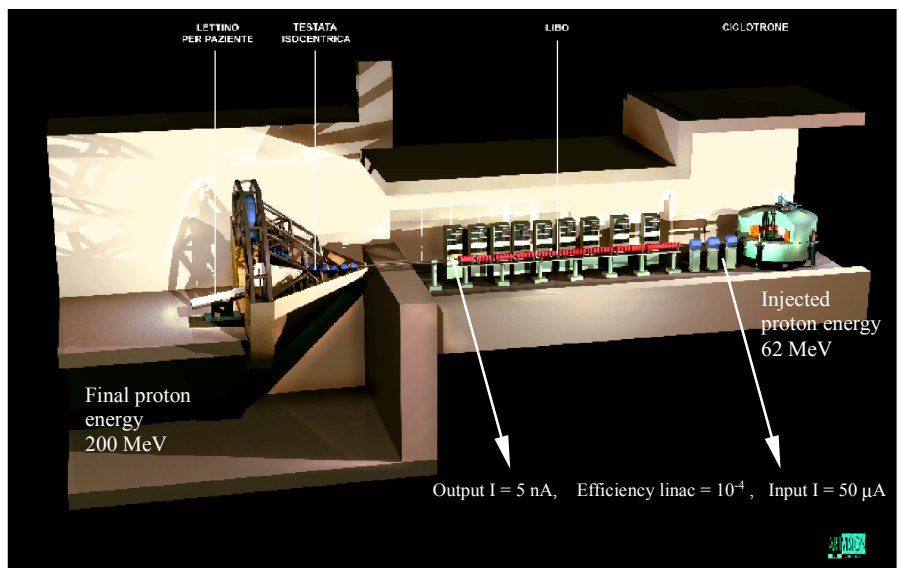

**Figure 2.5** *General scheme of a LIBO facility. The injected current is of the order of 50 μA and the global efficiency is $10^{-4}$, so the output proton current is about 5 nA, enough for cancer therapy.*

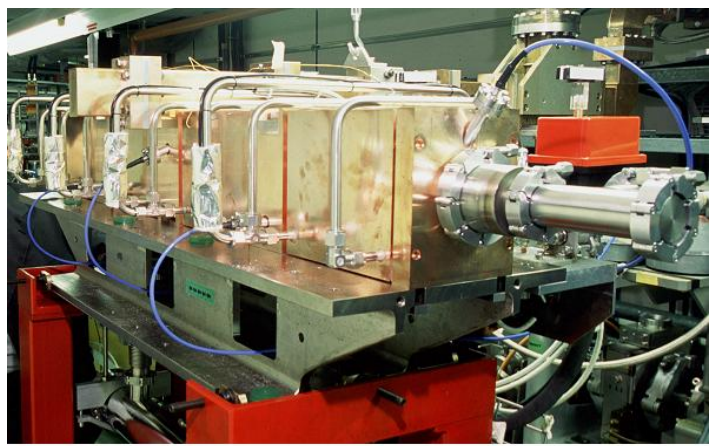

**Figure 2.6** *First LIBO-62 module built at CERN. The design, construction and test of this prototype are the main subjects of this thesis [A.2, A.4, A.5, A.6, A.7].*

## 2.2 Clinical requirements of the therapeutic proton beams

The accelerator design for hadrontherapy must follow stringent physical constraints, dictated by clinical requirements. In the next three paragraphs these clinical requirements for hadrontherapy and the relative minimum physical specifications of a compact proton accelerator [B.7] are presented in detail as well as the performance reached by LIBO.

❑ *Beam range*

In order to satisfy all therapeutic requests, for superficial and deep-seated tumours, the beam range must be performed as follows:

1) a variable range between 2 cm and 20 cm for deep treatments,

2) a variable range between 2.5 cm and 3.2 cm for ocular treatments,

3) a variable range between 2 cm and 10 cm for head and neck treatments.

❑ *Bragg peak modulation and range adjustment*

The Bragg peak modulation is defined on a depth dose curve as the distance between the proximal and the distal points corresponding to the 90% of the peak height. Normally the expression range adjustment refers to the possibility of translating the SOBP in depth. The modulation of the Bragg peak should be variable in steps of 0.05 cm for extension ≤ 5 cm and in steps of 0.1 cm for extension > 5 cm.

❑ *Dose rate*

For ocular treatments, which require very short irradiation time (≤ 20-30 sec), high dose rates in the SOBP are necessary (about 30-45 Gy/min), due to the necessity to guarantee a greater accuracy in the dose deposition. These parameters are necessary for cancer volumes smaller than 50 cm$^3$. For target volumes between 50 and 2000 cm$^3$ the required dose rates in the SOBP are 2-10 Gy/min.

❑ *Field size*

For fixed horizontal beams the required field size ranges from 2 x 2 cm$^2$ to 15 x 15 cm$^2$ at the entrance surface of the patient. For coupled horizontal and vertical fixed beams as well as the gantry beam, the required field size ranges from 2 x 2 cm$^2$ to 20 x 20 cm$^2$ at the isocentre. The dimensions of the irradiated region can be increased by translation of the treatment table. In all cases the field dimensions must be adjustable with steps of 1 mm (± 0.05 mm).

❑ *Field homogeneity and symmetry*

The homogeneity along the beam axis is defined as:

$$R_l = \frac{P_{\max}}{P_{\min}} \cdot 100$$

where $P_{max}$ is the maximum dose absorbed everywhere in the irradiation field, averaged on a surface not larger than 1 cm$^2$, and $P_{min}$ is the minimum absorbed dose in the homogeneity region, averaged on a surface not larger than 1 cm$^2$. Considering a plane transverse to the beam axis, the field symmetry is defined as the maximum value (expressed in %) of the ratio between the highest and the lowest adsorbed dose, averaged on a surface not larger than 1 cm$^2$, for every pair of symmetrical positions with respect to the beam axis inside the homogeneity region. The transversal ($R_t$) and longitudinal ($R_l$) homogeneity of the field must be as: $R_t \leq 105\%$, $R_l \leq 111\%$. The field symmetry is less than 105%.

❑   *Lateral penumbra*

Consider a dose profile measured along one of the main axis of the field. The lateral penumbra is defined as the distance between the points of 80% and 20% absorbed dose. 100% corresponds to the beam axis position. Lateral penumbra of the beam should be not more than 2 mm for each side.

❑   *Distal dose fall-off*

Considering a depth dose curve along the beam axis, the distal dose fall-off is defined as the distance between the points of 80% and 20% absorbed dose along the beam axis beyond the Bragg peak. 100% corresponds to the maximum in the SOBP. The distal dose fall-off due to the momentum dispersion of the beam should be less than 2-3 mm in addition to the intrinsic distal fall-off of the beam.

*__Table 2.3__ Clinical requirements for hadrontherapy [B.4].*

|  | Clinical requirement | has influence on |
|---|---|---|
| *Beam range in patient* | 2-20 cm<br>continuously adjustable | Energy: 180MeV + 20 MeV accounting for energy loss in the spreading devices to obtain a 20 x 20 cm$^2$ field with a throw of 3 m. |
| *Field size* | 20 x 20 cm$^2$ | Beam spreading – beam intensity |
| *Range modulation* | continuously adjustable | |
| *Range adjustment* | continuously adjustable | |
| *Dose rate* | | |
|    *large fields* | >2 Gy/min | Beam intensity |
|    *small fields* | >> 10 Gy/min | Beam intensity |
| *Field homogeneity* | | |
|    *Longitudinal* | <= 111% | Beam spreading |
|    *Orthogonal* | <= 105% | Beam delivery system |
| *Field symmetry* | <= 105% | Beam spreading – beam delivery system |
| *Lateral penumbra* | < 2 mm at the entrance | Multiple scattering spreading devices<br>Source site – source to axis distance |
| *Distal fall-off* | < 2 mm | Range straggling (energy) – energy spread<br>Source to axis distance |

## 2.3      Minimum physical specifications of a compact accelerator for proton therapy*

❑   _Beam characteristics: energy, range modulation and range adjustment_

Range modulation and adjustment are two important aspects of a proton therapy facility. Different solutions can be used to obtain a range in patient or to spread the Bragg peak so to obtain a Spread Out Bragg Peak (SOBP). The beam range in patient between 2 and 25 cm is achieved by an energy interval between 60 and 200 MeV. However the effective range available for treatment is less due to the energy loss suffered by the beam in all the material positioned between the extraction port of the accelerator and the patient (especially when a passive method is used for beam delivery). When a passive beam spreading system is used with a 3 m throw to obtain the required maximum 20 x 20 cm$^2$ field size, the energy loss in the scatterers lowers the energy to 175 MeV, which will ensure the range in patient of 20 cm required for deep seated tumours.

The energy precision of the linac must be within ± 0.4 MeV to satisfy the depth accuracy of the Bragg peak. The precision of the energy variation must allow steps of 0.1 cm or 0.05 cm for ranges smaller than 5 cm.

The energy spread affects the distal dose fall-off. The clinical specifications require that the excess distal dose fall-off (between the 80% and 20% levels) should not be larger than 2 mm. This excess is generated by two causes: the energy spread due to the accelerator (ΔW) and the energy spread δW due to the passive system that creates the required field size (20 x 20 cm$^2$). It can be approximated by [A.1]:

$$excess\ distal\ dose\ fall-off = 1.3 \cdot \sqrt{\Delta R_{beam}{}^2 + \Delta R_{RS}{}^2}\quad [cm],$$

where

$$\Delta R_{beam} = 1.65 \cdot R(W) \cdot \frac{\Delta W}{W},$$

$$R(W) = 25 \cdot (\frac{W}{200})^{1.65},$$

$$\Delta R_{RS} = 0.012 \cdot \Delta X,$$

R(W) is the range in cm in the tissue, W is the output energy of the accelerator in MeV.  The quantity $\Delta R_{RS}$ is the straggling generated in the scatterer material with depth ΔX (cm). Using the formulas mentioned above one can obtain the energy spread of the proton beam (ΔW) at different energies in order to have a well-defined distal dose fall-off in the tissue (water equivalent).

According to [A.1, B.12] the energy spread of the beam should be < 2.2% up to 65 MeV and < 0.75% between 100 MeV and 200 MeV in order to obtain the clinical request for the excess distal dose fall-off (between the 80% and 20% levels) of 2 mm or less.

The transverse emittance has to be as small as possible since it determines the size of the magnets in the beam transport system and, as a consequence, the gantry weight. Moreover it influences the size of the beam spot at the isocenter. For this reason and using the specifications of the Massachusetts General Hospital facility in Boston [A.1 and paragraph 1.5], a transverse emittance less than 5 π mm mrad at 200 MeV has been fixed as physical specification of compact accelerators for protontherapy.

❑   *Field size, dose rate and beam intensity*

As shown in table 2.3 the maximum field size for the compact accelerator is 20 x 20 cm$^2$. At the end of the beam transport system, particle beams have a too small cross-section to be used in comparison to the large treatment areas that are needed in radiation oncology. It is then necessary to spread these narrow beams: this aspect is guided by the beam delivery system. Dose rate has a strong dependence on beam intensity and, as a consequence, on the characteristics of the beam spreading devices and construction characteristics of the accelerator. The required dose rates for a proton facility vary with the type and size of the target volume. Large fields (20 x 20 cm$^2$) can be treated with dose of 2 Gy/min. For ocular and stereotactic treatments, where a short treatment time may be required, dose rates ranging between 30 and 45 Gy/min are necessary.

The need to deliver to the appropriate volume, uniformly within a few percent, the correct dose defines the beam intensity of the accelerator. For the application in radiotherapy the average current is important, that is proportional to the number of protons per second. For example, in order to irradiate a 20 x 20 x 10 cm$^3$ tumour (4 litres) with 2 Gy in one minute, one must deliver 8 J. Using a lucite modulator, the average contribution of one proton to the Spread-Out-Bragg-Peak is about 10 MeV/cm [A.1]. This means that this proton will contribute to the total tumour dose (for 10 cm in depth) with about 100 MeV. So a number of 8 J/100 MeV = 5 $10^{11}$ protons will be necessary, which corresponds to 1.2 nA for one minute. Taking into account an overall efficiency of 50% and several safety factors, a current of about 3 nA is necessary that corresponds to 2 $10^{10}$ protons per second. Using the same approach, for eye melanoma treatment at 65 MeV the average contribution of one proton to the SOBP is about 20 MeV/cm [A.1]. This means that this proton will contribute to the total tumour dose (for 1 cm in depth) with about 20 MeV. Considering a typical area of 3.5 cm of diameter, including the tumour target, one has a number of about 0.45 J/ 20 MeV = 1.4 $10^{11}$ protons which corresponds to 0.37 nA for one minute. With the assumption made, to deliver these doses to the target volume and considering the different beam delivery techniques, the accelerator has to be set, before the treatment, to have a current between 0.1 nA and 5 nA. Of course if the accelerator meets this specification with passive spreading, it will easily meet it in the scanning mode due to the increased efficiency.

## 2.4    Physical performance of LIBO-62

LIBO-62 is a compact accelerator designed to reach the physical specifications requested by the clinical constraints for proton therapy. Below follows a brief summary of its performance.

❑  *Clinical specification: range in the patient between 2 cm and 20 cm*

   *Clinical specification: range modulation and adjustment: steps of 0.1 cm for range > 5 cm*

   *Clinical specification: range modulation and adjustment: steps of 0.05 cm for range < 5 cm*

   *Minimum physical specification: Maximum energy 200 MeV*

   *Minimum physical specification: Energy accuracy ± 0.4 MeV*

   *Physical specification of LIBO-62: maximum energy of 200 MeV and continuous energy variation*

   *between approximately 100 and 200 MeV*

As already mentioned, the maximum energy has been fixed to 200 MeV in order to have at least 20 cm range in water. The energy increase above this value means adding some cavities at high energy. This is feasible without changing the design due to the natural modularity of this accelerator. The simplest solution for the energy variation is to fix the accelerator energy and change this by external system of absorbers. In case of passive technique, output energy has to be set to the value corresponding to the maximum depth of the tumour, while the SOBP is produced by a range modulator placed in the nozzle. At the end of the treatment, beam output energy and gantry magnets must be re-set. In case of active technique the energy has to be changed in a very short time and the beam transport system must be designed accordingly. LIBO is composed of separate modules, which can be considered as single RF units. Switching off a number of them results in a continuous of change the output energy, controlling the power of the last fed module. The control system is able to determine the power and so the beam energy. LIBO fits perfectly for the use of an active scanning system. In fact with this structure the energy can be varied also from one pulse to the next. LIBO output beam is practically unaffected by this operation. The subdivision in several modules is an advantage in satisfying the energy resolution requests.

❑  *Clinical specification: distal dose fall-off (80% - 20%) < 2 mm behind the intrinsic straggling*

   *Minimum physical specification: Energy spread < 0.75% above 100 MeV, < 2.2% at 65 MeV*

   *Physical specification of LIBO-62: Energy spread < 0.75% above 100 MeV, < 0.23% at 65 MeV*

In order to have a distal fall-off smaller than 0.2 cm, the energy spread of LIBO must be lower than the values mentioned before. For example, by using the formulas mentioned above, one could find that $\Delta R_{RS}$, at 200 MeV and for a 5 cm ($\Delta X$) water equivalent material of the scatterer (typically used for a field of 20 x 20 $cm^2$), is about 0.6 mm. At 200 MeV the calculated energy spread of the proton beam, for a distal fall-off smaller than 0.2 cm, is then 0.34%. Paragraph 4.2.1 will show that the energy spread of the output beam is well below these minimum physical specifications for all the LIBO energy range. Moreover simulations show energy spreads, due to the RF modulation, inside of the minimal physical specifications (chapter 4).

❑ *Physical specification of LIBO: norm. transverse emittance of the injector cyclotron $\cong$ 3.7π mm mrad,*
   *normalised transverse acceptance of LIBO $\cong$ 4.2π mm mrad*

LIBO can produce beams with emittances far smaller than other accelerators. This is generated by the fact that the use of high frequency at low beam current allows reducing the bore hole for the beam passage. As a consequence it is possible to reduce also the magnet apertures and the gantry size.

❑ *Clinical specification: Dose rate: 45 Gy/min for eye melanoma treatment and*
                          *2 Gy/min for deep seated tumours (field size 20 x 20 cm$^2$)*
   *Minimum physical specification: between 0.1 and 5 nA at different energies*
   *Physical specification of LIBO-62: between 0.1 and 10 nA at different energies*

The beam intensity is reached by LIBO easily at several energies by acting on the source or on the beam delivery system. The linac accelerator is designed for a current ranging between 0.1 nA and 10 nA, enough for eye melanoma and deep proton therapy.

❑ *Physical specification of LIBO-62: beam time structure and the active scanning application*

Regarding the time structure, LIBO is a 3 GHz radiofrequency (RF) accelerator and therefore the beam is delivered in bunches every 0.33 nsec. This micro-time structure is compatible with both patient treatments and accuracy dosimetry. However attention must be paid for the macro-time structure. The ideal accelerator should generate a continuous beam because it facilitates the use of any standard dosimetry and of passive delivery systems. Nevertheless, if the beam pulse is not too short, it is possible to use standard dosimetry and, having special care in the synchronisation and control of the current stability, also an active technique can be used. The performances of a linear accelerator as LIBO depend on the pulse repetition frequency. The error due to ionisation chamber saturation effects are of the order of 1% at 100 Hz [A.1], so at 400 Hz, the LIBO repetition rate, no problems arise. With the pulse length of 5 μsec and a repetition rate of 400 Hz, an active scanning system can be implemented by using a single shot for each voxel. In fact, for the largest 20 x 20 cm$^2$ area, and with a 0.7 cm FWHM dimension of the spot inside the target, 1000 spots are required; they can be delivered in 2.5 sec. In a 50 sec session, 20 slices can be treated by changing the energy (the slowest variable) and the current of the accelerator. It is evident that LIBO must be capable to vary not only the transversal position of the beam in x-y plane, but also the energy slice by slice. This is possible, above 100 MeV, changing the power of the RF generators in the last modules. This operation can be implemented on a very short time (less than 1 sec). During the changing energy the accelerator is temporary switched off. With the above considerations it is evident that LIBO can therefore fulfil the needs for a scanning system also for the largest fields. As to passive modulating and spreading systems, the modulator disk, which produces the SOBP, has to be driven in steps and synchronised to the 400 Hz repetition frequency.

# Chapter 3

# Basics on Radio Frequency

# and beam dynamics physics of linear accelerators

*"My little machine was a primitive precursor of this type of accelerator which today is called a "linac" for short. However I must now emphasise one important detail. The drift tube was the first accelerating system which had earthed potential on both sides, i. e. at both the particles' entry and exit, and was still able to accelerate the particles exactly as if a strong electric field was present. This fact is not trivial. In all naivetè one may well expect that, when the voltage on the drift tube is reversed, the particles flying within would be decelerated, which is clearly not the case. After I had proven that such structures, earthed at both ends, were effectively possible, many other such systems were invented."*

*[Rolf Wideroe, 1928]*

## 3.1      Brief introduction to linear accelerators

A linear accelerator (generally called linac) produces particles accelerated in a straight line using electromagnetic field. The RF linac is a periodic structure divided into cells that contain sinusoidal varying electromagnetic fields, usually in the ultra high frequency and microwave frequency range [C.6, C.8]. An extensive theory on these structures has been developed in the past years. In this chapter the main concepts on RF cavities will be presented as well as the general nomenclature used for a linac design [C.1].

The history of linear accelerators starts with G. Ising in 1924, when he proposed a system of sequentially pulsed drift tubes, where the particles would be accelerated between gaps. In 1928 R. Wideroe [C.1] suggested the use of RF voltage between consecutive drift tubes, and successfully tested such a device, followed also by D. H. Sloan and E. O. Lawrence [C.1] in 1931. It was, however, not until the end of the Second World War that the development of linear accelerators really started. The availability of RF power sources in the MW range and at frequencies of hundreds of MHz, developed for radar purposes, was a great asset. In 1946, L. Alvarez built in Berkeley his famous proton drift tube linac of 32 MeV, operating at 200 MHz [C.1]. At the same time the development of a 3 GHz electron linear accelerator at SLAC was under way. Since then, many linear accelerators have been developed and built, and the theory and practice have progressed considerably, and are still doing so nowadays.

Generally speaking, the main advantages of linear accelerators are their capability to produce high energy, high intensity charged particle beams of excellent quality in terms of size (small diameter) and low energy spread. The main reasons for these characteristics can be summarised as follows. Acceleration to high energies is not limited by electrical breakdown as in dc accelerators. Electric breakdown limits the maximum energy for charged ions in dc accelerators to several tens of MeV. Strong focusing can easily be provided for containment of high intensity beams. The beam passes through the structure in a single pass, and therefore repetitive exposure to error conditions, causing destructive beam resonances, as in a circular machine, does not occur. This aspect increases the current limit for acceleration of high intensity beams. Moreover due to the fact that the beam travels in a straight line, there is no power loss from synchrotron radiation, normally presents in high-energy electron beams in circular machine.

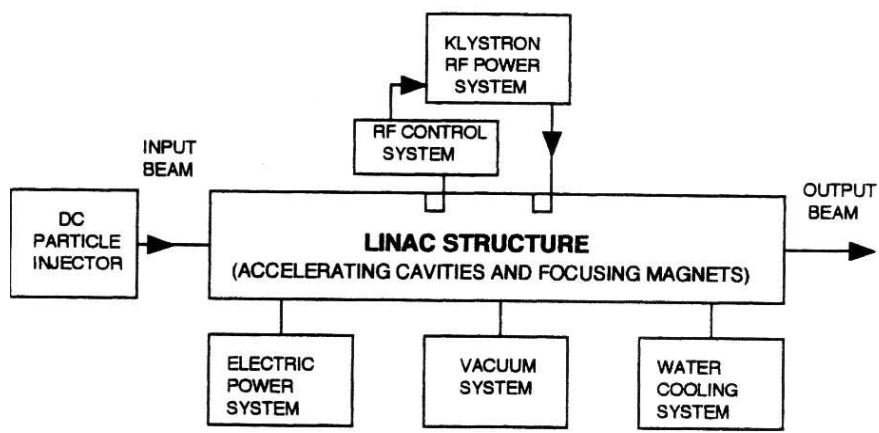

**Figure 3.1** *Simplified block diagram of a linac.*

Injection and extraction are simpler than in circular accelerators: the natural orbit of the linac is open at each end and therefore, although the beam must generally be bunched before injection, special techniques for these beam transfers are unnecessary. Figure 3.1 shows a simplified block diagram of a linac structure with accelerating cavities and focusing magnets, and supplied with electromagnetic energy by an RF power system energy through a transmission line or waveguide. Beam is injected from a dc injector system. A vacuum system is required for good beam transmission, while a cooling system removes the heat generated by the resistive wall losses.

In RF linac, which uses a sinusoidal varying electric field for acceleration, particles can either gain or lose energy depending on their phase relative to the crest of the wave. Therefore the accelerator must provide an electric field with the correct phase relationship relative to the beam to maintain acceleration along the structure. Normally the beam is grouped into bunches [C.2], that are separated longitudinally by one period of the electromagnetic field oscillations. Two types of linear structures are normally considered:

- the travelling wave structures,
- the standing wave structures.

In both the particles and the electromagnetic waves are in synchronism.

The travelling wave structures are conceived as a waveguide in which the injected electromagnetic wave travels with a phase velocity equal to the beam velocity along the structure. The particles gain energy continuously from the longitudinal electric field of the wave, depending on their phase relative to the crest.

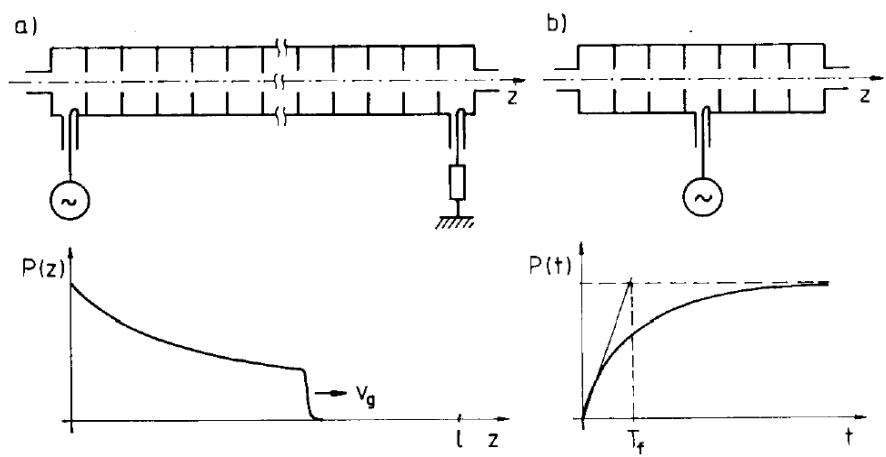

**Figure 3.2** *Periodic RF structures can be operated in two different ways, as a travelling-wave (TW) accelerator or a standing-wave (SW) accelerator. In a TW-structure, figure a, the fields build up in space with the wave front travelling with the group velocity. The output of the structure is matched to a load where the left-over energy is dissipated. In a SW-structure the fields build up in time, figure b. The incoming wave is attenuated along z, reflected at the output end and experiences another attenuation. Back at the input end the wave is partially reflected and partially transmitted through the input coupler. If the length of the structure is a multiple of half-wavelengths, the wave reflected at the input will be in phase with the incoming wave. The process of reflection at both ends will continue until equilibrium is reached. At the equilibrium none of the backward power will pass out of the input coupler if the coupler is correctly matched [C.11].*

The wave travels with attenuation due to the ohmic losses in the structure walls, and beam loading in case of high current, and after a certain distance the remaining power is absorbed in a load, with an impedance matched to the characteristic impedance of the guide. The typical travelling wave structure is disk-loaded as designed for electron linacs [C.1].

Standing wave structures consist in resonant cavities containing several accelerating cells. If the frequency of the injected wave matches the resonant frequency of the cavity, a characteristic standing wave pattern is built up, as a result of reflections of the injected wave from the ends of the cavity. A correct relationship of the field relative to the beam in terms of phase, must be achieved. Also here the longitudinal electric field component provides the accelerating force. For this application several types of multicell structures have been invented for optimum application over specific range of beam energy. The standing wave pattern mentioned above also can be described as a linear combination of two equal amplitude travelling waves, but with opposite directions. In this scenario the forward wave is synchronous with the beam and generates acceleration. In many cases the backward wave is not synchronous and does not produce a net average force on the beam. An example of standing wave is given in figure 3.3. The type of the electromagnetic wave and mode is TM$_{010}$, normally used in linacs. Another example of a linac cavity is shown in figure 3.4, where the phase difference of the electromagnetic wave between adjacent cells is $\pi$ (structure mode).

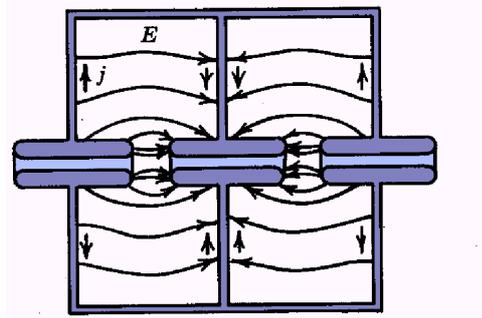

**Figure 3.3** *Electric field lines and wall currents of TM$_{010}$ modes in two cells of standing wave linear accelerator. The accelerating field is concentrated in the gap. The wall currents are the main origin of the power dissipation [C.11].*

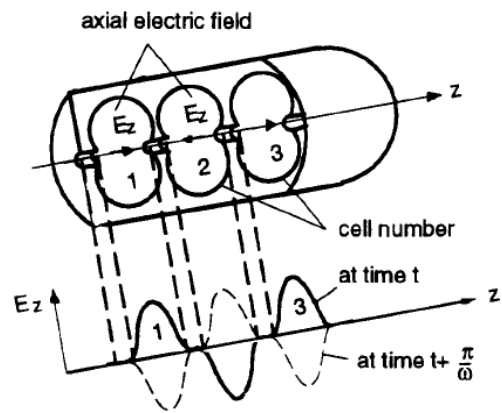

**Figure 3.4** *Standing wave accelerator operating in the $\pi$ structure mode [C.6, C.11].*

### 3.2      Radio frequency cavities for proton linear accelerators

### 3.2.1    General considerations for RF cavity design

❑   _Main cavity parameters_

There are several parameters that characterise the physics of the linac structure; the fundamental ones are:

▪   Transit time factor T:

$$T = \frac{\left| \int_0^L E_z(z,0) \cdot e^{-j(\omega z / v_p)} \cdot dz \right|}{\int_0^L E_z(z,0) \cdot dz}$$

The transit time factor T shows how well a field $E_z$ in the longitudinal direction accelerates particles. The denominator is equal to $E_0 \cdot L$, where $E_0$ is the average field in a cell, and L is the cell length. The transit time factor measures the ratio of the energy gained in the time varying RF field to that in a dc field of voltage $E_0 \cdot L \cdot \cos\phi$. It is a measure of the reduction in the energy gain caused by the sinusoidal time variation of the field in the gap ($\phi$ is the phase of the RF wave, when the particles pass in the mid gap).

▪   The shunt impedance is:

$$r_s = \frac{V_0^2}{P} \qquad [\text{M}\Omega].$$

It is a figure of merit that is independent of the excitation level of the cavity and measures the effectiveness of producing an axial voltage $V_0$ for a given dissipated power P. The peak energy gain of the particle occurs when $\phi=0$ and is:

$$\Delta W_{\phi=0} = q \cdot V_0 \cdot T.$$

Consequently, we define an effective shunt impedance of a cavity as:

$$r_{seff} = \left( \frac{\Delta W_{\phi=0}}{q} \right)^2 \cdot \frac{1}{P} = \frac{(V_0 \cdot T)^2}{P} = r_s \cdot T^2.$$

This parameter measures the effectiveness per unit power loss for delivering energy to a particle. It is also convenient to introduce the shunt impedance per unit length as:

$$Z = \frac{r_s}{L} = \frac{E_0^2}{P/L}.$$

Similarly the effective shunt impedance per unit length $Z_{seff}$ is:

$$Z_{seff} = Z \cdot T^2 = \frac{\left( energy \ \ gain / metre \right)^2}{power \ \ dissipated / metre} = \frac{r_{seff}}{L} = \frac{(E_0 \cdot T)^2}{P/L} \qquad [\text{M}\Omega \ \text{m}^{-1}]$$

The effective shunt impedance indicates the efficiency of the acceleration for a given dissipated RF power P.

▪ Quality factor Q

$$Q = \omega \cdot \frac{U}{P}$$

where U is the cavity stored energy, P the average power dissipated, ω the angular frequency. The stored energy is defined as:

$$U = \frac{\varepsilon_0}{2} \int E^2 \cdot dV + \frac{\mu_0}{2} \int H^2 \cdot dV$$

In the steady state for any resonant mode, the electric and magnetic stored energies oscillate in time, 90° out of phase. The total stored energy U remains constant. The time averaged electric and magnetic energies are equal. To calculate U it is easy to use either the electric or magnetic term alone with the peak value of the field in time. Another useful parameter is the ratio of effective shunt impedance to Q, often called r over Q:

$$\frac{r_{seff}}{Q} = \frac{(V_0 \cdot T)^2}{\omega \cdot U} \, .$$

This ratio measures the efficiency of acceleration per unit stored energy at given frequency. One may wish instead to quote the $\dfrac{Z \cdot T^2}{Q}$. Either of these ratios is useful, because they are a function only of the cavity geometry and are independent of the surface properties that determine the power losses.

▪ Duty factor

The duty factor shows the fraction of time during which the accelerator is in active mode. Linacs normally operate in a pulsed mode where the pulse is repeated a specific number of times per second. The duty factor is the product of the pulse length and the repetition rate.

❑ *Scaling factors of a RF linear accelerator*

During an analysis for the accelerator design it is also important to consider the relation between the structure parameters and the frequency. Suppose we scale all cavity dimensions with wavelength or as $f^{-1}$. The cell lengths of a multicell structure are proportional to $\beta \cdot \lambda$ and also would scale as $f^{-1}$. The total length for a given energy gain is independent of frequency if $E_0$ and the energy gain ΔW are fixed. Therefore the diameter, gap, radial aperture and cell length are scaled as $f^{-1}$. If $E_0$ and ΔW are fixed then all other fields are fixed, then:

$$T \propto f^0, \quad E \propto f^0, \quad B \propto f^0, \quad \Delta W \propto f^0, \quad L \propto f^0, \quad Surface \propto f^{-1},$$

$$Cavity \;\; Volume \propto f^{-2}, \quad U \propto f^{-2}$$

Surface resistance and power loss scaling depends on the type of the accelerating structure. For normal conducting structure the RF surface resistance $R_s$ scales as $f^{1/2}$, while the RF power dissipation scales as:

$$P = \frac{R_s}{2} \cdot \left| \frac{B}{\mu_0} \right|^2 \cdot dA \propto f^{-1/2}$$

This shows that higher frequency gives reduced power losses. The Q factor scales as $f^{-1/2}$ and the effective shunt impedance per unit length as $f^{1/2}$. Also here, the shunt impedance improves at higher frequencies. The ratio $\dfrac{Z \cdot T^2}{Q}$ scales as $f^1$.

❑   *Notes on the equivalent circuit for a resonant cavity system*

We consider a cavity coupled with an RF generator (figure 3.5). The unloaded Q factor is indicated as $Q_0$ and it can be defined as the stored energy U divided by the energy dissipated in the cavity:

$$Q_0 = \frac{\omega_0 \cdot U}{P_c}$$

where $P_c$ is the power dissipated in the cavity. If $P_{ext}$ is the power dissipated in the external circuit, the loaded Q factor is defined in terms of the total power dissipation in the system, that is external circuit plus cavity:

$$Q_L = \frac{\omega_0 \cdot U}{P_c + P_{ext}}.$$

For a cavity coupled to a waveguide it is convenient to define a general parameter that is the measure of the waveguide to cavity coupling strength, known as the parameter $\hat{\beta}$. This is defined by:

$$\hat{\beta} = \frac{P_{ext}}{P_c} = \frac{Q_0}{Q_{ext}}, \quad \text{with } Q_{ext} = \frac{\omega_0 \cdot U}{P_{ext}}.$$

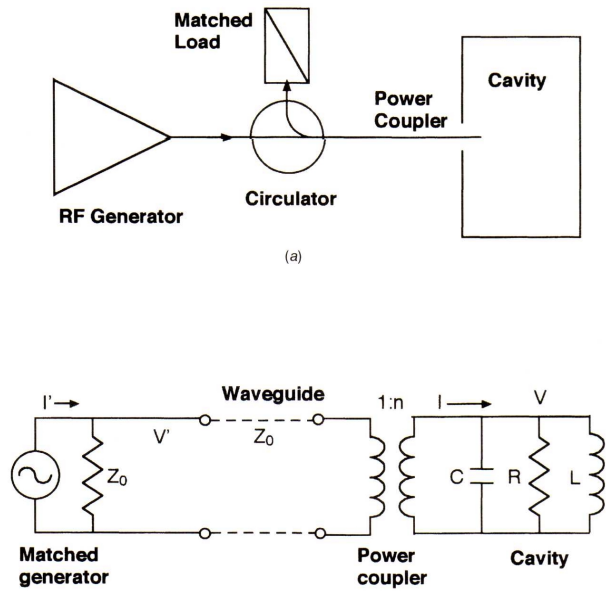

**Figure 3.5** *Block diagram of RF system components of a linac accelerator composed by the accelerating cavity, the RF generator and the coupler apparatus. The equivalent circuit of the global system is also visible. The cavity is represented as an equivalent electric circuit, where the capacitance (C), inductance (L) and resistance (R) are indicated.*

Substituting the above equations, one can obtain:

$$Q_L = \frac{Q_0}{1 + \hat{\beta}}.$$

When $\hat{\beta}$ <1 the waveguide and cavity are said to be under-coupled, for $\hat{\beta}$ >1 the waveguide and cavity are said to be over-coupled, when $\hat{\beta}$ =1 we have a critical coupling and it is the normal condition for a match, or maximum power transfer from the external circuit to the cavity.

The time dependence of the resonator fields during the turn-on transient is defined by the filling time constant. It defines the time needed for the electromagnetic energy to fill the cavity as:

$$T_f \propto \frac{Q_L}{f_0}.$$

### 3.2.2    Standing wave high energy structures for protons

❑  *Standing wave modes for RF structures*

We consider a resonant cavity, which consists of a linear and periodic sequence of accelerating cells [C.28]. We will focalise the attention on the properties of a $TM_{010}$ standing wave mode [C.14].

Two standing wave modes are normally used for linacs: the 0 mode and the $\pi$ mode. In 0 mode the field has the same sign in all gaps and there is one accelerating gap per axial period L (where L is the field period). In the $\pi$ mode the fields of two sequential accelerating gaps have different signs and there are two accelerating gaps per axial period 2L (2L is the field period). The π/2 mode is included in this family, due to the fact that in the accelerating structures there is still a phase shift of $\pi$ between two consecutive accelerating cells. These modes are shown in the schematic figure 3.6. To treat both cases, one must assume a field period of 2L. The electric field component along the beam axis can be expressed as [C.6]:

$$E_z(r,z,t) = \left[ \frac{a_0(r) \cdot \cos \omega t}{2} + \sum_{m=1}^{\infty} \frac{a_m(r)}{2} \cdot \left[ \cos\left( \omega t - \frac{m \cdot \pi \cdot z}{L} \right) + \cos\left( \omega t + \frac{m \cdot \pi \cdot z}{L} \right) \right] \right].$$

This equation describes the standing wave pattern as a sum of forward and backward travelling waves. These waves are also called space harmonics. The wave numbers of the mth wave are $K_m = \pm \frac{m \cdot \pi}{L}$, where the signs represent the forward and backward waves. The corresponding phase velocities are:

$$\beta_m = \frac{\omega}{K_m \cdot c} = \pm \frac{2 \cdot L}{m \cdot \lambda}.$$

In a linac one of these waves will be synchronous with the beam. The effect of all other waves on the particles averages to zero, due to the fact that they do not maintain a specific phase relationship with the beam.

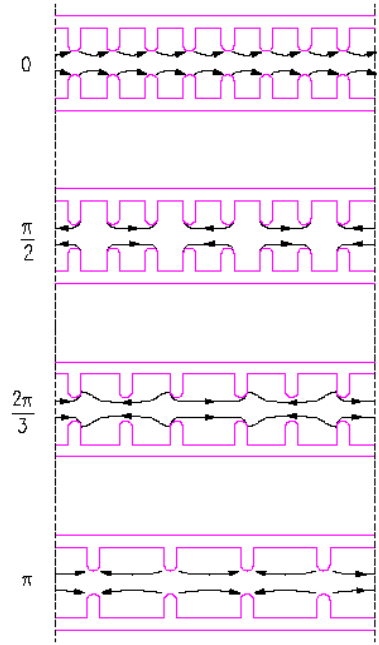

**Figure 3.6** *Instantaneous electric field configurations for different structure modes between accelerating gaps.*

We now consider $\pi$ and $\pi/2$ modes, where one complete field period contains two accelerating gaps with an alternating sign of the field. Examination of the symmetry of the $a_m$ coefficients, reveals that $a_m = 0$ for all even m values. The period of the field is 2L and all odd harmonics can be present. The wavelength of the mth harmonic in the Fourier expansion is $\lambda_m = \dfrac{2 \cdot L}{m}$. Therefore the terms with m odd integers are the only nonzero terms. Let $n = \dfrac{m-1}{2}$, the nonzero terms are n=0, 1, 2, …., ∞. The phase velocity of the nth harmonic is $\beta_n = \dfrac{\omega}{K_n \cdot c} = \pm \dfrac{L}{(\dfrac{n+1}{2}) \cdot \lambda}$. If we require the phase velocity $\beta_n$ of the nth travelling wave to equal the particle velocity $\beta$, we must have $L = \dfrac{n+1}{2} \cdot \beta \cdot \lambda$. This is possible for all values of n (n=0, 1, 2,..). *The choice of synchronous wave for the biperiodic coupled cavity structure is n=0, so we have* $L = \dfrac{\beta \cdot \lambda}{2}$ *and* $\beta_n = \pm \dfrac{\beta}{2n+1}$.

❑  *Specific properties of the* $\dfrac{\pi}{2}$ *mode for a biperiodic system and the side coupled (SCL) structure*

It has been proven that the $\pi/2$ normal mode of a chain of coupled oscillators has unique properties (see annex 3.1). In fact one can use this normal mode with a suitable geometry, satisfying a synchronous condition for the particles and resulting in good acceleration efficiency (high shunt impedance). To provide the synchronism in the $\pi/2$ normal mode in a periodic array of cavities, a $\dfrac{\beta \cdot \lambda}{2}$ length is fixed between sequential excited cavities (figure 3.7 left). So it provides two excited accelerating cells per $\beta \cdot \lambda$ (as in the $\pi$ mode structure) and includes two unexcited cells.

However this configuration does not generate an efficient accelerating structure. In fact the field is concentrated in only half of the cells for the $\pi/2$ mode, then the net power dissipation is larger than that for the $\pi$ mode, where all cells can produce the same energy gain. To solve this efficiency problem one can use a biperiodic chain, in which the accelerating cavities geometry are chosen to increase shunt impedance, and the unexcited cavities (called coupling cavities) have different shape and are tuned at the same uncoupled resonant frequency. The two most popular geometries have been 1) the on axis coupled structure [C.6, C.32], where the coupling cavities are reduced in length, and 2) the side coupled cavities, where the coupling cells are completely removed from the beam line. For both geometries the coupling cells are coupled magnetically to the accelerating cells.

The side coupled linac (SCL) is then a standing wave, with a <u>biperiodic structure</u> (consisting in accelerating and coupling cells) operating in the $\pi/2$ mode. For protons it usually operates at frequencies below 1 GHz, and is used for beam energies from 100 MeV on. Figure 3.7 (right) shows a cutaway drawing of a typical 800 MHz cavity of the side coupled type as used in the Los Alamos Meson Facility linac [C.31] developed and built in the 70s. LIBO-62 intends to use this kind of structure, operating for the first time at a frequency of 3 GHz in an energy range between 62 and 200 MeV.

While the perturbation analysis is helpful in showing the properties of general modes in a cavity chain, it is fundamental to illustrate the properties of the $\pi/2$ mode itself. A more restricted analysis is presented below.

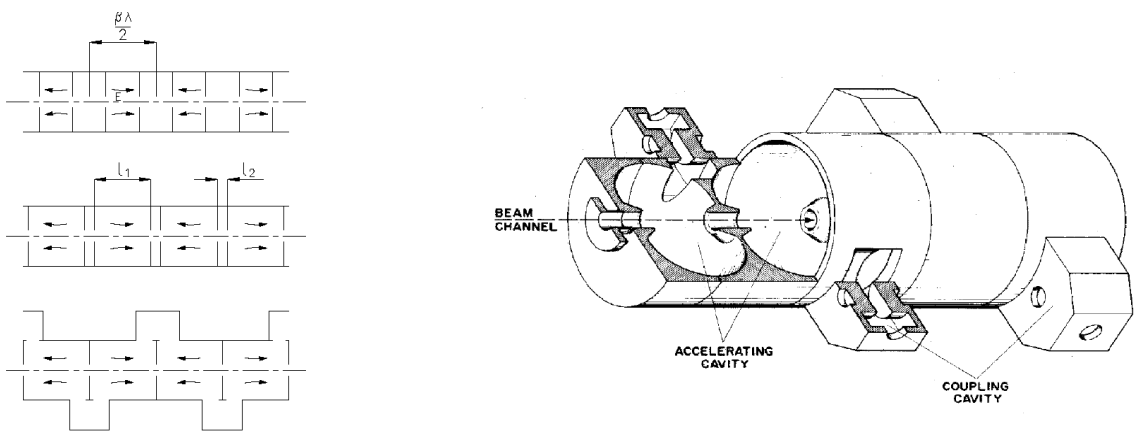

**Figure 3.7** _On the left is shown the biperiodic structure and the origin of the side coupled structure concept operating with $\pi/2$ normal mode. This kind of structure shows good stability and high efficiency. On the right the 800 MHz side coupled structure (SCL) typically designed for protons with $\beta > 0.4$._

For a biperiodic system with the stopband closed, and considering only the nearest neighbour coupling k between the accelerating (a) and coupling (c) cells (see annex 3.1 and chapter 5), the coupled resonator equations yield for the $\pi/2$ mode (f = $f_{\pi/2}$) the field levels $X_n$ in the cells as:

$$X_0 \cdot \left( \frac{1}{jQ_0} \right) + k \cdot X_1 = 0, \qquad X_{2n} \cdot \left( \frac{1}{jQ_{2n}} \right) + \frac{1}{2} \cdot k \cdot \left( X_{2n-1} + X_{2n+1} \right) = 0$$

We number the excited cavities (or accelerating cells) 2n, the unexcited cavities (or coupling cells) 2n+1, $Q_{2n}$ = $Q_a$, $Q_{2n+1}$ = $Q_c$. The total number of cells must be odd. Then, by assuming no drive in the region between 0 and 2m<2N and solving these equations stepwise, we find for n<m and an initial amplitude $X_0$ in cell 0:

$$X_{2n+1} = (-1)^n \cdot \frac{j(2n+1)}{k \cdot Q_a} \cdot X_0, \text{ and } X_{2n} = (-1)^n \cdot \left( 1 + \frac{2 \cdot n^2}{k^2 \cdot Q_a \cdot Q_c} \right) \cdot X_0$$

where third and higher order terms in $\frac{1}{k \cdot Q}$ have been neglected. These expressions give accurate predictions of the variation of fields in the coupling and accelerating cells in a biperiodic $\pi/2$ mode cavity chain. The coupling cell fields are orthogonal to the fields in the accelerating cells, have small amplitude and tilt linearly away from the drive. The above equations show also that losses do not produce any phase shift in the fields of the excited cavities down the chain (a result unique to the $\pi/2$ mode). While a small lossdependent amplitude change may occur, for most cases of interest (high Q values) it is negligible.

If frequency errors are present in the chain, then the general expression become very complex. The coupled resonator equations with frequency errors $\Delta f_{0n}$ in the cell n yield:

$$\left( -\frac{2 \cdot \Delta f_{0n}}{f_0} - \frac{j}{Q_n} \right) \cdot X_n + \frac{1}{2} \cdot k \cdot \left( X_{n-1} + X_{n+1} \right) = 0,$$

defining $\frac{4 \cdot \Delta f_{0n}}{k \cdot f_0} = \Delta_n$, and again solving stepwise one obtains the field levels for the coupling cells:

$$X_{2n+1} = X_0 \cdot (-1)^n \cdot \left[ \frac{1}{2} \cdot \Delta_0 + \Delta_2 + \dots\dots + \Delta_{2n} + \frac{j(2n+1)}{k \cdot Q_a} \right]$$

where third order terms have been neglected, and the field levels for the accelerating cells:

$$X_{2n} = X_0 \cdot (-1)^n \cdot \left[ 1 + \frac{2 \cdot n^2}{k^2 \cdot Q_a \cdot Q_c} + terms \quad \Delta_a \Delta_c, \frac{j\Delta}{k \cdot Q} \right]$$

Here again the results of the above equations show that to first order the fields in the excited cavities are independent of cavity errors. The second order terms generate a tolerance to errors for N cells of:

$$\left\langle \frac{\Delta X}{X} \right\rangle_{rms} \cong N \cdot \left\langle \Delta_a \right\rangle_{rms} \cdot \left\langle \Delta_c \right\rangle_{rms}$$

These results are then valid for a biperiodic structure operating in $\pi/2$ mode, so they will be extensively used for the design of the SCL LIBO structure (see chapter 5).

❑   *Main properties of the side coupled  structure (SCL)*

We have seen that the most obvious and efficient way to build a π/2 mode resonant accelerator structure is to operate with a side coupled structure. Figure 3.7 indicates placement of the unexcited cavities completely off the beam line to the side of the excited cavities. It has many technical advantages not present in the other schemes. Among these are the almost complete freedom to design the on line cavity to maximise efficiency. For the coupling may be accomplished through a small slot near the outer wall of the cavity; the beam aperture is not used for energy transfer. Also the accessibility of both the excited and unexcited cavity for cavity chain tuning and field measurement is excellent. We have seen that the SCL combines high efficiency and good electromagnetic stability. The cells in a coupled cavity linac are half as long as drift tube linac cells at the same velocity, so for the same diameter one has about half the inductance. The capacitance must be twice as large, which means the gap spacing is reduced. The smaller gap is positive due to the fact that it raises the transit time factor T. This can compensate another negative property: unlike drift tube linac, the cells in a coupled cavity linac have end walls, on which additional power is dissipated. The effective shunt impedance increases as function of beam energy. For low energy the wall power loss effect favours the more open drift tube linac structure, but at higher energy the transit time factor effect favours the coupled cavity linacs. Figure 3.8 shows the transition between these two regimes for the TOP linac designed for cancer therapy [B.12], and composed by sections of side coupled drift tubes and sections of side coupled linacs and where LIBO is the second part.

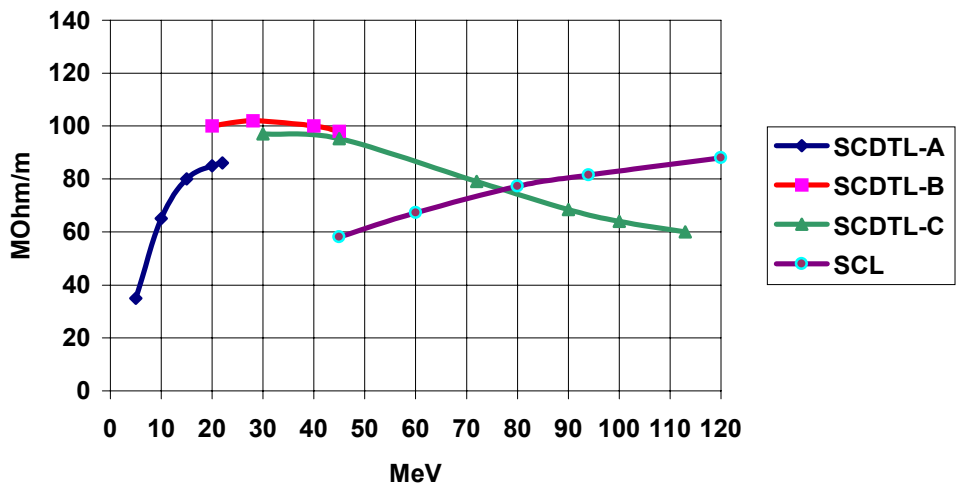

**Figure 3.8** *Example of shunt impedance values for a chain of different types of linear structures. The figure is refereed to the 3 GHz TOP linac for hadrontherapy (see chapter 2) designed by ISS in Rome. It is composed by different sections (A, B and C) of side coupled drift tube linacs (SCDTL) for energies ranging between 5 MeV and 70 MeV, followed by a side coupled linac (SCL) from 70 MeV up to 200 MeV. The figure shows how the efficiency of the two structures changes at different energy levels.*

## 3.3    Basics on beam dynamics of a proton linear accelerator

### 3.3.1    Electromagnetic field in a RF accelerating cavity and electric field in a gap

The general expression for the z-component (figure 3.9) of the electric field in a RF accelerating cavity takes the form [C.6]:

$$E_z(r,z,t) = \left[ \frac{A_0 \cdot J_0 \cdot (2\pi r / \lambda)}{2} + \sum_{m=1}^{\infty} A_m \cdot I_0 \cdot (K_m \cdot r) \cdot \cos \frac{\pi \cdot m \cdot z}{L} \right] \cdot \cos \omega t$$

and, on average, the particles will see the field of the synchronous wave, which is:

$$E_Z(r,z,t) = \frac{A_0}{2} \cdot I_0(K_0 r) \cdot \cos(\omega \cdot t - \frac{2\pi \cdot z}{\beta \cdot \lambda})$$

where the different terms are reported in [C.6]. The other field components can be written as follows. Deriving from Gauss's law in the absence of charge, one can obtain:

$$E_r(r,z,t) = \left[ \sum_{m=1}^{\infty} A_m \cdot \frac{m \cdot \pi}{K_m \cdot L} \cdot I_1(K_m r) \cdot \sin \frac{\pi \cdot m \cdot z}{L} \right] \cdot \cos \omega \cdot t$$

Deriving from Ampere's law for zero current, one can obtain:

$$B_\vartheta(r,z,t) = \left[ -\frac{\omega}{c^2} \cdot \sum_{m=1}^{\infty} \frac{A_m}{K_m} \cdot I_1(K_m r) \cdot \cos \frac{\pi \cdot m \cdot z}{L} \right] \cdot \sin \omega \cdot t$$

The remaining components can be shown to be zero. In fact $B_z = 0$ because we assume a TM solution. This fact together with requiring that the solutions is finite at r=0, proves that $B_r$ and $E_\vartheta$ are zero.

Starting with $E_z$, $E_r$, $B_\vartheta$ fields in the TM$_{010}$ mode and writing the equations of motion one can obtain the energy gain of the particles of charge q travelling over a distance L (cell length) [C.9]:

$$\Delta W = q \cdot E_0 \cdot T \cdot L \cdot \cos \phi \quad \text{with} \quad E_0 = \frac{1}{L} \cdot \int_{-L/2}^{L/2} E(0,z) \cdot dz$$

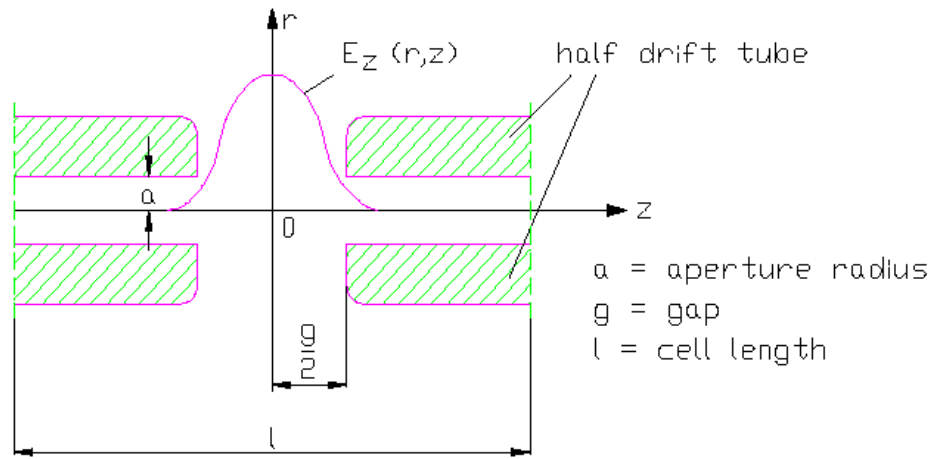

**Figure 3.9** _Electric field distribution across the accelerating gap. Z is the longitudinal direction of the beam, r is the radial (or transversal) direction. The transversal plane can be also represented with orthogonal co-ordinates x, y._

where $E_0$ is the spatial average of the accelerating field, calculated by the expression of the longitudinal electric field seen by a particle on axis passing through an accelerating gap [C.6]. The phase $\phi$ is 0 if the particle arrives at the origin when the field is at the crest. It is negative if the particle arrives earlier than the crest and positive if it arrives later. This phase convention differs from one normally used for circular accelerators, where the crest corresponds to 90°. Maximum acceleration occurs when $\phi = 0$, which is often used for relativistic electrons, but for non-relativistic protons, where the phase focusing is important, it is necessary to operate with negative phases. $E_0 \cdot L$ is the voltage gain experienced by a particle passing through a dc field equal to the field in the gap at time t=0. The particles in an accelerating gap experience radial forces due to the radial RF field components. In particular when the particles enter in the gap they see both longitudinal accelerating and radial focusing electric fields. Viceversa at the end of the gap the radial component becomes defocusing. For longitudinal stability $\phi_s$ must be negative, which means that the field is rising when the synchronous particle is injected. This creates higher defocusing than focusing field. In conclusion the RF gap produce a net radial defocusing that must be considered when the beam dynamics in the transversal plane is analysed (see paragraph 3.3.3).

### 3.3.2    Longitudinal dynamics

A proton linear accelerator is designed to produce a given energy gain per cell. Particle with correct initial velocity will continue to gain the right amount of energy to match the speed of the wave. For a field amplitude above a certain threshold, there will be two phases for which the velocity gain is equal to the design value, one earlier and the other later than the crest. One stable point at a certain phase exists and it is called synchronous phase. This aspect is due to the fact that particles near to this stable point arrive earlier than the synchronous phase and so they see a smaller accelerating field. Particles that arrive later will experience a larger field. This provides a longitudinal restoring force that keeps the particles oscillating about the stable phase, providing phase focusing or phase stability [C.1, C.2].

The particle with the correct initial velocity at the stable phase is called the synchronous particle, and it maintains exact synchronism with the RF wave. More precisely one can imagine a range for stability, within which the particles oscillate about the synchronous phase, at a frequency slower than the RF frequency of the electromagnetic field. Stable particles are then grouped in bunches, which on the average maintain synchronism with the wave and gain energy at the design rate. This stable region is called "bucket". Some particles are not captured in the buckets and they do not experience a net acceleration and fall behind in phase. These particles are lost radially before they arrive at the end of the linac.

We consider now an array of standing wave accelerating cells. We call the phase, energy and velocity of an arbitrary particle in the nth cell as $\phi_n, W_n, \beta_n$. For the nth cell we can also define a particle with synchronous phase $\phi_{sn}$, synchronous energy $W_{sn}$ and synchronous velocity $\beta_{sn}$. The particle phase is defined as the RF field phase when the particle is at the center of the nth cell (center of the gap), and the particle energy is the value at the end of the nth cell. From gap n-1 to gap n, the particle has a constant velocity $\beta_{n-1}$. The RF phase advance of the particle is:

$$\phi_n = \phi_{n-1} + \omega \cdot \frac{2 \cdot l_{n-1}}{\beta_{n-1} \cdot c}$$

where $l_{n-1} = \frac{1}{2} \cdot \beta_{s,n-1} \cdot \frac{\lambda}{2}$. For synchronism the cell length is $L_n = \frac{1}{2} \cdot (\beta_{s,n-1} + \beta_{s,n}) \cdot \frac{\lambda}{2}$, and for an arbitrary particle one can write $\phi_n = \phi_{n-1} + \pi \cdot \frac{\beta_{s,n-1}}{\beta_{n-1}}$.

The phase change of the particle through a cell relative to the synchronous particle is given by [C.6]:

$$\Delta(\phi - \phi_s)_n = \Delta\phi_n - \Delta\phi_{s,n} = \pi \cdot \beta_{s,n-1} \cdot (\frac{1}{\beta_{n-1}} - \frac{1}{\beta_{s,n-1}}) = -\frac{\pi \cdot (W_{n-1} - W_{s,n-1})}{m \cdot c^2 \cdot \gamma_{s,n-1}^3 \cdot \beta_{s,n-1}^2}$$

Remembering the energy gain for an arbitrary particle in the nth cell, the energy gain in cell n relative to the synchronous particle is:

$$\Delta(W - W_s)_n = q \cdot E_0 \cdot T \cdot L_n \cdot (\cos\phi_n - \cos\phi_{s,n})$$

To study the stability of the motion, it is convenient to convert the difference equations to differential equations. In this approximation we replace the discrete action of the standing wave field by a continuous field. Defining n as the cell number, one can treat the problem with a continuous variable. Using

$n = \dfrac{s}{\frac{1}{2} \cdot \beta_s \cdot \lambda}$ one can change variables from n to the axial distance s. The equation for the longitudinal

motion can be written as:

$$\gamma_s^3 \cdot \beta_s^3 \cdot \frac{d(\phi - \phi_s)}{ds^2} + 3 \cdot \gamma_s^2 \cdot \beta_s^2 \cdot \left[\frac{d}{ds}(\gamma_s \cdot \beta_s)\right] \cdot \frac{d(\phi - \phi_s)}{ds} + 2 \cdot \pi \cdot \frac{q \cdot E_0 \cdot T}{m \cdot c^2 \cdot \lambda} \cdot (\cos\phi - \cos\phi_s) = 0 .$$

Using the nomenclature and integrating the above equation (keeping $\beta_s$ constant), one can find:

$$W = \frac{W - W_s}{m \cdot c^2}, \quad A = \frac{2 \cdot \pi}{\beta_s^3 \cdot \gamma_s^3 \cdot \lambda}, \quad B = \frac{q \cdot E_0 \cdot T}{m \cdot c^2} \quad \Rightarrow \quad \frac{A \cdot W^2}{2} + B \cdot (sin\phi - \phi \cdot \cos\phi_s) = H_\phi$$

The quantity $H_\phi$ is a constant of integration and this result is a statement of energy conservation for the longitudinal motion. We identify $H_\phi$ as the Hamiltonian. The first term is the kinetic energy and the second is the potential energy. Figure 3.10.b shows the potential energy $U_\phi$ versus $\Delta\phi$ ($\phi$ - $\phi_s$); a potential well exists for particles around $\phi_s$. Particles within the potential well will perform stable oscillation about the synchronous particle. One can find a stability limit for the phase motion. The limiting stable trajectory is called the separatix. For small $|\phi_s|$ the stable phase region is $\cong 3 \cdot |\phi_s|$. At $\phi_s = -90°$, the phase acceptance is maximum, extending over the full 360° (no acceleration). At $\phi_s = 0$ the phase acceptance vanishes. In order to maintain synchronism, the linac is designed to provide an acceleration rate:

$$\frac{dW_s}{ds} = q \cdot (E_0 \cdot T)_{design} \cdot \cos\phi_s .$$

Finally we find the energy half-width of the separatix which occurs for $\phi = \phi_s$ [C.6, C.9]:

$$\frac{\Delta W_{max}}{m \cdot c^2} = \sqrt{\frac{2 \cdot q \cdot E_0 \cdot T \cdot \beta_s^3 \cdot \gamma_s^3 \cdot \lambda}{\pi \cdot m \cdot c^2} \cdot \left(\phi_s \cdot \cos\phi_s - sin\phi_s\right)}$$

If the stable region is defined and known, one can estimate the trapped efficiency between the injection and the first tank of a linac, as mentioned in chapter 4. Moreover, starting from the equation for small oscillations relative to the synchronous phase, one can write the period of the longitudinal oscillations as:

$$L_l = \frac{2 \cdot \pi}{\sqrt{\dfrac{2 \cdot \pi \cdot q \cdot E_0 \cdot T \cdot sin(-\phi_s)}{m \cdot c^2 \cdot \beta_s^3 \cdot \gamma_s^3 \cdot \lambda}}}$$

which can be compared with the RF period $\beta \cdot \lambda$. The longitudinal oscillation frequency is typically <10% of the RF frequency. At bigger energies the longitudinal oscillations approach zero.

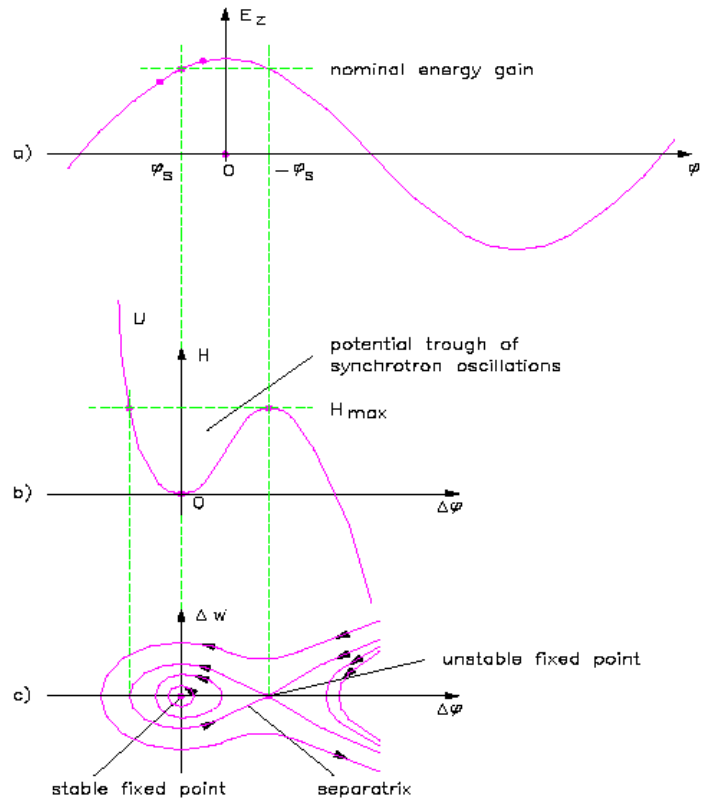

**Figure 3.10** *Figure a shows the accelerating field as function of phase. Figure b is the effective potential well, and figure c presents the phase-plane trajectories, including the separatix.*

### 3.3.3    Transverse dynamics

For a positive particle moving into the linac parallel to the beam axis, with velocity β and transverse co-ordinates x and y, the equations of motion with axial position s as the independent variable are [C.3, C.6]:

$$\frac{d^2 x}{ds^2} + K^2(s) \cdot x - \frac{k_{i0}^2}{2} \cdot x = 0, \qquad \frac{d^2 y}{ds^2} - K^2(s) \cdot y - \frac{k_{i0}^2}{2} \cdot y = 0$$

where $K^2(s) = \frac{q \cdot G(s)}{m \cdot c \cdot \gamma \cdot \beta}$ and $G = \frac{B_x}{y} = \frac{B_y}{x}$ is the quadrupole gradient. The effect of the RF defocusing in the gaps may be approximately included as a continuous force and it is represented by the term:

$$k_{i0}^2 = \frac{2 \cdot \pi \cdot q \cdot E_0 \cdot T \cdot sin(-\phi)}{m \cdot c^2 \cdot \lambda \cdot (\gamma \cdot \beta)^3}.$$

The quadrupoles are arranged in a regular lattice (the FODO array), which is periodic to allow the increasing cell lengths as the beam is accelerated. Because of the periodic force the equation of motion have the form of Hill's equation [C.10], for which the solutions are well known in the accelerator field[1]. The main results of these solutions are briefly mentioned in figure 3.11. The effect of the RF defocusing can be approximated by a thin-lens impulse at the center of the gap, where its focal length is given by:

$$\frac{1}{f_g} = \frac{\Delta x'}{x} = \frac{\pi \cdot q \cdot E_0 \cdot T \cdot sin(-\phi) \cdot L}{m \cdot c^2 \cdot \lambda \cdot (\gamma \cdot \beta)^3}$$

L is the length of the integrated RF pulse. The well-known relation gives the quadrupole focal length:

$$\frac{1}{f_Q} = \frac{\Delta x'}{x} = \pm \frac{q \cdot G \cdot l}{m \cdot c \cdot \gamma \cdot \beta}$$

where l is the length of the quadrupole.

To study the stability properties of Hill's equation, it is convenient to introduce a matrix solution, where each element is represented by a transfer matrix.

---

[1] The solution of the Hill's equation describes the particle motion, where a periodic focusing structure is present. This solution is written as $x(s) = \sqrt{\varepsilon_1 \cdot \tilde{\beta}(s)} \cdot cos(\phi(s) + \phi_1)$, where $\tilde{\beta}(s)$ and $\phi(s)$ are called the amplitude and phase functions, and $\varepsilon_1$ and $\phi_1$ are constants determined by the initial conditions. The solutions of Hill's equation are described by the Twiss parameters ($\tilde{\alpha}(s)$, $\tilde{\beta}(s)$ and $\tilde{\gamma}(s)$) [C.10]. The co-ordinates x and x' of the particles satisfy the equation $\tilde{\gamma}(s) \cdot x^2 + 2 \cdot \tilde{\alpha}(s) \cdot x \cdot x' + \tilde{\beta}(s) \cdot x'^2 = \varepsilon_1$, an ellipse centred at the origin of x-x' phase space with area $\pi\varepsilon_1$. The general ellipse in phase-space x, x' is shown in figure 3.11. A particle beam at any location in a focusing channel consists of a large sample of particle co-ordinates each on its own ellipse, each enclosing its own area, and each with its own initial phase. Each ellipse changes periodically as a function of s, and each area remains constant. The area of the ellipse is called emittance of the beam in the generic phase space x-x'. The beam is univocally determined by the phase spaces x-x' and y-y' for the transverse and ΔW-Δϕ for the longitudinal motion [C.10].

The matrices can be multiplied together in the proper order to obtain the total transfer matrix over a period:

$$P = F_{1/2} \cdot L \cdot G \cdot L \cdot D \cdot L \cdot G \cdot L \cdot F_{1/2}$$

where the focusing and defocusing quadrupoles are represented by F and D respectively, the RF gap by G and the drift space d between the end of the quadrupole and the accelerating gap by L. These matrices are:

$$F_{1/2} = \begin{bmatrix} \cos(\sqrt{K} \cdot l/2) & \dfrac{sin(\sqrt{K} \cdot l/2)}{\sqrt{K}} \\ -\sqrt{K} \cdot sin(\sqrt{K} \cdot l/2) & \cos(\sqrt{K} \cdot l/2) \end{bmatrix}$$

$$D = \begin{bmatrix} \cosh(\sqrt{|K|} \cdot l) & \dfrac{sinh(\sqrt{|K|} \cdot l)}{\sqrt{|K|}} \\ \sqrt{|K|} \cdot sinh(\sqrt{|K|} \cdot l) & \cosh(\sqrt{|K|} \cdot l) \end{bmatrix}$$

$$G = \begin{bmatrix} 1 & 0 \\ \dfrac{1}{f_g} & 1 \end{bmatrix}, \quad L = \begin{bmatrix} 1 & d \\ 0 & 1 \end{bmatrix}$$

From the smooth approximation, which averages over the rapid flutter associated with the individual focusing lenses, we obtain to lowest order an expression for the phase advance per period [C.6]:

$$\sigma_0^2 \cong \left( \frac{q \cdot G \cdot l \cdot L}{m \cdot c \cdot \gamma \cdot \beta} \right)^2 - \frac{\pi \cdot q \cdot E_0 \cdot T \cdot sin(-\phi) \cdot (2 \cdot L)^2}{m \cdot c^2 \cdot \lambda \cdot (\gamma \cdot \beta)^3}$$

The first is the quadrupole term, and the second is the RF defocus impulse. For stability the quadrupole term must be larger than the RF defocusing term. We see that the RF defocusing term falls off with $\beta \cdot \gamma$ faster than the quadrupole term, so that the RF defocusing is relatively more important at low velocities. In the smooth approximation the trajectory is sinusoidal, corresponding to simple harmonic motion, and the local effect of the individual lenses in the lattice is averaging out. The stability considerations [C.1] for a FODO structure foresee that the phase advance is limited to $\sigma_0 < \pi$. Finally there is a resonance between the radial and phase motion when the longitudinal oscillation frequency is twice the transverse oscillation frequency. The design of the accelerator must be performed to avoid this resonance.

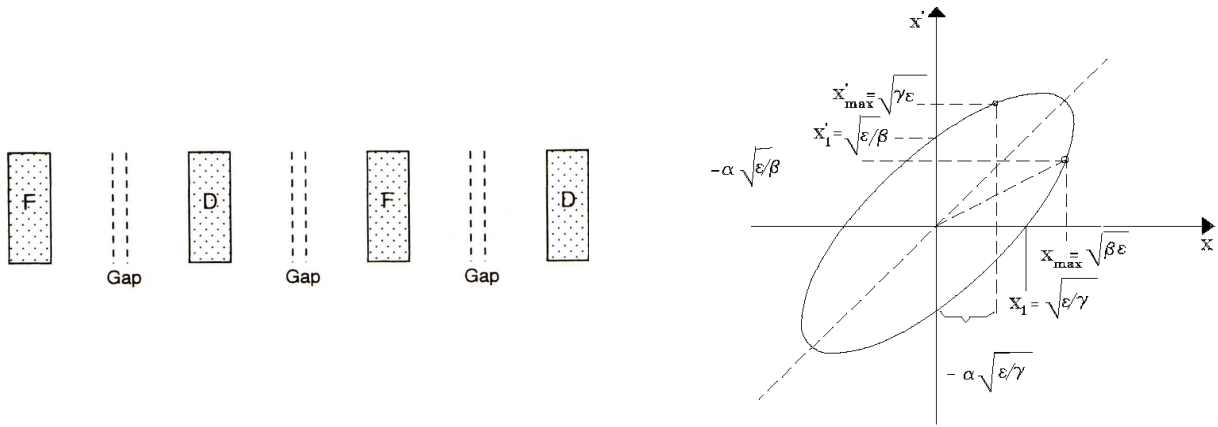

**_Figure 3.11_** _FODO quadrupole lattice with defocusing RF accelerating gaps in a linear accelerator. The figure shows the general trajectory ellipse and its parameters [C.1, C.2, C.6]._

*Chapter 4*

**The Side Coupled Linac (SCL) structure of LIBO-62 facility**

## 4.1     General description

LIBO is a standing wave Side Coupled Linac (SCL) where the accelerating cells operate on the axis in the $\pi$ mode and the coupling cells are placed off axis. In this configuration the coupling cells are unexcited, maintaining, as mentioned in chapter 3, the whole structure extremely stable with respect to mechanical and tuning errors. With this configuration (accelerating plus coupling cells) the structure operates in the $\pi/2$ mode, where the effect of adjacent perturbing modes is minimised.

The stability of the accelerating fields is especially important in proton structure applications, where, due to the change of $\beta$, the maintenance of phase stable operation requires rigid tolerance limits on field phase and amplitude.

The SCL of LIBO-62 facility is a structure capable to increase the proton energy from 62 MeV to 200 MeV. It consists of nine separate modules. The accelerating structure of each module is composed by three basic elements (see figures 4.1 a and b), all in highly pure copper: the "half-cell-plate", the bridge coupler and the end half-cell. All the above elements are brazed together and connected to vacuum pumps and cooling system to guarantee the thermal stability.

As shown, the accelerating and coupling cells are grouped to form an accelerating tank; each tank contains 13 accelerating cells and 12 coupling cells. Permanent magnetic quadrupoles (PMQs) are placed in the bridge couplers and end half-cells, forming a typical FODO structure, in order to focus the beam. Four tanks form then a module, fed through the central bridge coupler by a klystron. The electromagnetic energy passes from one accelerating cell to the next via a coupling slot and from tank to tank via the bridge coupler.

This system provides then an extremely efficient transfer of electromagnetic energy into particle energy in the velocity range from $\beta=0.3$ to 1.

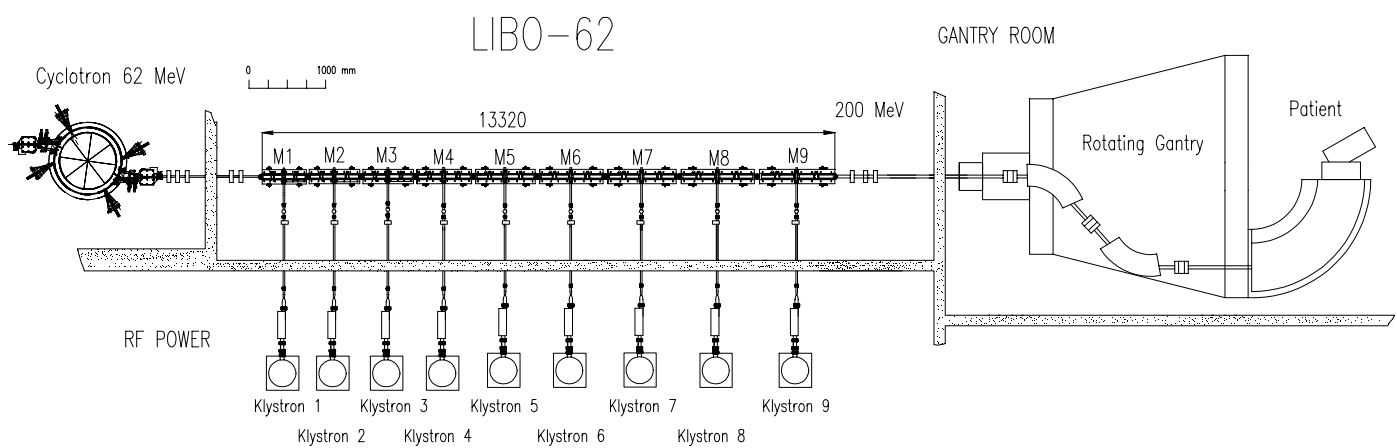

***Figure 4.1.a*** *Scheme of full LIBO-62 facility [A.4, A.13].*

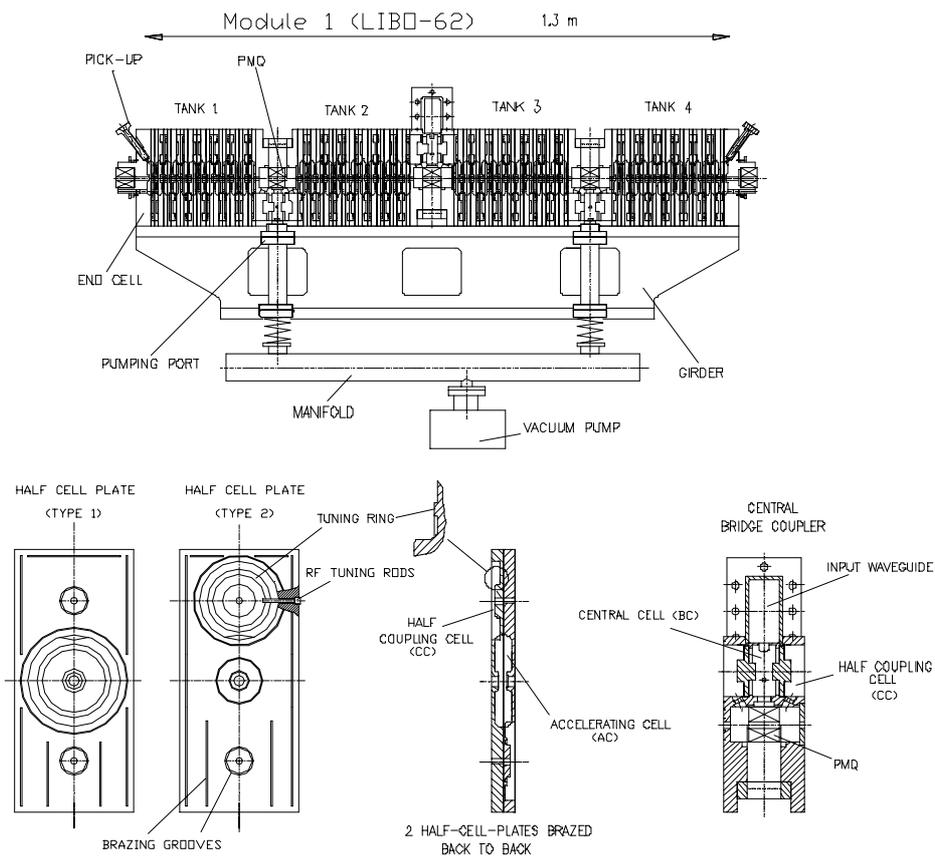

***Figure 4.1.b*** *The first module of LIBO-62. It is about 1.3 m long and composed by 102 half-cell plates (grouped in four accelerating tanks), three bridge couplers, and two end half-cells. The resonant cavities can be tuned at the correct frequency, changing the cell volume by machining of rings or by tuning rods insertion, as shown in figure.*

Table 4.1 shows the main parameters of LIBO-62 facility, which uses the Clatterbridge cyclotron as injector, as mentioned in chapter 2 [A.1]. The configuration shown in figures is a compromise between the needs to reduce the number of klystrons and the possibility to vary the final energy of the accelerator between 100 and 200 MeV for the therapy of deep-seated tumours. In this chapter the main design aspects of a full LIBO-62 will be presented in brief, and they will be used to extrapolate important considerations on the design and construction of LIBO prototype in chapters 5 and 6.

**Table 4.1** *Main parameters of LIBO-62 facility.*

| *Clatterbridge Cyclotron Specification* | |
|---|---|
| Proton energy [MeV] | 62.0 |
| Proton energy resolution (FWHM [%] | < 0.2 |
| Output energy spread [%] | 0.25 |
| Internal beam current [μA] | < 100 |
| External beam current [μA] | 75 |
| Transverse emittances (normalised) [mm mrad] | ~ 3.65 π |
| Cyclotron RF frequency (first harmonic) [MHz] | 25.1 |

| *LIBO (SCL) Specification* | |
|---|---|
| Operating frequency [MHz] | 2998 |
| Input energy [MeV] | 62 |
| Maximum output energy [MeV] | 200 |
| Aperture radius [mm] | 4 |
| Number of cells/tank | 13 |
| Number of tanks/module | 4 |
| Number of modules | 9 |
| Total number of tanks | 36 |
| Number of klystrons | 9 |
| Number of PMQs | 36 |
| Quad gradient [T/m] | 160 |
| Total length [m] | 13.32 |
| Synchronous phase angle [degree] | - 19 |
| Peak RF power [MW] | 32.8 |
| RF duty cycle [%] | 0.2 |
| Beam duty cycle [%] | 0.18 |
| Repetition rate [Hz] | 400 |
| Transverse acceptance (normalised) [ mm mrad] | 4.2 π |
| Trapped cyclotron beam [%] | 9.6 |

## 4.2    Project criteria

### 4.2.1    Preliminary design

The design of a linear accelerator is an iterative process, as the optimisation depends on several parameters, which are interrelated. First of all one must start from the physical specifications of the accelerator dictated by the clinical constraints for a cancer therapy. Before to start the design in detail, one has to define the accelerator itself in general terms. The following preliminary considerations are of interest in this sense.

❑  *Choice of operating RF frequency*

The 3 GHz frequency has been chosen for the following reasons.

-   The technology is well developed and used in the majority of the conventional electron linacs around the world.

-   At high frequency the accelerator is small and this aspect is compatible with the specification of a compact accelerator for medical applications.

- The effective shunt impedance $Z_{seff}$ is proportional to $f^{1/2}$ [C.6]: high frequency implies less RF power;
- The breakdown limit is higher, because the breakdown field, according to Kilpatrick's law, increases approximately as $f^{0.4}$.

❑ *Approximate linac length estimation*

To calculate the length of LIBO one can use, at the first stage, the simple formula [C.11]:

$$L = \frac{W_{fin} - W_{in}}{q \cdot E_0 \cdot T \cdot \cos\phi_s}$$

where q is the charge of the proton, $E_0 \cdot T$ is the effective accelerating field, $\phi_s$ is the synchronous phase, and L is the approximate accelerator length.

Selecting a reference value of $E_0 = 15.4$ MV/m, an indicative transit time factor of 0.86 and a synchronous phase $\phi_s$ of $-19°$, for a total energy gain of 138 MeV (62 MeV – 200 MeV), the approximate active accelerator length is about 12 m (not including the inter-tank spaces).

❑ *Approximate RF  power estimation*

The peak RF power required to feed the accelerator is calculated applying the following formula:

$$P_{RF} = \frac{(E_0 \cdot T)^2}{Z_{seff}} \cdot L$$

where $Z_{seff}$ is the effective shunt impedance per unit length in MΩ/m. The additional power which is delivered to the beam being accelerated is neglected due to the very small beam current required for cancer therapy.

The average RF power is obtained by multiplying $P_{RF}$ by the RF duty factor (repetition rate * RF pulse length). Consequently using the above relation one can have a rough estimation of the RF peak power needed to feed the accelerator of about 35 MW. The nominal average RF power is then about 70 kW (for 400 Hz and 5 μsec).

❑ *Energy spread of the proton beam at the exit of LIBO*

According to the physical specification, the radiotherapy requires a distal fall-off ΔR due to momentum spread in the beam dp limited to 2 mm.

The range R in the tissue (water equivalent) at beam energies between 100 and 200 MeV is given by the approximate formula [A.1]:

$$R \approx 100 \cdot (\beta \cdot \gamma)^{3.3} [cm] \quad \rightarrow \quad \frac{\Delta R_{beam}}{R} = 1.65 \cdot \left(\frac{\Delta W}{W_{out}}\right).$$

Reaching 25 cm in the tissue at 200 MeV ($W_{out}$), one allows energy spread ΔW ≤ 1 MeV as a physical requirement.

From the beam dynamics one has[*]:

$$\frac{\Delta W_{out}}{\Delta W_{in}} = \left( \frac{(\beta \cdot \gamma)_{out}}{(\beta \cdot \gamma)_{in}} \right)^{3/4}$$

and assuming the following approximate input values:

$(\beta \cdot \gamma)_{200} = 0.68$,  $(\beta \cdot \gamma)_{60} = 0.37$,  $\Delta W_{in} = 200 keV$ , one can estimate the energy spread of the accelerator as $\Delta W_{out} \cong 0.37$ MeV, in accordance with the physical requirements (chapter 2).

❑  *Main physical parameters*

In order to simplify the design and consequently the construction process, it has been established to maintain the length of the accelerating cells constant over the entire tank. This length increases proportionally to the velocity of the particles and changes from tank to tank. The length of each accelerating cell is $\frac{\beta \cdot \lambda}{2}$ (where λ is the wavelength calculated at 3 GHz and β is the relativistic factor) and the distance between the tanks must be equal to $\frac{(2n+1) \cdot \beta \cdot \lambda}{2}$ maintaining the synchronism of the beam with the RF accelerating field. In principle the distance between two consecutive modules could be fixed arbitrary due to the fact that the phases can be adjusted via phase shifter, although a particular care must be taken for the stability in the transverse plane. However for the design of LIBO-62 it has been decided to have a reasonable value of the inter-tank, as well as of the inter-module, fixed to $\frac{5 \cdot \beta \cdot \lambda}{2}$, enough for the bridge coupler and end cell construction constraints.

At 3 GHz the wavelength λ is about 10 cm, while the relativistic factor β is calculated by the well-known relations:

$$\beta = \sqrt{1 - \frac{1}{\gamma^2}} \qquad \text{and} \qquad \gamma = \frac{W + m \cdot c^2}{m \cdot c^2}$$

where W is the proton beam energy.

---

[*] If the acceleration rate is small, the parameters of the phase-space ellipses for small oscillations vary slowly. The area of the ellipse describing the small-amplitude oscillations is an adiabatic invariant. The constant area of the ellipse for any particle is:

$$area = \pi \cdot \Delta\phi_0 \cdot \Delta W_0 = \pi \cdot \Delta\phi_0^2 \cdot \sqrt{q \cdot E_0 \cdot T \cdot m \cdot c^2 \cdot \beta_s^3 \cdot \gamma_s^3 \cdot \lambda \cdot sin(-\phi_s)/2\pi}$$

If the accelerating field and the synchronous phase are fixed, we have:

$$\Delta\phi_0 = \frac{const}{(\beta_s \cdot \gamma_s)^{3/4}}$$

and since the phase-space area is an adiabatic invariant, we conclude from the above equations that:

$$\Delta W_0 = const \cdot (\beta_s \cdot \gamma_s)^{3/4}$$

This result describes a decrease of the phase amplitude and an increase in the energy amplitude of phase oscillations during acceleration in a linac (called phase damping).

### 4.2.2    Brief considerations on Radio Frequency and beam dynamics

❑  *Choice of basic LIBO design parameters*

The linac design procedure starts from the selection of few fundamental parameters, connected by the physical constraints for cancer therapy. These parameters are: the average accelerating field on axis $E_0$; the focusing gradient G of the quadrupoles; the synchronous phase $\phi_s$; the transverse phase shift per period $\sigma_0$ (betatron oscillations), the number of cells per tank $n_c$.

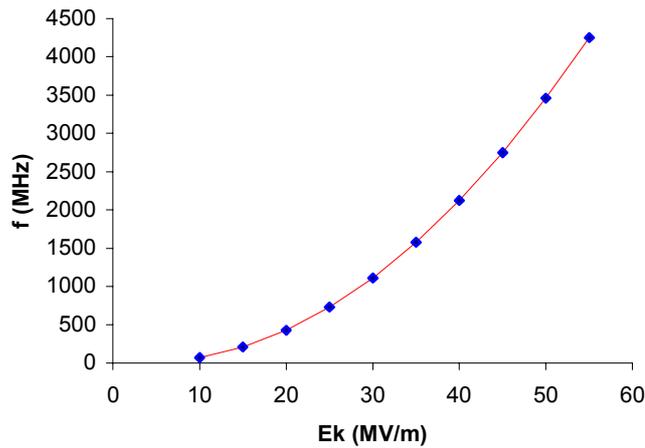

**Figure 4.2** *Kilpatrick limit as function of frequency. The Kilpatrick criterion is used in cavity design for RF electric breakdown analysis [C.13].*

One of the most critical parameter is the choice of the average accelerating field in the tanks, $E_0$.

Due to the fact that linacs are a single pass device, the accelerator length and the ohmic power consumption in the cavity walls become a critical issue. To shorten the accelerator, it is necessary to raise the longitudinal electric field.

The peak surface electric field and magnetic field are important in cavity design because very high peak surface electric field can bring electric breakdown. The Kilpatrick criterion is often used as a design guideline for room temperature cavity using the  frequency dependent "Kilpatrick" formula [C.1, C.13]:

$$f(MHz) = 1,64 \cdot E_k^2 \cdot e^{-8,5/E_k}$$

where $E_k$ is called Kilpatrick field (MV/m) and f is the RF frequency of resonant cavities. Inverting the formula, one finds that $E_k \propto f^{0,4}$ for frequency range which interests us. The peak surface electric field can exceed $E_k$ if particular care is taken for a clean vacuum system and short pulses (<1 msec). A high $E_0$ makes the accelerator short (interesting for a hospital), but the required RF power increases, being proportional to $E_0^2$. Moreover $E_0$ can not be too big due to the fact that the defocusing components of the RF field is proportional to $E_0$. The Kilpatrick breakdown limit $E_k$ is, at 3 GHz, 47 MV/m (figure 4.2), and as this limit is very conservative, one usually allows for the maximum surface field $E_s \leq 2 E_k$. A reasonable compromise is to select for the first tentative analysis an average axial accelerating field gradient $E_0 = 15.4$ MV/m.

Concerning the quadrupole gradient, a maximum value of 160 T/m has been chosen, that can be obtained by using permanent quadrupoles with an external diameter of 40 mm and a beam hole radius of 5 mm.

Remembering the considerations mentioned in chapter 3 on beam dynamics, the synchronous phase $\phi_s$ is linked to the energy gain per cell ($E_0 \cdot T \cdot L \cdot \cos\phi_s$) and the longitudinal acceptance (that for small oscillations is $\approx 3\phi_s$). A compromise between a high energy gain and a large acceptance is $\phi_s$ = -19°. The choice of $\sigma_0$ fixes the maximum value of the Twiss coefficient $\tilde{\beta}_{max}$ (the transversal acceptance is proportional to $1/\tilde{\beta}_{max}$) and then the gradient of quadrupoles. For a FODO structure (not including the defocusing RF) this value shows a large minimum for $\sigma_0$ = 76.3° [B.12]. The beam envelope is $\sqrt{\varepsilon_x \cdot \tilde{\beta}_x}$ in the horizontal plane and $\sqrt{\varepsilon_y \cdot \tilde{\beta}_y}$ in the vertical plane, and the non-normalised emittance $\varepsilon_{x,y}$ decreases as function of $\dfrac{1}{\beta \cdot \gamma}$. For the chosen quadrupole gradient of 160 T/m, $\sigma_0$ varies between 55° and 58°.

The choice of $n_c$ (number of cells in a tank), which determinates the length of the single tank, must be compatible with the quadrupoles focusing. A good compromise has been obtained using 13 full accelerating cells in each tank.

❑  *Design procedures*

Once the above parameters have been fixed, one can start to design the linac. The basic design procedure can be summarised as follows [A.1, C.11].

i)    A structure program as SUPERFISH [C.15] determines the cell geometry corresponding to the frequency of 3 GHz and to different particle velocities. The program also computes the field distribution in the cell, the transit-time factor T, the wall losses, etc (see annex 4.1). Output parameters of SUPERFISH are widely used for the engineering design of the cavity, such as the geometrical definition and the thermal behaviour of the structure under full RF power.

ii)   The cells computed by SUPERFISH are grouped in tanks and analysed by the program DESIGN. The program computes the energy gain of each tank, and determines the required focusing strength of magnetic quadrupoles.

iii)  The last step is made with a beam simulation program, called LINAC, which simulates the beam by a few thousand particles and follows them through the linac.

### 4.2.3    RF cavity design of a SCL structure

The cavity design is performed by using SUPERFISH computer program that solves numerically Maxwell's equations with specific boundary conditions. The geometry of the cells of the SCL is dictated by the following considerations. The main task is to maximise the effective shunt impedance of a single cavity. Starting with a simple pillbox cavity for reference, nose cones centred on the beam axis are added in order to reduce the gap and raise the transit time factor. The nose creates also a region of more localised electric field concentrated near the beam axis and a region of magnetic field near the outer part of the cavity ($TM_{010}$ mode). The cavity optimisation is performed in steps. First, the gap and nose cones geometry can be chosen to maximise the $r_s/Q$ parameter in order to increase the energy gain per unit stored energy. Then the outer walls geometry is adjusted to maximise Q, thus minimising the power dissipation per unit stored energy. In principle the optimum leads to a spherical shape for the outer wall, which corresponds to the smallest surface area for a given volume. The aperture radius is usually chosen to satisfy the beam dynamics requirements for good transmission. It also can cause larger peak surface electric field. The effective shunt impedance of a linac depends greatly on the dimension of the beam hole. This aperture is a function of the beam emittance, of the margin for alignment tolerance, and of the distance between focusing elements. A small equatorial "flat" on the nose has been also foreseen in the design in order to reduce the peak surface electric field and to allow a cavity tuning during manufacture (at the end this tuning facility has not been used). Figure 4.3 shows for example the shape of the accelerating cells for two different values of β, corresponding to the input and output energy of LIBO.

The accelerating cells have been designed varying the ratio g/L (figure 4.4) maintaining constant the diameters in all 36 tanks, for simplicity of manufacture, and the resonance frequency at 2998 MHz.

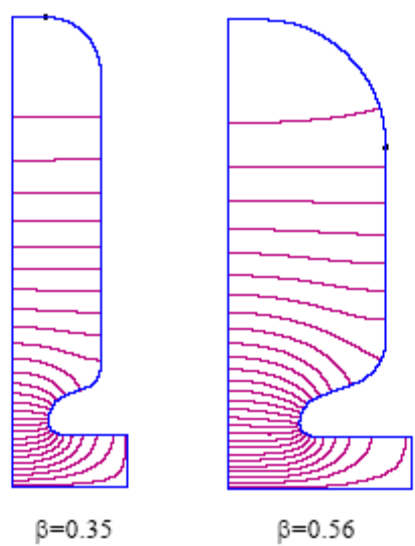

β=0.35          β=0.56

**Figure 4.3** *Shape of the accelerating cells of LIBO computed at input (62 MeV) and output (200 MeV) energies. It is evident that the accelerating electric field (in red) is concentrated on axis in the nose region [D.57].*

This choice is a compromise between the request of high shunt impedance and the geometrical configuration of the cavity. Bigger diameters generate bigger gaps and consequently lower transit time factors and shunt impedance values. Once these main parameters have been fixed, the cavity shape has been optimised in order to have the maximum efficiency compatible with a reasonable construction complexity. The hole diameter is inversely proportional to the structure efficiency and the maximum surface electric field $E_S$. An 8-mm hole is a good compromise between the needs to increase the structure and to reduce the electric field at the surface. The web thickness must be reduced as much as possible, to reduce the RF power dissipation in the cavity, maintaining at the same time a good mechanical stability. The angle $\alpha$ of the noses reduces the Kilpatrick factor and increases at the same time the shunt impedance. An optimised value has been found at 20°.

A general study of the accelerating cells has been performed to understand the RF properties useful for the design optimisation [J. Stovall and M. Weiss, A.4 and A.5]. The program SUPERFISH is used and is limited to rotationally symmetric structure. Neglecting at first the coupling slot, each cell type becomes rotationally symmetric and can be designed separately. Due to symmetry, only a quarter of the cell is analysed, saving thus mesh points, which are denser in region where a greater precision is required. The effect of cell diameter and web thickness on the shunt impedance ($ZT^2$), peak surface electric field ($E_{peak}$) and tuning sensitivity have been investigated. Figures 4.5 and 4.6 show that a 7 cm diameter is a good compromise between the shunt impedance $ZT^2$ (ranging between about 45 and 80 M$\Omega$/m for $\beta = 0.35$ and $\beta = 0.56$) and the peak surface electric field $E_{peak}$ (about 1.6 times the Kilpatrick limit). For the web thickness one would necessitate a relatively thick septum for mechanical stability, but this affects negatively the shunt impedance and increases the peak surface field ($E_{peak}$).

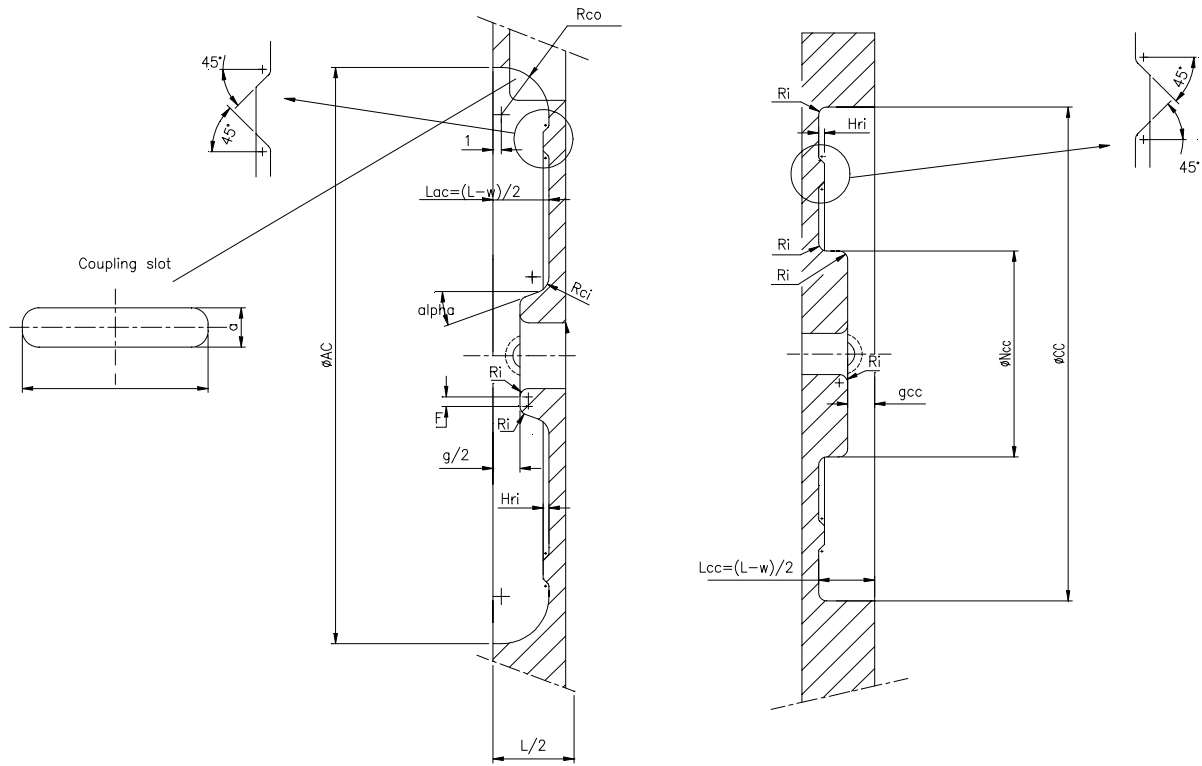

***Figure 4.4** Basic accelerating (left) and coupling (right) geometrical cell parameters of LIBO-62.*

These negative effects, which are most important at low β, are shown (for β = 0.35) in figure 4.5. In this figure the notation $ZT^2_{5\%}$ indicates a shunt impedance calculated considering also the effect of a 5% coupling slot between the accelerating and coupling cell. For the final design a more conservative coupling coefficient of 4.5% will be considered. Figure 4.7 shows the average power dissipated per cell, at a RF duty factor of 0.2%, as a function of beam energy. In these figures the degradation in shunt impedance that will result from coupling slots has been ignored. It is evident how the SCL becomes less efficient at lower energies, because at low energies the cells are very short and there is an important energy loss per unit length on the walls. Figure 4.8 shows peak surface electric fields as function of β: the conservative design assures a good safety margin for the breakdown limit. SUPERFISH program has also calculated the electric (E) and magnetic (H) field distribution in the accelerating cavity, as well as the frequency change with respect to a change of cavity volume. As expected the electric field is highly peaked on the inner nose where acceleration takes place.

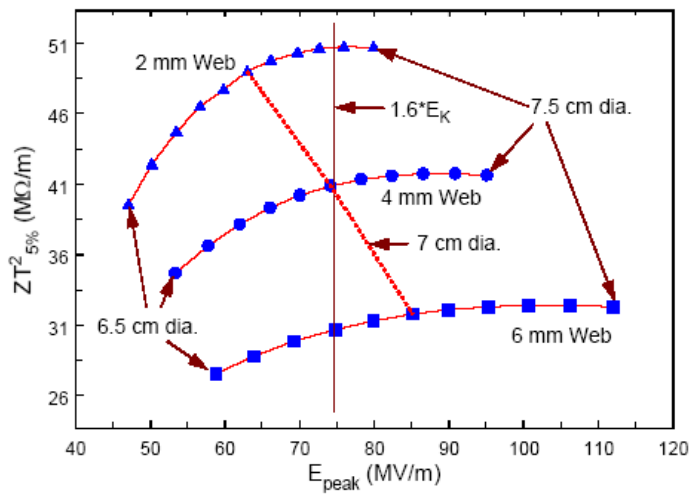

**Figure 4.5** *General RF results of LIBO cells (frequency: 2998 MHz, β: 0.35, diameter: 70 mm, average electric field on axis $E_0$: 15.4 MV/m). In figure is evident the relation between the shunt impedance and peak surface electric field in the first accelerating cells (β = 0.35) as a function of its diameter and web thickness.*

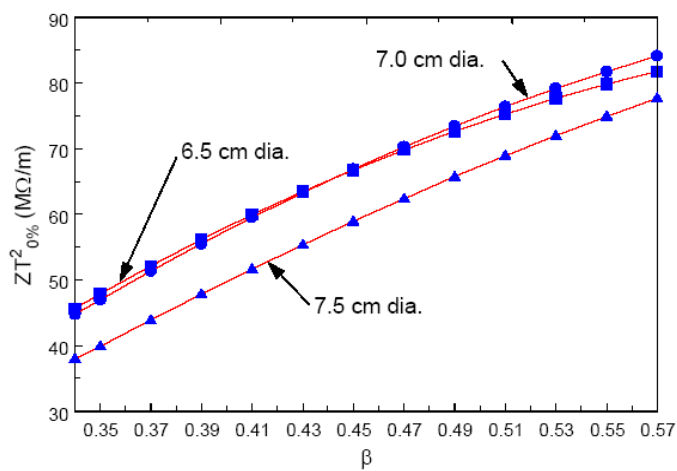

**Figure 4.6** *Shunt impedance as function of energy for three accelerating cell diameters.*

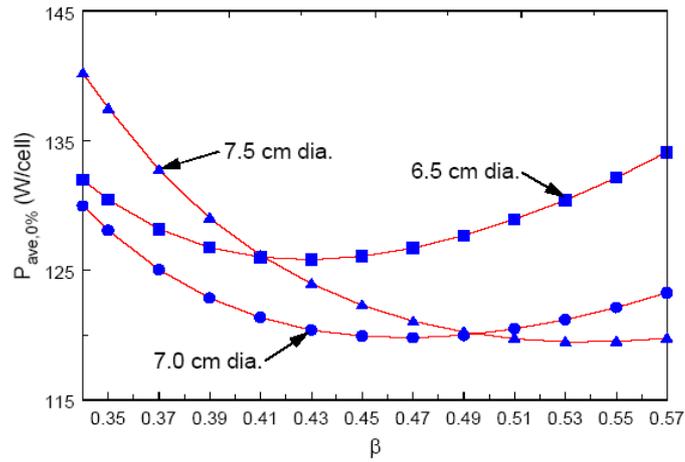

***Figure 4.7*** *Average power dissipated per cell as a function of proton energies for three accelerating cell diameters.*

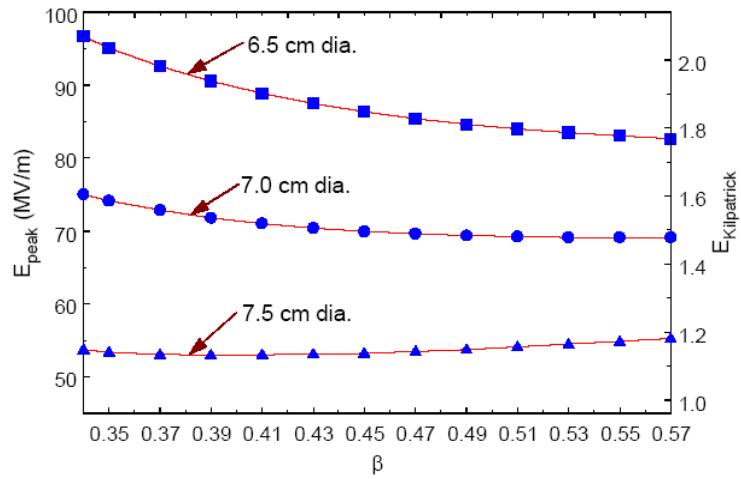

***Figure 4.8*** *Peak surface electric field as a function of proton energies for three accelerating cell diameters.*

#### 4.2.4    Basic design of LIBO-62

❑   _Injection efficiency from cyclotron to LIBO_

From the beam dynamics point of view, the injected beam coming from the cyclotron operates at lower frequency (25.7 MHz) than the linac frequency (3 GHz), so it is not necessary to establish a synchronism between the two RF structures.

Different aspects generate the main characteristic of LIBO current. First of all the cyclotron beam emittance is considerably larger than the LIBO emittance. One can estimate the normalised cyclotron beam emittance, the geometric emittance multiplied by the relativistic factors βγ, as 3.7 π mm mrad. The normalised transverse acceptance of the SCL is [A.1, A.4, A.12]:

$$A_n = \frac{r^2}{\tilde{\beta}} \cdot \beta \cdot \gamma$$

where r is the aperture radius (4 mm) for the cavity beam hole, $\tilde{\beta}$  the Twiss parameter (1.9 m) and βγ are the relativistic factors (βγ=0.365). With the above configuration the normalised acceptance of LIBO is about 4.2 π mm mrad. So with a beam hole radius of 4 mm the transverse acceptance of LIBO contains easily the cyclotron emittance. Longitudinally, each cyclotron beam bunch fills several LIBO buckets, but only part of the 25.7 MHz cyclotron beam falls into the 2.998 GHz LIBO buckets (see figure 4.9)[†]. Here the bucket length is about three times the synchronous phase angle ($\phi_s$). For simplicity no attempt was made to match the beam using a buncher.

❑   _Length of the accelerating tanks_

The basic accelerating unit of the SCL is the multicells tank [C.28]. Due to the fact that for "high" energy sections of proton linacs the velocity change in each tank is small, one can simplify the construction by designing each tank at the same constant cell length. The length of each cell is $\beta_s\lambda/2$, where $\beta_s$ is the velocity of the reference particle at the center of the tank. This construction changes the dynamics of the reference particle. In fact with constant length cells, the synchronous phase must be $\phi_s = -\pi/2$, which corresponds to no acceleration for the true synchronous particle and no increase in cell lengths. However the reference particle is injected at a specific phase near the crest of the wave. The reference particle injection phase is chosen so

---

[†] We have seen in chapter 3 that the equation of motion can be written in Hamiltonian form: the term $H_\phi$ is the energy invariant. The canonically conjugate variables are φ and $p_\phi= -w =(W - W_s)/m$ $c^2$, and the second order differential equation of motion is derived from the Hamilton's equations:

$$p_\phi' = -\frac{\partial H_\phi}{\partial \phi}, \qquad \phi' = \frac{\partial H_\phi}{\partial p_\phi}.$$

It can be proven that the area in the phase space occupied by the particles is constant for a system described by a Hamiltonian that is independent of time. This is the Liouville's theorem, and it is valid for systems with no dissipation that satisfy the continuity equation in phase-space. Normally the dissipation term is present only when there is acceleration. When the acceleration occurs slowly, the beam dynamics parameters change adiabatically. The phase-space area can be shown to be an adiabatic invariant, and therefore is still approximately constant. If the beam is accelerated slowly, the shape of the area enclosed by the particle trajectory changes but the area remains constant. This is a good approximation for the energy levels presented into LIBO.

that the phase of the reference particle advances to a maximum positive value in the first half of the tank and decreases back to its initial injection value in the second half. The total energy gain is the sum of the two equal contributions of the two halves of the tank. The energy gain of the particle in the tank is [C.6]:

$$\Delta W_r = q \cdot E_0 \cdot T \cdot \cos \phi_r \cdot N_c \cdot \beta_s \cdot \frac{\lambda}{2}$$

where $N_c$ is the number of the cells in the tank and $\phi_r$ is the mean value of the reference particle phase. The approximate maximum phase excursion of the reference particle through the tank, usually no more than a few degrees, is [C.6]:

$$\phi_0 - \phi_1 = \frac{\pi \cdot N_c}{8} \cdot \frac{\Delta W_r}{\gamma_s^3 \cdot \beta_s^2 \cdot m \cdot c^2}$$

The length of LIBO tanks is a compromise among several factors. For protons the requirement for focusing usually sets a limit on the tank length, due to the fact that focusing elements are foreseen at the end of each tank and the space between focusing elements are guided from the overall focusing strength. Another aspect to be considered in setting the tank length is that the power flow droop effect in the π/2 mode must not affect significally the electric field distribution.

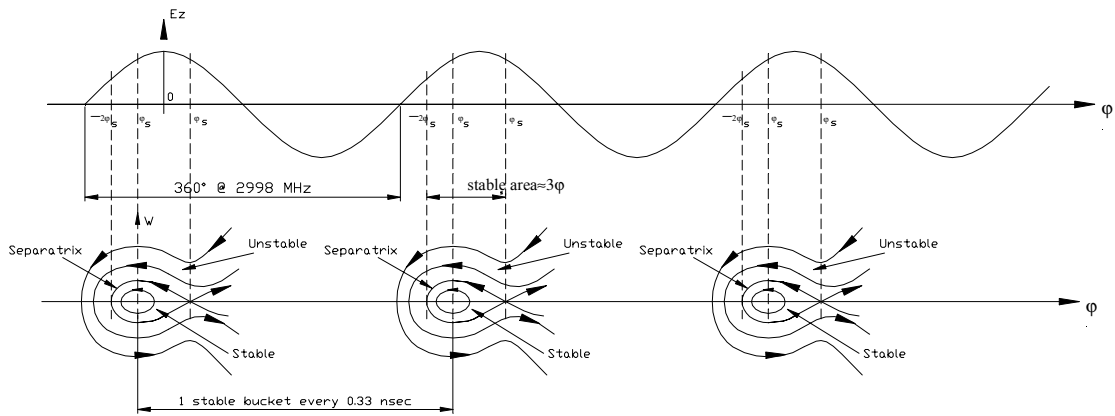

*Figure 4.9 Longitudinal phase plane: LIBO stable buckets.*

The DESIGN program is used to calculate the best configuration of the accelerating tanks. The main beam dynamics simulations are due to K. Crandall and M. Weiss that first have analysed the physics of proton beams for LIBO [A.4, A.12]. Table 4.2 gives some results of the linac design: the SCL accelerator is about 15 m long (including inter-tank distances), contains 36 tanks and consumes about 33 MW of peak RF power. The quadrupoles placed between the tanks (the inter-tank distance is 2.5 βλ) are 40 mm long and their gradients are maintained constant over the whole structure at the value of 160 T/m (the specific transverse betatron phase advance per focusing period (σ₀) ranges between about 55° at 62 MeV and 58° at 200 MeV) [A.1]. Transverse focusing is designed also to avoid synchro-betatron parametric resonances [D.50].

**Table 4.2** *Main physical parameters of LIBO-62.*

| Tank number | Tank length [mm] | Total length [mm] | Energy [MeV] | $Z_{seff}$ [MΩ/m] | Power/tank [MW] |
|---|---|---|---|---|---|
| 1 | 227.61 | 314.3 | 64.86 | 48.2 | 0.948 |
| 2 | 232.24 | 634.9 | 67.78 | 49.8 | 0.940 |
| 3 | 236.84 | 962.0 | 70.77 | 51.3 | 0.933 |
| 4 | 241.41 | 1295.4 | 73.82 | 52.8 | 0.927 |
| 5 | 245.94 | 1635.0 | 76.92 | 54.2 | 0.921 |
| 6 | 250.43 | 1980.9 | 80.09 | 55.6 | 0.917 |
| 7 | 254.89 | 2333.0 | 83.32 | 57.0 | 0.913 |
| 8 | 259.31 | 2691.2 | 86.61 | 58.3 | 0.910 |
| 9 | 263.69 | 3055.5 | 89.95 | 59.6 | 0.907 |
| 10 | 268.03 | 3425.8 | 93.36 | 60.9 | 0.905 |
| 11 | 272.34 | 3802.0 | 96.82 | 62.1 | 0.903 |
| 12 | 276.60 | 4184.2 | 100.34 | 63.3 | 0.901 |
| 13 | 280.83 | 4572.2 | 103.91 | 64.4 | 0.900 |
| 14 | 285.01 | 4966.1 | 107.54 | 65.5 | 0.899 |
| 15 | 289.15 | 5365.7 | 111.22 | 66.6 | 0.899 |
| 16 | 293.25 | 5770.9 | 114.96 | 67.6 | 0.899 |
| 17 | 297.31 | 6181.8 | 118.76 | 68.7 | 0.899 |
| 18 | 301.33 | 6598.3 | 122.60 | 69.6 | 0.899 |
| 19 | 305.31 | 7020.2 | 126.5 | 70.6 | 0.899 |
| 20 | 309.24 | 7447.7 | 130.45 | 71.5 | 0.900 |
| 21 | 313.13 | 7880.5 | 134.45 | 72.4 | 0.900 |
| 22 | 316.98 | 8318.7 | 138.50 | 73.3 | 0.901 |
| 23 | 320.79 | 8762.1 | 142.60 | 74.1 | 0.902 |
| 24 | 324.55 | 9210.8 | 146.75 | 74.9 | 0.904 |
| 25 | 328.28 | 9664.6 | 150.94 | 75.7 | 0.905 |
| 26 | 331.95 | 10123.5 | 155.19 | 76.4 | 0.906 |
| 27 | 335.59 | 10587.5 | 159.48 | 77.1 | 0.908 |
| 28 | 339.19 | 11056.5 | 163.81 | 77.8 | 0.909 |
| 29 | 342.74 | 11530.3 | 168.2 | 78.5 | 0.911 |
| 30 | 346.25 | 12009.4 | 172.62 | 79.2 | 0.913 |
| 31 | 349.72 | 12492.7 | 177.09 | 79.8 | 0.915 |
| 32 | 353.14 | 12981.0 | 181.60 | 80.4 | 0.916 |
| 33 | 356.52 | 13474.0 | 186.16 | 81.0 | 0.918 |
| 34 | 359.87 | 13971.6 | 190.76 | 81.5 | 0.920 |
| 35 | 363.17 | 14473.8 | 195.39 | 82.1 | 0.922 |
| 36 | 366.43 | 14980.6 | 200.07 | 82.6 | 0.924 |

❑  *Brief description of error analysis and final constraints for LIBO construction*

A Monte Carlo error analysis has been also performed with the program LINAC along the SCL [A.12], considering for this study the following combination of errors:

▪   quadrupole displacement errors ±0.1 mm;

▪   quadrupole gradient errors ±1%;

▪   quadrupole rotation errors ±1°;

▪   tank displacement errors ±0.1 mm.

With these constraints the beam loss along the structure and the output emittances have been computed. In particular several runs have been simulated, where, for each run, an independent random error selected in an uniform distribution has been considered. The analysis shows that the most critical error is on quadrupole displacements, but always inside a safety margin. For the above mentioned alignment tolerances and with a bore radius of 4 mm, there is a 90% probability for a transmission of the beam in the SCL bigger than 10%, and 50% probability for a transmission bigger then 11% (figure 4.10).

Concerning the electric field errors, the analysis shows that uniformity variation of electric field from cell to cell and from tank to tank must be less than 5%. With this constraint one can obtain the same transmission result mentioned above. The electric field distribution, one of the main critical parameters to be tested on the first prototype module, will be widely discussed in chapters 6 and 7.

With these considerations, one can conclude *that LIBO-62 accelerates about 10 % of the input beam for therapy*, the rest leaves with energies below 70 MeV. *To obtain an average LIBO current of 8-9 nA, necessary for cancer therapy, and with a beam duty cycle of 0.18%, the peak intensity in the pulsed cyclotron beam must be about 50 μA*. To avoid unnecessary irradiation to the copper structure, the cyclotron source will be pulsed at 400 Hz, so that the extracted beam comes in synchronism with the 5 μs LIBO pulses.

❑   *LIBO output beam*

LIBO-62 is a modular structure composed by nine separate modules, each fed by own klystron: this aspect permits to consider each module as a unique RF unit. According with protontherapy specifications, one should be able to change the beam energy between 100 and 200 MeV, with small energy spread, in order to reduce the distal dose fall-off to 2 mm or less.

With LIBO this energy variation could be achieved with two different adjustment methodologies: the gross variation and the fine variation. The first can be obtained switching off some RF modules, the second changing the power of the klystron of the last operating module.

After a certain number of simulations [M. Weiss and A.1], it has been seen that the output beam energy varies with the amplitude of the RF field in the last switched on module. Moreover is visible that the energy spread is constant in all energy variation ranges and fits very well the radiotherapy specifications. In any condition the beam will come out from the accelerator structure without losses. In this context the general scheme of RF power distribution shown in figure 4.12 can cover the above physical requirements.

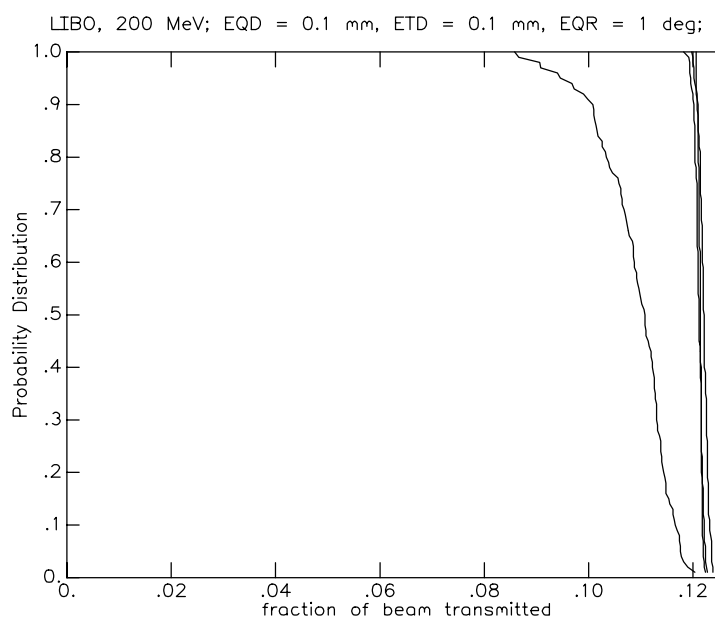

**Figure 4.10** *Transmission probability distribution of LIBO-62: the figure shows a 90% probability for a beam transmission >10% and a 50% probability for a beam transmission >11% [M. Weiss and A.1].*

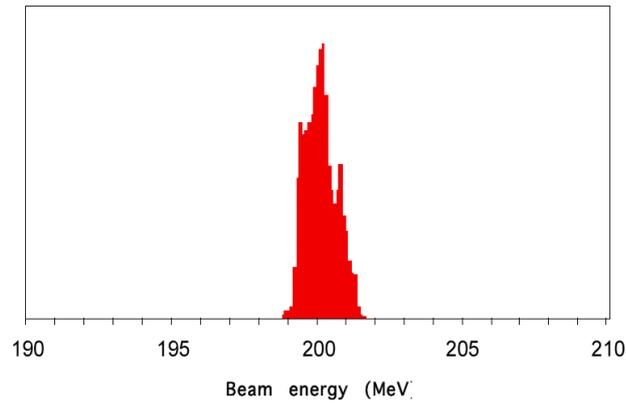

*Figure 4.11* Output energy distribution of LIBO-62 [A.2].

❑  *Main technical details on RF power system*

Different schemes have been looked at in order to provide pulsed RF power to the accelerating cavities. The klystrons operate with a frequency of 2998.5 MHz (the standard European S-band frequency). The RF pulse width measured at flat-top of the klystron output, and applied to each tank, is about 5 μs, while the nominal repetition rate of the system is 400 Hz.

Constraints on the design will be mainly: 1) capital cost investment for the purchase and installation, 2) spares holding costs for components with long delivery delays (klystrons), 3) required operational flexibility and availability in the medical environment, 4) integration into any existing controls system to avoid equipment duplication, 5) on-line and breakdown maintenance and adjustments by medical technicians, 6) yearly running costs including maintenance and technical staff training. System performance, installation cost and reliability are the principal design features that would determine the acceptability of one klystron-modulator scheme over another.

An extensive study for klystron-modulator system to be used for LIBO has been performed by TERA during the period 2002-2005 [P. Pearce]. Several solutions have been investigated, where a single high-power klystron-modulator can feed one or more modules in parallel or a single modulator can be fitted with one or two klystrons (figure 4.12).

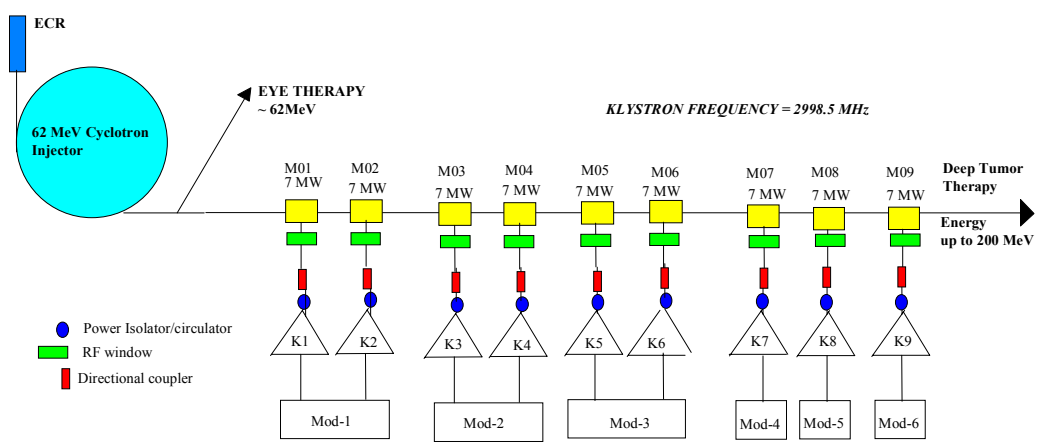

*Figure 4.12* Example of a possible Klystron distribution for LIBO-62 facility [P. Pearce].

In all cases the power has to be divided equally between the modules (for example by using a set of 3 dB power splitters and power attenuators). Also to be able to compensate for the change in phase along the linac, each module will be equipped with a phase shifter. Vacuum to $SF_6$ windows are used to separate the machine vacuum conditions from the waveguide environment, permitting the removal of a klystron without letting the linac up to atmospheric pressure and probable contamination.

The klystrons have a power isolator in their output circuit for protection from any high-power breakdowns in the modules or waveguide components. These breakdowns cause reflected power generation with risks for the klystron output window, so the loads of the protection isolator are there to absorb this reflected energy. Moreover studies have been performed in order to understand the use of the klystron-modulator system in a medical environment. The total stability of the peak klystron power over the useful pulse width will ultimately determine the proton beam stability in LIBO, and an optimal figure of ± 0.1 % seems to be necessary.

# Chapter 5

# Physics and engineering design

# of the first LIBO-62 prototype module

*"Quelli che s'innamoran di pratica senza scienza*
*son come l'nocchier ch'entra in navilio senza timone o bussola,*
*che mai ha certezza dove si vada"*
*[Leonardo Da Vinci ≈ 1500]*

## 5.1    Introduction

In 1998 TERA has promoted a collaboration, chaired by Dr. M. Weiss, with CERN, Universities and INFN sections of Milan and Naples, to realise and test a prototype of LIBO, which would then be considered as a "proof of principle". For this purpose the first LIBO-62 module has been logically chosen as test: it represents, as all modules, a basic unit with its own RF power supply. The full construction of a linear accelerator is a complex problem where several aspects must be considered, including beam dynamics, Radio Frequency physics, vacuum technology, material science, machining and brazing technology. In particular one should design the accelerating structure maintaining the following items:

- Controllability of electric filed levels and, as a consequence, the acceleration rate and particles motion.
- Efficiency in transferring RF power into particle energy.
- Freedom from electrical breakdown problems within the cavities.
- Mechanical tolerance requirements to obtain the needed physical performances.
- Possibility to operate at high vacuum levels.
- Low costs for fabrication, using as much as possible a standard technology.

Figure 5.1.a shows the engineering design of the prototype module, which contains 102 half-cells (grouped into four tanks of 25 [tank 1 and 4] or 26 [tank 2 and 3] plates), 3 bridge couplers and 2 end half-cells.

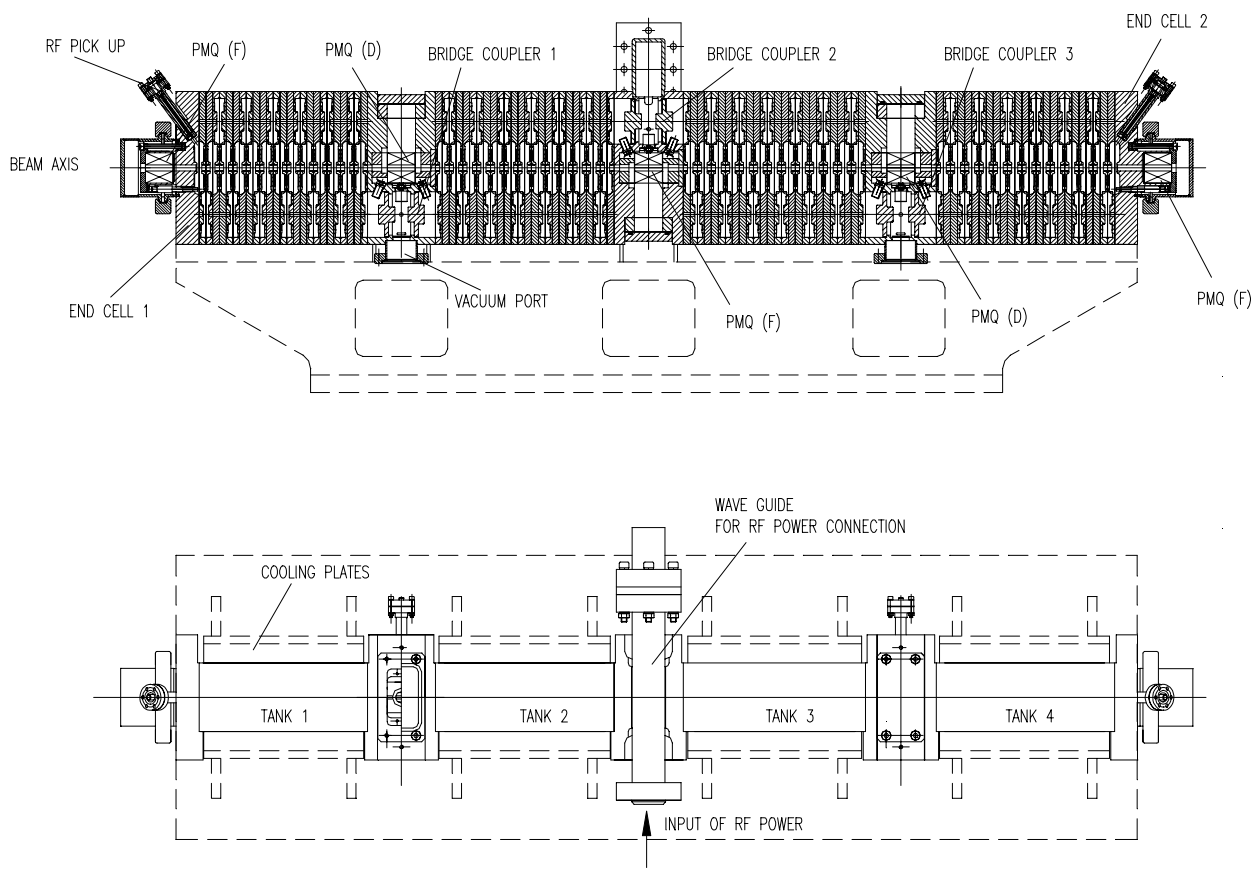

***Figure 5.1.a*** *LIBO prototype module: the engineering design [D.49].*

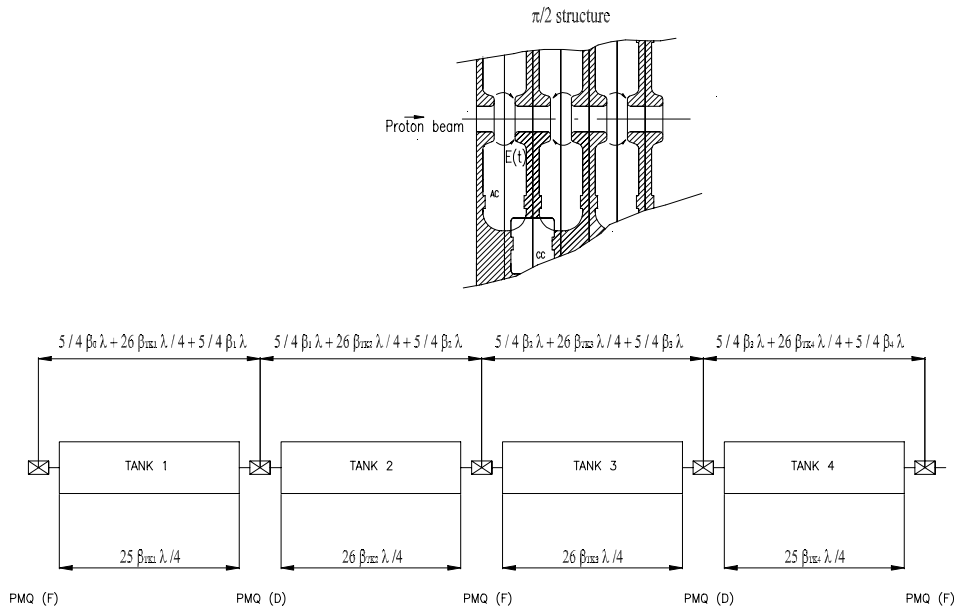

***Figure 5.1.b*** *LIBO prototype module. π/2 mode configuration of the accelerating electric field and the FODO system for beam focusing are shown.*

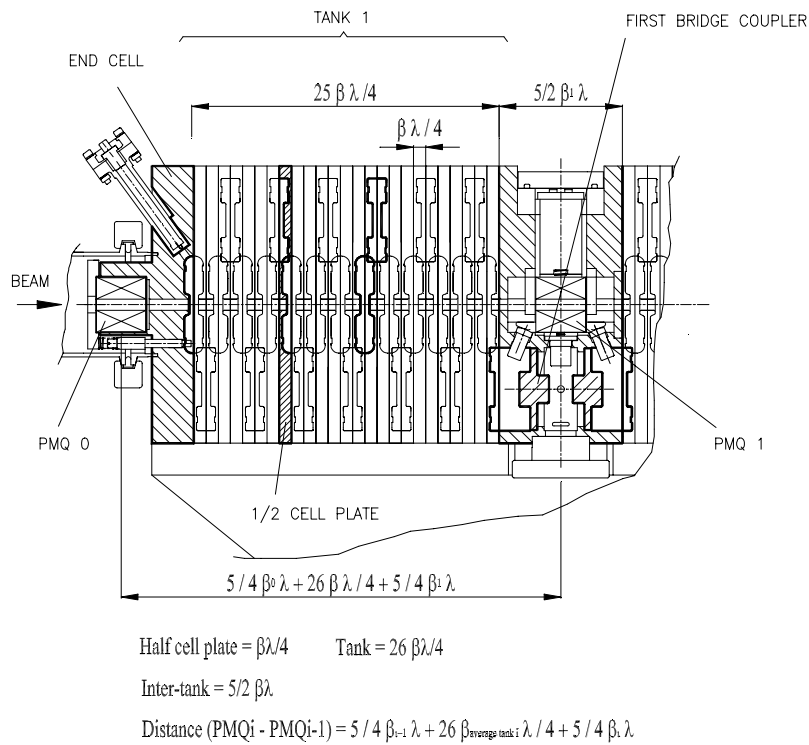

***Figure 5.1.c*** *LIBO prototype module. The accelerating tank and its physical parameters are shown.*

All single elements are machined separately and then brazed together in vacuum at high temperature to form an accelerating module. Connections in stainless steel to the beam line at both ends of the module and to the vacuum manifold are available while cooling plates are brazed on the lateral faces of tanks. A permanent magnet quadruple (PMQ) for beam focusing, a pick-up loop for RF field measurements, and tuners for frequency corrections are housed in each bridge coupler and end half-cell after the module has been brazed. The finished copper structure is mounted on a rigid girder, onto which an external reference is fixed for alignment purposes.

## 5.2    Materials selection

❑    _General considerations_

The choice of the materials for the prototype construction must take into account the following constraints:

- compatibility with vacuum levels of about $10^{-6}$ mbar and high electric fields (very low impurities levels),

- compatibility with the brazing alloys and the high temperature thermal cycles,

- high electric (IACS 101%) and thermal conductivity,

- grain size as small as possible (brazing heat treatment increases grain growth),

- stability of the material when machining and brazing.

One of the main problems for a correct material selection is due to the different steps of high temperature treatment used for brazing and stress relieves. These treatments change drastically the crystallographic structure and consequently the mechanical properties of the materials, and they must be taken into account during the design process. An extensive analysis of these problems will be presented in chapter 6.

❑    _Summary of materials adopted for the prototype_

The following materials have been adopted for the construction of the LIBO-62 prototype:

- _Half-cell plate_: OFHC copper, laminated in one preferential direction. Pre-machining of the half-cells with 0.2 mm of over dimensions, followed by heat treatment for stress relieve (2h x 230°C in air).

- _Bridge coupler and end cell_: OFHC copper, forged (in three directions). No heat treatment is applied.

- _Cooling plates:_ OFHC copper plates with stainless steel (316LN) pipes brazed.

- _Flanges and inserts for fixation brazed on the copper structure_: stainless steel (316 LN).

- _Other stainless steel pieces:_ stainless steel (304 L).

- _RF pick-ups:_ copper beryllium, stainless steel (316 LN), OFHC copper.

- _Tuners:_ ETP copper.

**Table 5.1** _Material selection for the prototype._

| | |
|---|---|
| _OFHC copper_ | ASTM C 10100, High conductivity, Cu 99.99%, $O_2$ 0.0005% 85-115 HV, $\sigma_r$: 280-330 N/mm$^2$ , Δl (A5): 8% min Conductivity: 101% IACS |
| _ETP copper_ | Cu 99.9%, $O_2$ 0,005-0.04% 65 HV10 min, $\sigma_r$: 250 N/mm$^2$, Δl (A5): 15% Conductivity: 98% IACS min, 57 m/Ω mm$^2$ |
| _Stainless steel 316 LN_ | Cr Ni 18.12 Mo N, 190 HB, $\sigma_r$: 600-800 N/mm$^2$, Δl (A5): 35% |
| _Stainless steel 304 L_ | Cr Ni 18.10 180 HB, $\sigma_p$: 175 N/mm$^2$, Δl (A5): 50% |
| _Copper beryllium_ | CuBe$_2$, Cu 98%, Be 2% 100 HV10 min, $\sigma_r$: 450 N/mm$^2$, Δl (A5): 35% |

### 5.3    Half-cell plates

❑    *General considerations*

The half-cells of the prototype have been conceived as a rectangular copper plates where two types of half-cells are present (forming a biperiodic structure): the accelerating (AC) or "a" and the coupling (CC) or "c" cells (figure 5.3 and 5.4). These are generated by machining on both sides and grouped together to form a sequence of full accelerating and coupling cells, called tanks. The form of the half-cells does not change along the structure, except for the thickness, which increases from tank to tank proportionally to the beam energy. All accelerating cells in a tank have the same length, while the length of the coupling cells does not vary over the whole module. The cells are magnetically coupled, via the coupling slot, which remains the same in the module. This is then taken into account later with MAFIA program [C.16 and R. Zennaro] and RF measurements on half-cell models.

To describe accurately the behaviour of the actual cavity chain, the simple resonant circuit model mentioned in chapter 3 can be used. For the correct final design of a biperiodic structure, one has to establish the nearest neighbour coupling factor k in a tank as well as the influence of the next (or second) nearest neighbours.

The coupling factor k depends on the slot geometry, the fields and energies in each cell, determining many aspects as group velocity, distance between resonant frequencies of various modes, sensitivity to various errors, etc. The next nearest neighbour coupling between accelerating cells $k_a$ (or $k_1$), and the next nearest neighbour coupling between coupling cells $k_c$ (or $k_2$) both affect the resonant frequency of the $\pi/2$ mode. From the coupled circuit theory, and numbering the even or excited cavities as 2n, and the odd or unexcited cavities as 2n+1, we remember the field (X) equations for the $\pi/2$ mode configuration:

$$X_{2n} = A \cdot \cos 2n\phi \ \text{ and } \ X_{2n+1} = B \cdot \cos(2n+1)\phi$$

where $\phi = \dfrac{\pi \cdot q}{2 \cdot N}$, q=0,1,.....2N. A and B are reported in annex 3.1, while for the next we will indicate $f_1=f_a$ and $f_2=f_c$ as the resonant frequencies of the accelerating and coupling cells respectively.

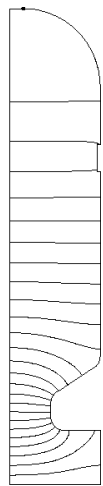

**Figure 5.2** *Final shape of the accelerating half-cell of the prototype. The ring for tuning by machining is also visible. The nose presents a evident flat region reducing the peak electric field.*

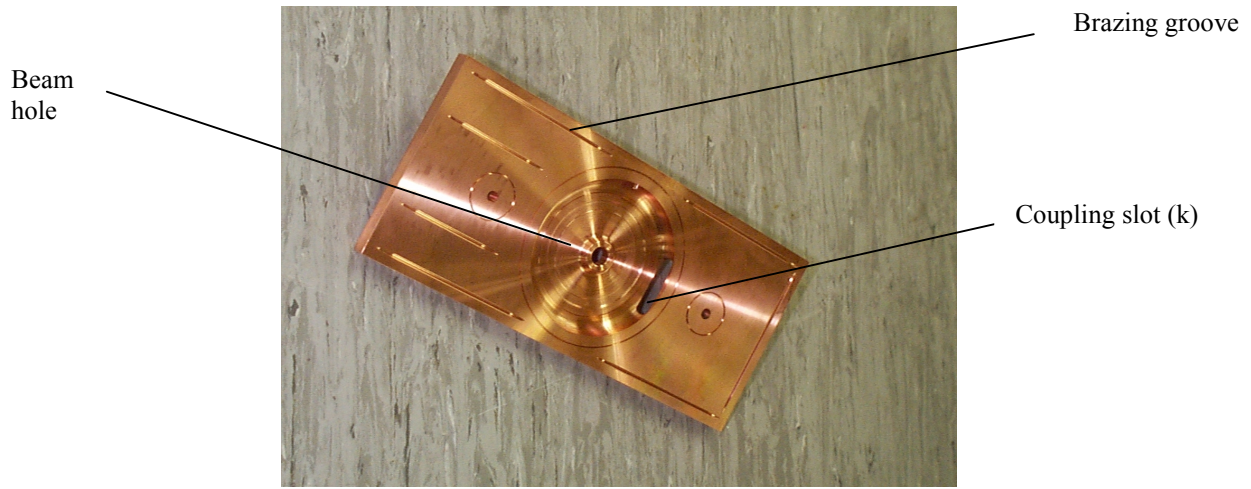

*Figure 5.3* *Half-cell plate of the first LIBO module: the accelerating cell as well as the coupling slot (k) are visible.*

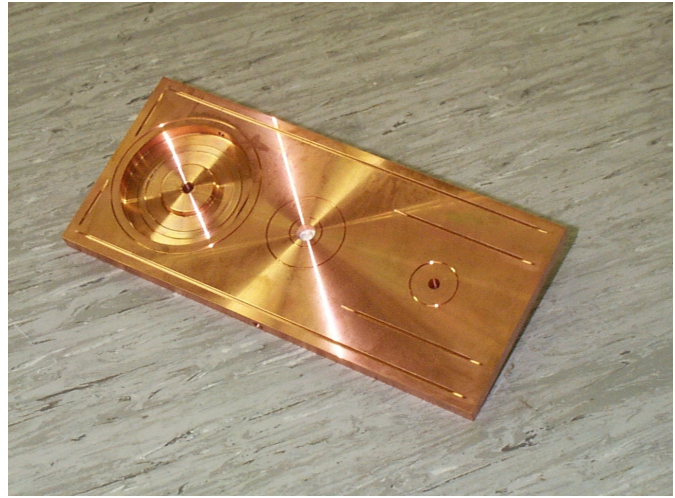

*Figure 5.4* *Half-cell plate of the first LIBO module: the coupling cell is visible.*

The dispersion relation is then:

$$k^2 \cdot \cos^2 \phi = \left[ 1 - \frac{f_a^2}{f^2} + k_a \cdot \cos 2\phi \right] \cdot \left[ 1 - \frac{f_c^2}{f^2} + k_c \cdot \cos 2\phi \right]$$

This reduces to dispersion relation for periodic structure [C.11, C.28] when $f_a \to f_c$ and $k_a = k_c = 0$.

For $\phi = \pi/2$, two solutions exist:

$$f_0 = f_{\frac{\pi}{2}} = \frac{f_a}{\sqrt{1 - k_a}} \quad and \quad f_0 = f_{\frac{\pi}{2}} = \frac{f_c}{\sqrt{1 - k_c}}$$

where $f_0$ is the wanted operational frequency for the $\pi/2$ mode. If the two $f_0$'s calculated from these relations are not the same, there is a discontinuity at this point in the dispersion relation, and the region between the two solutions is the "stopband" (SB). There are no solutions between the two $f_0$'s defining the stopband. To have a stable structure, it is necessary to satisfy both conditions in order to avoid a stopband in the middle of the passband. The next nearest neighbour factors are of the order of 1/10 of the coupling coefficient k, and they will be measured on half-cell test models. It can be proven that the nearest neighbour couplings do not appreciably affect the stability properties of the $\pi/2$ mode, so at the first stage one can assume that $k_a = k_c = 0$, so one has that $f_{\pi/2} = f_a = f_c$. By solving the resultant coupled resonator equations one can find, for initial amplitude $X_0$ in cell 0, the following equations as mentioned in chapter 3:

$$X_{2n+1} = (-1)^n \cdot \left\{ \frac{[j(2n+1)]}{k \cdot Q_a} \right\} \cdot X_0 \text{ and } X_{2n} = (-1)^n \cdot \left\{ 1 + n^2 \cdot \frac{2}{k^2 \cdot Q_a \cdot Q_c} \right\} \cdot X_0$$

where $Q_{2n} = Q_a$ is the Q factor of the excited or accelerating cells, and $Q_{2n+1} = Q_c$ is the Q factor of the unexcited or coupling cells. For our general considerations, the third and higher order terms in $1/k \cdot Q$ have been neglected. The first equation indicates that in $\pi/2$ mode the fields in the odd cells are in quadrature with the excited cavity fields and have very small amplitude (no electric field in the coupling cells). In order to have a preliminary estimation of the coupling k to be used in the tank, one can use the above formula for excited cell. The droop of the accelerating field in the tanks as function of k is then:

$$\frac{\Delta X}{X} = \frac{2N^2}{k^2 \cdot Q_a \cdot Q_c}$$

where N is the number of accelerating cells. Allowing for example a droop of about 1% of the electric field and introducing the Q values computed by SUPERFISH ($Q_a$=8500 and $Q_c$=5000), one obtains a coupling factor k of about 5%.

The advantages by using N coupled resonators operating in $\pi/2$ mode respect to the $\pi$ mode are different. First, one can prove that the mode separation is N/$\pi$ times as large for the $\pi/2$ mode as for the $\pi$ mode[1]. The mode separation is an important aspect. In fact if the natural width of the normal mode (calculated by $\Delta f = f/Q$) is larger than the spacing to the nearest normal mode, the normal modes will overlap and the klystron generator will excite both modes rather than only the desired one. Moreover perturbation analysis shows that for a given set of oscillator errors, modes that are closer together cause larger perturbations to the accelerating fields than modes that are farther apart. All this implies also that the number of coupled cavities

---

[1] This aspect can be shown here for simplicity for a periodic chain of N coupled resonators by a coupling factor k. In this case the bandwidth, defined as the frequency difference between the highest ($\pi$) and the lowest (0) normal mode frequencies, is:

$$f_\pi - f_0 = \frac{f_r}{\sqrt{1-k}} - \frac{f_r}{\sqrt{1+k}} .$$

where $f_r$ is the resonant cell frequency. For k << 1 is $f_\pi - f_0 \approx k \ f_r$. The mode separation in the middle of the band (around the $\pi/2$ mode) and at the $\pi$ mode for N large are [C.6]:

$$\frac{\Delta f_{\pi/2}}{f_r} \cong \frac{k \cdot \pi}{2 \cdot N}, \qquad \frac{\Delta f_\pi}{f_r} \cong \frac{k \cdot \pi^2}{4 \cdot N^2} .$$

This shows that for the $\pi/2$ mode configuration the closest mode spacing is greater than the other normal modes.

is much more severely limited in the π mode than in the π/2 mode, and for the π mode configuration the number of feed points has to be increased correspondingly, with a problem of costs and complexity.

As a consequence of these aspects it is possible to show that a π/2 mode can predict rather large construction tolerances of the cavities with little effect on performance. An estimate may be made from the solution of the resonator equations: $\left\langle \Delta X / X \right\rangle_{rms} \cong N \cdot \left\langle \Delta_a \right\rangle_{rms} \cdot \left\langle \Delta_c \right\rangle_{rms}$, where $\Delta_a \propto \dfrac{\Delta f_a}{k \cdot f_a}$, $\Delta_c \propto \dfrac{\Delta f_c}{k \cdot f_c}$. The errors are assumed random and the effect of the sums is lumped in a statistical factor N [C.31]: for $\Delta_a$ and $\Delta_c$ of the order of $10^{-2}$ (Δf/f = $10^{-2}$), $\left\langle \Delta X / X \right\rangle_{rms} \cong 10^{-4} N$. For about 100 cell module, one should expect about 1% field fluctuations. With general mechanical tolerances of the cells in the range of ± 0.02 mm, one should expect a spread in frequency in the range of 3-5 MHz (see also results in chapter 6), then better of the error mentioned above. This is a proof that the frequency tolerances of RF cavities can be achieved using standard machining techniques, where numerically controlled machines can assure repetitive errors even smaller than ± 0.01 mm at relatively low costs.

❑  *Construction and low power RF measurements of test half-cells*

The design of the cells is a complex and iterative process. In 1999 it has been established to build a six half-cells block at CERN Central Workshop, as a test, for the final definition of the cavities geometry, materials, as well as the procedures adopted for machining, brazing and RF tuning (see also chapter 6). This was a key step, before the construction of the prototype module, due to the difficulty to achieve the required precision of the resonators: machining in the workshop has to go on in parallel with RF measurements for frequency tuning and field adjustment.

Once the general geometry has been defined with SUPERFISH, one must optimize the structure, as function of RF tuning corrections and measurement results, changing the gap g (figure 4.4). The goal, for the final design, is to have, for the complete module, $f_{\pi/2}$ = 2998 MHz and stopband SB=0.

The tentative frequencies of the test cells calculated by SUPERFISH and produced for the RF measurements are $f_a$ = 3040.71 MHz and $f_c$ = 3048.76 MHz. The first measurements have been performed without tuning rods and with a $f_{\pi/2}$ = 2957 MHz. The tuning rods insertion in the accelerating cells has generated an increase of frequency, closing at the same time the stopband ($f_{\pi/2, a}$ - $f_{\pi/2, c}$). In this condition $f_{\pi/2}$ = 2968.8 MHz.

In order to have a correct reference value for the final frequency, one must also take into account different effects. The contribution of air on the measurements can be estimated equal to −0.8 MHz [D.50, D.51], the frequency decreasing for a long chain of cells (the complete module is a sequence of 52 accelerating cells) is −3.34 MHz [D.50, D.51], and the contribution due to the presence of the end cells with a lower frequency (in order to avoid possible bumps in the field) is about −0.19 MHz [D.50, D.51].

Considering also the tuning adjustment after production, one can definitely design the cells with the criteria reported in tables 5.2 and 5.3. With these measurements the coupling slot has also been investigated. Generally speaking, the presence of the coupling slot lowers the resonant frequency and the quality factor Q, and the frequency shift depends mainly on the ratio of the slot length towards the cell size. Using the six half-cell plates an experimental relation between the coupling factor k and the geometrical dimensions of the slot has been found. Starting with the assumption that [C.21, D.50]:

$$k \propto l^3 \cdot \frac{H_a}{\sqrt{U_a}} \cdot \frac{H_c}{\sqrt{U_c}}$$

where l is the slot length, $H_a$ and $H_c$ the magnetic fields in the middle of the coupling slot, $U_a$ and $U_c$ the stored energies in accelerating and coupling cells respectively, the following empirical relation has been obtained:

$$k = \frac{1}{722,7} \cdot l^3 \cdot \frac{1}{\sqrt{L_c}} \cdot \frac{1}{\sqrt{L_a}}$$

where $L_c$ and $L_a$ are the coupling and accelerating half-cell lengths measured in mm.

With a slot length of 28 mm, a mechanical tolerance of ± 0.1 mm, and an accelerating and coupling half-cell length of 6.75 mm, one has a coupling factor k of 4.5% (± 1%), adopted then for the final design of the prototype. The stored RF energy is proportional to $E_0L$, and from the general rule $k^2U$=const., one obtains that $kE_0$=const. An important advantage to have the same shape of the cavities over the full module is the small spread of k, with a consequent small spread in $E_0$, a crucial aspect of the LIBO design.

Final value of the next nearest neighbour coupling between accelerating cells $k_a$ is about –0,74%, while the next nearest neighbour coupling between coupling cells $k_c$ is about 0 [D.50].

Extrapolating these measurements performed on a simple six half-cells block, the full set of cell parameters of the first module has been found, as reported in table 5.4.

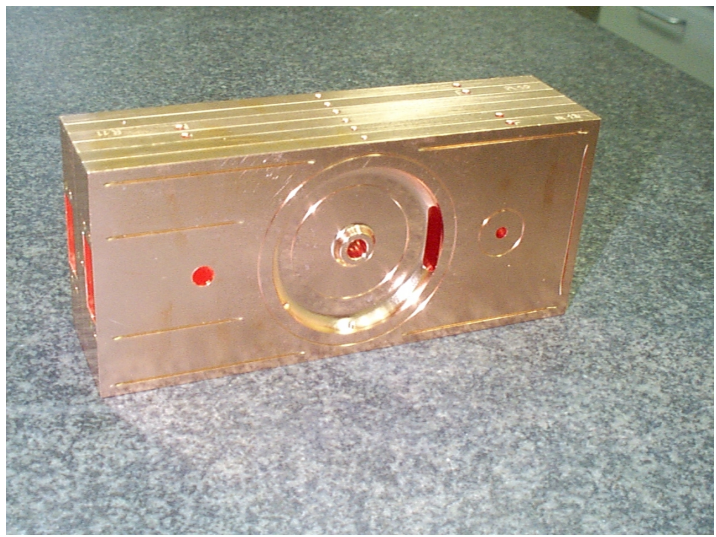

**_Figure 5.5_** _Six half-cell plates produced before full production for mechanical and RF tests. Coupling slot is visible._

**Table 5.2** *Accelerating cell frequency definition.*

| | |
|---|---|
| Frequency calculated by SUPERFISH for test cells | 3040.71 MHz + |
| $f_{\pi/2}$ goal | 2998 MHz - |
| $f_{\pi/2}$ measured for SB=0 | 2968.8 MHz + |
| Frequency variation to obtain SB=0 | 11.8 MHz + |
| Frequency variation for long cell stack | 3.34 MHz + |
| Frequency variation for the end cell presence | 0.19 MHz - |
| Frequency variation for air correction | 0.8 MHz - |
| Half frequency variation with all tuning facilities[*] | 11.5 MHz |
| ***Final AC frequency*** | ***3072.68 MHz*** |

(* See table 5.6)

**Table 5.3** *Coupling cell frequency definition.*

| | |
|---|---|
| Frequency calculated by SUPERFISH for test cells | 3048.76 MHz + |
| $f_{\pi/2}$ goal | 2998 MHz - |
| $f_{\pi/2}$ measured for SB=0 | 2969 MHz + |
| Frequency variation for air correction | 0.8 MHz - |
| Half frequency variation with all tuning facilities[*] | 13.32 MHz |
| ***Final CC frequency*** | ***3063.64 MHz*** |

(* See table 5.6)

❑   *Final design parameters for the prototype module construction*

The first prototype of LIBO-62 is capable to accelerate proton beams from 62 MEV to 73.8 MeV, so with a net energy gain of 11.8 MeV. The final average value of the electric field on axis $E_0$ has been fixed to 15.3 MV/m. The Kilpatrik limit at 2998 MHz is $E_k = 46.77$ MV/m, and the maximum surface electric field is $E_s = 1.43\,E_k = 66.9$ MV/m. The values of the surface electric fields in the tanks of the prototype are reported in table 5.5. The peak power needs are of the order of about 1 MW for each tank, with an effective shunt impedance $Z_{seff} \approx 59$ MΩ/m. The shunt impedance of the four tanks is then about 55.2 MΩ, calculated as:

$$r_{seff} = Z_{seff} \cdot \frac{L}{2} \cdot 2 \cdot N_a$$

where *L/2* is the average accelerating half-cell length, N is the number of accelerating cells in the module. The ratio $r_{seff}/Q$  is then 7.27 kΩ. These parameters have been verified in depth during the RF power tests, and they will be briefly presented in chapter 7.

From the construction point of view the half-cell plates are grouped to form four accelerating tanks and are identical, except for three parametric longitudinal dimensions: the thickness of the plate, the depth of the nose g and the accelerating cell. These aspects have been taken into consideration in order to reduce the complexity during construction.

The overall shape of the cells are produced with a tolerances of ± 0.02 mm, except for few critical dimensions, such as the noses of the accelerating and coupling cells where a more conservative tolerance of ± 0.01 mm has been used. Precise lateral contour is adopted as an external alignment reference during production and RF measurements [D.49]. Grooves on the face of each cell are also machined to allow the positioning of the brazing wires (see chapter 6). Particular care must be foreseen for the surface roughness of the cell as well as for all the brazing surfaces. The roughness in the cavity is strictly linked with the breakdown limit and the ohmic losses present in the accelerating structure.

Due to the fact that the RF currents and power losses are confined within about a skin depth at the surface[2], high roughness generates, at high power levels, risks of discharge and high power losses. The typical roughness values for the half-cell plates are: 0.4 μm (Ra) for the cavity and 0.8 μm (Ra) for the brazing surfaces [D.49].

**Table 5.4** _Final geometrical cell parameters of the LIBO-62 prototype module (see figure 4.4)._

| *Accelerating cell* |
|:---:|
| Energy module 1: 62 – 73.8 MeV |
| Length (L = βλ/2): 17.5 (TK1), 17.86 (TK2), 18.22 (TK3), 18.56 (TK4) mm |
| Average electric field on axis: 15.3 MV/m |
| φAC: 70 mm; $R_{co}$: 5.75 mm; $R_{ci}$: 2 mm |
| g/2: 3.26 (TK1), 3.39 (TK2), 3.51 (TK3), 3.63 (TK4) mm |
| Lac : 6.75 (TK1), 6.93 (TK2), 7.11 (TK3), 3.63 (TK4) mm; |
| α: 20.0 °;  F: 1.2 mm; $R_i$: 1 mm; |
| coupling slot: l x a: 28 x 6 mm |

| *Coupling cell* |
|:---:|
| Energy module 1: 62 – 73.8 MeV |
| φCC: 60 mm; $R_i$ : 1 mm |
| φNcc: 25 mm (TK1, TK2, TK3, TK4) |
| gcc:  3.24 mm (TK1, TK2, TK3, TK4) |
| Lcc:   6.75 mm (TK1, TK2, TK3, TK4) |

---

[2] If dS is the area element on the cell walls, the average power dissipation per cycle is [C.6]:

$$P = \frac{R_s}{2} \int H^2 \cdot dS$$

where P is the average power over an RF cycle, H is the peak magnetic field at the surface in ampere per meter, dS is the surface area element in meter squared, and $R_s$ is the RF surface resistance in ohms/metre. Defining ω=2πf the RF angular frequency, for normal conducting metals:

$$R_s = \frac{1}{\sigma \cdot \delta}$$

where σ is the dc conductivity and δ is the skin depth, given by: $\delta = \sqrt{\dfrac{2}{\sigma \cdot \mu_0 \cdot \omega}}$ .

***Table 5.5*** *RF parameters of the first LIBO prototype module.*

| |
|---|
| Energy gain: $W_{out} - W_{in}$ =73.8-62 = 11.8 MeV |
| Average electric field on axis: $E_0$ = 15.3 MV/m |
| Kilpatrik limit at 2998 MHz: $E_k$ = 46.77 MV/m |
| Surface electric field tank 1: $E_s$ = 1.43 $E_k$ |
| Surface electric field tank 2: $E_s$ = 1.41 $E_k$ |
| Surface electric field tank 3: $E_s$ = 1.40 $E_k$ |
| Surface electric field tank 4: $E_s$ = 1.39 $E_k$ |
| Peak RF power / tank: $P_{rf} \approx$ 1MW |
| Coupling coefficient between AC and CC: 4.5 % |
| Quality factor (averaged over 4 tanks): Q = 7600 |
| Effective shunt impedance : $Z_{seff} \approx$ 59 M$\Omega$/m |
| Transit time factor : T = 0.846 |
| $r_{seff}$ (Shunt imp. of ensemble of 4 tanks) = 55.2 M$\Omega$ |
| $r_{seff}$/Q = 7.27 k$\Omega$ |

❑  *RF tuning solutions on LIBO cavities*

One of the main goal of the module construction is the homogeneity of the accelerating field $E_0$ in order to maintain the synchronism in phase of the proton beam. From the perturbation analysis one can find that the average value of the square root of frequency shift of the cavity is proportional to the average accelerating field $E_0$ (see paragraph 6.3). So it is evident that a crucial issue of the cell production is to reduce as much as possible the final spread in frequency, maintaining at the same time under control the complexity, and consequently the costs, of the construction design.

There are two variables that determine the primary frequency of any resonant cavity. The first is the physical size: in general, the smaller the cavity, the higher its resonant frequency. The second is the shape of the cavity. Both are linked to the stored energy that can be inserted or removed from a cavity.

Finally the resonant frequency of a cavity can be varied by changing any of three parameters: cavity volume, cavity capacitance, or cavity inductance.

Changing the frequencies of a cavity is known as tuning. Tuning systems can be summarised as follows.

▪ Varying the volume it will result in a new resonant frequency because the inductance and the capacitance of the cavity are changed by different amounts. If the volume is decreased, the resonant frequency will be higher. The resonant frequency will be lower if the volume of the cavity is made larger.

▪ Capacitive tuning consists in an adjustable slug or screw, placed in the area of maximum E lines. As the slug is moved in, the distance between the two plates becomes smaller and the capacitance increases. The increase in capacitance causes a decrease in the resonant frequency. As the slug is moved out, the resonant frequency of the cavity increases.

▪ Inductive tuning is accomplished by placing a nonmagnetic slug in the area of maximum H lines. The changing H lines induce a current in the slug that sets up an opposing H field. The opposing field reduces the total H field in the cavity, and therefore reduces the total inductance. Reducing the inductance, by moving the slug in, raises the resonant frequency. Increasing the inductance, by moving the slug out, lowers the resonant frequency.

***Table 5.6*** *Tuning facilities of accelerating (AC) and coupling (CC) cells.*

| Total frequency correction of tuning rings (2 per cell) | |
|---|---|
| AC | 19.1 MHz |
| CC | 21.7 MHz |
| Total frequency correction of lateral tuning rods (2 per cell) | |
| AC | 4 MHz |
| CC | 5 MHz |

A good compromise between the needs requested by RF constraints and the construction processes is to machine the cavities with conventional CNC (numerically-controlled) machines with tolerances ranging between at least 0.01 and 0.02 mm, adjusting then cell by cell with different tuning facilities, to be used during the module assembly.

For the half-cell plates two tuning systems have been adopted. The first consists in a tuning ring (0.7 mm high and 2 mm wide) on the flat face of both accelerating and coupling cells. The tuning of the small LIBO cells is done in parallel with the production, in a series of steps. First of all the cells are measured with a special tools (annex 6.1), in order to control all the RF parameters, and then machined accordingly, to reduce the spread in frequency for the full assembly. Remachining this ring brings the individual half-cell frequency inside the nominal value (±0.5 MHz) with conventional machining tolerances. The sensitivity of the frequency to the machining is 1.6 and 2 MHz per 0.1 mm of the ring height for half accelerating and coupling cells respectively [D.51 and table 6.4].

To compensate for residual frequency errors due to the cell brazing and to tune the complete module to the correct operational frequency, each half-cell has a second tuning facility, that consists of a lateral hole where a small tuning rod is inserted (figures 4.1.b and 5.6), adjusted to the correct length, and then brazed at the last brazing step.

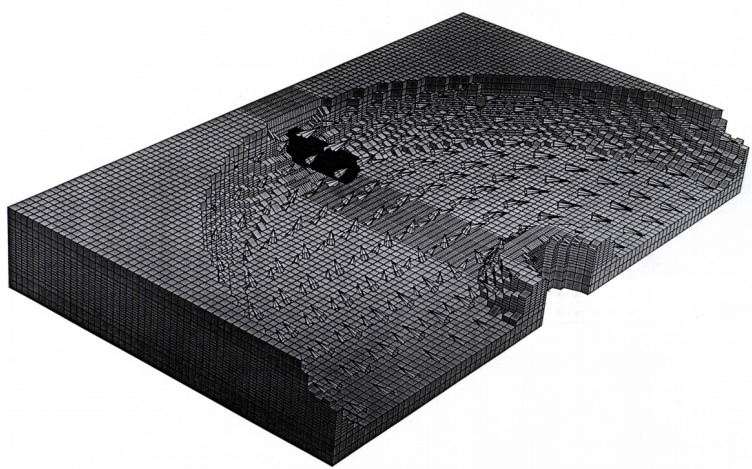

***Figure 5.6*** *RF tuning rod facilities of the accelerating cell. In total two tuning rods are inserted in <u>each half-cell</u> (one in the coupling and one in the accelerating cell). The circular magnetic field near the tuning rod is visible ($TM_{010}$ mode) [R. Zennaro].*

## 5.4    *Bridge coupler bodies*

The elements used to give RF continuity to the different consecutive tanks in a module, maintaining at the same time the possibility to have enough space to house the permanent quadrupole magnets for transverse beam focusing, are called bridge couplers. The use of such a RF component is extremely important due to the fact that an adequate phase control between all the tanks driven from a common source could require a too high power phase shifter between the power splitter and the tank or very precise resonant frequency control. The bridge couplers are conceived as a 3-cells magnetically coupled cavities system: the two coupling cells (CC) and the central cell (BC). The two types of bridge coupler bodies present in the prototype are shown in detail in figures 5.7 and 5.8. They are essentially identical, except for the lengths that change as function of β. Each module counts three bridge couplers. The central bridge is used to connect the accelerator to the RF power source through an iris and a waveguide. The waveguide is tangent to the central cell and is terminated by a short-circuit at 5 λ/4 from the iris. The iris dimensions determine the matching factor ($\hat{\beta}$) between the klystron and the module. The other bridge couplers are called lateral bridges and they are also used for vacuum pumping. The RF power flows between the cells of the bridges through the coupling slots machined in the two disks. The two coupling cells in the bridge coupler (CC) are connected with the adjacent half coupling cells of the half-cell plates to form a complete coupling cell. The bridge couplers are designed trying to dissipate as little power as possible while accomplishing the bridging function.

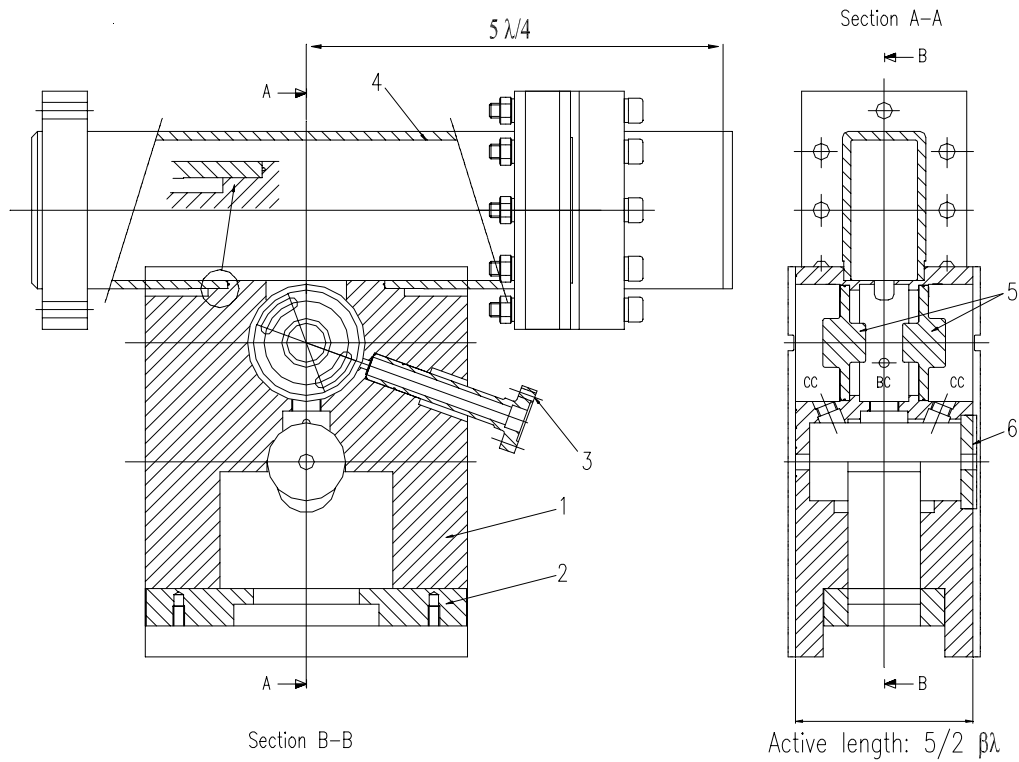

**Figure 5.7** *Central bridge coupler body full assembled with the all pieces brazed on. (1) Bridge coupler copper body, (2) stainless steel flange from the bottom, (3) stainless steel pick-up body for RF field measurement in the cavity, (4) wave guide for RF feeder in copper, (5) copper central disks (RF noses), (6) copper plug.*

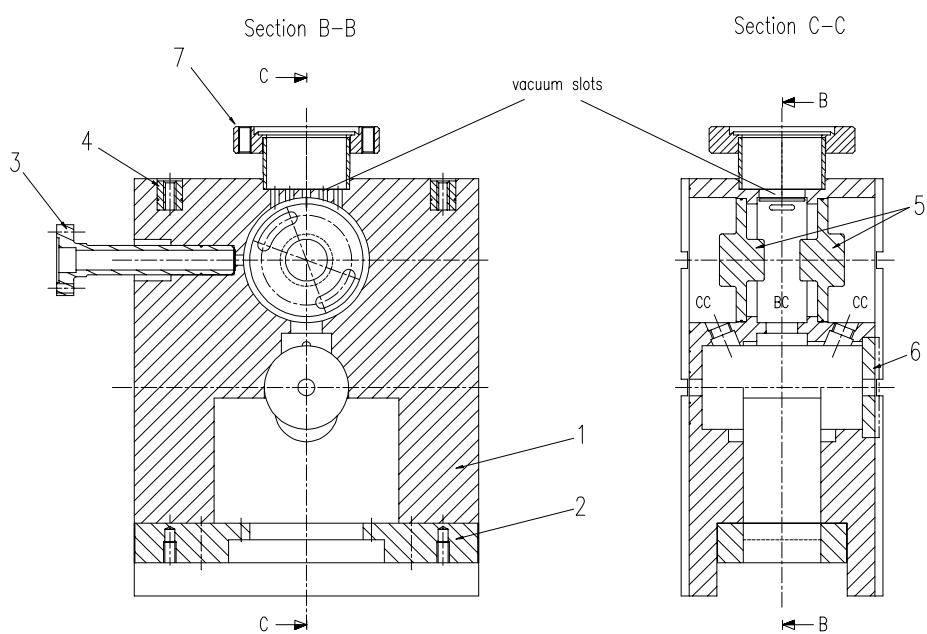

***Figure 5.8*** *Bridge coupler with vacuum pumping port full assembled with the all pieces brazed on. (1) Bridge coupler copper body, (2) stainless steel flange from the bottom, (3) stainless steel RF pick-up body, (4) inox inserts for fixation, (5) copper central disks (RF noses), (6) copper plug, (7) stainless steel flange for vacuum pumping (CF type).*

An in depth analysis [C.18, D.51] shows that by increasing the coupling from coupling cell to the bridge cavity over that from coupling to accelerating cell, the stored energy in the bridge coupler can be reduced substantially. The slots into the bridge are thus cut larger than those into the accelerating cell: their final dimension being set by a trade off between losses in the bridge and ease of power coupling into the bridge tank. The coupling cells (CC) of the bridges are longer than in the tanks, to provide sufficient space for the PMQ, but at the same time they are strictly connected to the frequency of the coupling cells in the tanks. Similarly the central cell (BC) is connected to the frequency of the accelerating cells. The frequencies design are 3002.75 MHz for the central cavity (BC) and 2995.35 MHz for the coupling cells (CC). The methodology performed by TERA specialists to reach the final operating frequency, starting from the design frequencies are briefly discussed in chapter 6, including the RF tuning procedures.

From the engineering point of view the bridge couplers are composed of several pieces: the body, the disks with the slots to form three coupling cells, a copper plug to close PMQ opening after machining, stainless steel flanges, inserts and pick up body. Two brazing steps and intermediate machining are needed to produce a complete bridge coupler. The three bridge couplers are similar, except for two parametric dimensions: the thickness of the body and the depths of the nose in the central cavity (BC). Geometrical tolerances of the overall shape of bridge cavities are ± 0.02 mm as well as the main tolerances connected to RF, beam dynamics and brazing aspects. As for the half-cell plates, brazing grooves are used also for the bridge couplers for the brazing between the body and the adjacent tanks. The bridge couplers are also the main alignment reference of the different module components: the lateral contours of the bodies are used as an external reference in the jig on the machine tool and for the brazing. The PMQ's houses, machined in the bridge coupler, have this precise lateral contour as a reference. After final brazing of the entire module, the transversal alignment of the PMQ geometrical centres must be precise to 0.1 mm, with respect to theoretical axis, in accordance to the beam dynamics specifications (see figures 7.2, 7.3, and 7.4). Technical details for the construction of the bridge couplers are widely presented in [D.49].

In the bridge couplers are foreseen also RF tuning facilities, to be used for field adjustment between two consecutive tanks (see figure 6.24). This frequency adjustment procedure was based on variable screw tuners in copper, equipped in the three cavities of the bridge couplers. This facility could be used also after the final brazing (see figure 5.10). Once the tuner position is fixed and the cavity measured, it is blocked with copper beryllium spring and the bridge is closed under vacuum.

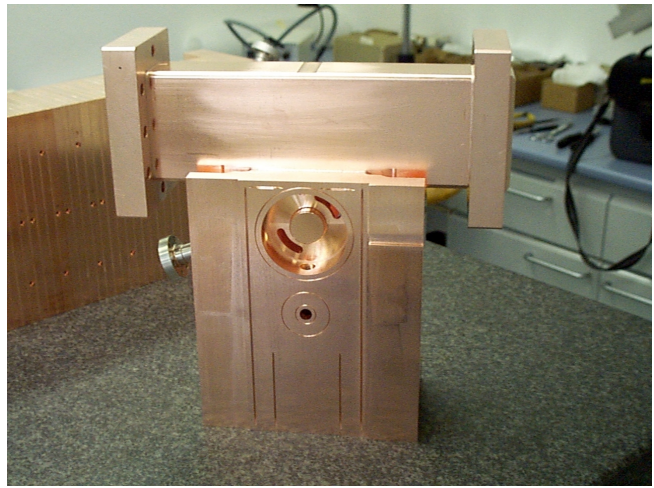

*Figure 5.9* Central bridge coupler of the first LIBO-62 prototype module.

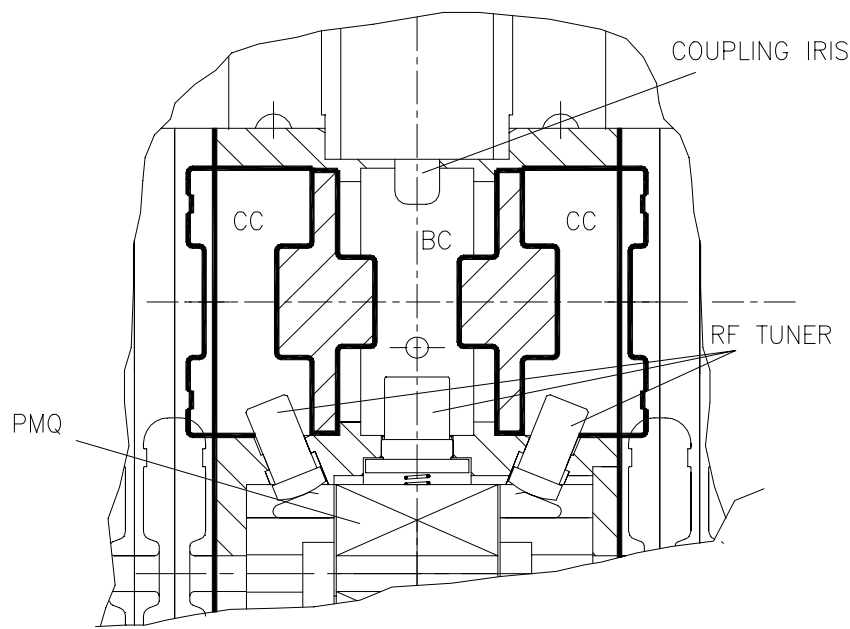

*Figure 5.10* RF tuning facility for the bridge couplers in the LIBO-62 prototype. Once the tuner positions are fixed, they are blocked with copper-beryllium springs, and then the PMQ is housed and blocked. At this stage the bridge coupler is ready for vacuum pumping.

## 5.5    End cell plates

The end cell bodies have the function to close the half accelerating cells of the tanks positioned on both ends of the module. Moreover they must be connected mechanically to the adjacent modules, housing at the same time the PMQs for the continuity of the FODO focusing system.

End cells contain also special stainless steel flanges through which one passes magnetic probes in order to excite and measure the electromagnetic field in the cavity: the RF pick-up.

The design of the half accelerating cell in the end cell body is performed with SUPERFISH program, with the difference, respect to the normal accelerating cells of the tank, that here is present only one open slot for RF power connection. In order to have electric field homogeneity over the full module, one must design the end cell and the other accelerating cells following the relation [B.12]:

$$f_e = f_a \cdot \frac{\sqrt{1 - \dfrac{k_a}{2}}}{\sqrt{1 - k_a}}$$

where $k_a$ is the coupling coefficient between two consecutive accelerating cells.

As already mentioned, it is possible also to prove that:

$$f_a = f_{\frac{\pi}{2}} \cdot \sqrt{1 - k_a}$$

Remembering that $f_{\pi/2}$ is 2998 MHz, and from RF measurements of $k_a$ (-0,74%), one can find the frequencies of the accelerating cells and, as a consequence, the frequency of the end cell to maintain the structure correctly tuned ($f_e$ = 3004 MHz).

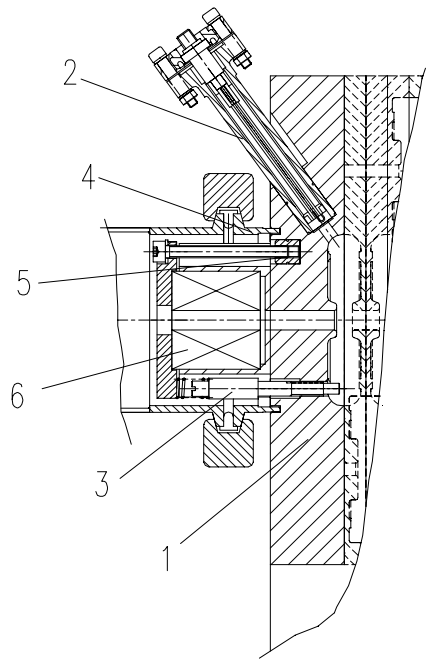

**Figure 5.11**  End cell body full assembled with the all pieces brazed on. (1) End cell copper body, (2) stainless steel RF pick-up body, (3) RF tuner, (4) stainless steel flange for vacuum connection, (5) inox insert, (6) PMQ.

The final design of the end cell bodies foresees also RF tuning facilities using three adjustable screws in the longitudinal direction.

Mechanically speaking, the end cells are composed of several pieces: the copper central body, inserts, vacuum flange and RF pick-up body in stainless steel. The end half-cells of the module are identical, except for three parametric dimensions: the depths of the nose and the cavity in the AC with respect to the brazing surface and the longitudinal PMQ position, according to the above considerations on frequencies and FODO structure. Geometrical tolerances of the overall shape of cell range between 0.01 and 0.02 mm,

Lateral contour is also present, using the same approach of the bridge coupler, as an external alignment reference in the jig on the machine tool and for the brazing. The PMQ's houses, machined in the end cell bodies, have this precise lateral contour as a reference. After final brazing, the transversal alignment of the PMQ geometrical centres must be precise to 0.1 mm, with respect to theoretical axis. Roughness (Ra) varies between 0.4 μm for the RF cavity and 0.8 μm for brazing surfaces [D.49].

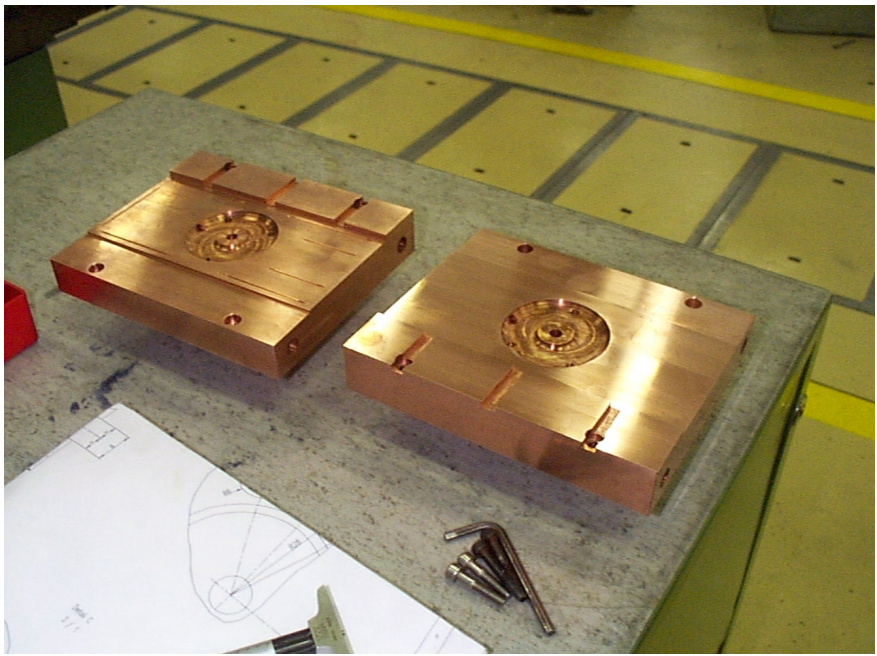

**Figure 5.12** *Two end cell bodies of the first LIBO-62 prototype module.*

## 5.6    Permanent Magnet Quadrupoles (PMQs) for beam focusing

To guarantee the correct beam focusing, maintaining at the same time the reduced space requests of a medical accelerator, Permanent Magnet Quadrupoles (PMQs) have been considered. This technology was developed for linear accelerators and beam transport lines approximately 15 years ago and the fundamental theory was described by K. Halbach and based on anisotropic samarium-cobalt permanent magnet materials [C.10]. Today such device is starting to be attractive for several applications ranging from drift tube linac to various isotope production and neutron generation accelerators.

PMQs consist of a small cylinder where eight blocks of samarium-cobalt 2-17 alloy are inserted, generating the correct magnetic field distribution. The geometrical axis of the aluminium house can be made to coincide with the magnetic axis within ± 0.002'' (50 μm) and then the PMQs are precisely lodged in the bridge coupler and end cell bodies.

The PMQs can not be exposed to more than 250 °C, well below any hard-brazing temperature. For this reason they are inserted in the module once the module is fully assembled and brazed.

Typical requirements for the PMQs used for the prototype are, for a bore radius of 5 mm:

- gradient of 160 T/m,
- length of 36 mm,
- quadrupole displacement error of 0.1 mm,
- quadrupole rotation error of 1°,
- quadrupole gradient error of 2%.

In a real quadrupole one should take into consideration for the beam dynamics the natural harmonics generated by the segmentation of the samarium-cobalt 2-17 blocks (M=8) as well as the other harmonics generated by imperfections of production.

Consider the ring shaped area ($r_1 < r < r_2$) which represents the cross-section of a PMQ. The magnetic polarization of the inserted blocks has a fixed amplitude $B_r$ (remanence) and the pattern of magnetization generates a field within the aperture $r < r_1$ which is:

$$B(z) = B_r \cdot \sum_{n=0}^{\infty} \left( \frac{z}{r_1} \right)^{N+n \cdot M - 1} \cdot \frac{N + n \cdot M}{N + n \cdot M - 1} \cdot \left[ 1 - \left( \frac{r_1}{r_2} \right)^{N+n \cdot M - 1} \right] \cdot K_n$$

and

$$K_n = \frac{M \cdot \cos^{N+n \cdot M}(\pi / M) \cdot sin[\pi \cdot (N + n \cdot M)/M]}{\pi \cdot (N + n \cdot M)}$$

with z = x + i y, $r_1$ internal radius (5 mm), $r_2$ external radius (20 mm). N is the order of the multipole, N+nM is the order of the different harmonics generated by the fact that the multipole is produced by M elements (in our case M=8).

For a quadrupole (N=2) one can write:

▪ n = 0 fundamental harmonic

$$B_x + i \cdot B_y = 2 \cdot B_r \left[ \frac{z}{r_1} \right] \cdot \left[ 1 - \frac{r_1}{r_2} \right] \cdot K_0$$

with $\quad K_0 = \left[ M \cdot \cos 2(\pi / M) \cdot \sin(2 \cdot \pi / M) \right] / 2 \cdot \pi$

▪ n = 1 first harmonic

$$B_x + i \cdot B_y = \frac{2+M}{1+M} \cdot B_r \cdot \left[ \frac{z}{r_1} \right]^{1+M} \cdot \left[ 1 - \frac{r_1}{r_2} \right]^{1+M} \cdot K_1$$

$$K_1 = \left[ M \cdot \cos^{2+M}(\pi / M) \cdot \sin(\pi \cdot (2+M) / M) \right] / (\pi \cdot (2+M)) \,.$$

We can write the above equations in the cylindrical co-ordinates as:

$$B_x(n=0) = A_0 \cdot \left( \frac{r}{r_1} \right) \cdot \cos\phi, \qquad B_y(n=0) = A_0 \cdot \left( \frac{r}{r_1} \right) \cdot \sin\phi,$$

$$B_x(n=1) = A_1 \cdot \left( \frac{r}{r_1} \right)^{1+M} \cdot \cos[(1+M) \cdot \phi], \qquad B_y(n=1) = A_1 \cdot \left( \frac{r}{r_1} \right)^{1+M} \cdot \sin[(1+M) \cdot \phi]$$

$$A_0 = 2 \cdot K_0 \cdot B_r \cdot \left[ 1 - \frac{r_1}{r_2} \right], \qquad A_1 = K_1 \cdot B_r \cdot \left[ 1 - \left( \frac{r_1}{r_2} \right)^{1+M} \right] \cdot \frac{2+M}{1+M} \,.$$

Then one should compare the first harmonic with the fundamental one for the maximum value at φ=0. According with the above formulas, for $B_x$ this ratio is explicated as:

$$\frac{B_x(n=1)}{B_x(n=0)} = \frac{A_1}{A_0} \cdot \left( \frac{r}{r_1} \right)^M \,.$$

The first allowed error harmonic of the PMQs (certified by the constructor) is in the range of 0.15% of the quadrupole field at r = 5 mm, less then of the analytic value calculated with the above formulas.

To understand the contribution of the imperfections of manufacture one should estimate, at the first order, also the contributions of dipole components respect to the fundamental quadrupole field.

A dipole component is equivalent to a difference between the magnetic and the geometric axis of the PMQ[3], and this comparison can be used to understand the importance of these components on the particle motion.

---

[3] The electromagnetic fields encountered by the particles can be derived from a potential function V(x, y, z), which is determined by the well-known Laplace equation [B.34, C.7]:

$$\nabla^2 V = 0,$$

where

$$V(r, \phi, z) = -\frac{c \cdot p}{q} \cdot \sum_{n>0} \frac{1}{n!} \cdot A_n(z) \cdot r^n \cdot e^{i \cdot n \phi}$$

To demonstrate the types of field errors generated by magnet misalignments we use the co-ordinate system fixed in the displaced magnet respect to the reference path. The magnet potential of the nth order expressed with respect to the beam center is:

$$V_n(r, \phi) = -\frac{c \cdot p}{q} \cdot \frac{1}{n!} \cdot A_n \cdot (r - \delta r)^n \cdot e^{i \cdot n \cdot (\phi - \delta\phi)}$$

From the beam dynamics we have the maximum allowed tolerance of the PMQ as 0.1 mm (see previous paragraphs) and for this value we calculate the intensity of the generated dipole component. By using the nomenclature used in [B.12, C.43] and equating the potential of the two different systems of co-ordinates as:

$$V(z) = \sum_n c_{m,n} \cdot \left(\frac{z_m}{r_1}\right)^n = \sum_n c_{c,n} \cdot \left(\frac{z_m}{r_1} + d\right)^n \qquad \text{with} \qquad \begin{array}{l} z_c = z_m + d \cdot r_1 \\[2mm] c_{m,n} = c_{c,n} + n \cdot d \cdot c_{c,n+1} + \dfrac{n \cdot (n+1)}{2} \cdot d^2 \cdot c_{c,n+2} \end{array}$$

n=1 corresponds to the dipole, while n=2 to the quadrupole component. d is the normalized displacement of the quadrupole defined as the ratio between the displacement of the axis and the internal radius of the quadrupole. This can be explained also as the distance between the two different systems of co-ordinates, one perfectly fixed with the PMQ ($z_c$) while the second ($z_m$) coincides with the line where the field hence the dipole component equal to zero. At the first order one has:

$$d = \frac{c_{c,1}}{2 \cdot c_{c,2}} \cdot$$

Then a displacement of 0,1 mm for the PMQ, with an internal radius of 5 mm, generates a ratio between dipole ($c_{c,1}$) and quadrupole ($c_{c,2}$) components < 4%. The PMQ constructor guarantees the errors created by magnet imperfections < 1% of the quadrupole field [C.8], then less, for example, of the errors generated by the misalignments of 0.1 mm of the focusing elements during final assembly.

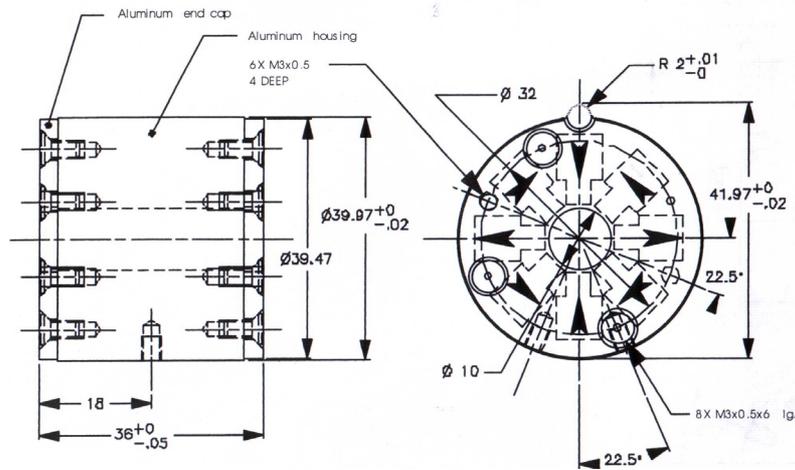

**_Figure 5.13_** _Permanent Magnet Quadrupoles (PMQs): the patterns of magnetisation are also indicated._

From this one can then write:

$$V_n(r,\phi) \approx -\frac{c \cdot p}{q} \cdot \frac{A_n}{n!} \cdot \left[ r^n \cdot e^{i \cdot n \cdot \phi} - i \cdot n \cdot \delta\phi \cdot r^n \cdot e^{i \cdot n \cdot \phi} + \sum_{j=0}^{n-1} \binom{n}{j} \cdot \delta r^{n-j} \cdot r^j \cdot e^{i \cdot n \cdot \phi} \right] + \Theta(2) \cdot$$

The effects of magnet misalignments become obvious. The first term in the square bracket shows that to first order the original nth order fields have not been changed due to magnet misalignments. The second term demonstrates the appearance of a rotated multiple component of the same order nth with the strength n δφ being linearly proportional to rotational misalignment. Transverse misalignments (δr) lead to the third term, which is the sum over a series of lower order field components. In practice a displaced quadrupole generates, at the first order and on the beam axis, a dipole field component.

### 5.7        Thermal stabilisation for frequency control

The cooling system design of a linac is connected to the resonance frequency stability requirements and the magnitude of the RF losses. For a cavity, driven by a fixed frequency oscillator, any deviation of the resonant frequency from the nominal value results in a reduction of the field amplitude. This effect is large, especially in high Q cavity. Thus, for a system of several cells, it will be necessary to maintain the resonant frequency at or very near to the frequency of the driving RF system. Consequently, very precise temperature control is required. It seems feasible, either to demand that the cooling system maintains a constant temperature for frequency stability, or alternatively, that the cooling system is used effectively for the servo stabilisation of the cavity resonant frequency.

So after the injection of the RF power into the module through the central bridge coupler, a cooling is foreseen in order to stabilise the copper structure, maintaining the cavity tuned. As calculated by simulations with SUPERFISH and ANSYS codes [C.35], the expected rate of cell detuning resulting from RF heating will be corrected by cooling the structure bringing the cavity back into resonance.

### 5.7.1    Finite element thermal analysis for RF cavity considerations

❑    *General aspects*

In order to perform dynamic beam scanning, LIBO must work at high repetition rates, so it is designed to operate at 400 Hz pulse repetition frequency with an RF pulse length of 5 μsec, resulting in an RF duty factor of 0.2%. With this duty factor, and assuming a high cell-to-cell coupling coefficient of 5% (conservative design value), each cell in the first tank will dissipate nominally 147 W. The RF power, dissipated on the cavity walls, heats them and causes a change of dimensions, with a consequent change in frequency. A cooling system must therefore be foreseen and designed to keep the resonant frequency of the various cells at reasonable values. This aspect is extremely critical in low beta structure respect to the most SCLs due to the fact that they operate at a higher β and so they have a higher $ZT^2$ and then less power dissipation than LIBO.

In π/2 mode operation the coupling cells are practically empty and the electromagnetic field is concentrated in the accelerating cell, so the analysis of cooling will deal only with the accelerating cells. It is possible to evaluate the frequency thermal detuning, due to the heat created in the cavity, using the Slater perturbation theory [C.6, C.14]. Applying Slater's formula presented in annex 6.2, one can estimate the change in frequency due to local perturbations to the cavity's surface. This "detuning sensitivity" has been plotted for LIBO cells designed at β = 0.35. In this analysis it is evident that the detuning sensitivity is dominated by the electric fields on the nose. To calculate the expected thermal deformation of the cavity the finite element code ANSYS has been used. The principle of this numerical tool is briefly described in annex 5.2.

The axisymmetric shape of the cell is generated by SUPERFISH including the average surface power density dissipated on each surface element, based on the operating RF duty factor. The power dissipated per unit area is proportional to the square of local magnetic field.

The model with ANSYS uses two types of elements for thermal and structural analysis respectively: PLANE55 and PLANE42 elements. The first is used as a 2-D plane element with a two-dimensional thermal conduction capability. It has four nodes with a single degree of freedom, the temperature, at each node. In the same way PLANE42 is used for 2-D modelling of solid structures. This element can be used either as a plane element or as an axisymmetric element and it is defined by four nodes having two degree of freedom at each node: the translation in the nodal x and y directions. The model describes then an axisymmetric half-cell in which all nodes are constrained axially, but are free to move radially at the median plane of the cell. In addition there is a symmetry constraint applied at center and the web which couples all nodes, so that they have the same longitudinal displacement independent of their radius. Otherwise the cell is free to deform both radially and axially. The material properties correspond to annealed copper. In this study an isothermal surface Tsink has been defined at 5 mm from the outer diameter of the cavity. This cooling system will be easily reached by lateral cooling plates presented in figure 5.1.a.

In this analysis the cavity's thermal deformation, resulting from a fixed RF dissipation, has been studied while cooling the outer cavity wall to various temperatures [D.53, D.57]. Figure 5.14 shows how the initial LIBO cavity, with a 5 mm outer wall thickness deforms in two cooling scenarios. In these figures the initial unpowered cavity shape is plotted for reference. In figure 5.14.a the initial temperature of the outer wall of the accelerator Tsink (used as a constraint) corresponds to the ambient value. In the calculations, this initial Tsink is given a value of 0 °C (for any ambient temperature).

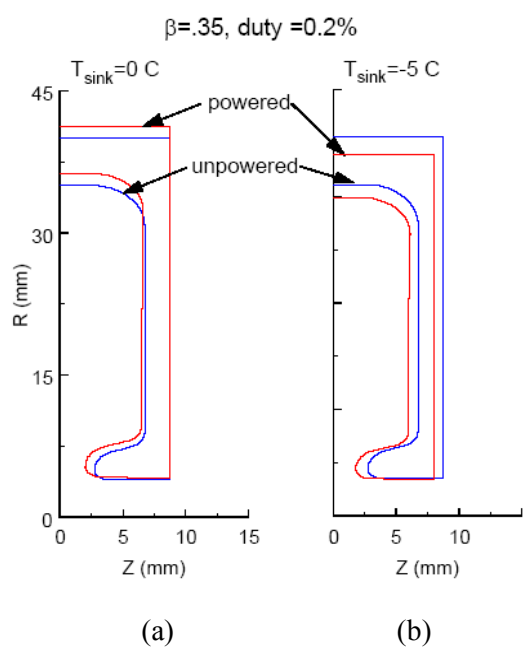

(a)                    (b)

**Figure 5.14** _Geometry modification of the LIBO accelerating cells as a consequence of RF heating for two temperature values of the water cooling system. In the powered case the deformations are magnified by a factor of 1000 to put in evidence the thermal distortion._

Because the temperature of the outer wall is held constant the cell does not expand longitudinally under RF power. At the same time the cell diameter increases and the nose lengthens but there is relatively little growth in the web thickness. In figure "b" the outer wall of the linac has been cooled by 5 °C below ambient which forces a reduction in both the cavity's diameter and length. In this case the cooling has brought the cavity back into resonance. Figure 5.15 shows some properties of the thermal performance of the accelerating cell at $\beta = 0.35$ under RF power.

The first picture shows the final temperature profile. The arrows along the cavity contour indicate the thermal load that is being applied on the cavity surface, while the arrows along the outer wall that its temperature is being constrained to Tsink. A large temperature gradient (about 8 °C) in itself does not represent a particular liability. The principal stresses vector $\sigma_{princ} = (\,\sigma_I\,,\,\sigma_{II}\,,\sigma_{III}\,)$ is then calculated from the stress components[4].

---

[4] Let consider an infinitesimal cubic element with a stress configuration applied in all directions (see figure).

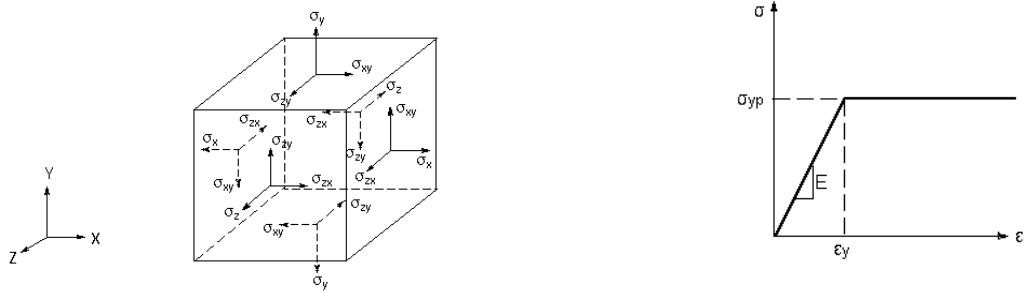

The relation between stress (N/mm$^2$) and strains (a-dimensional) in the elastic domain is [C.54, C.55]:

$$\{\sigma\} = [D]\cdot\{\varepsilon^{el}\}$$

where the stress vector, defined for the infinitesimal cubic element, can be written as:

$$\{\sigma\} = \begin{bmatrix} \sigma_x & \sigma_y & \sigma_z & \sigma_{xy} & \sigma_{yz} & \sigma_{xz} \end{bmatrix}^T$$

The elasticity matrix is:

$$[D]^{-1} = \begin{bmatrix} \dfrac{1}{E_x} & -\dfrac{v_{xy}}{E_y} & -\dfrac{v_{xz}}{E_z} & 0 & 0 & 0 \\[2mm] -\dfrac{v_{yx}}{E_x} & \dfrac{1}{E_y} & -\dfrac{v_{yz}}{E_z} & 0 & 0 & 0 \\[2mm] -\dfrac{v_{zx}}{E_x} & -\dfrac{v_{zy}}{E_y} & \dfrac{1}{E_z} & 0 & 0 & 0 \\[2mm] 0 & 0 & 0 & \dfrac{1}{G_{xy}} & 0 & 0 \\[2mm] 0 & 0 & 0 & 0 & \dfrac{1}{G_{yz}} & 0 \\[2mm] 0 & 0 & 0 & 0 & 0 & \dfrac{1}{G_{xz}} \end{bmatrix}$$

The resulting thermal stress intensity $\sigma_{int}$ must be compared with the elastic limit of the copper $\sigma_{yp}$, reduced by a safety factor $\eta$ (see annex 5.1). Below follows the definition of the resistance criteria as:

$$\sigma_{int} = Max \; \left(\left|\sigma_I - \sigma_{II}\right|, \; \left|\sigma_{II} - \sigma_{III}\right|, \; \left|\sigma_{III} - \sigma_I\right|\right) < \frac{\sigma_{yp}}{\eta} = k \quad \text{(see note}^5\text{)}$$

The distribution of thermal stress intensity is shown in the second picture. The yield strength of soft copper is about 50 N/mm$^2$ (elastic limit) so, in the worst case, with a Young module E of 129000 N/mm$^2$, a Poisson ratio $\nu$ of 0.343 and the thermal expansion coefficient $\alpha$ of 17 $\mu$m/$^\circ$K m, the maximum stress represents only 20% of its yield strength.

---

where typical terms, defining the material properties, are: $E_i$ Young's module in the i direction, $\nu$ Poisson's ratio and G shear modulus. $\sigma_{yp}$ is defined as the plastic limit. Under this limit one has a linear dependence between stress and strain. Above the plastic limit, the material has a non-linear dependence. For the heat-treated pure copper the plastic limit has been found to be equal to 50 N/mm$^2$.

The strain vector is defined as:

$$\{\varepsilon^{el}\} = \{\varepsilon\} - \{\varepsilon^{th}\}$$

where, defining $\Delta T$ the difference in temperature of the material respect to the reference value, $\alpha$ the thermal expansion coefficient, the strain vector due to the thermal gradients is:

$$\{\varepsilon^{th}\} = \Delta T \cdot \begin{bmatrix} \alpha_x & \alpha_y & \alpha_z & 0 & 0 & 0 \end{bmatrix}^T$$

and the total strain vector is:

$$\{\varepsilon\} = \begin{bmatrix} \varepsilon_x & \varepsilon_y & \varepsilon_z & \varepsilon_{xy} & \varepsilon_{yz} & \varepsilon_{xz} \end{bmatrix}^T$$

From the strain calculation one can obtain the stress components. The principal stress vector ($\sigma_{princ} = \sigma_I, \sigma_{II}, \sigma_{III}$) is then calculated from these stress components by the cubic equation:

$$\begin{vmatrix} \sigma_x - \sigma_{princ} & \sigma_{xy} & \sigma_{xz} \\ \sigma_{xy} & \sigma_y - \sigma_{princ} & \sigma_{yz} \\ \sigma_{xz} & \sigma_{yz} & \sigma_z - \sigma_{princ} \end{vmatrix} = 0$$

$^5$ The resistance criteria used in this analysis is associated to the names of Tresca and Guest and considers, as a reference, the maximum value of the tangential stress ($\tau_{max}$) in the material, to be compared with its strength limit k. The above relation is equivalent to the following six equations, where the principal stresses are used to calculate the safety domain:

$$-k \leq \sigma_\alpha - \sigma_\beta \leq k, \qquad \alpha, \beta = I, II, III$$

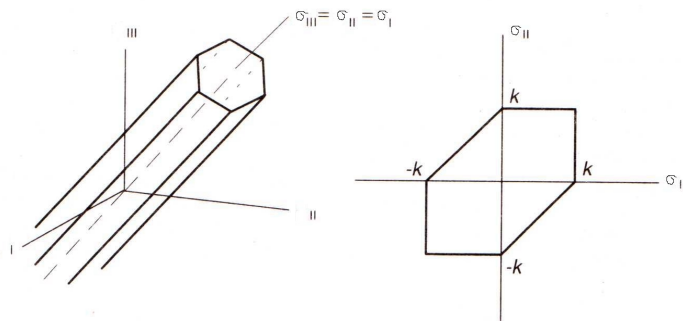

These equations represent in the principal stress-space a hexagon: all the stress configurations inside this hexagon are then considered below the strength limit of the material.

The third picture shows the thermal distortion in the longitudinal direction. The overall cavity extension is constrained by the outer body temperature and never exceeds 1 mm. The forth picture shows the integrated thermal distortion which is primarily in the radial direction.

Once the basic thermal behaviour of the cavities has been studied, a 3D analysis has also been performed with ANSYS, in collaboration with INFN and Milan University, in order to understand in detail the temperature distribution and to verify the results of the 2D analysis. In particular the 3D analysis allows checking the intrinsic asymmetric structure of the accelerator and the effects of the coupling cell and the related slot. Also for this model, one assumes the material properties corresponding to annealed copper. No thermal analysis on the slot has been considered due to the low RF power in the cells at a duty cycle of 0.2%. The average RF power density dissipated at each surface area, generated by SUPERFISH, are the input data of the problem along the cavity contour. Figure 5.16 shows the main co-ordinates and the relative RF power density. Maximum temperature gradient (between nose and external walls) is under 7 °C, while, considering the copper thermal expansion coefficient α, the thermal expansion of the nose (the most critical area for detuning sensitivity) is about 10 μm, which is of the same order as the mechanical tolerances. Figures 5.17 and 5.18 put in evidence these results. Here again it has been assumed that coupling cells are practically empty, as it is visible from the temperature distribution.

Given the deformed shape of the cell generated by ANSYS, one can calculate the surface displacements resulting from RF heating and uniform external cooling of the cavity. At the same time given the fields at each surface element it is possible to calculate the cavity detuning from Slater's formula.

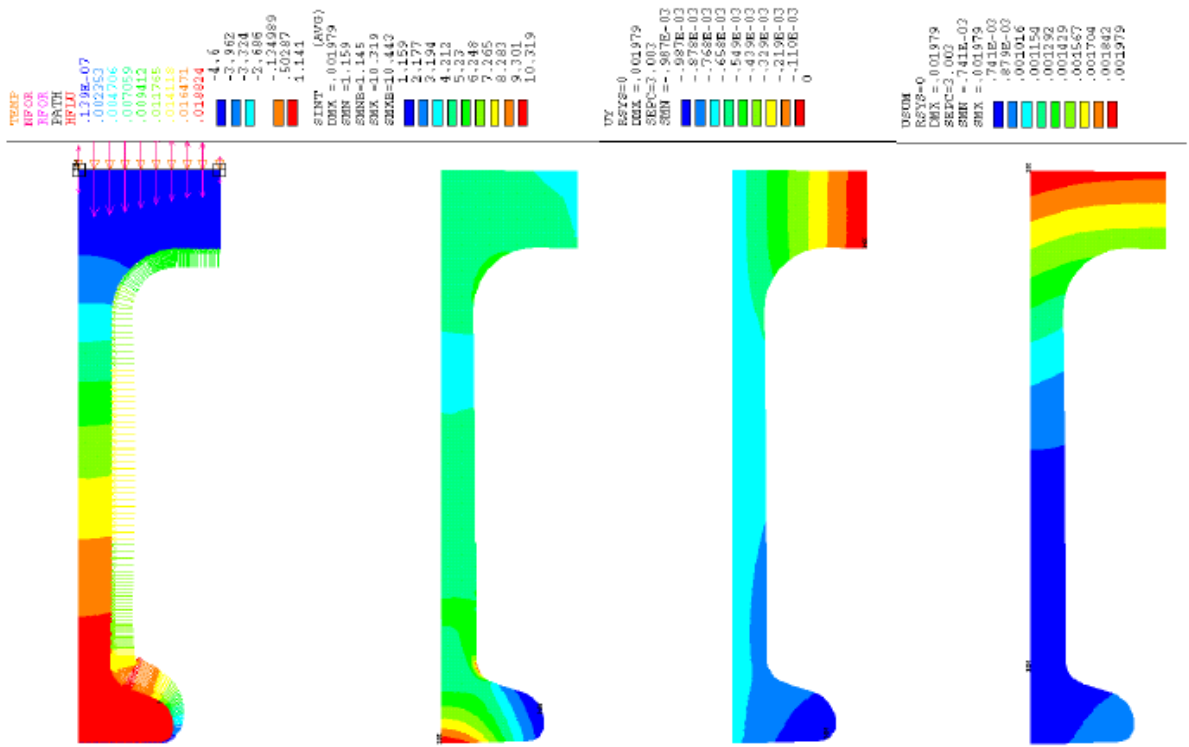

**Figure 5.15** *Thermal stress analysis of the first accelerating cell of the LIBO-62 prototype. From the left to the right are shown: temperature distribution, resulting thermal stress intensity, thermal distortion in the longitudinal direction and the integrated thermal distortion. The analysis foresees the use of the material properties for annealed copper.*

The expected rate of cavity detuning per unit length along its contour resulting from RF heating has been calculated as well as the subsequent cooling to bring the cavity back into resonance. Summing the contribution from each surface element one obtains the total expected cavity detuning.

The tuning sensitivity for the first LIBO cavity is plotted in Figure 5.19. In this analysis it is visible that, for an uncooled configuration where the outer wall of the cell is held at ambient while the nose temperature increases by about 7 °C, the cavities detune by about -250 kHZ [A.4]. Cooling the outer wall to 5 °C below ambient brings them back into resonance. To verify the results generated from the Slater's relation an inverse analysis has been also performed using the thermally deformed shapes generated by ANSYS and computing the resonant frequency of the cavity with SUPERFISH. The expected changes in frequency agree within 2% using the two methods [J. Stovall and A. Catenaccio].

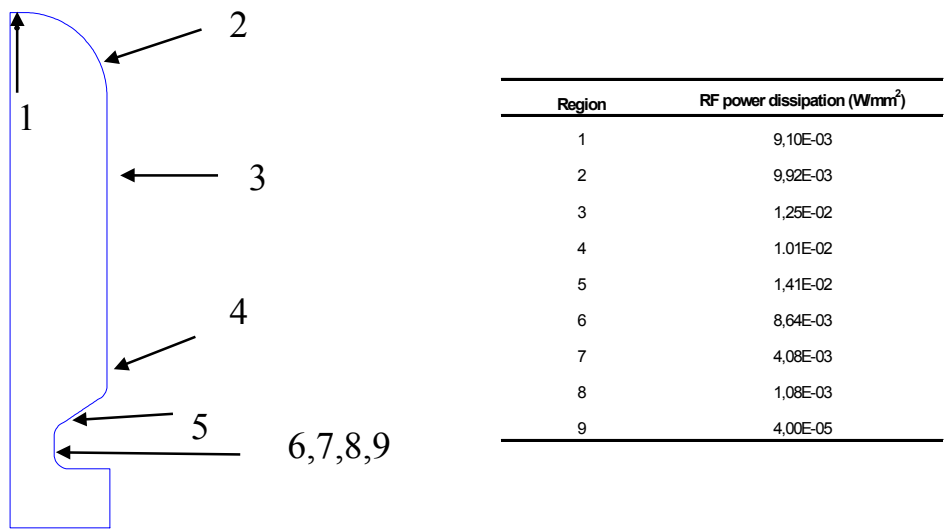

| Region | RF power dissipation (W/mm$^2$) |
|--------|---------------------------------|
| 1 | 9,10E-03 |
| 2 | 9,92E-03 |
| 3 | 1,25E-02 |
| 4 | 1.01E-02 |
| 5 | 1,41E-02 |
| 6 | 8,64E-03 |
| 7 | 4,08E-03 |
| 8 | 1,08E-03 |
| 9 | 4,00E-05 |

**Figure 5.16** *Power density distribution on the LIBO accelerating cell (duty cycle 0.2%).*

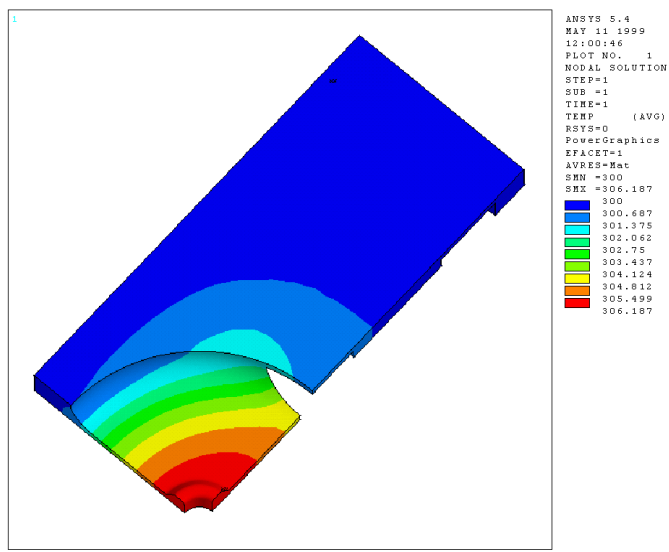

**Figure 5.17** *3D temperature distribution (°K) in the accelerating cell of LIBO-62 prototype.*

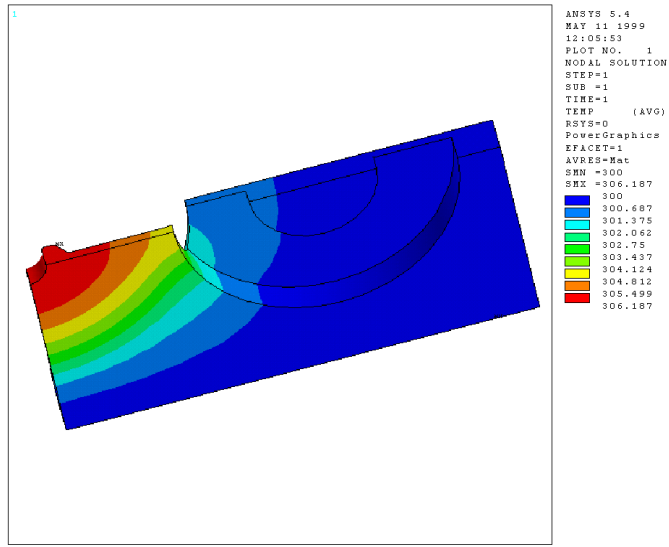

**_Figure 5.18_** _3D temperature distribution (°K) in the coupling cell of LIBO-62 prototype._

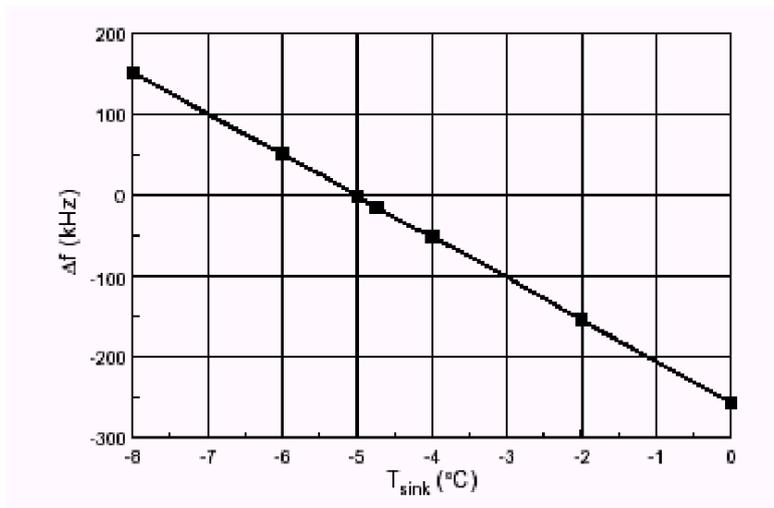

**_Figure 5.19_** _Cavity detuning respect to the temperature of the cooling system._

Finally one finds that the change of frequency of an accelerating cell is proportional to changes in its linear dimensions. If one further assumes that the cavity temperature can be controlled isothermally then it is possible to obtain the following relation:

$$\frac{\Delta f}{\Delta T} = -\alpha_{Cu} \cdot f$$

where $\alpha_{Cu}$ is the thermal expansion coefficient for copper. The rate of frequency change with the temperature, for an isothermally controlled cavity, is then between −50 and -60 kHz/°C (calculated in the 2 and 3D analysis).

❑  *Conclusions on the thermal analysis*

The final conclusions of these analyses can be summarised with the following points [A.4].

1. The peak surface electric field and the power dissipation are in the range of a good safety margin.

2. The expected rate of cavity detuning with temperature, once a steady state temperature has been established, can accurately be described by assuming an isothermal cavity model.

3. Temperature distribution, thermal stresses and mechanical deformations of the cell are computed and they will not affect irreversibly the cavity during the operation at full power (no permanent plastic deformations). The temperature gradient between the nose (cell centre) and the water cooling temperature $T_{sink}$ is found to be about 7°C.

4. The expected cavity deformation and relative thermal detuning under full RF power are within a reasonable range (between -50 and -60 kHz/°C).

5. By lowering $T_{sink}$, the original cavity can be mechanically deformed in the opposite sense. So the thermal detuning can be corrected via temperature control. The cooling system will be designed accordingly.

### 5.7.2   Design of cooling circuits*

To maintain the precise dimensions of the linac during operation, temperature control is then required. Without this, the working frequency may be varied to correct the effects of input power or ambient temperature changes. The RF power loss determines the amount of heat generated in the copper structure and occurs in a thickness of the cavities walls known as the skin depth. The temperature rise in the skin depth during the RF pulse is generally low for typical operating power levels. The volume-averaged cavity wall temperature determines the cell dimensions. This temperature depends upon the coolant temperature and the temperature drops across the coolant film (inducted by convection phenomena) and across the cavity wall (inducted by conduction phenomena) (figure 5.20).

The efficiency of the heat removal system is determinate by the intimacy of the contact between the coolant and the accelerator.

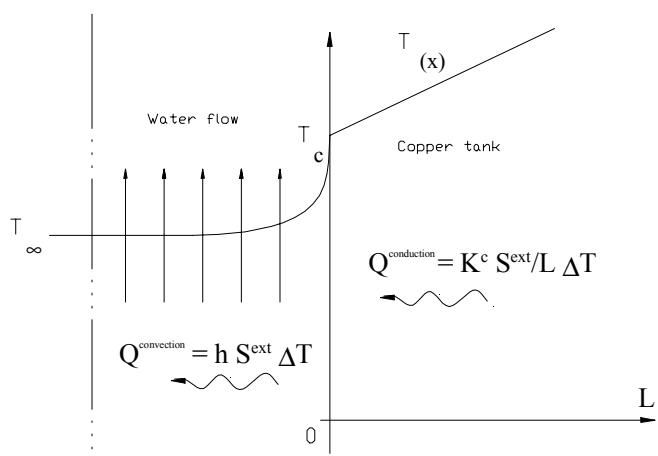

***Figure 5.20*** *Temperature profile in the water channel of cooling plates.*

The coolant flow rate is fixed by the need for temperature uniformity, the flow configuration, the cost of pumping, corrosion and erosion. A high coolant flow rate will give a small coolant temperature rise. A high velocity will yield high heat transfer coefficients, resulting in a small temperature drop between the surface being cooled and the coolant. On the other hand high flow rates increase the pumping and heat exchanger costs. The most desiderable and reliable heat transfer characteristics occur with turbulent flow, which generally means high velocities (see note 6). The acceptable coolant velocity range depends on the coolant and materials used for the construction of the cooling channels. For water, flowing over a relatively soft copper material, it is desirable to keep velocities below 5 m/sec to minimise possible erosion. For LIBO water is naturally used as a coolant for economical reasons.

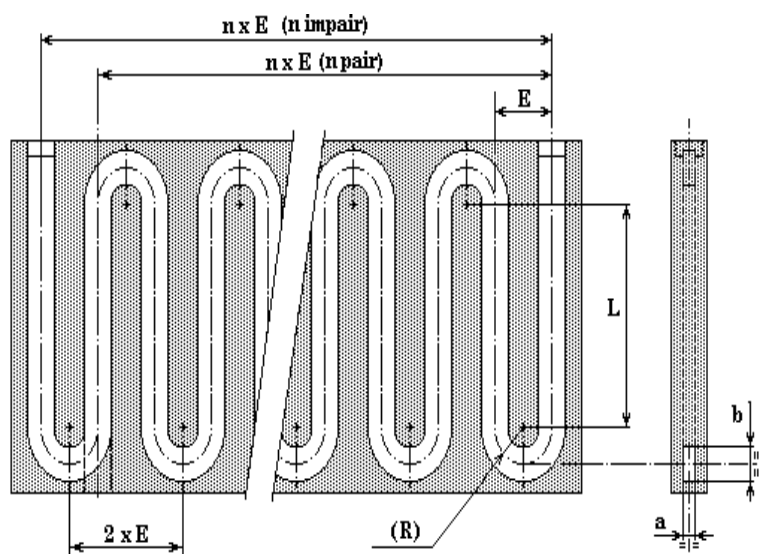

***Figure 5.21*** *Cooling channels of LIBO tank.*

In the prototype, the cooling system is composed by cooling plates, brazed on the lateral sides of the LIBO tanks (see figure 5.1.a). Each plate is divided in two parts: one presents a long rectangular channel, pre-machined by a CNC milling machine and then brazed on the second one.

It has been found that temperature drops between the entrance and the exit of the cooling system of 0,7 °C is enough for the thermal stabilisation connected with the frequency shift errors. All the calculations and results referee to the prototype and they are foreseen at the nominal RF peak power of about 4.2 MW. With 5 μsec pulse length and a repetition rate of 400 Hz, the average power dissipated in the copper structure is then 8400 W. So, to dissipate the average power of 2100 W per tank, a water flux of about 45 l/min per tank is needed.

After an optimisation process of pump and circuit parameters as well as cost analysis considerations, it has been decided that four cooling circuits in parallel can assure the correct cooling of the prototype. Rectangular channels compose the cooling system, where a turbulent flow is created for a good heat exchange. The calculated Reynolds number is 38250, while total pressure drop in each cooling plate is about 4 bar.

Once the water starts to flow in the channels, a convection heat exchange generates a temperature drop between the channel axis and the copper wall (figure 5.20), according with the basic law:

$$\Delta(T_c - T_\infty) = \frac{P}{h \cdot S}$$

where P is the power exchanged, S is the internal channel surface and h is the forced convection coefficient. We remember the Dittus-Boelter experimental correlation [C.53]:

$$Nu = 0,023 \cdot \text{Re}^{0,8} \cdot \text{Pr}^{0,4}$$

where Re, Pr and Nu are the Reynolds, Prandtl and Nusselt dimensionless parameters, that express univocally the forced convection phenomena[6]. From the above correlation one can estimate the transfer convection coefficient of the LIBO cooling system as:

$$h = \frac{0,023 \cdot \gamma^{0,8} \cdot v^{0,8} \cdot d^{-0,2} \cdot K_w^{0,6} \cdot c_w^{0,4}}{\mu^{0,4}}$$

where $\gamma$ is the water density, $v$ the water speed, d the hydraulic diameter, $K_w$ the water thermal conductivity, $c_w$ the water specific heat and $\mu$ the water viscosity. The transfer convection coefficient h is then equal to 13900 Watt/(m$^2$ K), showing a good capacity of the coolant to extract the RF power dissipated in the cavities. The final cooling configuration is reported in table 5.7, with the main parameters calculated in steady state condition at the nominal RF power.

---

[6] We remember that the dimensionless parameters of internal flows in convection condition are:

- Prandtl number $\text{Pr} = \dfrac{c_w \cdot \mu}{K_w}$, that provides a measure of the relative effectiveness of momentum and energy transport by diffusion in the velocity and thermal boundary layers.

- Reynolds number $\text{Re} = \dfrac{\gamma \cdot v \cdot d}{\mu}$ may be interpreted as the ratio of inertia to viscous forces in the velocity boundary layers. The Reynolds number determines the existence of laminar or turbulent flow.

- Nusselt number $Nu = \dfrac{h \cdot d}{K_w}$, that shows the dimensionless temperature gradient at the surface.

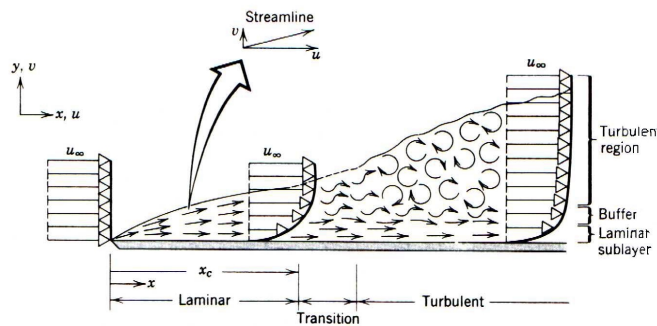

Two types of boundary layers (see figure) determine the convection problem: the laminar and turbulent condition. In laminar condition fluid motion is highly ordered and with low heat transfer capacity. In contrast fluid motion in the turbulent boundary layer is highly irregular with velocity fluctuations that enhance the transfer of momentum and energy, and increase surface friction as well as convection transfer rates.

***Table 5.7*** *Cooling system of LIBO-62 prototype.*

**Applied peak power: 4.2 MW**
**Duty cycle: 0.2% → Average power inside LIBO: 8400 W**

*Fluid channel:*
Section: 100 mm$^2$ (a x b = 10 x 10 mm)
E = 17.5 mm, R = 8.75 mm
Number of turns: n = 11
Length of the fluid circuit: Lcs = 1.98 m
Two parallel circuits for each tank.

*Thermal parameters:*
Power in a tank: 2100 W
$\Delta T_{water}$ (in-out) = 0.7 K
Water flux: Qv = 45.9 l/min
Water velocity: 3.825 m/sec
Heat exchange: forced convection system in turbulent regime (Re = 38250)
Transfer convection coefficient: h = 13900 W/(m$^2$ K)
Heat exchange surface of the fluid: S$_{est}$ = 0.079 m$^2$
$\Delta T_{Fluid\text{-}External\ wall}$ $(T_c - T_\infty)$ = 1.01 K
Estimated by 2 and 3D ANSYS computations: $\Delta T_{External\ wall\text{-}Nose}$ = 7 K

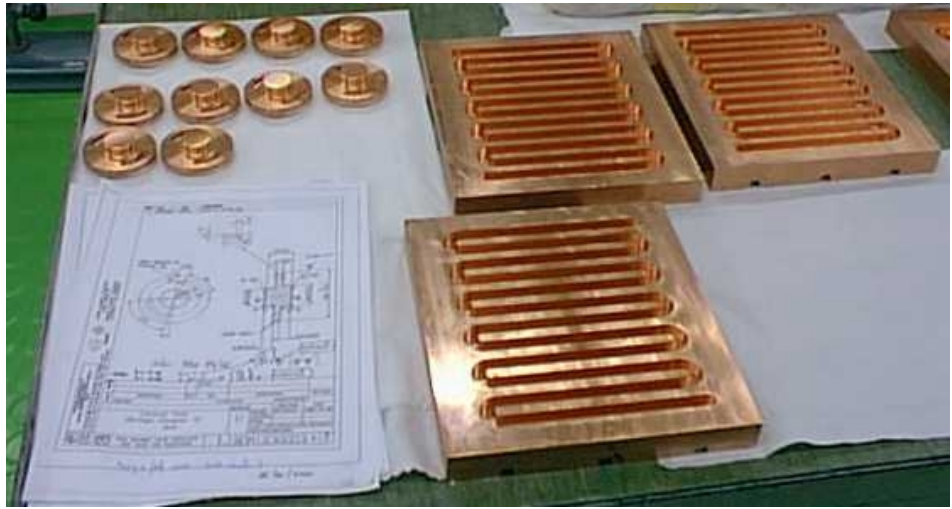

***Figure 5.22*** *Cooling plates of LIBO-62 prototype.*

### 5.7.3    Thermal stability of the prototype

❑   *General considerations of a simple model*

In order to study the thermal transient conditions of the tank under full RF power an analytic model of the global system has been developed. This paragraph is dedicated to present this model and it will be used to understand the thermal behaviour and the frequency dependence with the temperature of the cavity.

The model is valid under the following conditions:

- high conductivity of the monolithic copper structure;
- sub-systems, such as tanks and cooling plates, can thermally communicate and they can be considered as points (point model), where thermal signals can be exchanged [C.56, C.57, C.58].

Starting with the assumption that a tank is thermally connected to the cooling channels through a conduction process, one can write the basic equations of a system (tank-cooling water), as described in figure 5.23:

$$\begin{cases} P - K * \cdot (T_c - T_w) = c_c \cdot M_c \cdot \dfrac{dT_c}{dt} \\ K * \cdot (T_c - T_w) - c_w \cdot \Gamma \cdot (T_o - T_i) = c_w \cdot M_w \cdot \dfrac{dT_w}{dt} \\ T_w = \dfrac{T_i + T_o}{2} \end{cases}$$

where the different parameters are defined in table 5.8. $K*$ indicates the heat transfer coefficient water-copper. One can define the different time delays of the physical phenomena as:

$$\begin{cases} \tau_c = \dfrac{c_c \cdot M_c}{K *} \\ \tau_w = \dfrac{c_w \cdot M_w}{K *} \\ \tau_0 = \dfrac{M_w}{\Gamma} \end{cases}$$

where $\tau_c$ and $\tau_w$ represent the rapidity of copper tank and cooling water to exchange heat, while $\tau_0$ shows how rapid is the heat transfer, generated by the RF power and dissipated into LIBO, from the cooling channels to the external system.

**Table 5.8** *Nomenclatures for the analysis of the prototype thermal stability.*

| |
|---|
| P: RF power injected in the module |
| $\Gamma_i$: water flow in tank i |
| $T_i$: input temperature of cooling water |
| $T_c$: temperature of copper tank |
| $T_o$: output temperature of cooling water |
| $T_w$: average temperature of cooling water, calculated in the middle position of the tank |
| $K_c$: copper thermal conductivity |
| $K_w$: water thermal conductivity |
| $c_w$: water specific heat coefficient |
| $c_c$: copper specific heat coefficient |
| $\mu$: water viscosity |
| $M_c$: copper mass of accelerating tank |
| $M_w$: water mass in a cooling channel |
| h: convection heat transfer coefficient |

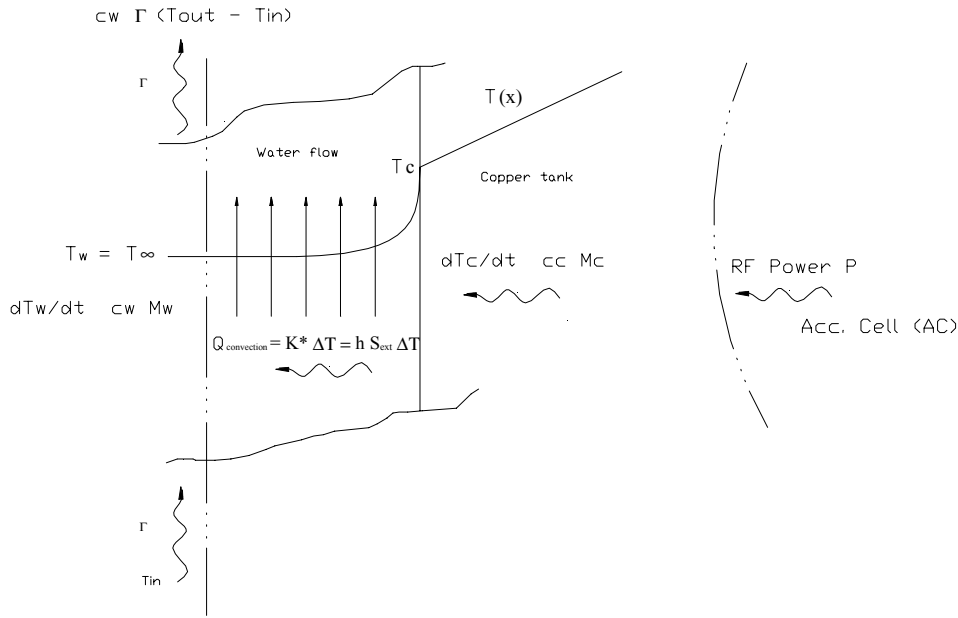

**Figure 5.23** *Scheme of the thermal behaviour of LIBO cavity under full RF power. The power is generated inside of the accelerating cell (AC) and transferred with cooling water through the lateral channel.*

Then the above system can be written including the effects of the appropriate time delays as:

$$\begin{cases} \dfrac{dT_c}{dt} = \dfrac{P}{K * \cdot \tau_c} - \dfrac{T_c - T_w}{\tau_c} \\ \dfrac{dT_w}{dt} = \dfrac{T_c - T_w}{\tau_w} - 2 \cdot \dfrac{T_w - T_i}{\tau_0} \end{cases}$$

We assume that K* is expressed as h · $S_{ext}$, with h= 13900 W/m$^2$K, $S_{ext}$= 0.079 m$^2$ (convection external surface), $\Gamma$ = 0.765 kg/sec (to be divided in two circuits in parallel), $M_w$=0.2 kg, $c_w$= 4183 J/kg K, $c_c$= 400 J/kg K. Typical values of time delays for operating condition are: $\tau_c$= 5 sec, $\tau_w$= 0.7 sec, $\tau_0$= 0.5 sec. It means that the dominant part during transient analysis is the time delay of copper [C.53].

❏   *Thermal stability of the prototype*

To have a better understanding of the above equations, we write the system in terms of input-output signals.

In control theory, functions called transfer functions are commonly used to characterise the input-output relationships of components or systems that can be described by linear, time invariant, differential equations. The transfer function is defined as the ratio of the Laplace transform[7] of the output (response function) to the Laplace transform of the input (driving function). The input-output system is represented in figure 5.24. It follows therefore that if δP is the change in power of the system operating in the steady condition, one can estimate the changes in temperature of the copper structure ($T_c$) and the cooling water ($T_w$).

---

[7] We remember that for s = σ + iω ∈ C, $F(s) = L[f(t)] = \displaystyle\int_0^{+\infty} e^{-st} \cdot f(t) \cdot dt$ is the Laplace transform of f(t) [C.57].

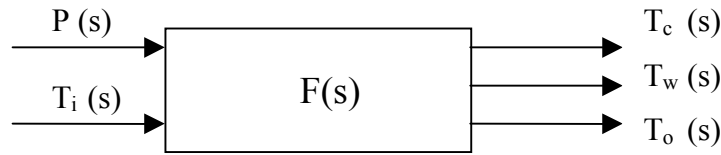

**Figure 5.24** _Input-output system with the relative transfer function can describe the thermal model of the accelerating tank. Input variables are the RF power (P) injected into LIBO and the input temperature of the cooling water ($T_i$). Output variables are the copper tank temperature ($T_c$), the cooling water temperature ($T_w$) (calculated in the middle of the cooling plates) and the output water temperature ($T_o$)._

We suppose that, if input parameters are applied under equilibrium conditions, the different variables can slowly change with small perturbations. These perturbations can be represented by the above equations using the Laplace transforms as:

$$\begin{cases} \delta\ P(s) - K*\cdot\delta\ T_c(s) + K*\cdot\delta\ T_w(s) = K*\cdot s\cdot\tau_c\cdot\delta\ T_c(s) \\ K*\cdot\delta\ T_c(s) - K*\cdot\delta\ T_w(s) - c_w\cdot\Gamma\cdot\big(\delta\ T_o(s) - \delta\ T_i(s)\big) = K*\cdot s\cdot\tau_w\cdot\delta\ T_w(s) \\ \delta\ T_w(s) = \dfrac{\delta\ T_i(s) + \delta\ T_o(s)}{2} \end{cases}$$

or, in terms of time delays, as:

$$\begin{cases} \delta\ T_c(s) = \dfrac{\delta\ P(s) + K*\cdot\delta\ T_w(s)}{K*\cdot(1 + s\cdot\tau_c)} \\ \delta\ T_o(s) = \dfrac{K*}{c_w\cdot\Gamma}\cdot\delta\ T_c(s) - \dfrac{K*\cdot(1 + s\cdot\tau_w)}{c_w\cdot\Gamma}\cdot\delta\ T_w(s) + \delta\ T_i(s) \\ \delta\ T_w(s) = \dfrac{\delta\ T_i(s) + \delta\ T_o(s)}{2} \end{cases}$$

In this case it is possible to compare the different heat exchange behaviours of water and copper. The above equations can be represented also in terms of signals diagram [C.56], where input and output variables are indicated, showing then how the sub-systems (tank and cooling plate) can thermally communicate.

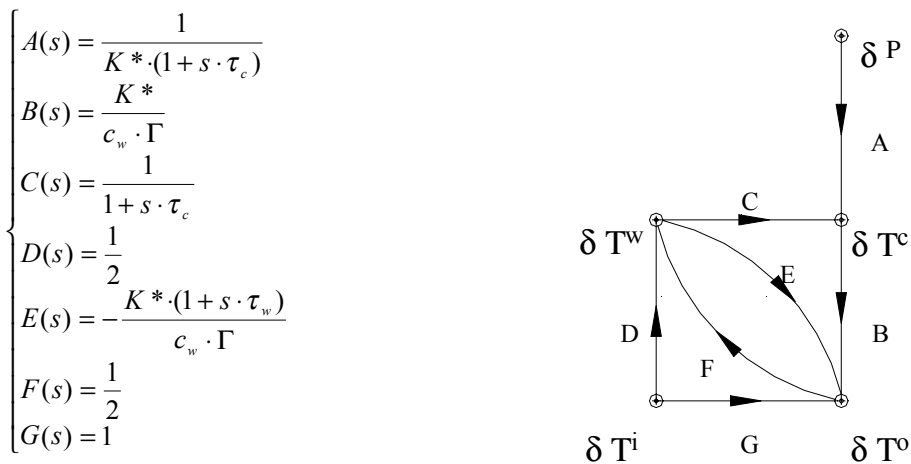

$$\begin{cases} A(s) = \dfrac{1}{K*\cdot(1 + s\cdot\tau_c)} \\ B(s) = \dfrac{K*}{c_w\cdot\Gamma} \\ C(s) = \dfrac{1}{1 + s\cdot\tau_c} \\ D(s) = \dfrac{1}{2} \\ E(s) = -\dfrac{K*\cdot(1 + s\cdot\tau_w)}{c_w\cdot\Gamma} \\ F(s) = \dfrac{1}{2} \\ G(s) = 1 \end{cases}$$

**Figure 5.25** _Signal-flow diagram (for small perturbations) can describe the thermal behaviour of a tank between the input and output variables [C.57]._

These relations, involved in the kinetics of the RF structure and including the feedback, are represented schematically by the signal-flow diagram of figure 5.25. Each branch linking two nodes represents the transfer function linking the corresponding variables.

The above equations represent a typical linear dynamic system[8], written in general terms as:

$$\begin{cases} \vec{x}(t) = A \cdot \vec{x}(t) + B \cdot \vec{u}(t) \\ \vec{y}(t) = C \cdot \vec{x}(t) + D \cdot \vec{u}(t) \end{cases} \quad \text{where} \quad \vec{u}(t) = \begin{bmatrix} P \\ T_i \end{bmatrix}, \quad \vec{x}(t) = \begin{bmatrix} T_c \\ T_w \end{bmatrix}, \quad \vec{y}(t) = \begin{bmatrix} T_c \\ T_w \\ T_o \end{bmatrix}$$

We must note that the approach mentioned above is applied only to a single tank/cooling sub-system, where a linear behaviour is present. However the method is more general and it should be used to describe also different tanks in series. This generates a typical non-linear dynamic system, which could be solved by linearization[9].

---

[8] A linear dynamic system of n variables is a process that can be described with the following equations [C.57]:

$$\begin{cases} \vec{x}(t) = A \cdot \vec{x}(t) + B \cdot \vec{u}(t) \\ \vec{y}(t) = C \cdot \vec{x}(t) + D \cdot \vec{u}(t) \end{cases}$$

where $\vec{x}(t)$ is the state variables vector, $\vec{u}(t)$ is the input vector and $\vec{y}(t)$ is the output vector.

Fixing the boundary conditions $\vec{x}(0) = \vec{x}_0$, then the solution can be expressed as:

$$\vec{x}(t) = e^{A \cdot t} \cdot \vec{x}_0 + \int_0^t e^{A(t-\xi)} \cdot B \cdot \vec{u}(\xi) \, d\xi \; .$$

Laplace transforming the above system, it is also possible to describe the dynamic system in terms of input-output parameters in the complex domain ($s = \sigma + i\varpi$). In fact by using the well known properties of the Laplace transform, one has:

$$L[x(t)] = X(s), \quad L[y(t)] = Y(s) \rightarrow \begin{cases} s \cdot X(s) - x(0) = A \cdot X(s) + B \cdot U(s) \\ Y(s) = C \cdot X(s) + D \cdot U(s) \end{cases}$$

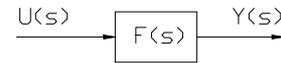

$$Y(s) = \left[ C \cdot (sI - A)^{-1} \cdot B + D \right] \cdot U(s) = \frac{N(s)}{D(s)} \cdot U(s) = F(s) \cdot U(s)$$

where F(s) is defined as the transfer function of the dynamic system.

[9] The non-linear input-output system can be described with the following equations:

$$\begin{cases} \vec{x}(t) = f(\vec{x}(t), \vec{u}(t)) \\ \vec{y}(t) = g(\vec{x}(t), \vec{u}(t)) \end{cases}$$

We define $\vec{x}_0$ as the initial stable configuration, and $\vec{x}_{eq}$ as the equilibrium point vector of the system after a small perturbation. We suppose to apply a perturbation $\vec{x} = \vec{x}_0 + \delta\vec{x}, \quad \vec{u} = \vec{u}_0 + \delta\vec{u}$, and we analyse the system behaviour into the iper-sphere $|\vec{x} - \vec{x}_0| = \eta$ around $\vec{x}_0$.

We write then the above system as:

$$\begin{cases} \delta\vec{x}(t) = f(\vec{x}_0 + \delta\vec{x}, \vec{u}_0 + \delta\vec{u}) = f(\vec{x}_0, \vec{u}_0) + \dfrac{\partial f}{\partial \vec{x}} \bigg|_{\substack{\vec{x}=\vec{x}_0 \\ \vec{u}=\vec{u}_0}} \cdot \delta\vec{x} + \dfrac{\partial f}{\partial \vec{u}} \bigg|_{\substack{\vec{x}=\vec{x}_0 \\ \vec{u}=\vec{u}_0}} \cdot \delta\vec{u} \\[4ex] \delta\vec{y}(t) = g(\vec{x}_o + \delta\vec{x}, \vec{u}_0 + \delta\vec{u}) = g(\vec{x}_0, \vec{u}_0) + \dfrac{\partial g}{\partial \vec{x}} \bigg|_{\substack{\vec{x}=\vec{x}_0 \\ \vec{u}=\vec{u}_0}} \cdot \delta\vec{x} + \dfrac{\partial g}{\partial \vec{u}} \bigg|_{\substack{\vec{x}=\vec{x}_0 \\ \vec{u}=\vec{u}_0}} \cdot \delta\vec{u} \end{cases}$$

We define $A_0$, $B_0$, $C_0$ and $D_0$ as the Jacobians of f and g in $\vec{x}_0$ and $\vec{u}_0$ respectively.

Anyway the above system for small perturbations and in terms of input-output variables is:

$$\delta \; \vec{y}(s) = F(s) \cdot \delta \; \vec{u}(s) = \left[ C \cdot (s \cdot I - A)^{-1} \cdot B + D \right] \cdot \delta \; \vec{u}(s)$$

where F(s) is the transfer function, while the matrixes of the system are described by:

$$A = \begin{bmatrix} -\dfrac{1}{\tau_c} & \dfrac{1}{\tau_c} \\ \dfrac{1}{\tau_w} & -\left(\dfrac{1}{\tau_w} - \dfrac{2}{\tau_0}\right) \end{bmatrix}, \quad B = \begin{bmatrix} \dfrac{1}{K * \tau_c} & 0 \\ 0 & \dfrac{2}{\tau_0} \end{bmatrix}, \quad C = \begin{bmatrix} 1 & 0 \\ 0 & 1 \\ 0 & 2 \end{bmatrix}, \quad D = \begin{bmatrix} 0 & 0 \\ 0 & 0 \\ 0 & -1 \end{bmatrix}$$

The equilibrium condition can be found analytically, imposing equal to zero the first equation of the system:

$$\dot{\vec{x}} = A \cdot \vec{x} + B \cdot \vec{u} \;\; \rightarrow \;\; 0 = A \cdot \vec{x}_{eq} + B \cdot \vec{u}_0$$

The equilibrium points vector is then:

$$\vec{x}_{eq} = -A^{-1} \cdot B \cdot \vec{u}_0 = \begin{bmatrix} \dfrac{P_0}{K *} \cdot \left(1 + \dfrac{\tau_0}{2 \cdot \tau_w}\right) + T_{i0} \\ \dfrac{P_0 \cdot \tau_0}{2 \cdot K * \tau_w} + T_{i0} \end{bmatrix} \quad (5.1)$$

Explaining the dynamic phenomena, with the relative input and output vectors and the matrixes, one finds:

$$\begin{bmatrix} \delta \; T_c(s) \\ \delta \; T_w(s) \\ \delta \; T_o(s) \end{bmatrix} = \underbrace{\frac{\begin{bmatrix} \left(s + \dfrac{1}{\tau_w} + \dfrac{2}{\tau_c}\right) \cdot \dfrac{1}{K * \tau_c} & \dfrac{2}{\tau_0 \cdot \tau_c} \\ \dfrac{1}{K * \tau_c \cdot \tau_w} & \left(s + \dfrac{1}{\tau_c}\right) \cdot \dfrac{2}{\tau_0} \\ \dfrac{2}{K * \tau_c \cdot \tau_w} & \left(s + \dfrac{1}{\tau_c}\right) \cdot \dfrac{4}{\tau_0} - \det(sI - A) \end{bmatrix}}{\det(sI - A)}}_{F(s)} \cdot \begin{bmatrix} \delta \; P(s) \\ \delta \; T_i(s) \end{bmatrix} \quad (5.2)$$

The above system could be then written by a linear form as:

$$\begin{cases} \delta\dot{\vec{x}}(t) = A_0 \cdot \delta\vec{x}(t) + B_0 \cdot \delta\vec{u}(t) + \dot{\vec{x}}_0(t) \\ \delta\vec{y}(t) = C_0 \cdot \delta\vec{x}(t) + D_0 \cdot \delta\vec{u}(t) \end{cases} \Rightarrow \begin{cases} s \cdot \delta\vec{x}(s) = A_0 \cdot \delta\vec{x}(s) + B_0 \cdot \delta\vec{u}(s) + \vec{X}_0 / s \\ \delta\vec{y}(s) = C_0 \cdot \delta\vec{x}(s) + D_0 \cdot \delta\vec{u}(s) \end{cases}$$

that can be represented with the following signal-flow diagrams, simplified by the well-known Mason rules [C.57]:

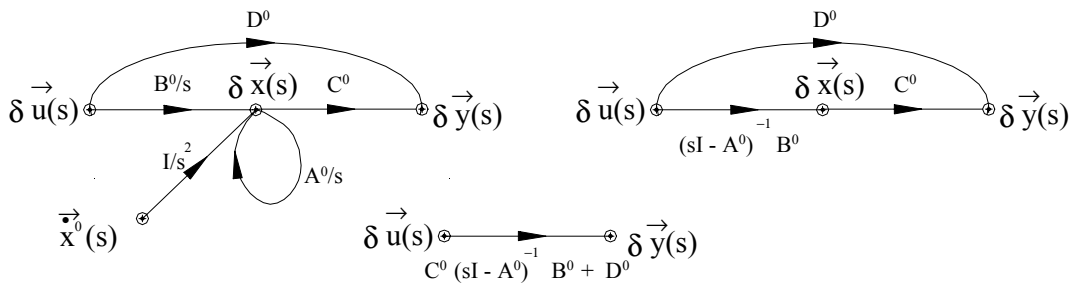

Finally we have the same results found for the linear system, but with the transfer function F(s) written with the Jacobians in a very general term. This approach is then valid inside of any iper-sphere η and for any small perturbation around a point vector $\vec{x}_0$ (Enrico Fermi Institute for Nuclear Studies, Politecnico Milan).

The stability of the system[10] can be analysed [C.57] from the poles of the transfer function F(s) and from the behaviour of the feedback transfer function G(s)[11]. In fact the basic characteristic of the transient response of a close loop system is related to the location of the close-loop poles. If the system has a variable loop gain, then the location of the closed-loop poles depends on the value of the loop gain g chosen. It is important therefore to know how the close-loop poles move in the s plane as the loop gain is varied. This can be known using the root-locus method, in which the roots of the characteristic equation are plotted for all values of a system gain. The poles of F(s) are defined by:

$$\det(sI - A) = 0 \quad \rightarrow \quad \left(s + \frac{1}{\tau_c}\right) \cdot \left(s + \frac{1}{\tau_w} + \frac{2}{\tau_0}\right) - \frac{1}{\tau_w \cdot \tau_0} = 0$$

while the feedback transfer function G(s) is:

$$G(s) = \frac{1}{\tau_c \cdot \tau_w} \cdot \frac{1}{\left(s + \frac{1}{\tau_c}\right) \cdot \left(s + \frac{1}{\tau_w} + \frac{2}{\tau_0}\right)}$$

---

[10] Let define a dynamic system in general terms as: $\vec{x}(t) = f(\vec{x}(t), \vec{u}(t))$. The input vector is $\vec{u} = const$ and the equilibrium point vector $\vec{x}_{eq}$ is defined as $f(\vec{x}_{eq}, \vec{u}) = 0$. Different basic stable configurations of the equilibrium point in the phase space are then reported below as a reference [Liapunov and C.57, C.58].

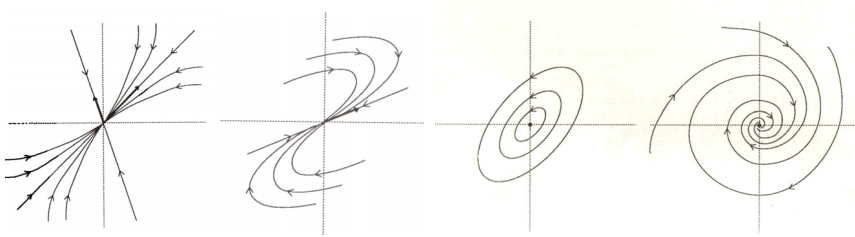

[11] Let consider a dynamic system, described in terms of its transfer function F(s):

$$F(s) = \frac{G(s)}{1 + G(s)}$$

and a feedback loop inside, as illustrated in the following figure.

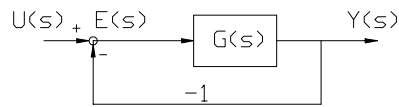

$G(s)$ is defined as the feedback transfer function:

$$G(s) = g \cdot \frac{(s - z_1) \cdot (s - z_2) \cdot \ldots \ldots \cdot (s - z_m)}{(s - p_1) \cdot (s - p_2) \cdot \ldots \ldots \cdot (s - p_n)} .$$

The general problem for stability of the system is connected to the poles of the transfer function F(s) [C.57].
One can prove that the critical value $g_{cr}$, where the system is unstable, is the point where the root locus cuts the axis j$\omega$, penetrating in the right quadrants of the complex variable s ($\sigma$+i$\omega$). The $g_{cr}$ value can then be calculated by:

$$|G(s)| = 1; \quad \rightarrow |g_{cr}| = \frac{|\overline{s} - p_1| \cdot |\overline{s} - p_2| \cdot \ldots \ldots |\overline{s} - p_n|}{|\overline{s} - z_1| \cdot |\overline{s} - z_2| \cdot \ldots \ldots |\overline{s} - z_m|} .$$

The system is stable for $|g| < |g_{cr}|$.

The root locus is then shown in figure 5.26. If the real part of the poles of the feedback transfer function G(s) is positive (i.e. if the poles are roots in the right-half plane of the plot of Im(s) versus Re(s)), then the temperatures of the system will grow exponentially with time, thus indicating an unstable response to the applied RF power perturbation. If, on the other hand, the real part of the poles are negative (i.e. they lie in the left half-plane), the roots will produce a contribution to the changes of temperatures that decay with time. The stability of the system is then calculated from the gain analysis of the root locus [C.57, C.58]:

$$g_{cr} = \frac{1}{\tau_c} \cdot \left( \frac{1}{\tau_w} + \frac{2}{\tau_0} \right) \; \rightarrow \; g_{cr} = g + \frac{2}{\tau_c \cdot \tau_0} \; \rightarrow \; always \; g < g_{cr}$$

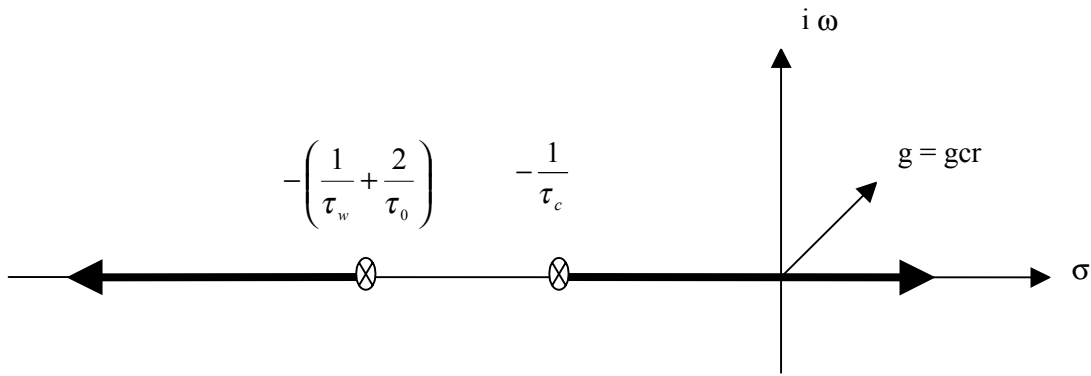

**_Figure 5.26_** _Root locus describing the thermal behaviour of the accelerating tank._

Due to the fact that the effective gain g for LIBO is $< g_{cr}$, then the system (which describes the thermal behaviour of the tank under operation) is stable in any conditions, also varying the input parameters, such as RF power P and the input water temperature $T_i$.

Measuring the output parameters (temperatures of copper body $T_c$, and cooling water at the input and output position $T_i$ and $T_0$), with simple thermocouples positioned on the prototype, it is then possible to measure indirectly the exact injected RF power in the accelerating cavity, also without the help of RF pick-up loops. Moreover, if thermal stable conditions can be easily reached, one can confirm the possibility to maintain at different stable temperatures consecutive accelerating tanks. This then analytically proves that it is possible to obtain fine RF tuning of accelerating cavities by changing the cooling water flows in each single tank.

To verify all these results, extensive measurements of LIBO thermal behaviour under RF power will be presented in chapter 7.

*Chapter 6*

*Fabrication of the LIBO-62 prototype and*

*low power RF measurements*

## 6.1 Introduction

Prototyping is mandatory to transform a conceptual design into a successful industrial product. However in this context it is difficult that a single external firm can provide technical support, especially when small quantities do not financially justify industrial interest. For these reasons, and at this stage, CERN has played a crucial role, representing a distinctive asset and providing full traceability of the fabrication steps in a flexible environment. Figure 6.1 shows the main sequences for the prototype construction performed at CERN in the period 1998-2000. Before to start the full production an R&D has been performed to qualify processes such as machining sequences, RF measurements, brazing thermal cycles as well as final design of the RF cavities and the material selection. Most of these engineering problems are further aggravated by the extensive use of OFHC copper with its very poor mechanical properties. Moreover, due to the difficulty to achieve the required precision of the high frequency resonators, detailed RF measurements and tuning have been performed in parallel to the full module construction. High accuracy in cavity frequency will assure high uniformity of the accelerating field on beam axis, in accordance to the design specification.

Finally the development of the prototype is both challenging and unique from the physics and engineering standpoint and it can be considered the first accurately fabricated SCL at 3 GHz studied so far.

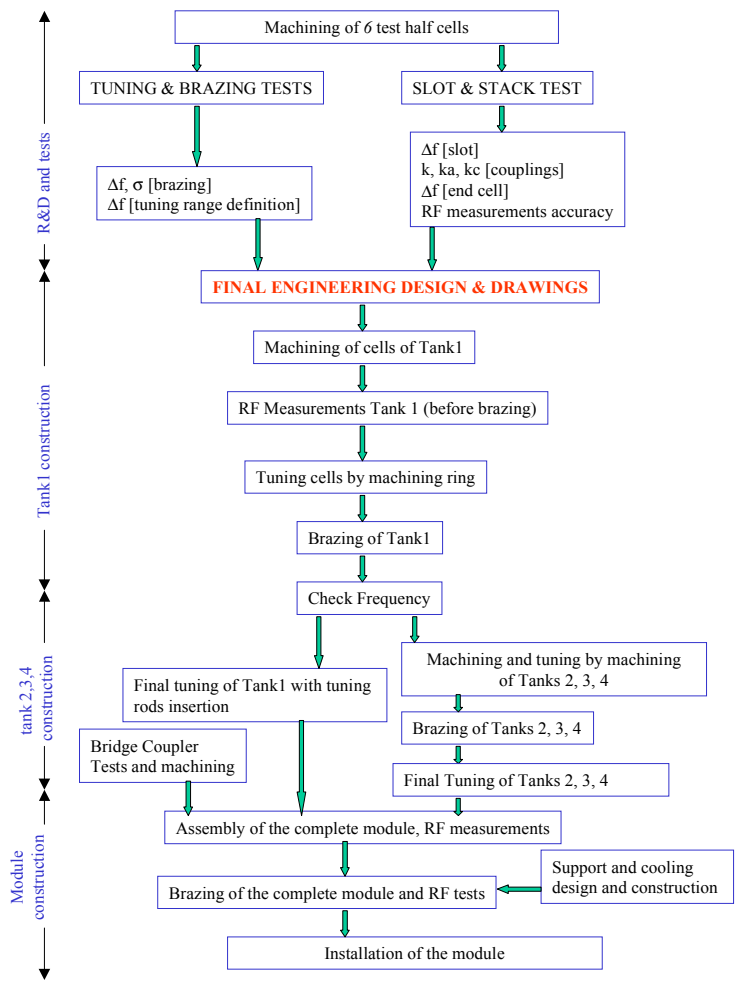

**Figure 6.1** *Main construction sequences of LIBO prototype built at CERN in the period 1998-2000.*

## 6.2    Fabrication of the LIBO-62 prototype[1]

### 6.2.1    Materials used for the prototype construction*

❑    _Copper material_

Copper is the highest electrical conductor of the commercial metals [C.49, C.50]. For this reason conductivity of materials are based on an IACS (International Annealed Copper Standard) rating. A 100% IACS rating is assigned to annealed copper with a volume resistivity of 0.017241 Ohm-mm$^2$ per meter at 20°C. Moreover copper is also the highest thermally conductive metal. It has good fabricability and it is easily brazed and soldered. This material is primarily chosen for the electrical conductivity. Since impurities impair the electrical conductivity, these are kept to a minimum. Nonetheless some impurities are intentionally added to achieve certain characteristics. High purity copper is a very soft metal. Cold and hot working are performed to increase the physical characteristics: cold working increases both tensile strength and yield strength while hot working is normally used to increase ductility above the recrystallization temperature. Work-hardened metal can be returned to a soft state by heating or annealing, where deformed and highly stressed crystals are transformed into unstressed crystals by recovery, recrystallization and grain growth. Heat treatment processes are also applied to achieve stress relieving after pre-machining.

The reference material used for LIBO is the high pure oxygen free copper (OFHC), indicated with C10100 and C10200 [C.49, C.50]. It is composed by 99.99% of copper and 0.0005% of oxygen and can be obtained by melting under nonoxidizing atmospheres or by adding deoxidizers such as phosphorus to the melt. Normally the OFHC copper is used for high conductivity needs coupled with ductility, low gas permeability, freedom from hydrogen embrittlement and low out gassing tendency.

Where possible (especially for small pieces brazed on OFHC copper) the electrolytic tough pitch copper (ETP) is also used for economical reasons. Its normal processing leaves about 0.04% oxygen in combination with copper as CuO. This level of impurity does not measurably lower the conductivity (it still exhibits a conductivity of 100% IACS), but it lowers a little bit the ductility.

Where pure copper is required with good resistance and very high ductility (for example for RF contacts or pick up loops) then copper-beryllium alloy is selected. This is grouped in the heat-treatable precipitation-hardening alloys, and in particular in the red and gold alloys that contain from 0.2 to 0.7%, and from 1.6 to 2.0% beryllium, respectively. Normally they have yield strengths from 170-200 N/mm$^2$ to 550-700 N/mm$^2$ with no heat treatment, and more than 900-1350 N/mm$^2$ after aging.

Optical microscopy is used for the examination, identification and classification of metallographic structures. Image analysis is used when simple visual observation is insufficient.

---

[1] The engineering subjects of LIBO have been performed in co-operation with Dr. Balazs Szeless at CERN in the period 1998-2002.  Thanks again.

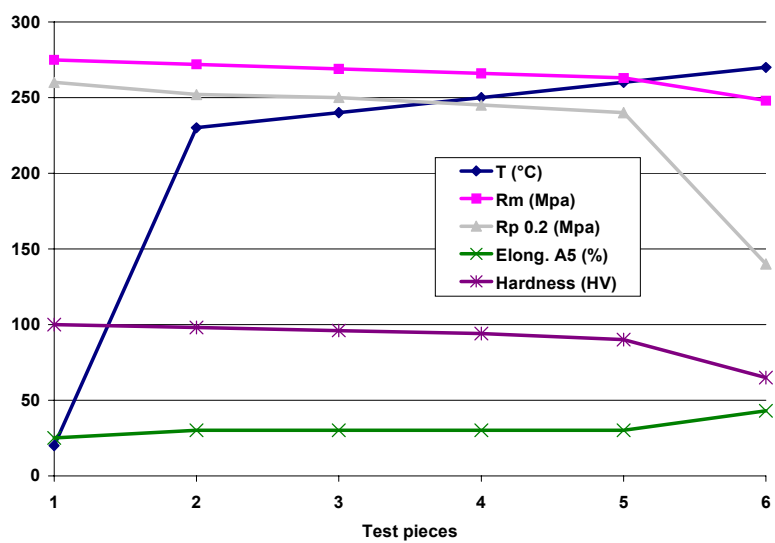

**Figure 6.2** *Summary of mechanical properties of OFHC copper obtained on six test pieces. 1) not heat treated, 2) heat treated at a temperature T of 230°C, 3) 240°C, 4) 250°C, 5) 260°C, 6) 270°C. The changes of the mechanical properties under stress relieve heat treatment are relevant and they must be taken into account for further machining (Rm: ultimate strength, Rp 0.2: elastic limit, A5: elongation (%), Hardness in HV-Vickers unit [C.54]) (CERN Surface and Material Laboratory).*

The purest form copper is a relatively simple structure, consisting of one phase only, except for the oxide presence. Normally oxygen is present in small amounts in the form of cuprous oxide ($Cu_2O$) which forms with copper a eutectic with a reduced melting point (1065°C) positioned at grain boundaries. Commercial copper, which contains impurities, is of a more complex nature. A considerable quantity of some of these elements, such as iron, nickel, etc. can be held in solid solution, and consequently their presence is not apparent on the microscopic examination. The solid solubility in copper of some other elements is negligible or extremely limited and their presence results in a separate phase, which can be detected microscopically. While the structure of copper as cast is comparatively coarse, mechanical working results in refinement and worked copper in the soft condition usually has small equiaxial grains. Cold working results in distortion of this equiaxial structure, so that the grains are elongated in the direction of working. For these reasons grain orientation must be defined for raw material selection. The amount of oxidation which occurs when copper is heated in air (as the stress relieve) is largely a function of temperature and it has been visible that while oxidation is slight at low temperature, there may be rather serious scaling at high temperatures. On the other hand, under certain conditions the oxides formed are adherent and so result in a diminished rate of attack. Two principal types of oxides are formed upon copper during heating: the black scale is cupric oxide (CuO), and the underlying adherent oxide, which has a characteristic red colour, is cuprous oxide ($Cu_2O$). All these aspects must be taken into consideration to prepare the copper surface correctly and in accordance to vacuum and RF requirements of the accelerating cavity.

As already mentioned to reduce possible deformations of pre-machined copper, stress relieve treatment in air must be performed. The changes of mechanical properties under this heating have been analysed at CERN at different temperatures and summarised in figure 6.2. Consecutive high temperature treatments, such as stress relieve and brazing cycles change the form and size of grains, as illustrated in figures 6.4 and 6.5. Short-time mechanical properties at elevated temperatures are available in literature and reported in figure 6.3 as a reference. These show how the tensile strength and grain sizes are strongly connected to the temperature. It is also evident that at high temperature the mechanical properties of copper can change drastically. Several tests have been also performed at CERN to check these properties under heating in vacuum ($10^{-5}$ mbar) in order to simulate stress relieve and high temperature brazing. The results show that the crystallographic structure of copper at 220-230 °C does not present drastic changes, so this temperature range can be used for final production to avoid deformation during machining. However above a temperature of about 550 °C the grains are irreversibly distorted and grown up to 1 mm [C.48]. Special silver-copper alloy has been also studied to reduce the grain size under high temperature treatment, but without success (figure 6.6).

As a consequence of these results, *particular care is then mandatory during the cavity design, trying to reduce as much as possible the brazing steps performed on each sub-component of the accelerator.*

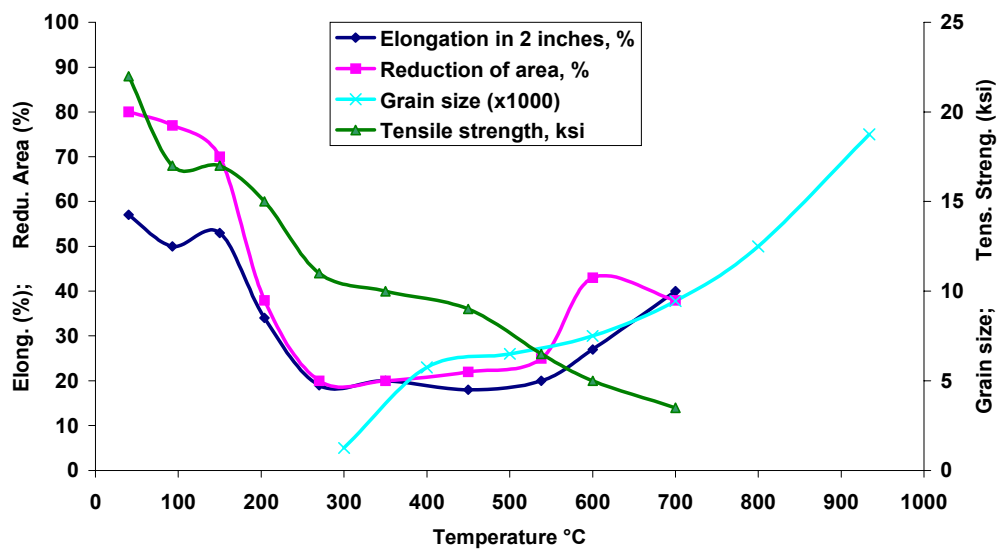

***Figure 6.3** Mechanical properties of OFHC copper under high temperature heat treatments.*

❑   *Stainless steel material*

Stainless steel is characterised by at least 10.5% chromium addition to iron. Normally the maximum chromium content is about 30% and the minimum iron content is 50%. The carbon is normally present for less than 0.03% to a maximum of 1.2%. Stainless steel characteristic arises from the formation of an invisible and very adherent chromium rich oxide surface film. This puts the steel in a passive state and when the film is breached, it immediately heals when oxygen is present. The standard AISI  (American Iron and Steel Institute) designation is the older and more widely used and consists of three-digit number. Generally speaking the 300 grades are the austenitic stainless steel. The most common austenitic stainless steel is the 304 grade.

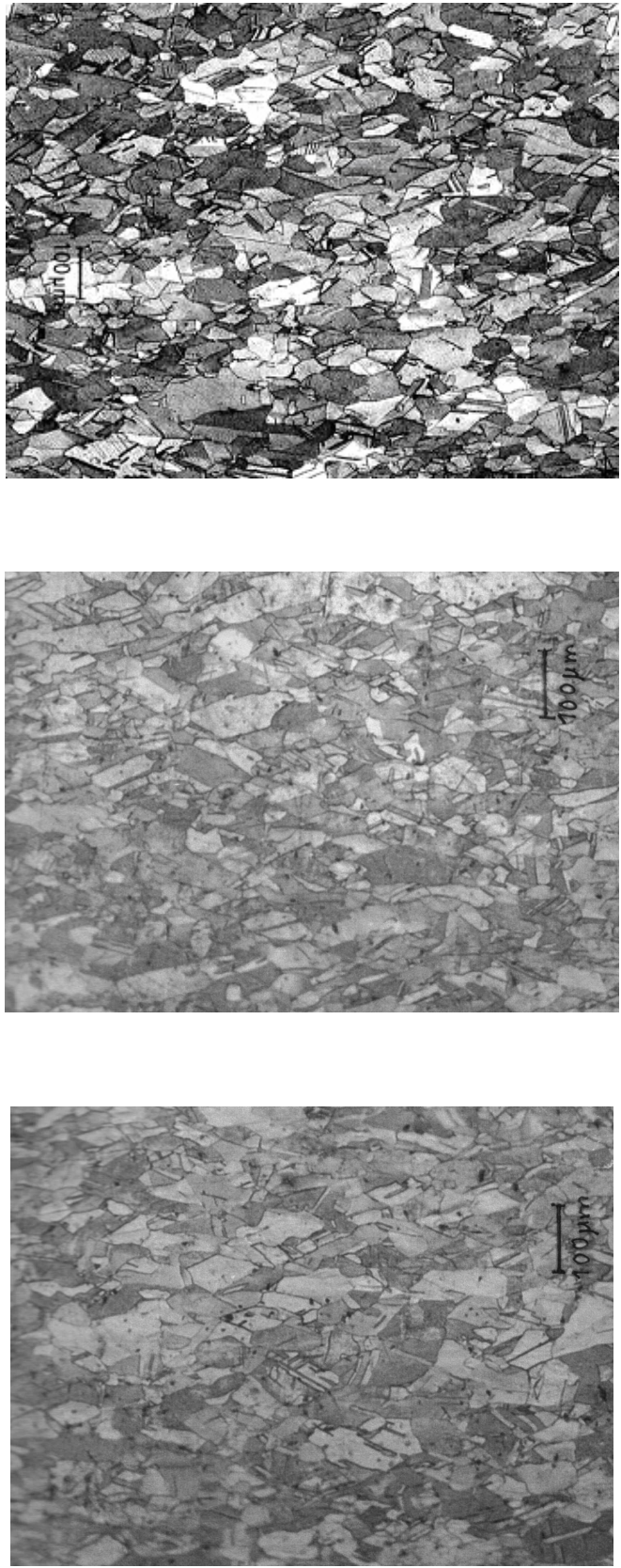

**Figure 6.4** *Crystallographic analysis of laminated oxygen free copper, after heat treatments at 180, 200 and 220 °C, used to simulate the stress relieve during production of LIBO prototype (CERN Surface Laboratory). Form the pictures the grain size, as a function of the temperature, is visible as well as the regularity of the crystallographic structure of high pure copper.*

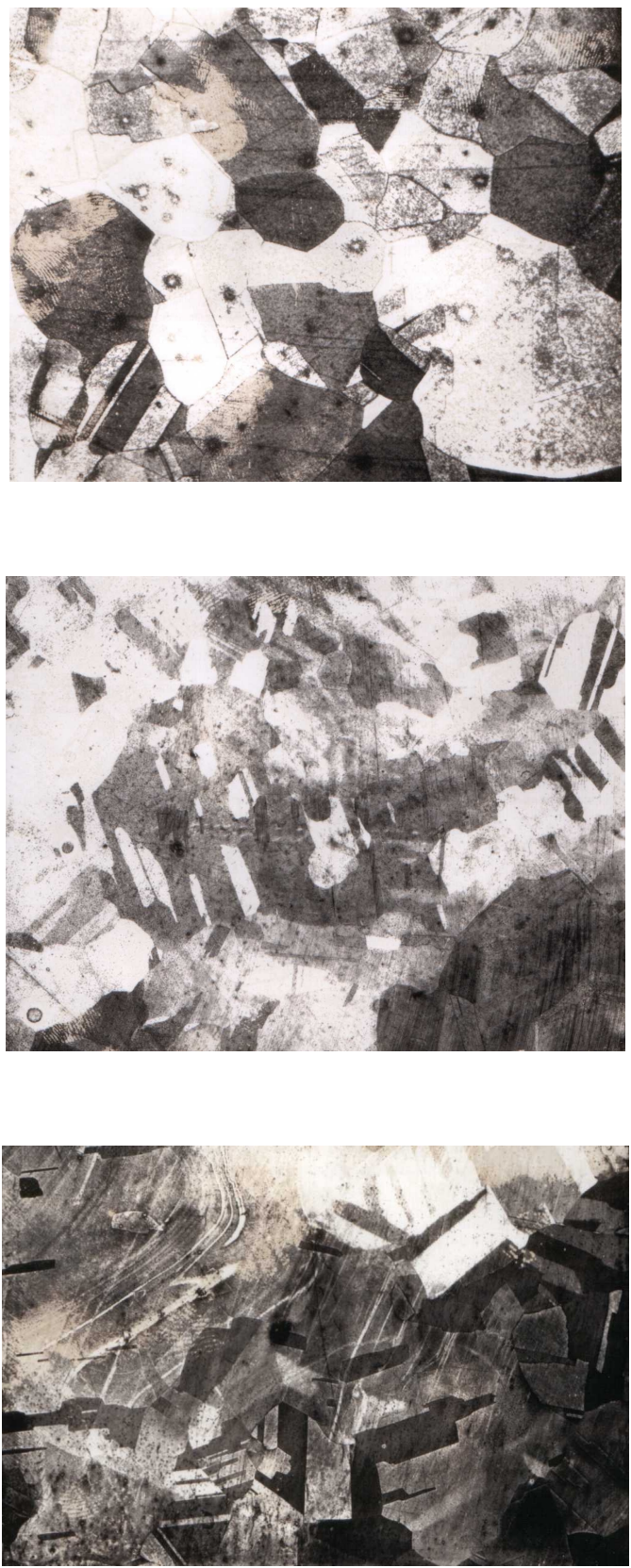

**Figure 6.5** *Crystallographic analysis of laminated oxygen free copper after three heat treatments at 750, 800 and 850 °C, used to simulate the brazing treatments on LIBO prototype (CERN Surface Laboratory).*

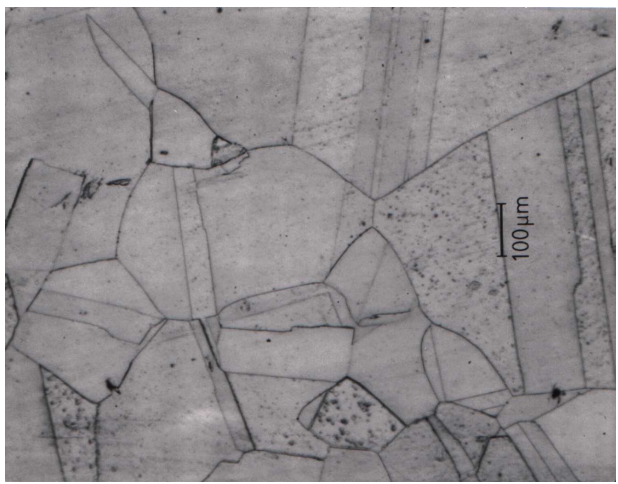

***Figure 6.6***  *Special silver-copper alloy studied for LIBO to reduce the grain sizes under high temperature brazing. In general terms high pure copper has regular crystals, with the borders well defined; few impurities stay on the edge of the grain. Figure shows the results after three treatments at 750, 800 and 850 °C. It is evident that, at high temperature, the crystallographic structure, also for this special alloy, is drastically compromised (CERN Surface Laboratory).*

This has the face centered cubic (FCC) structure at room and high temperature. Nickel and manganese, also present as added elements, are the principal austenite formers, although carbon and nitrogen are also used because they dissolve readily in austenite. A wide range of corrosion resistance can be achieved by balancing the stabilisers, such as chromium and molybdenum, with the austenite stabilisers. Molybdenum is added to type 304 to produce type 316 for enhanced pitting corrosion resistance and more stability at high temperature. Modifications of these grades are normally performed to achieve particular properties. Lower carbon grades (indicated with the L suffix) prevent intergranular corrosion. The time-temperature-carbide precipitation (sensitisation) curves of austenitic stainless steel shows that the L grade with a maximum 0.03% C has its carbide precipitation curve pushed to the right and therefore allows the austenite to be cooled slow without being sensitised (no carbide precipitation).

For practical application 304 L stainless steel is used for all the high vacuum components not brazed on the copper pieces of the LIBO prototype. 316 L stainless steel is more resistant and stable during high temperature thermal cycles and then is used for all vacuum flanges brazed on LIBO.

❑   *Brazing process: metallurgical aspects of the filler metal alloys and their correlation with OFHC copper and stainless steel*

Copper is probably the most easily brazed metal available, and a wide range of brazing filler metals can be used. Generally speaking the selection of the brazing process and the filler metal depends on the alloy or material composition, the shape and dimension of the parts to be joined. Moreover electrical conductivity is an important factor, and the physical characteristic across the joint must be controlled. Therefore, the finished metallurgical structures, grain sizes, and mechanical properties of the specific copper alloy must be carefully considered when brazing is utilised.

The metallurgical structures of brazed joints are largely those of the brazing filler metal, enriched with copper from the parent material. Because some surface melting of the copper alloys being joined occurs, the interfaces usually show primary dendrites of a copper-rich phase growing from the copper alloy and braze metal interface. The brazing filler metal becomes more copper-rich, producing additional dendritic growth with increased brazing temperature and time.

Possible oxide films on the copper surfaces slow or prevent the flow of the brazing filler metal. For this reason, *cleaning operations are needed to remove the surface oxides prior to brazing*.

To prevent extra oxidation during heating the brazing operation is normally carried out in either a protective atmosphere or one that has essentially reducing conditions. *For LIBO, brazing in high vacuum has been selected as the most promising solution*. In fact vacuum brazing of copper has several advantages with respect to hydrogen atmosphere, such as its cleanness and brightness, providing at the same time an excellent surface for RF currents. Generally speaking at high temperature one must take into account also the interaction between the material and the external atmosphere, and the correspondent negative consequences. In fact brazing with hydrogen is normally above 700°C, which allows the dispersed copper oxides in the material to react with the atmosphere, creating a deleterious high-pressure steam within the solid metal. Due to the fact that in the past wide experience with hydrogen ovens has been developed for linac construction (especially at Los Alamos National Laboratory), today OFHC copper is universally used for these purposes: the main advantage is that it can be brazed even in hydrogen-containing atmospheres with no embrittlement.

The brazing of OFHC copper to stainless steel presents additional complications in the form of little wetability and stress problems due to their differences in thermal expansion. The stainless steel, the brazing alloy or the copper must yield to prevent failure during the cooling cycle or the reheat for the next brazing step. Moreover the filler metal must be compatible with both the steel and copper. In a hydrogen atmosphere the lowest temperature at which stainless steel oxides can be reduced with hydrogen is near to the upper limit to which copper may be heated safely, limiting the number of available alloys that can be used for brazing unless steps are taken to preclude the formation of stainless steel oxides. With vacuum brazing all these aspects are then solved.

Other main advantages of vacuum brazing are: increased strength of brazed joints, increased wetting and flow of brazing alloy, and cleaner assemblies. In order to increase the wetability on stainless steel, the surfaces are flashed with approximately few microns of nickel plating (see next paragraph).

In connection with these considerations different tests have been performed to control the joints of OFHC copper pieces after brazing. Tests in air show a good penetration with few imperfections in the brazing interface, while high vacuum generates excellent results in terms of uniformity and brazing alloy control with respect to the crystallographic structure of the copper (figure 6.7).

For the final design a variety of brazing alloys could be used, such as copper, silver and nickel-base alloys [C.51]. Each brazing alloy has its own particular metallurgical characteristics, such as wetability, sluggishness, blushing, tendency to liquate, flow and filleting nature.

The brazing alloys with different melting points, used for LIBO are then the following [D.52, D.54]:

- Type I: Ag 67.5 %, Cu 22.5 %, Pd 10 %; B-Ag67CuPd-834/840 (cf. ISO 3677); purity > 99.99%.
- Type II: Ag 68.4 %, Cu 26.6 %, Pd 5 %; B-Ag68CuPd-807/810 (cf. ISO 3677); purity > 99.99%.
- Type III : Ag 72 %, Cu 28 %; B-Ag72Cu-780 (cf. ISO 3677); purity > 99.99%.
- Type IV : Ag 61 %, Cu 24 %, In 15 %; B-Ag61CuIn-685/730 (cf. ISO 3677); purity > 99.99%.

More information about brazing of the prototype will be presented in paragraph 6.2.3.

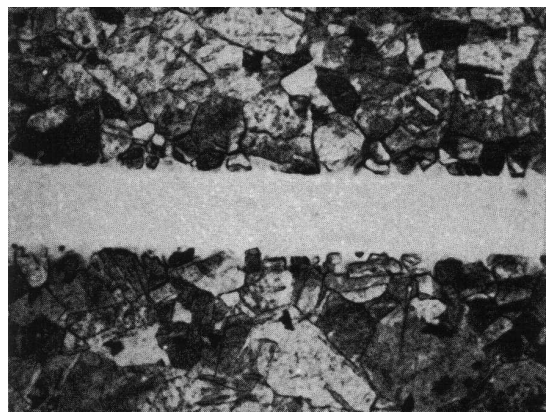

(a)

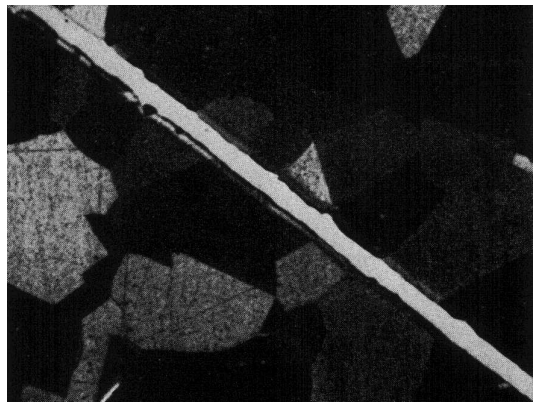

(b)

**Figure 6.7** *Crystallographic analysis of brazing joints on OFHC copper: figure (a) (x 200) shows joint for copper-copper (OFHC) brazing performed in <u>air</u> at 750 °C with copper-silver-phosphor alloy. In this case few imperfections are visible. Figure (b) (x 100) yields the joint results for brazing <u>under vacuum</u> at $10^{-5}$ mbar and 800 °C with silver-copper-palladium alloy. Here excellent results are obtained in terms of uniformity and controllability of the brazing alloy (CERN Surface Laboratory).*

### 6.2.2    Production of the LIBO-62 prototype

❑    *General aspects*

All LIBO components are produced with CNC milling and lathe machines as well as standard machines at CERN Central Workshop [D.52, D.55]. The main goal of LIBO construction is to use processes as conventional as possible, in order to facilitate the technology transfer to industry with a consequent cost reduction [C.46]. For the prototype no temperature control was used during production, while all the tolerances are applicable at workplace temperature of 20 °C. To maintain the same reference, independent from the ambient conditions, a step by step checking during production, especially for critical dimensions, was performed comparing the pieces on the machine with a reference precise pre-machined piece, which have passed a severe control of dimensions and roughness. The most critical parameter is the 0.02 mm planarity of the two brazing surfaces for the half-cell plates. This is important in order to assume an uniform distribution of brazing alloy, avoiding risks of vacuum leaks or discontinuity of electrical conductivity. For this reason pre-machining of the half-cell plates, with 0.2 mm of over dimensions, is followed by heat treatment for stress relieve. After several tests (table 6.1), it has been decided to use laminated OFHC copper for the half-cells production (for economical reasons) performing on the pre-machined pieces an heat treatment in air (chapter 4). This is a good compromise between an acceptable stress relieve and a reasonable crystallographic structure defined in terms of grain growth. After final machining a planarity better than 0.02 mm has been obtained.

**Table 6.1** *Machining test and planarity results on heat treated, laminated and forged half-cell copper plates (CERN).*

|  | Planarity of accelerating cavity surface [mm] | Planarity of coupling cavity surface [mm] |
|---|---|---|
| *OFHC forged copper* | 0.04 | 0.03 |
| *OFHC laminated copper heat treated at 180°C* | 0.10 | 0.07 |
| *OFHC laminated copper heat treated at 200°C* | 0.09 | 0.09 |
| *OFHC laminated copper heat treated at 220°C* | 0.07 | 0.11 |
| *OFHC laminated copper heat treated at 150°C* | 0.15 → 0.05 (after mach.) | 0.07 → 0.07 (after mach.) |
| *OFHC laminated copper heat treated at 250°C* | 0.02 → 0.03 (after mach.) | 0.10 → 0.05 (after mach.) |

For full production rapid steel and tools in widia were used as well as standard emulsion for copper and stainless steel. Ethyl alcohol is mandatory for any machining of pieces already cleaned or pre-brazed. The best-machined surface upon copper is obtained by means of high cutting speeds combined with light cuts. For turning and similar operations, speeds of about 250 to 800 ft per minute have been adopted. The most difficult constraints are the status of the heat-treated materials, especially in connection with the high quality surfaces. Generally speaking, copper in the work-hardened condition can be cut cleaner than the annealed material [C.47].

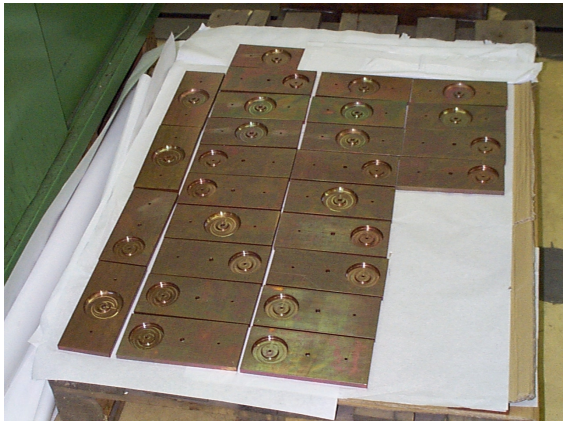 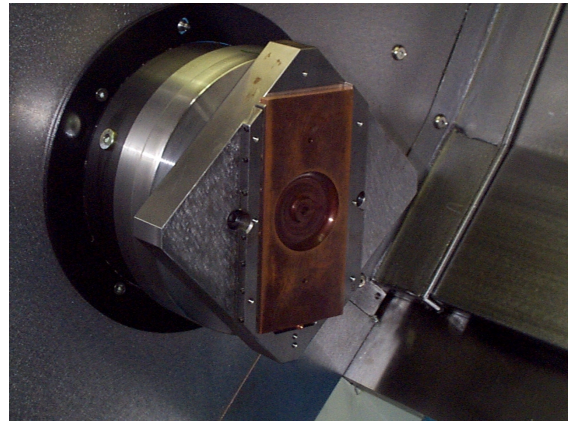

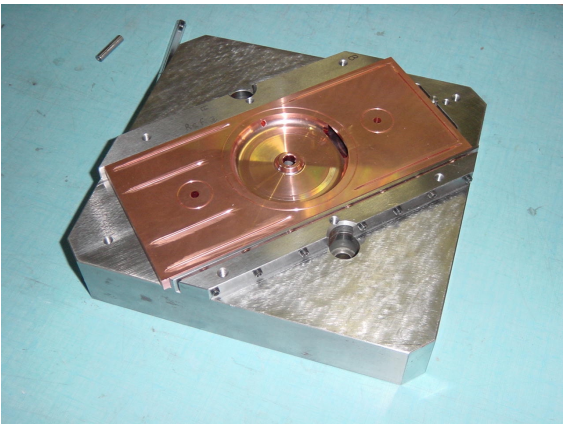 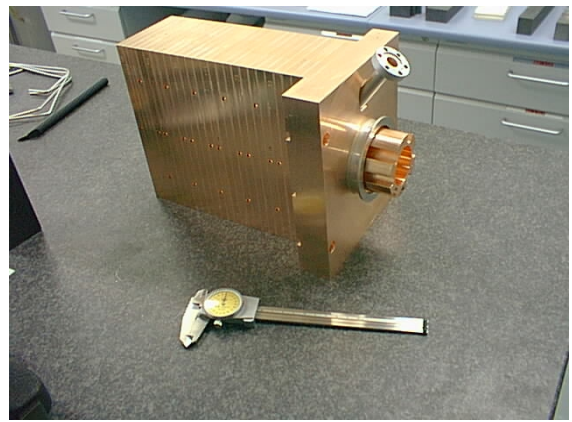

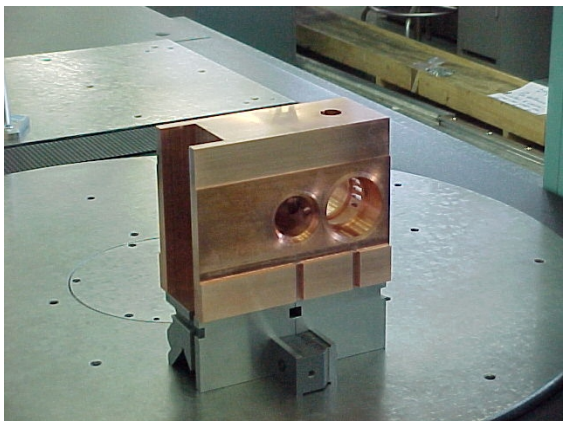 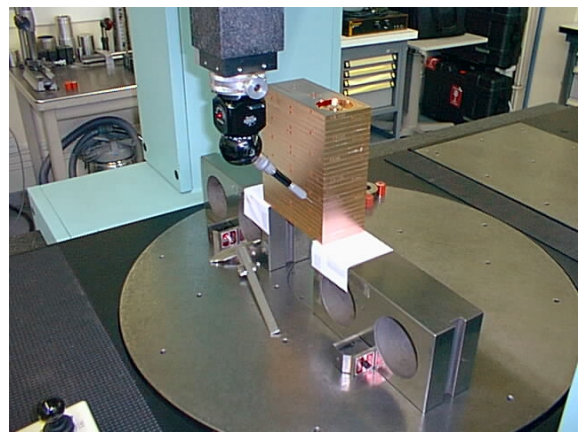

***Figure 6.8*** *Photo gallery of LIBO production at CERN Central Workshop in the period 1999-2001. For the prototype construction standard CNC lathe and milling machines have been used for cost reduction. Each part of the accelerator has passed a precise metrology control before and after RF measurements and brazing. From top to bottom and from left to right: full production of 150 pre-machined half-cell plates treated at 230 °C in air (note the surface oxidation of copper after treatment); half-cell plate on the lathe during production and after production on dedicate mask; tank 1 with the end cell already brazed and ready for RF measurements; bridge coupler body after production and ready for brazing; metrology tests of a tank after brazing [A.5, C.31, C.36, C.37, C.39, C.42, C.46, D.49, D.52].*

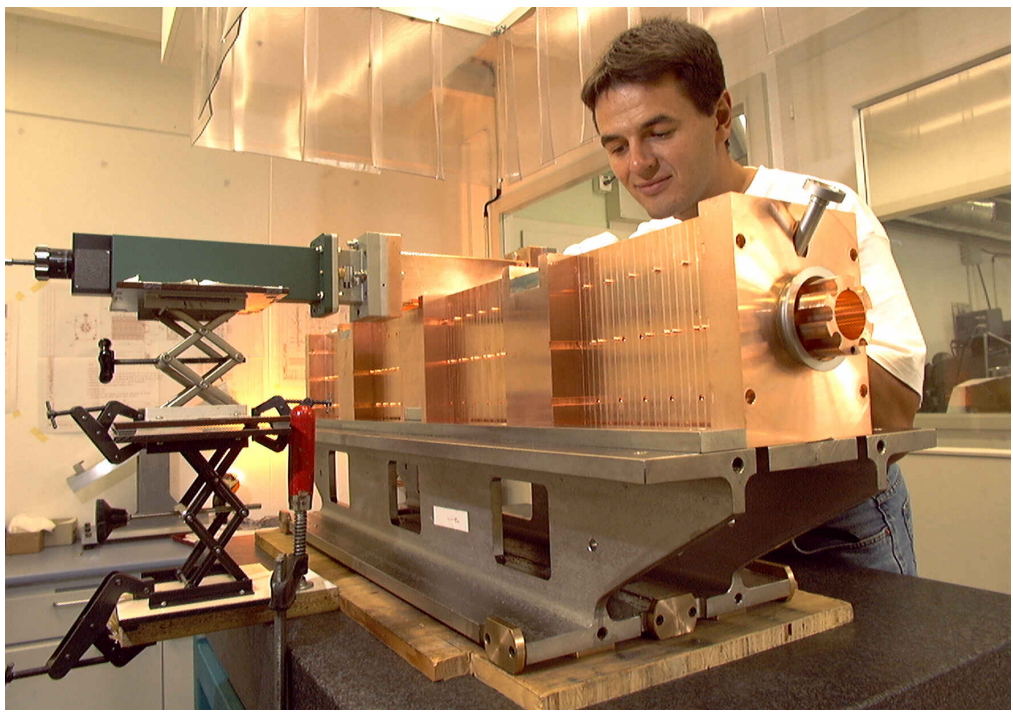

***Figure 6.9*** *LIBO-62 prototype pre-assembled on steel girder and ready for brazing.*

A key step during production was the tuning procedure of the cavity. Machining at CERN Workshop has been done in parallel with RF measurements for frequency tuning and field adjustment. The half-cell and end half-cell plates were put into a jig onto a lathe, after RF measurements, for cavities tuning. In particular the RF tuning is performed by machining of the tuning ring, according to RF measurements. A detailed description of production is reported in [D.49], while the full set of low power RF measurements are presented in paragraph 6.3. See also as a reference the minutes reported for the CERN Central Workshop meetings (references, D section). After production tests at CERN Metrology Laboratory have been made by using standard micrometers as well as a six-axis programmable machine. To declare ready the pieces for RF measurements and brazing assembly few microns out of tolerances have been chosen as the standard acceptance criteria.

❑ *Cleaning procedures\**

Chemical surface treatments are required in the fabrication process of much of the equipment used in ultra high vacuum, RF and high voltage systems [C.44]. The standard processes for cleaning use solvents or detergents. Normally speaking solvent cleaning can be performed in either the vapour or liquid phase, the difference being that in the vapour phase the cleaning agent is free of contamination from previous cleaning operations. In vapour phase solvent cleaning the part to be cleaned is suspended in a zone directly above a bath of boiling solvent. The solvent vapour condenses on the cold component dissolving any organic grease present on the surface. When aqueous detergents are used for cleaning, surface greases are removed and solubilised in water using fatty acid salts. At CERN ultrasonic agitation is applied to improve the cleaning efficiency. This produces a cavitation effect, which results in a mechanical action due to the implosion of microbubbles on the surface of the component to be cleaned. After these treatments, handling with gloves is mandatory and must be performed in a clean area, especially for the parts exposed to vacuum and RF fields. Several procedures have been developed at CERN for different materials (OFHC copper, stainless steel, copper-beryllium, ETP copper, etc.) [D.49, C.44][2]. All the processes are compatible with requirements of vacuum brazing ($10^{-5}$ mbar) for construction as well as RF fields (about 15 MV/m) and high vacuum (about $10^{-7}$ mbar) constraints for LIBO operating conditions.

❑   *Electrochemical procedures used for the prototype construction\**

Two main electrochemical processes have been used for LIBO construction: the metal plating and electrochemical polishing. The first is used for brazing between copper and stainless steel (see next section) in order to assure more wettability or electric contact between separate pieces inside the cavity not submitted to brazing welding. The general process is based on simple electrochemical plating by using the well-known galvanic processes. By proper design of the electrodes it is possible to obtain quite uniform layers on complex geometry. By using a pulsed current it is also possible to produce deposits ranging from few microns to millimetres, still keeping the grain structure of the plated layer under control. It is important to note that to obtain good adhesion of the plated metal it is necessary to apply complex preparatory treatments developed at CERN Surface Laboratory [C.44, D.49]. Electrochemical polishing is also used for surface preparation resulting in the dissolution of the metal surface, reducing the surface roughness and providing mirror like surface finishes. Electrochemical polishing is widely applied to prepare copper and stainless steel components to be used for high vacuum, such as the manifold and large vacuum tubes.

---

[2] For the construction, especially for high pure copper, the following chemical procedures are shown as a reference:
1.   Vapor degreasing in perchlorethylene ($C_2Cl_2$) at 121°C.
2.   Alkaline soak with ultrasonic agitation for 5 min in an alkaline detergent at 50°C (at CERN 1740S, supplied by Cleaning Technology, Nyon, Switzerland), and made up to a concentration of 10 g $l^{-1}$.
3.   Tap water rinse.
4.   Pickling in HCl (33%) 50% by volume with $H_2O$ 50% at room temperature for 1-5 min.
5.   Tap water rinse.
6.   At room temperature for 30-60 sec, acid etch in
     $H_2SO_4$ (96%) 42% by vol, $HNO_3$ (60%) 8% by vol, HCl (33%) 0.2% by vol, $H_2O$ to complete.
7.   Tap water rinse.
8.   Passivation at room temperature for 30-60 sec in $CrO_3$ 80 g $l^{-1}$, $H_2SO_4$ (96%) 3 $cm^3$ $l^{-1}$.
9.   Running tap water rinse.
10.  Cold demineralized water rinse.
11.  Dry with filtered air or dry $N_2$ and wrap in Al foils.

### 6.2.3   Brazing process and assembly sequences

Copper and stainless steel pieces are very difficult to join by any process other than furnace brazing [C.51], and the high vacuum condition has been selected as the best solution for LIBO (see previous paragraph). Generally speaking the design criteria is profoundly effected by the assembly process as well as the integration in the accelerating structure of several sub-components. For this reason one must pay close attention to joint design, sequence of brazing, alloy selection, compatibility to heating cycles, etc.

Assembly for vacuum brazing is prepared by using the pieces pre-cleaned with chemical degreasing. Where fixtures are required, the components are loaded in fixtures, and then the assemblies are placed in the vacuum chamber. Prior to heating the chamber is evacuated to prevent initial oxidation of the assembly. The complete cycle of evacuating, heating and cooling in vacuum takes between 10 and 24 hours.

Design clearances are primarily related to the brazing filler metal being used and their capillarity attraction. If the gap is too small, then the brazing filler metal will not flow into the joint. Moreover clearance can also influence the mechanical strength of the brazed joints. Typical joint clearances are in the order of 0.02 mm. Allowances should be made in joint design for relative expansion of the parts at the brazing temperature, especially where dissimilar metals are involved, such as brazing between copper and stainless steel. Penetration of the brazing filler metal through a joint to the opposite side of placement can be an effective measure of brazing filler metal flow, and a preferable joint design allows this observation to be made. Generally braze joints should be designed for an effective joint area of 90-95%.

The main features to be reached after brazing are the following:

- to know and control thickness of brazing,
- to prevent the brazing alloy from flowing inside the cavity and on other brazing surfaces (for this reason sharp angles on all contours are requested),
- to maintain a uniform distribution of the brazing alloy, over the entire wetted surface, for vacuum tightness and electrical conductivity.

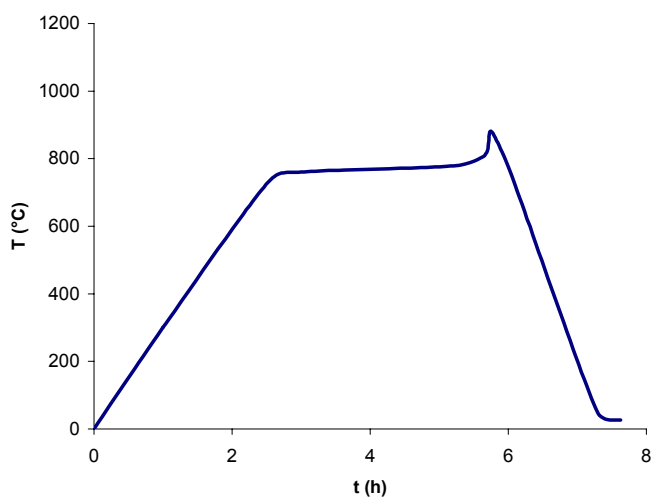

**Figure 6.10** _Typical thermal cycle performed on copper and stainless steel for high temperature brazing [A.5, D.54]._

❑ *Brazing sequences*

The full assembly of the module is divided in different brazing steps [D.49, and technical drawings], namely:

▪ Brazing of the four tanks (end cell not included) one by one.

▪ Brazing of all flanges and stainless steel inserts on the end half-cell bodies.

▪ Brazing of tank 1 and 4 with end half-cells.

▪ Brazing in two steps of all copper pieces, flanges and stainless steel inserts on the bridge coupler bodies. An intermediate machining, between the two steps was necessary in order to prepare the lateral surface of the bridge for the next brazing step.

▪ Brazing of the four tanks (with the end half-cells), the three bridge coupler bodies and the lateral tuners.

▪ Brazing of the lateral cooling plates.

The outer contour for alignment purpose has been chosen as reference as the most precise. The alignments of all the copper pieces are performed out of the oven on special ceramic and graphite plates, before brazing, with dedicate tools. This procedure is essential to guarantee the alignment for all sub-assembly-copper-pieces (tanks and module) in each direction with accuracy better than 0.02 mm. At the end the monolithic brazed module has been aligned so that the geometrical axis of the PMQ's houses, pre-machined in the bridge coupler and end cell bodies, stay within a circle of about 0.1 mm (see chapter 7).

❑ *Technical details for the LIBO prototype construction*

In the frame of high precision brazing, three different methods were considered for LIBO construction (in increasing order of cost and precision) [D.52, D.54]: 1) brazing with the foil, 2) brazing with wires in grooves, 3) brazing/diffusion bonding with a recess of 0.02 mm for brazing over the whole surface with exception of areas which need diffusion bonding to stop the flow of brazing material. The last method needs very precise machining and surface roughness of 0.1 μm.

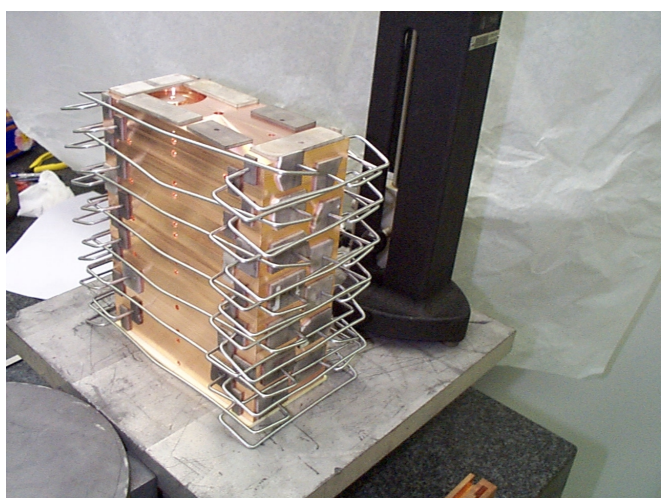

***Figure 6.11*** *Assembly of LIBO tank for brazing.*

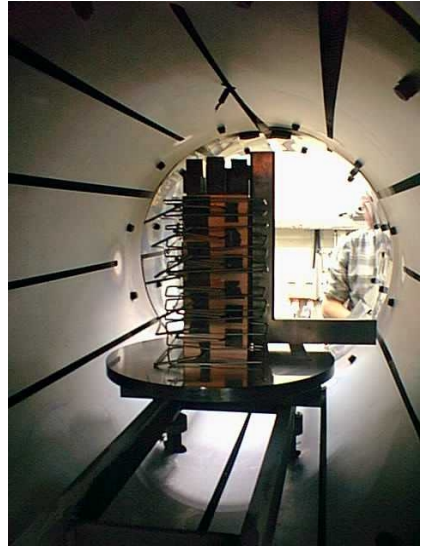

**_Figure 6.12_** _Tank assembled in the brazing oven._

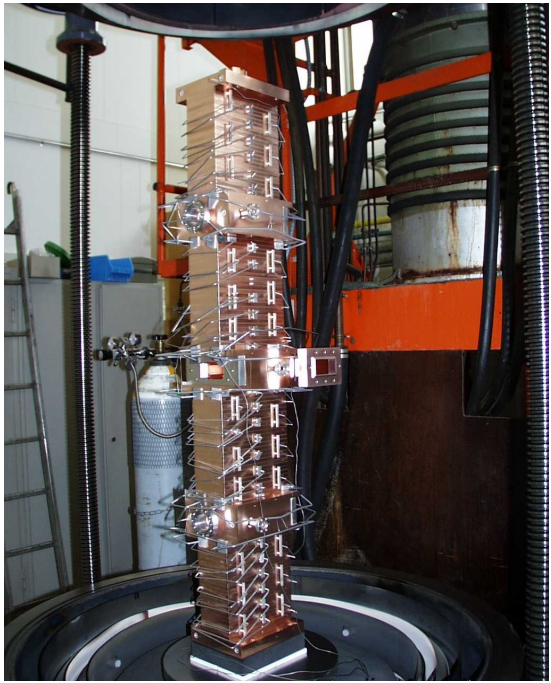 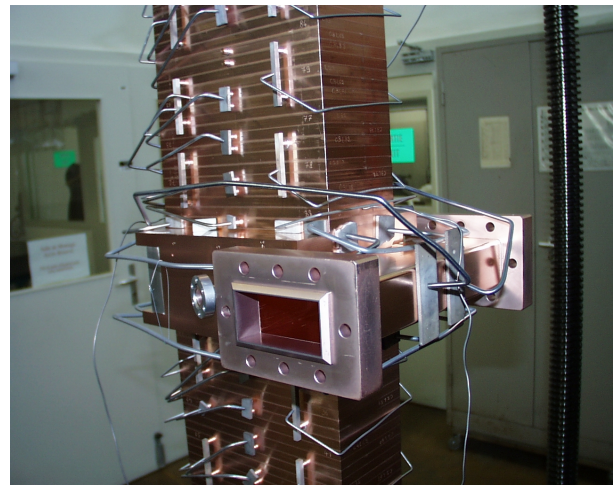

**_Figure 6.13_** _Assembly of the prototype before brazing and a detail of the wave-guide on the central bridge coupler._

According to CERN experience the brazing was planned and performed with LIBO upright and all flat brazing surfaces in horizontal position, as shown in figure 6.13. _Brazing by capillarity with wires has been then chosen_ for the half-cells, end plates and bridge couplers, except for few vacuum flanges in the bridges and the cooling plates, where foil technique has been adopted. For the different brazing steps, commercially available silver-base alloys are used, with a decreasing brazing temperatures ranging from 850 °C to 750 °C (see paragraph 6.2.1). This is mandatory for successive brazing thermal cycles executed on the same elements. These filler metal alloys are well known for their resistance to oxidation at high temperatures. All brazing operations are performed in all-metal vacuum furnaces at $10^{-5}$ mbar.

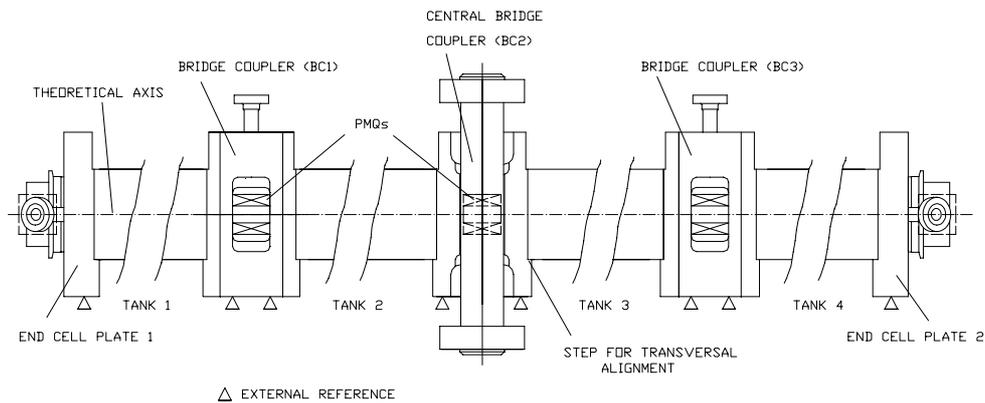

**Figure 6.14** *Alignment of LIBO prototype. The external references on the bridge couplers and end half-cell plates are shown. The half-cell plates are aligned and brazed to form an individual tank. Each tank is aligned one respect to the other with pre-machined steps on the bridge couplers and end cells. The five PMQs, in the bridge couplers and end cells, are also precisely positioned in machined houses, with their geometrical axis aligned within a circle of 0.1 mm in the transverse plane. All brazing operations must respect the alignment requirements.*

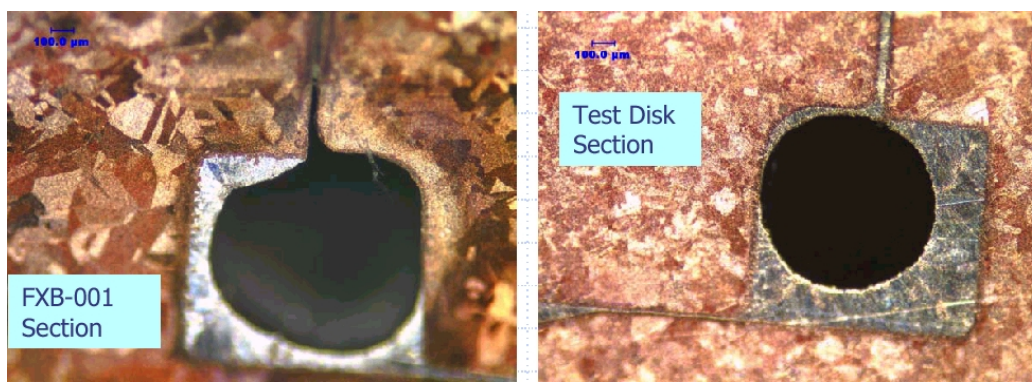

**Figure 6.15** *Example of brazing alloy in a groove pre-machined on copper cavities. On the left the braze material reached its melting point but did not have enough soak time to flow. On the right high temperature and longer soak time produce better results in terms of brazing penetration (Fermilab, USA).*

For the half-cell plates positions and dimensions of the brazing grooves, determining the quantity and the distribution of the brazing alloy, sample pieces have been machined, brazed and finally tested with an automatic total-immersion ultrasonic method. The aim was to check the uniformity of the alloy distribution on the brazing surface. The scheme of this test is shown in figure 6.17. It consists of the visualisation of the ultrasonic indications present in a controlled volume by projecting these onto a surface parallel to the control surface. By choosing a delay interval on the time base of the ultrasonic apparatus, it is possible to record only the defects present in a given layer of the material. This apparatus used for our tests is associated with a high frequency unit up to 100 MHz, and it is particularly adapted for testing of brazed joints. The results of these tests are reported in figure 6.18, where the brazing alloy between two half-cell plates is distributed uniformly. A deep cleaning of all elements has been made prior to brazing, and a careful insertion of the brazing alloy as well as the stepwise assembly were done in a clean room under laminar air flow.

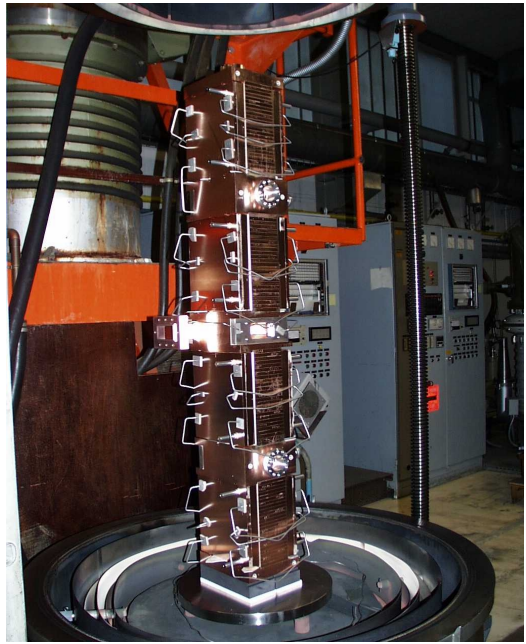

*Figure 6.16* Prototype with the cooling plates ready for vertical brazing.

In order to be declared ready for brazing, the structure has satisfied the following conditions: 1) the mechanical tolerances have met the requirements specified in the technical drawings, checked by metrology, 2) visual inspection has been performed in order to control possible damages or scratch of the RF cavities, external contours, brazing surfaces, etc., 3) the chamfer on copper and stainless steel pieces have been polished by hand, especially for the brazing grooves. The oven has been equipped with a programmable temperature control with accuracy of ± 5 °C, and under vacuum at < $10^{-5}$ mbar. The total residual gas content due to hydrocarbons or any other diffusion pump vapour has not exceeded 10%. During the realisation of LIBO, the steel girder was used to handle the module after brazing.

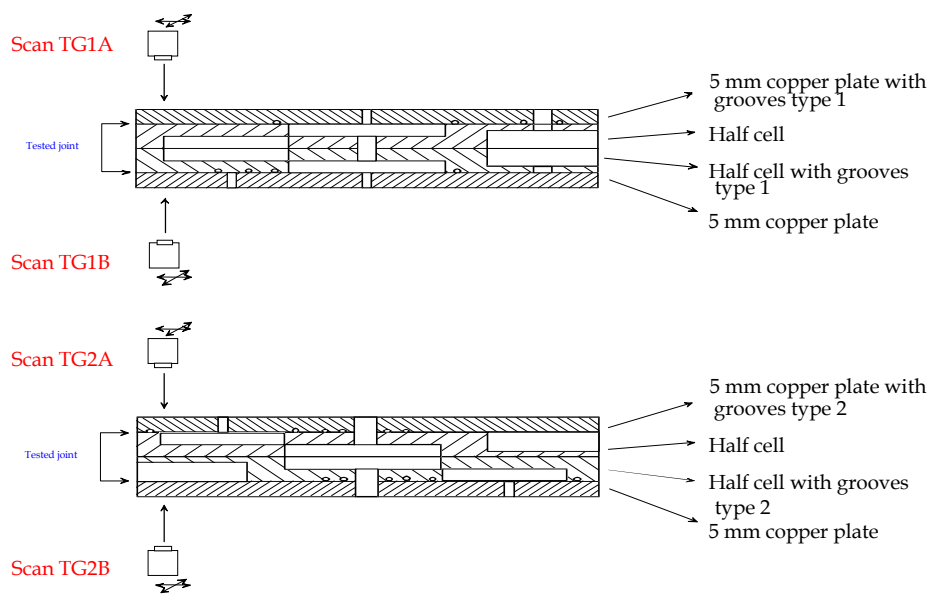

*Figure 6.17* Automatic total-immersion ultrasonic test for brazing alloy distribution analysis between RF cavities. Two test pieces have been produced to check the accelerating and coupling cell joints (CERN Surface Laboratory).

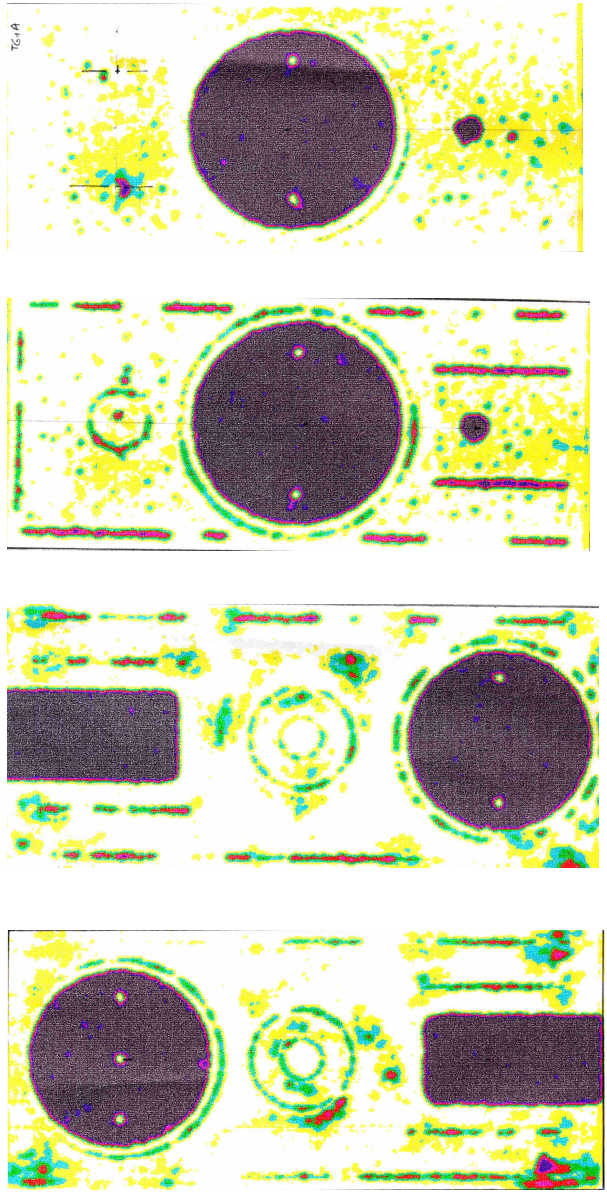

**Figure 6.18** *Ultra sound results on half-cell plates: the brazing alloy is well distributed on the wetted surface, allowing vacuum tightness and good electric contact between consecutive cells  (CERN Surface Laboratory).*

### 6.3  Basics on low power RF measurements and tuning procedures[3]

The main strategy for LIBO design is to fix relative large errors in the manufacturing of the cavities to reduce the production costs, using then different tuning solutions to reach the correct final resonant frequencies of the oscillating electromagnetic field [A.5, A.6, C.25, C.27, C.29, C.30]. Once this is obtained then the module is ready to accelerate the particle beam at the correct energy and phase.

In this paragraph the low power RF measurements made on the prototype module will be presented in brief to verify the soundness of the fabrication and the tuning procedures [D.50, D.51].

The main goal is to have no stopband at the $\pi/2$ mode in the dispersion curve for the final module, maximising the effective shunt impedance of the accelerating structure (table 6.2).

Generally speaking the results of these tests show that the predictions of the equivalent circuit analysis for $\pi/2$ mode  (the LIBO operating mode) have been verified. Final results of the low power measurements on the prototype indicate a final operating resonant frequency of $f_{\pi/2} = 2998$ MHz, a stopband (SB) of 150 kHz, a shunt impedance of 50 M$\Omega$/m and a maximum accelerating (electric) field variation less than 3%.

❑   *Summary of RF measurements on LIBO prototype*

▪ <u>Electric field flatness:</u> this measurement is the most important and it has been repeated under different settings and conditions such as tuners position and temperature.

▪ <u>Stop band (SB) and stability:</u> it has been repeated under all possible conditions in order to have ideally final SB = 0 kHz. The stability has been evaluated as function of temperature change.

▪ <u>Matching:</u> The ideal condition is to reduce as much as possible the reflected power at the RF entrance of the module through the waveguide and the iris machined in the central bridge coupler (see figure 5.10). In case of low matching the only possible solution is to operate with the piston tuner position of the central bridge in order to dismatch the BC cell and to lower the coupling with the waveguide. LIBO injection is designed to have an over-coupled condition.

▪ <u>Q value:</u> it is important to get this measurement with the definitive LIBO setting. Q values are important to define the frequency errors of the measurements and the efficiency of the assembled module.

*Table 6.2* *Tuning of LIBO-62 prototype module: final goals of the RF resonators [D.20].*

| Items | Final goal |
|---|---|
| *Overall final frequency* | ± 100 kHz (± 2°C water temperature) |
| *Stopband (SB) ($f_a - f_c$)* | < 500 kHz |
| *AC frequency error* | < 600 kHz |
| *CC frequency error* | < 200 kHz |
| *Electric field error on axis* | < 5% |

---

[3] The RF measurements and the mechanical tuning procedures of the cells have been performed in co-operation with the RF specialists Drs Mario Weiss (LIBO project leader) and Riccardo Zennaro.

❑   *RF measurement techniques*

The perturbation theory can be used to measure the electric field distribution on accelerator axis: this is done with the so-called bead pulling technique (annex 6.2).

In this standard RF method one can measure the RF electric (or magnetic) fields versus position using a small bead and moved through the cavity using the set-up as shown in figure 6.19. The bead is attached to a thin nylon line, which is driven through the cavity by a motor.

The perturbation at any bead position causes a frequency shift, which can be calculated using the Slater perturbation theorem. The general result for a spherical perfectly conducting and diamagnetic bead is [C.6]:

$$\frac{\Delta f}{f_0} = -\frac{3\Delta V}{4U}\left(\varepsilon_0 E^2 - \frac{\mu_0 H^2}{2}\right)$$

where $\Delta V$ is the volume of the perturbation, H is the magnetic field, $\mu_0$ is the permeability, E is the electric field, $\varepsilon_0$ is the dielectric coefficient, U is the total stored energy. From the above relation it is visible that the frequency variation is proportional to the square of the local field. The frequency shift can easily be measured by driving the cavity using a phase-locked loop that blocks the driving frequency to the resonant frequency of the cavity. The measurement determines then the ratio of squared field to the stored energy. The stored energy can be then determined from the relation U = QP/f, and from separate measurements of Q and the cavity power dissipation P, using standard microwave measurement techniques [C.6, C.14]. With these relations and the measurements of $\Delta f$ from the bead-pull, $ZT^2/Q$ can be also determined [D.51].

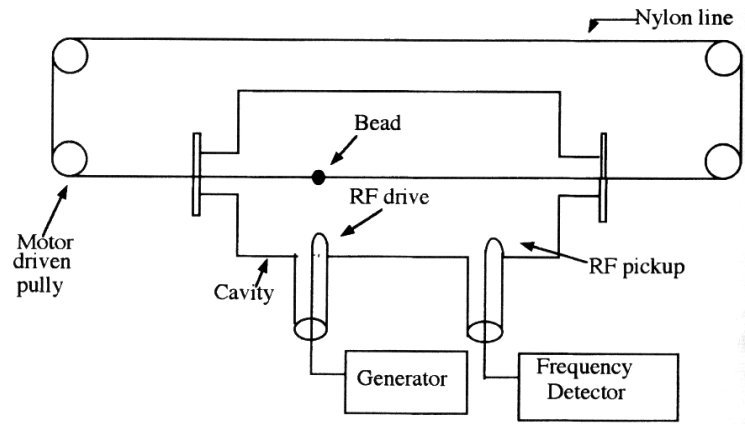

**Figure 6.19** *Bead pulling apparatus for accelerating electric field measurement on LIBO axis.*

The Q of a resonator is also a measure of the width $\Delta f$ of the resonance, defined as the bandwidth between the frequencies, where the stored energy has fallen to half its resonant value while being driven from a constant RF power source. This bandwidth is related to Q by:

$$\Delta f/f_0 = 1/Q.$$

The Q measurement is the most delicate because it depends so strongly on the state of the metallic cavity surfaces and their good contact (annex 6.1). This can be assured if the cavities are brazed together.

❑    *Tuning of accelerating tanks*

The main goal at the end of the module assembly is to have ideally $f_{\pi/2}$ = 2998 MHz and stopband SB=0 [D.50, D.51]. For the construction, 30 half-cell plates for each tank have been machined at the design frequency of 3072.68 MHz and 3063.64 MHz for accelerating and coupling cells respectively (tables 5.2 and 5.3). The cells have been measured one by one and full assembled, to check the $f_{\pi/2}$ and the stopband.

In order to calculate the correct final geometry of the cavities, it is necessary to estimate the influence of several aspects on the resonant frequency. The frequency error in the measurements is strictly connected with the Q value and then with the electric contact between consecutive half-cells. Once the cavities are definitely joined by brazing, an increasing of the resonant frequency is visible. Experimental data indicate also that there is a frequency increase due to the presence of high vacuum. Moreover the $\pi/2$ mode frequency of a multicell stack is lower than that of a single two half-cells system, due to the symmetry conditions at the reflecting plane since the two half-cells chain appeared to be an infinite chain with the coupling cavity on one side only. Considering the final frequency goal $f_{\pi/2}$ = 2998 MHz, then the quantitative values of all these effects are [D.51]: 1) brazing (+1 MHz), 2) air (-0.8 MHz), 3) frequency variation of the measurements from one tank to the full module (-1 MHz). With these values, the theoretical frequency goal of the tank $f_{\pi/2,\text{ tank}}$ is 2995.2 MHz. The main approach is then to measure the cells one by one, reducing the frequency spread as much as possible by machining the tuning rings in steps in the workshop, and then making measurements of $f_{\pi/2}$ and stopband on the assembled tanks. An example of frequency distribution of the cells during tuning with the ring is shown in figures 6.20.a and b for accelerating and coupling cells respectively. At the end the frequency spread between the cells is considerably reduced and the average frequency is centred close to the final frequency value. The tuning is based to the fact that resonant frequencies of individual cells change by altering the capacitive or inductive loading within that cavity. As mentioned in paragraph 5.3 and in accordance with these figures, cell frequency decreases if the ring is machined, while resonant frequency as well as electric field amplitude increase if the tuning rods are inserted in the cavity chain. The tuning of the coupling cells is essentially identical to the accelerating cells, taking into account the coupling constants.

*Table 6.3 RF tuning procedure for the prototype construction (see annex 6.4 for details).*

| | Step | Parameters to be measured | Tools for RF measurements |
|---|---|---|---|
| 1 | Measurements of parameters of single half-cell | $f_{\text{mean}}$, $\Delta f$ | Covers and short circuit |
| 2 | Measurements of stack parameters | $f_{\pi/2}$, SB | Covers |
| 3 | Machining of tuning ring of half-cells | Thickness of the tuning ring | |
| 4 | Measurements of stack parameters and definition of cell distribution in each tank | $f_{\pi/2}$, SB, field | Bead pulling |
| 5 | Meas. of tank and RF tuning with lateral rods | $f_{\pi/2}$, SB | Bead pulling |
| 6 | Measurements of the complete module before and after brazing | $f_{\pi/2}$, SB | Bead pulling |

**Table 6.4** _Frequency ranges for tuning of accelerating and coupling cells [D.29]._

|  |  | _Total range_ | _Sensitivity_ | _Mechanical accuracy_ | _Measurement error_ | _Residual spread_ |
|---|---|---|---|---|---|---|
| _Tuning ring_ | AC | ± 9.55 MHz | 3.2 MHz/0.1 mm | 0.02 mm | ± 500 kHz | ± 1.1 MHz |
|  | CC | ± 10.85 MHz | 4.0 MHz/0.1 mm | 0.02 mm | ± 500 kHz | ± 1.3 MHz |
| _Each tuning rod_ | AC | ± 2 MHz | 500 kHz/mm | 0.1 mm | ± 100 kHz | ± 150 kHz |
|  | CC | ± 2.5 MHz | 500 kHz/mm | 0.1 mm | ± 100 kHz | ± 150 kHz |

Coupling strength between cells determinates the frequency separation between 0 and $\pi$ modes. One must have strong coupling in order to have large bandwidth or more separation. The tuning procedure described earlier was used to flatten the tank, and also to move the tank frequency to the desired frequency $f_{\pi/2}$ of 2998 MHz. One of the main problems for low power RF measurements was the low quality of electric contact, generating a low Q value. Remembering the relation between frequency, relative error and Q, one can easily obtain, for a frequency of 2998 MHz, Q values of about 2800 for single cell and about 500 for the assembled tank (underline before brazing), an intrinsic frequency error for RF measurements of 1 MHz and 6 MHz respectively. Preliminary assembly (figure 6.21) has been obtained by joining the four tanks with the end cells already brazed (but without bridge couplers) and measuring, with a bead pulling, the accelerating field distribution on beam axis. Tuning of the end cells consisted of adjusting the end cell frequency $f_{\pi/2}$ so that it matched the $f_{\pi/2}$ of an interior cavity of the infinite chain. This is possible by using longitudinal piston tuners as shown in figure 5.6. A conclusive measurement result for the four tanks (already brazed but without the bridge couplers) is visible in figure 6.22 [D.51]. Here, with the help of lateral tuning rods, the error of the field distribution is less then 3%, with $f_{\pi/2}$ = 2997.5 MHz and a stop band SB = +300 kHz.

❑  *Tuning of bridge couplers*

When the bridge couplers have been inserted between tanks, the module has been tuned with the help of piston tuners in the bridge couplers. The tuning procedure of BCs is based on the fact that, for a coupled resonator system, the only direct measurable quantities are the resonant frequencies of the whole structure, which differ from the single cell values [D.50, D.51]. For a three-coupled-cell system like the 3-cell bridge (figure 5.10), the unknown single cell frequencies are ($f_1$, $f_2$, $f_3$), while the measured ones are ($f_-$, $f_0$, $f_+$). The tuning procedure consisted, first, in equalising the frequencies of the coupling cells (CC) of the bridge ($f_1$ = $f_3$) and then varying the central cell (BC) frequency $f_0$ to get a set of measured data. For $f_1 = f_3$, a linear relation applies between a combination of the sum and of the ratio of the measured frequencies [A.2]:

$$\left( \frac{f_0^2}{f_+^2} + \frac{f_0^2}{f_-^2} = \frac{1 - \frac{k_2}{2}}{1 - \frac{k_2}{2}} + \frac{f_0^4}{f_+^2 \cdot f_-^2} \cdot \frac{1 + \frac{k_2}{2}}{1 - \frac{k_2}{2} - \frac{k_1^2}{2}} \right)$$

By a linear fit to the data in that formula, it was possible to obtain the first and second coupling coefficients, $k_1$ and $k_2$, so the single cell frequencies could be found from the theory of coupled resonant cavities.

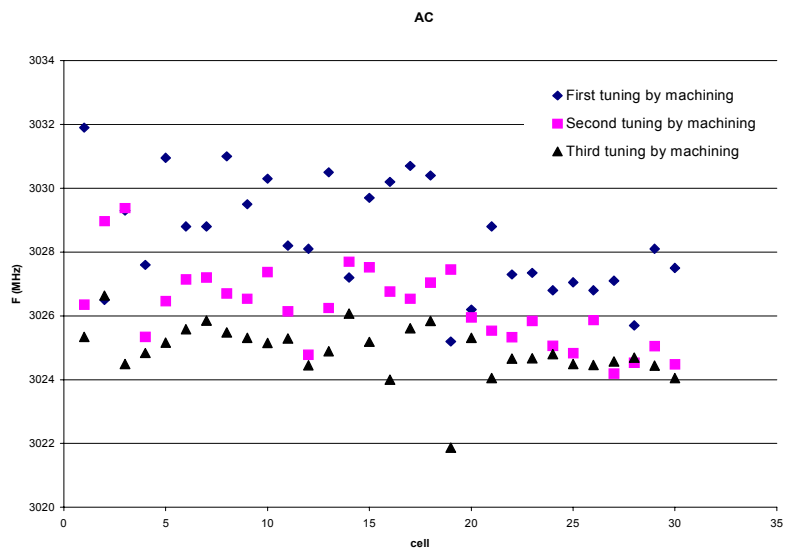

***Figure 6.20.a*** *Example of frequency distribution for the accelerating half-cells in the first tank. The different steps are generated by machining the tuning ring of each accelerating cell. The measurements have been performed in the RF laboratory while the mechanical tuning at the CERN Central Workshop. For each tank 30 half-cells have been produced, but only the best 25 cells (or 26 for tank 2 and 3) have been used to compose the prototype module.*

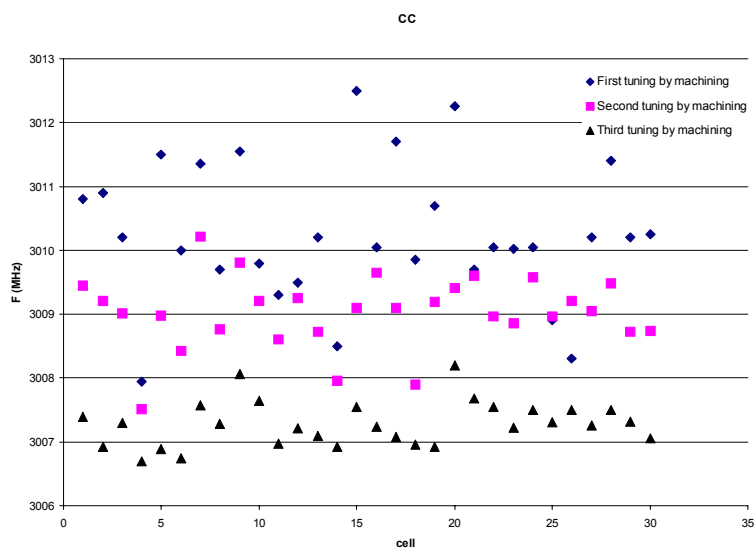

***Figure 6.20.b*** *Example of frequency distribution for the 30 coupling half-cells produced for the first tank. The different steps are generated by machining the tuning ring of each coupling cell.*

In this way, one has correctly tuned the cells in the bridge couplers, except for the central cell of the central bridge, which remained 6.5 MHz below the nominal value. This caused a small bump in the field level between tanks 2 and 3 (as visible in figure 6.24). The waveguide of the RF feeder line has been brazed tangentially to the central bridge coupler and terminated in a short-circuit at 5λ/4 from the elliptical coupling iris: the injection port in the central bridge coupler of the RF power (figure 5.7). The matching factor $\hat{\beta}$ measured finally was 1.14, corresponding to a reflected input power of 1.6% [D.50].

**Table 6.5**  _Real frequencies of the half-cells in the four tanks of the prototype. The different values reefer to the different tuning phases (ring machining or tuning rods insertion) [D.35, D.41, D.50, D.51]._

| | Tank 1 | | Tank 2 | | Tank 3 | | Tank 4 | |
|---|---|---|---|---|---|---|---|---|
| _Frequencies on the 1/2 cells (MHz)_ | AC | CC | AC | CC | AC | CC | AC | CC |
| **First measurement** | 3028.42 | 3010.22 | 3028.42 | 3009.66 | 3030.92 | 3010.28 | 3032.72 | 3010.29 |
| _Dispersion_ | 1.77 | 1.06 | 2.53 | 0.83 | 0.82 | 0.80 | 0.67 | 0.61 |
| **Second measurement** | 3026.28 | 3009.02 | 3024.70 | 3005.44 | 3025.38 | 3004.42 | 3025.56 | 3003.12 |
| _Dispersion_ | 1.27 | 0.57 | 0.45 | 0.27 | 0.26 | 0.25 | 0.26 | 0.38 |
| **Third measurement** | 3024.91 | 3007.29 | 3019.79 | 2999.80 | 3018.85 | 2997.96 | 3021.73 | / |
| _Dispersion_ | 0.85 | 0.35 | 1.18 | 0.47 | 0.36 | 0.26 | 0.30 | / |
| **Fourth measurement** | 3020.68 | 3001.38 | / | / | / | / | 3019.71 | 2999.39 |
| _Dispersion_ | 0.43 | 0.28 | / | / | / | / | 0.24 | 0.22 |
| **Final goal** | 3021.77 | 3002.82 | 3020.71 | 3000.95 | 3020.54 | 2999.08 | 3019.30 | 2999.50 |
| **Variation from the goal** | -1.09 | -1.44 | -0.92 | -1.15 | -1.69 | -1.12 | +0.41 | -0.11 |
| _π/2 mode frequencies_ | | | | | | | | |
| $f_{\pi/2}$ **(final values)** | 2994.64 | | 2993.1 | | 2992.3 | | 2994.75 | |
| $f_{\pi/2}$ **(goal)** | 2995.2 | | 2995.2 | | 2995.2 | | 2995.2 | |
| **Goal variation** | -0.56 | | -2.1 | | -2.9 | | -0.45 | |

**Table 6.6** _Frequency parameters of the tanks after brazing._

| | Tank 1 | Tank 2 | Tank 3 | Tank 4 |
|---|---|---|---|---|
| _Frequency $f_{\pi/2}$ (MHz)_ | 2996.66 | 2994.75 | 2994.78 | 2995.08 |
| _Stop band (MHz)_ | + 0.6 | + 0.6 | + 0.15 | - 0.37 |
| _Q_ | 1960 | 2546 | 2515 | 5890 |

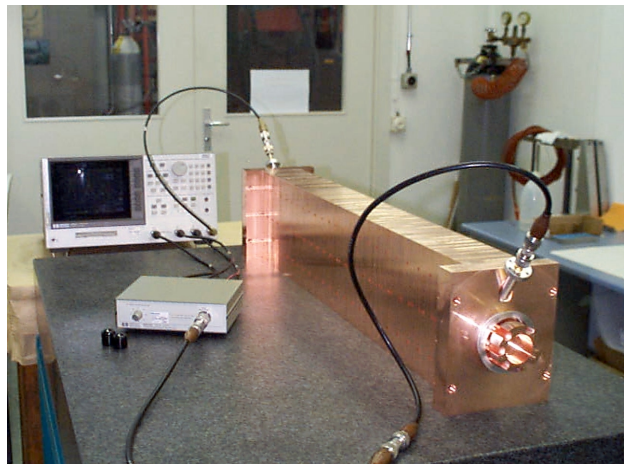

**Figure 6.21** _The four tanks (already brazed) are assembled and measured with the two end-cells, before final brazing with the bridge couplers. This measurement is used to check the accelerating field distribution on axis in the tanks._

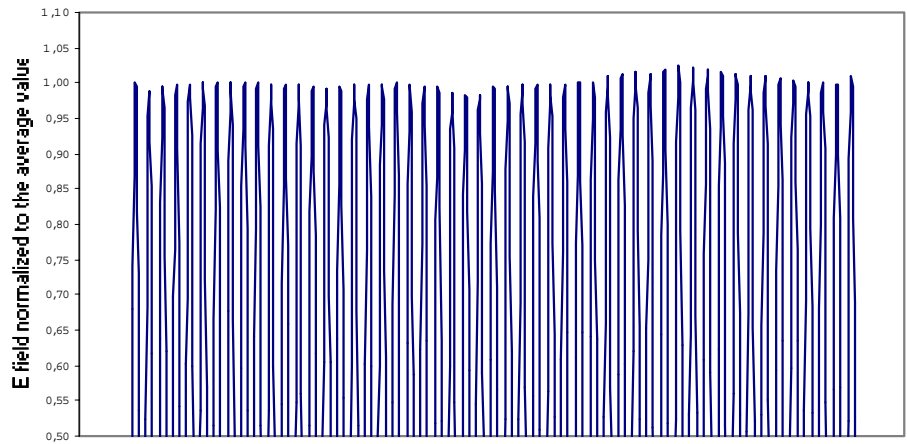

**Figure 6.22** *Electric field distribution on axis (produced with the bead pulling technique) in the four tanks before final brazing and without bridge couplers.*

❑   *Tuning of the full assembled module*

The complete module (tanks and bridge couplers) has been then assembled and measured before the penultimate brazing, and the length of the tuning rods in the accelerating cells (one per half-cell) has been adjusted to give the correct overall frequency and the required field flatness.

Finally to adjust the LIBO accelerating field to the correct value during the power tests, the total shunt impedance of the module has to be known. This impedance was determined prior to the tests by means of an innovative combination of perturbation measurements and SUPERFISH computations. The measured value of the total shunt impedance is 50 M$\Omega$, with an estimated error of $\pm$ 5% [R. Zennaro and A.5, A.6].

Acting in the same way on the coupling cells, the stopband was also adjusted to be +150 kHz. All the rod tuners in a tank, except in the accelerating end cells, are cut at the same length.

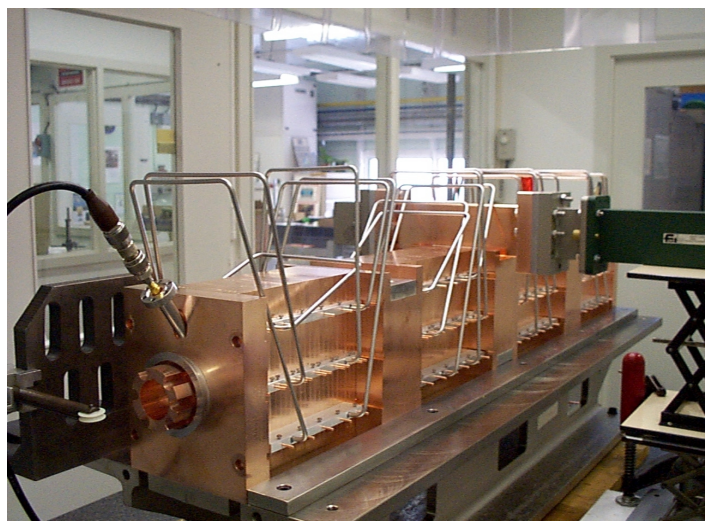

**Figure 6.23** *Full assembled module on the girder during low power RF measurements with the bead pulling apparatus.*

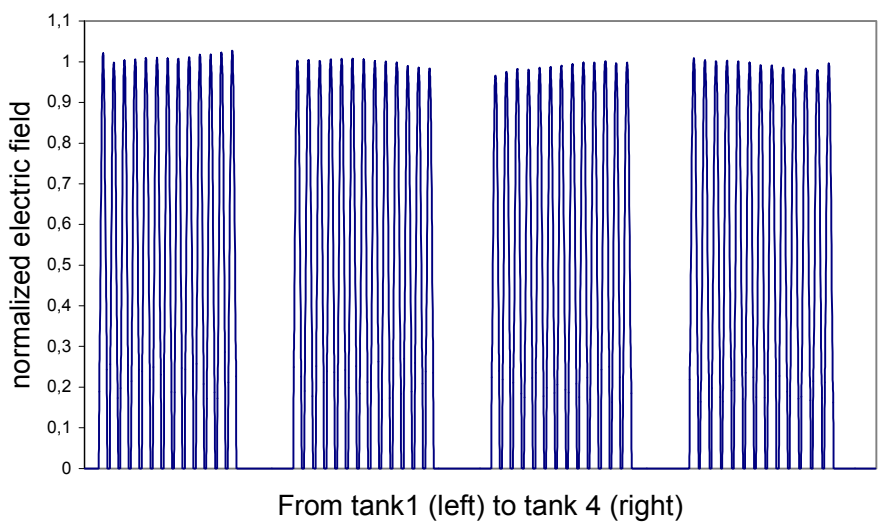

**Figure 6.24** *Final electric field distribution on axis of the full prototype module (after brazing).*

They were used essentially to compensate the differences in tank frequencies and to act on the stopband, but not to equalise the single cell frequencies. The frequency and field distribution, measured before and after vertical brazing, remained the same, as confirmed by bead-pulling measurements.

_Finally the accelerating field distribution in the four tanks of LIBO plus bridge couplers shows a uniformity around 3% (figure 6.24) [D.51]. With this configuration the final $f_{\pi/2}$ is 2997.28 MHz at 28.8 °C. Remembering the frequency variation with the temperature of about 50-60 kHz/°C (chapter 5), one can conclude that, to have $f_{\pi/2}$ = 2998 MHz, the prototype will operate at a temperature of 30.2 °C._

The point for further studies is connected to the possibility, under full-scale production, to prepare the tank frequencies very close to a predicted value during the machining stage, if high accuracy in construction is used. This could be done probably without extra costs if mechanical tolerances are maintained in the range of 5-10 μm. In this case one can envisage even to avoid final tuning with rods (that could generate risks for brazing), if some constraints in field flatness could be relaxed.

In the same way, if tuning procedures are maintained, one could also think to relax considerably the mechanical tolerances of the cavity in the range of 0.05-0.1 mm (except of course for the nose gap). Here one can use the machining of rings for frequency-spread reduction and rods insertion for final field adjustment, but with a good cost reduction for machining.

In parallel several studies have been performed to tune the cells using the so-called "dinging" technique developed at Los Alamos National Laboratory [C.42], but applied at a 3 GHz structure. These tests proved the physical feasibility of this tuning procedure (annex 6.3), but of course with a reduced tuning range (about 500 kHz), respect to the insertion of lateral tuning rods (2-2,5 MHz) (see table 6.4).

*Chapter 7*

**Successful tests of the LIBO-62 prototype**

## 7.1 Metrology and vacuum tests on LIBO-62 prototype

❑ *Metrology tests*

One of the main goals for the prototype construction is the alignment constraint of the PMQs inside module as well as the alignment of the accelerating gaps respect to the theoretical axis.

After final brazing, a 3D-metrology measurement, with an accuracy of 3 μm, has been performed at CERN Metrology Laboratory in autumn 2001. The main strategy was to map the contours of the bridge couplers and the end cells, as well as the shape of the PMQ houses, machined in the copper bodies (figures 7.1 and 7.2). Extrapolating these values with the single-piece-measurements performed before brazing, the theoretical axis of the accelerator has been determined. All the PMQ axes are inside a circle of 0.1 mm in x, y plane (transverse plane), in accordance with the beam dynamics specifications (figure 7.3). Moreover all the PMQs have been also measured and the geometrical axis distribution found once the PMQs have been inserted into the cavity. Also here the magnetic axis of PMQs are within the above-mentioned circle of 0,1 mm in x, y direction (figure 7.4).

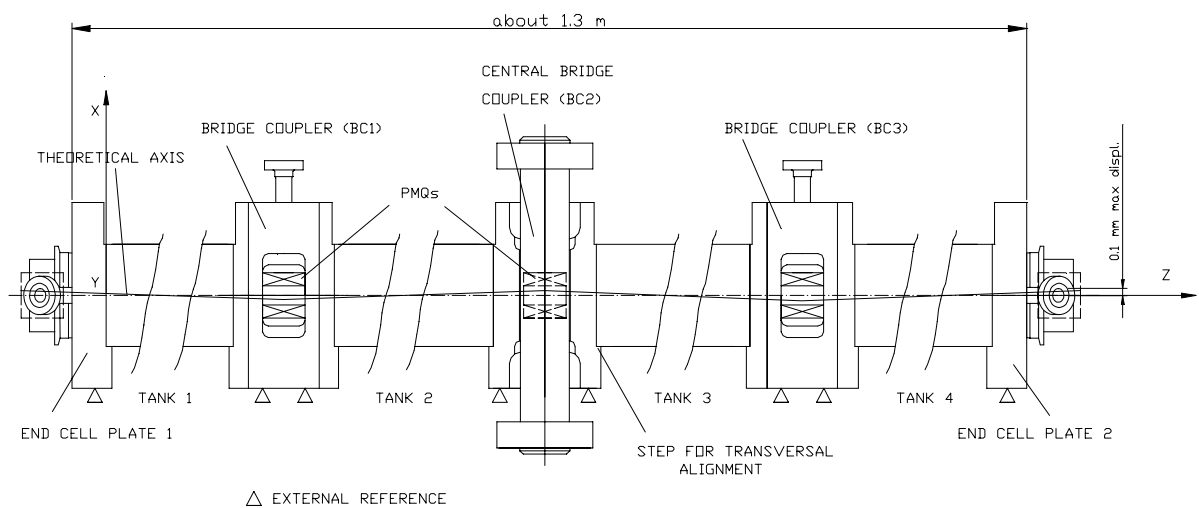

**Figure 7.1** *Reference system for module alignment. Metrology measurements on the prototype have been performed after final brazing to prove the perfect alignment of the module needed for good beam transmission (CERN Metrology Laboratory).*

❑ *Vacuum tests*

It is good practice in the accelerator technology to make individual leak detection before installation [C.44]. Once LIBO has been brazed and measured at low RF power, leak detection with helium has been performed at CERN Brazing Laboratory. Normally 60% of the full leak signal is obtained after 100 seconds. In figure 7.5 the set-up experiment is visible. A flow rate of $10^{-9}$ Atm cm$^3$/sec has been measured at a vacuum level of $10^{-6}$ mbar, showing the success of the test and the perfect tightness of the prototype after brazing.

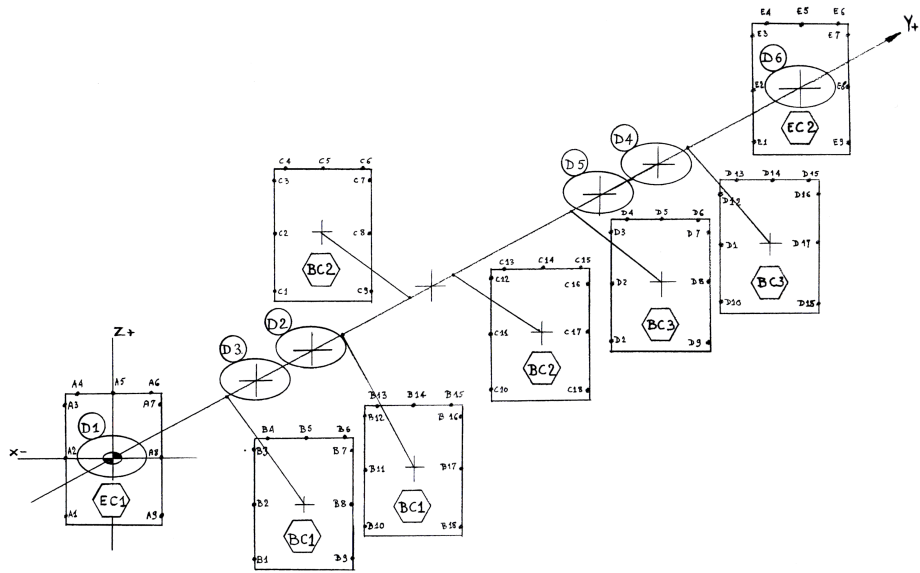

**Figure 7.2** *3D-metrology on the prototype after final brazing in 2000 at CERN Metrology Laboratory. Theoretical axis of the accelerator is calculated by direct measurements of different external contours of the bridge couplers and the end cells and then compared with the measurements performed on each piece before brazing. With this method it is possible to calculate the theoretical centres of the PMQ houses in the copper bodies even for the cases where access inside the module (after brazing) is denied for direct measurements.*

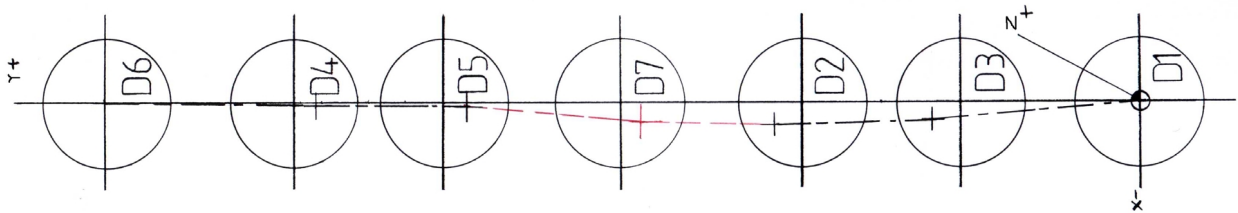

**Figure 7.3** *Metrology measurements on the prototype after brazing (CERN Metrology Laboratory). The figure shows the geometrical centres of the <u>PMQ houses</u> machined in the prototype. These centres are inside of a circle of 0.1 mm in x,y transverse plane.*

After leak detection, the module with the steel girder has been mounted on the support and tested at full vacuum in a dedicate laboratory, as shown in figure 7.6.

A primary pumping group, needed to reach a preliminary rough vacuum level ($10^{-2}$ mbar) inside of the cavity is part of the vacuum system. After this stage, turbo molecular pump starts to be operative, followed by the ion pump for high vacuum levels. All the tests have been controlled by dedicate valves and gauges, as mentioned in table 7.1. The copper structure is connected to the pumping system with stainless steel and polished manifold, through bellows, KF and CF standard vacuum flanges. After three days, levels of $5 \times 10^{-7}$ mbar and $5 \times 10^{-8}$ mbar have been easily reached in the module and in the vacuum manifold, respectively, as it is shown in figure 7.7. This is much better than the design requirement of $10^{-6}$ mbar, needed to accelerate the low beam current used for therapy.

Due to the care taken during the assembly process, it was decided not to bake out the module. In case of needs, one could proceed with the baking out by using hot water (~140 °C) at 8 bar in the cooling circuit.

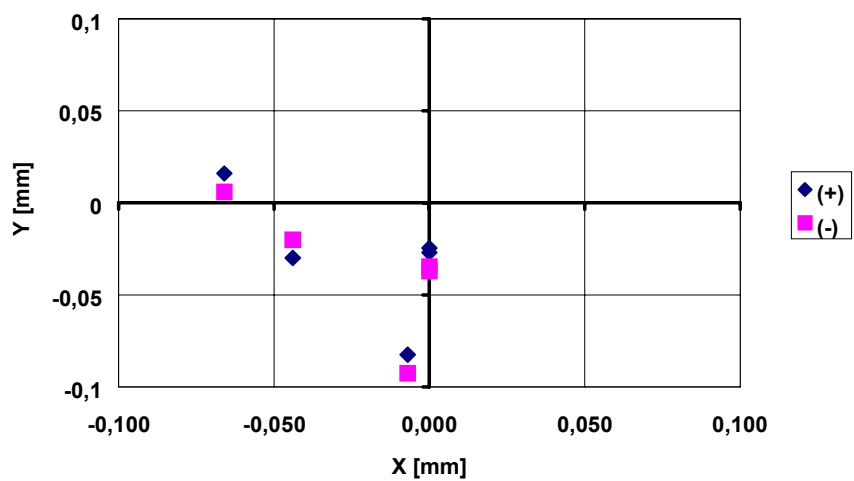

**Figure 7.4** *Geometrical centres distribution of the <u>PMQs</u> inserted in the prototype module. Each quadrupole is defined by a range of tolerances (maximum positive error is marked with +, and the maximum negative error is marked with -). Taking into account also the error of magnetic axis respect to geometrical axis of the PMQs (given by the constructor), the final results show the perfect alignment of the LIBO sub-components after brazing as well as the perfect alignment of the PMQs once they have been inserted into the module. All this is in accordance with the design specifications (± 0.1 mm).*

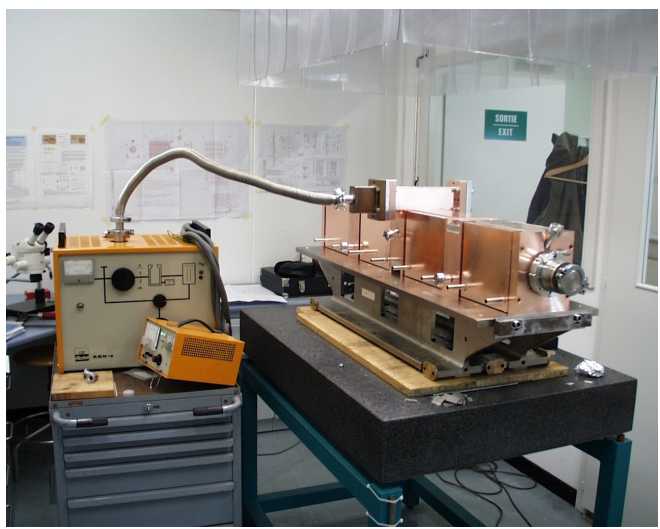

**Figure 7.5** *Helium leak detection performed on the prototype after final brazing. This test has shown successful results of brazing process. Details of this type of test are reported in [C.44].*

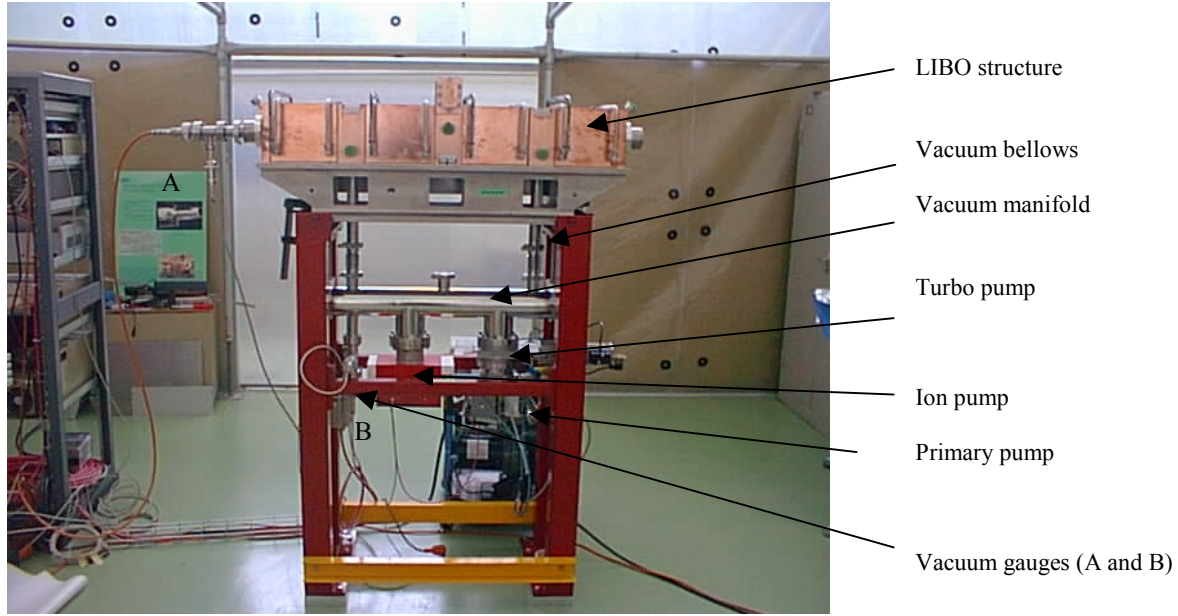

**Figure 7.6** *Full vacuum tests on the prototype before high power experiment. Pumping systems are visible (primary pumping group, turbo and ion pumps with valves and gauges). The pumping system is connected to the structure by using stainless steel manifold and two bellows through the lateral bridge couplers.*

*The vacuum levels are measured at the end cell position (gauge A) (defined as a reference for vacuum in the cavity) and in the vacuum manifold (gauge B). A brief description of the vacuum system is presented in annex 7.1 and table 7.1.*

**Table 7.1** *Vacuum system of the LIBO-62 prototype [C.44].*

| | |
|---|---|
| Ion pump (Vaclon plus 75, Varian) | Nominal pumping speed: 75 l/s<br>Maximum starting pressure: $< 10^{-3}$ mbar |
| Turbo pump (V250 Varian) | |
| Commande pneumatique (DN 100, Vat serie 10) ||
| IKR triaxiale (Penning) (DN 40 CF, Balzers) | Measurement domain: $5x10^{-3} – 10^{-8}$ mbar |
| UHV valve (DN 16 CF, ZCR20R) ||
| TPR 018 (Pirani) (DN 16 CF-F, Balzers) | Measurement domain: $100 – 10^{-2}$ mbar |
| 2 Balzers TPG 300 ||
| 3 Ion pump supply ||
| Extra-ion pumps for wave guide with CERN standard connections ||
| Primary vacuum group ||
| Vacuum manifold ||

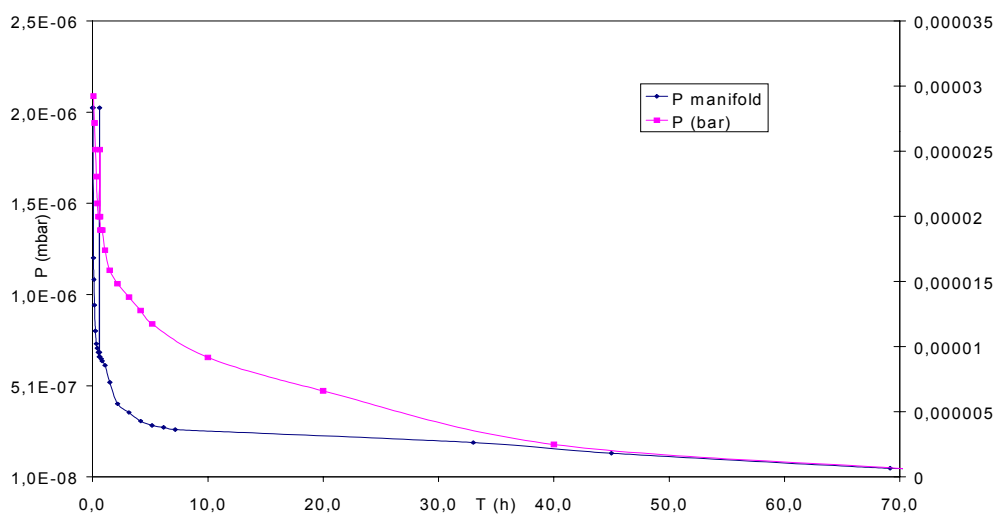

**Figure 7.7** *Vacuum levels for the prototype. The two curves referee at manifold and accelerating structure positions (right and left scales respectively). From the figure it is visible that the vacuum regime is reached after about 70 hours.*

## 7.2    RF high power test at CERN on LIBO-62 prototype

The prototype, tuned and vacuum tested, has been installed at CERN-LIL area, after the operation of LEP has stopped, for RF power tests [A.3, A.5, A.6]. The main goal was to prove that the module can support the peak power corresponding to an average electric field on axis, with the relative maximum surface field ($E_s$), needed to accelerate proton till to about 74 MeV, as required by design.

In this context the module behaviour has been investigated under power as well as the maximum RF limit supported by the module itself. In the following paragraphs the main aspects of these tests will be presented [D.46, D.47, D.48].

### 7.2.1    Installation facilities of the LIBO-62 prototype

The prototype module was transported at LIL tunnel in autumn 2000. During the September shut down the red LIBO support, equipped as much as needed, has been pre-installed in the tunnel, where, to shorten the final installation, precise marks have been drawn on the floor to define the horizontal position of the support, checking at the same time its height with respect to the feeder waveguide.

In November 2000, after the alignment constraints have been solved with respect to the existing RF power line, the module, on its steel girder, has been deposed on it and connected to vacuum equipment and thermostat-controlled water supply made available in the area and powered by the klystron-modulator. Figure 7.9 shows the module installed at LIL tunnel: one distinguishes the waveguide of the RF feeder line as well as valves and tubes supplying the water to the channels of the cooling plates brazed on either side of each tank. The vacuum manifold with the ion and turbo pumps is placed underneath the LIBO. A dedicate control system allows to read and record : 1) the cooling water temperatures,  2) the vacuum at the end cell position, 3) the status position of the 4 water valves regulating the water flow in the cooling channels, 4) the 5 pick up loop signals for RF field reading. The temperatures of the thermocouples, fixed on the top of the four tanks, are displayed by the thermocontrollers installed in the gallery, as well as a TV monitor connected to a CCD camera, installed near the module, allows to observe continuously the water valves.

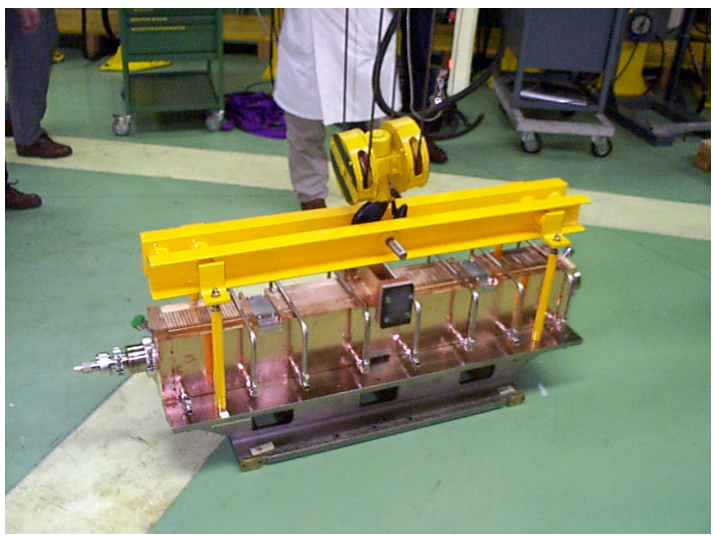

**Figure 7.8** *Installation of the prototype in LIL tunnel at CERN (November 2000). During the installation the module is under nitrogen at 0.1 bar and closed with vacuum flanges.*

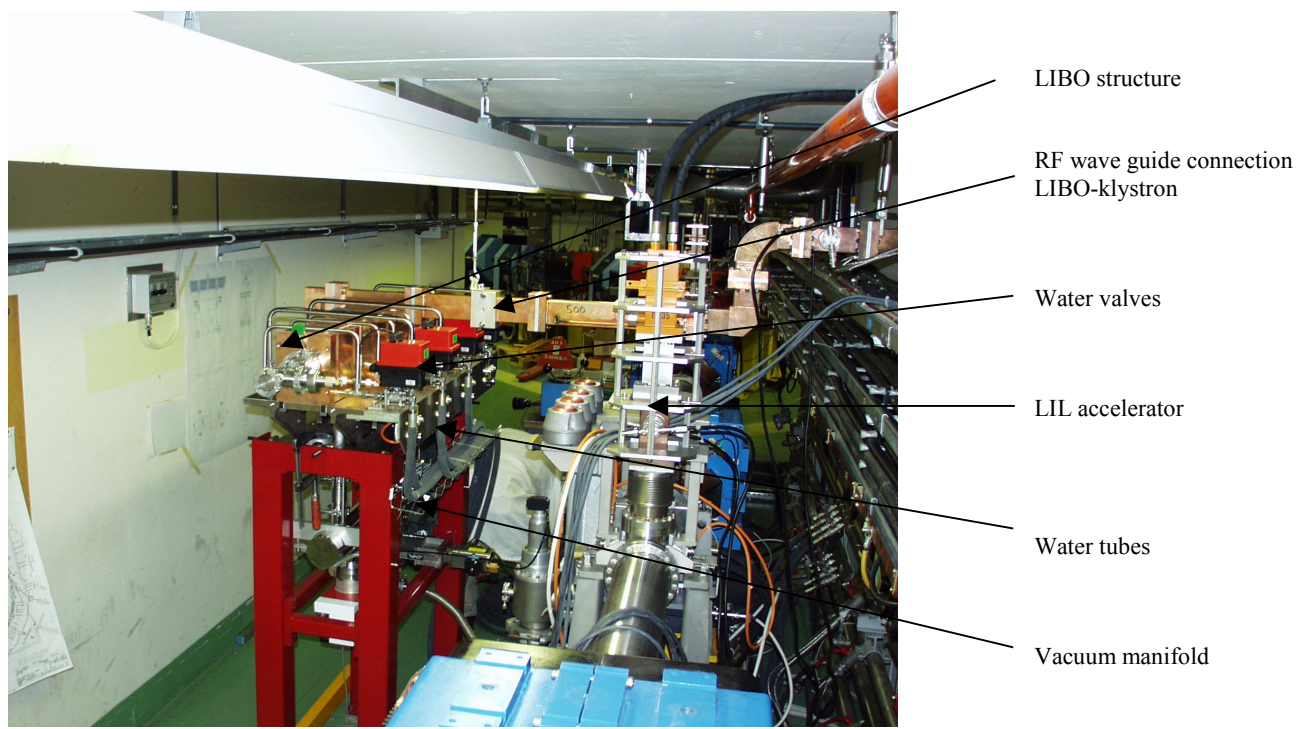

**Figure 7.9** *Prototype module of LIBO-62 at CERN. The figure shows the facilities used for the test. The vacuum and cooling systems as well as the RF power connections are visible.*

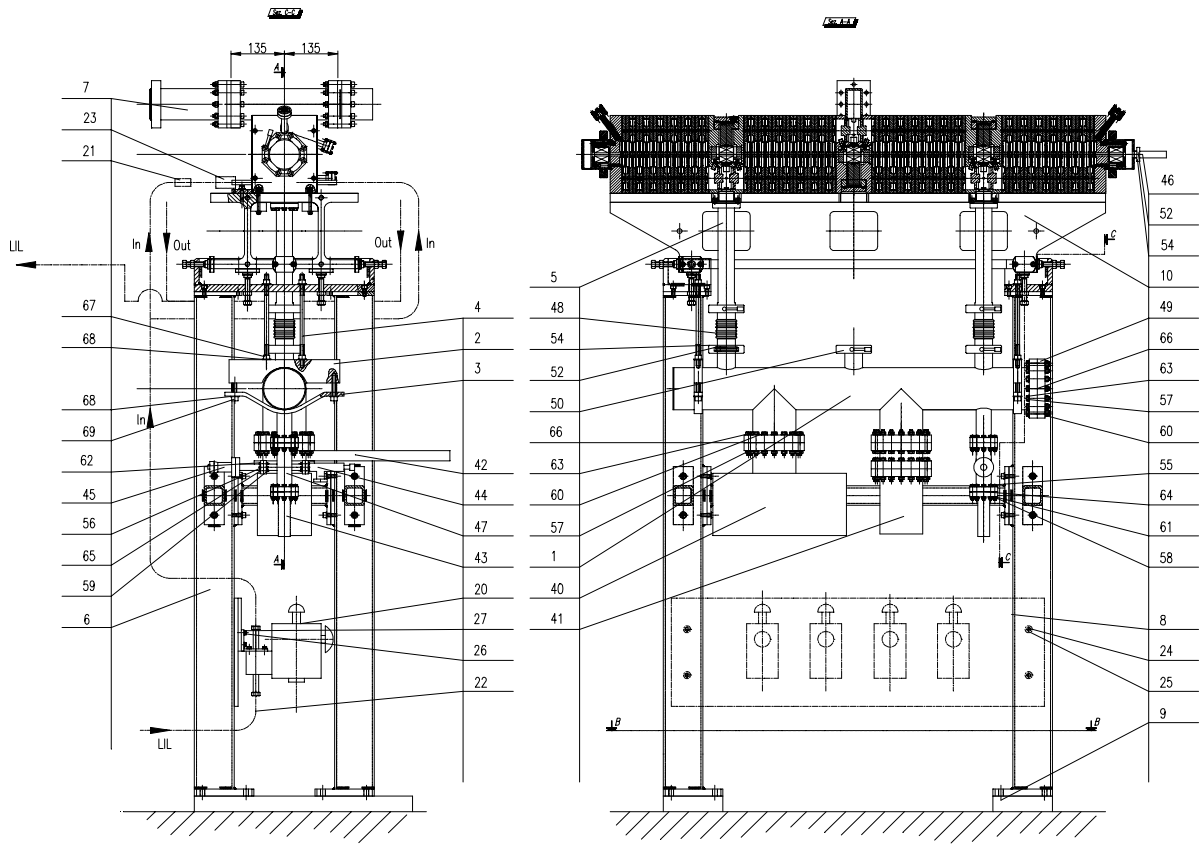

**_Figure 7.10_** _Engineering design of the prototype at CERN. Vacuum and cooling systems are visible (the nomenclature is shown in the technical drawing LIBOM1000000 [D.49])._

❑   *The klystron and the water cooling system*

For the power tests it has been decided to use an existing 3 GHz klystron of LIL tunnel at CERN. The main parameters of this system are summarised in table 7.2. A maximum power of 35 MW, with pulse length up to 5 μsec and repetition rate ranging between 100 and 200 Hz, can be achieved. The connection between the klystron and the cavity must be designed to limit the reflected power, in order to prevent deterioration of the performance. A method for protection is to use a ferrite device called circulator. It is a microwave component, which transmits RF power travelling from the RF amplifier to the cavity. Any power coming back from the cavity to the circulator is delivered to a matched load rather than back to the RF amplifier, where it could cause damage. A RF control system is used to maintain the phase and amplitude of the cavity field. A RF pick up loop signal is fed back to the field control electronics. Figure 7.11 shows a schematic view of the RF power system. For the preliminary tests the klystron uses high levels of peak power, but pulsed at a duty cycle of 5 x 10$^{-4}$ (100 Hz * 2 μsec).

The injected power must be removed by the de-mineralized water flowing in the cooling channel of the plates brazed laterally to the tanks. The main water line is composed by four circuits in parallel, for totally eight water channels (see figure 7.13). The water flow is regulated by special valves and checked by a flow meter positioned in the LIL tunnel. The water is thermostated by special heat exchangers positioned in the LIL gallery and maintained at constant temperature with an accuracy of ± 0.1 °C.

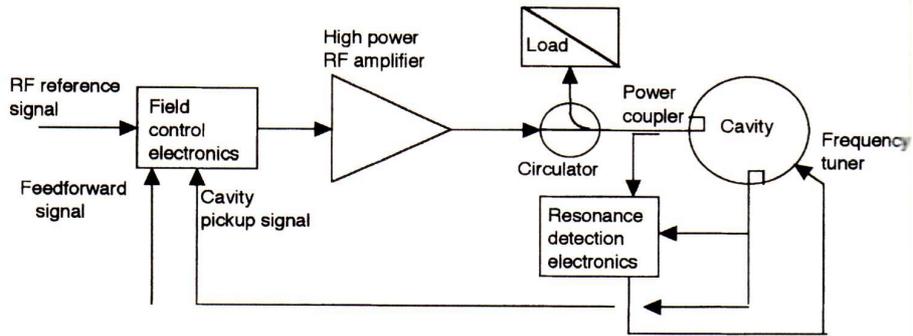

**Figure 7.11** *Schematic view of the RF power system [C.6].*

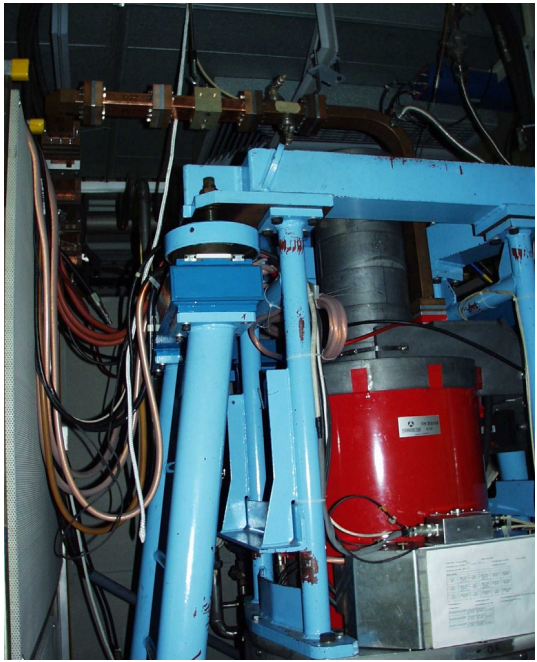

**Table 7.2** *Main RF parameters of the klystron used for power tests at LIL tunnel.*

| Peak output power | 35 MW |
|---|---|
| Pulse length (max) | 5 μs |
| Repetition rate | 100-200 Hz |
| Efficiency | 45% |

**Figure 7.12** *CERN klystron at LIL gallery. The facility has been used for high power tests of LIBO prototype.*

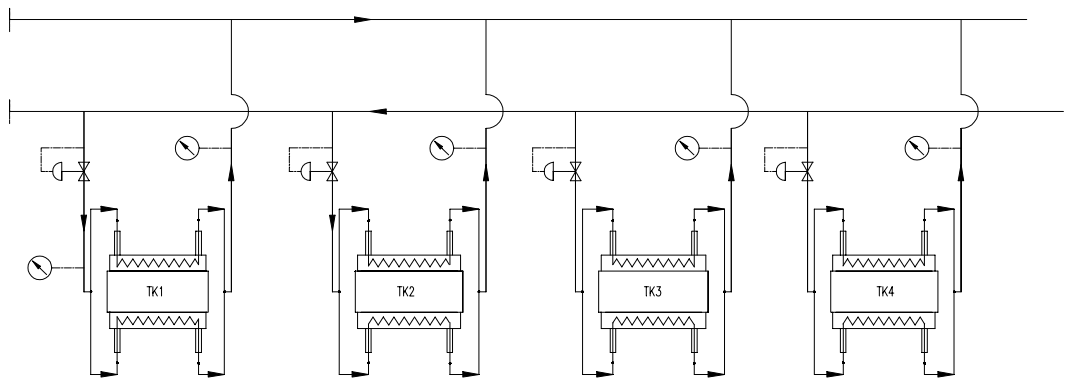

**Figure 7.13** *Scheme of cooling system for the prototype used at LIL tunnel. The four valves for water regulation and fine RF tuning are shown as well as the resistive probes (Pt100) for water temperature measurements. The input water line is taken from the main cooling system of the LIL accelerator.*

❑   *Control and Data Acquisition System*

The system has been designed in collaboration with Milan University and INFN specialists [D. Giove], according to the following scheme [C.43, C.44, C.57]:

1.   Safety system.

2.   Steering system (driving and maintaining the operating point).

3.   Data acquisition and analysis.

Safety system is devoted to the control of safe operation conditions of the main apparatus involved on LIBO, steering system will help to drive LIBO toward the required operating point and to maintain it against variations in control parameters while data acquisition and analysis system is devoted to acquire measures related to the RF power test.  The main choices for these levels have been summarised in table 7.3, where the times of intervention are indicated.

*Table 7. 3 Controls and Data acquisition systems for prototype power tests at LIL tunnel.*

| System | Controlled elements | Characteristic | Time scale |
|---|---|---|---|
| *Safety* | Vacuum | HW driven | FAST (msec) |
| | Klystron | no human interaction | FAST (msec) |
| | Cooling | | SLOW (sec) |
| | Electrical Power | | FAST (msec) |
| *Steering* | Temperature | HW signal conditioning | SLOW (sec) |
| | RF Instruments | SW control loops | SLOW (sec) |
| | Vacuum | SW human interaction | SLOW (sec) |
| *DAQ* | Vacuum | HW signal conditioning | SLOW (sec) |
| | RF instruments | IEEE-488 interface | SLOW (sec) |

*Table 7.4 Safety System Controlled elements at LIL tunnel.*

| Reading (input) | Characteristic | Reading comes from |
|---|---|---|
| Main Electrical Power Availability | Digital (ON/OFF) | |
| Pumps Electrical Power Availability | Digital (ON/OFF) | |
| LIBO Temperature monitor | Analog | 4 (Tc or Pt100) installed on LIBO |
| Water Flow Monitor | Analog | Available at LIL |
| LIBO Vacuum monitor alarm | Digital (ON/OFF) | Balzer TPG 300 in the Klystron Gallery |
| Klystron Ready to power | Digital (ON/OFF) | |
| Klystron power state | Digital (ON/OFF) | |
| *Actions (output)* | *Characteristic* | *Action works on* |
| Klystron shutdown | Digital (ON/OFF) | Klystron control unit |
| Vacuum pump Valves | Digital (ON/OFF) | Turbo valve |

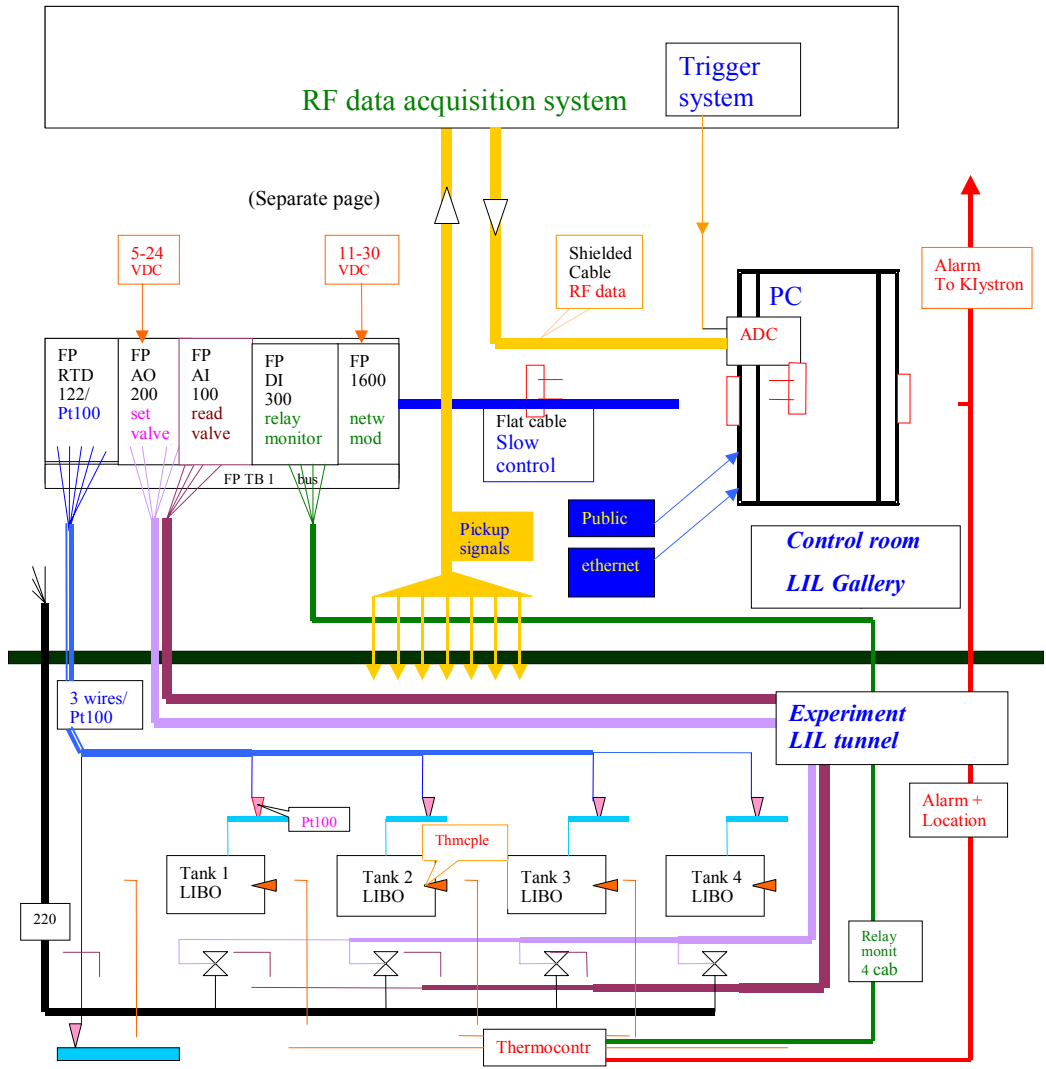

**Figure 7.14** *Controls and Data Acquisition for high power tests of LIBO-62 prototype at CERN [D.48].*

Figure 7.14 shows a general scheme of data acquisition and control system. Resistive probes (Pt100) are mounted on the stainless steel water tubes in order to check the cooling temperature, while thermocouples ($T_c$) are fixed on the tanks to control the temperature of the prototype under RF power. All these signals are managed by commercial data acquisition system (produced by National Instrument) and connected to a PC for recording. Signals from vacuum system are also generated and connected to the klystron source for safety reason. RF signals are generated by the five pick-up loops, positioned in the cavity, to control the electromagnetic field distribution in the accelerating tanks. These are connected to the RF data acquisition system positioned in the control room (LIL gallery). Tables 7.4 and 7.5 summarise the above mentioned systems.

**Table 7.5** *Details of Steering System Controlled elements.*

| *Steering System Controlled elements* | *Reading (input)* | *Characteristic* | *Reading comes from* | *Comments* |
|---|---|---|---|---|
| Temperature | 5 temperature readings: 4 related to the 4 tanks and 1 related to the water input circuit | Pt100 | Temperature sensors installed on LIBO | Signals use 5 cables (of 10 available) from LIBO test area up to the Klystron Gallery |
| Cooling | 1 digital alarm condition due to LIBO over temperature | Digital (ON/OFF) | Thermocontroller which controls the 4 Tcs or Pt100s in LIBO test area | It uses one of the 10 cables available from LIBO test area up to the Klystron gallery |
| | Motorised valves | 4 signals (4-20 mA) | Valve controls in LIBO test area | Signals use half conductors in 4 cables (of 10 available) from LIBO test area up to the Klystron Gallery |
| Vacuum | Vacuum level in LIBO | RS-232 | Balzer TPG 300 | The instrument is in Klystron Gallery |
| | *Actions (output)* | *Characteristic* | *Action works on* | *Comments* |
| | Motorised valves | 4 signals (4-20 mA) | Valve controls in LIBO test area | Signals use half conductors in 4 cables (of 10 available) from LIBO test area up to the Klystron Gallery |

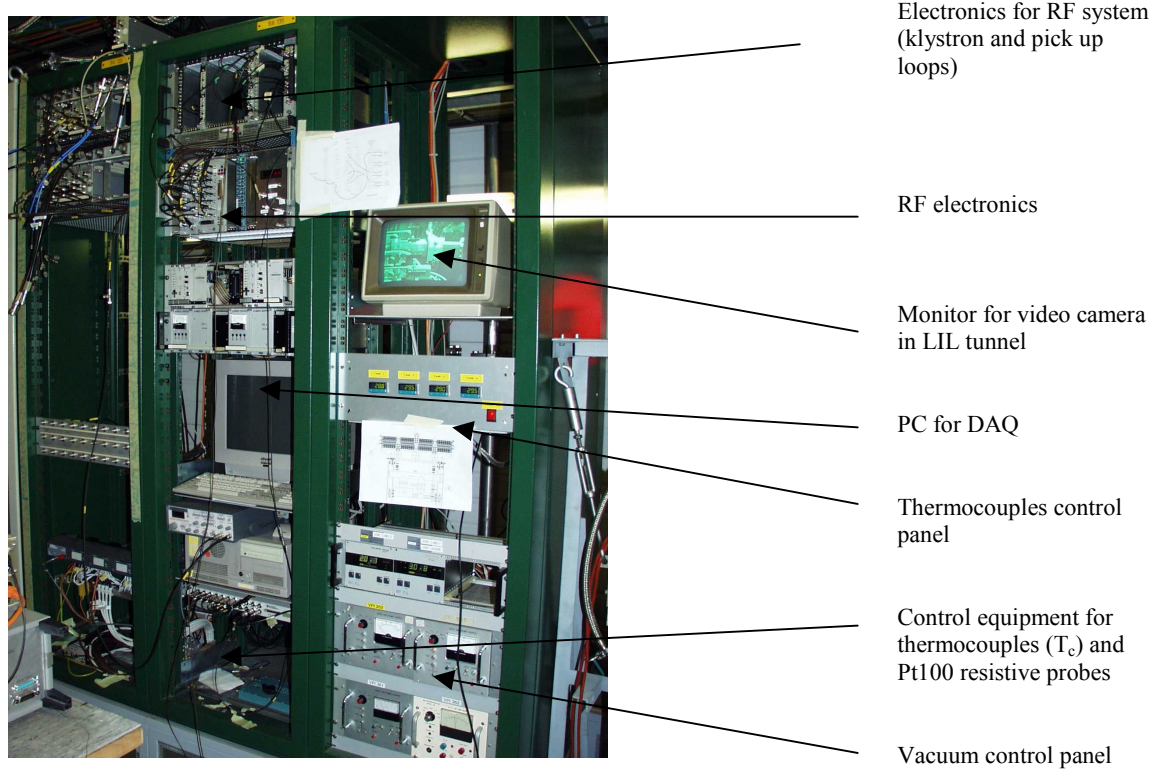

Electronics for RF system (klystron and pick up loops)

RF electronics

Monitor for video camera in LIL tunnel

PC for DAQ

Thermocouples control panel

Control equipment for thermocouples (T$_c$) and Pt100 resistive probes

Vacuum control panel

**Figure 7.15** *Control room for high power experiment at LIL gallery.*

### 7.2.2    Conditioning and RF tests of the LIBO-62 prototype

❑   *General considerations on high power behaviour of RF cavities*

The properties of the accelerating structure, which have been discussed till now, have been measured at low power levels, of the order of milliwatts. At this stage the structure behaves as a linear passive network. When high power, at levels of megawatts, is applied to the structure, interaction between the intense fields, the metal boundaries and the residual gas molecules begin to assume importance [C.1, C.6, C.13]. As the incident power level is raised from zero, the first phenomena which may be encountered is a resonant high frequency discharge known as the "multipactor" effect. This discharge can occur between two surfaces satisfying the requirement that an electron accelerated by one half cycle of the RF should produce more than one secondary electron at the right moment to be captured by the next half cycle and accelerated in the opposite direction. The process can be repeated at each half cycle, building up a space-charge limited discharge current. When this phenomena appears at a high impedance point, for example across the gaps, it can: 1) load the cavity for the duration of the RF pulse and completely prohibit build-up, or 2) it might load the cavity only for a short period at the beginning, producing jitter in the leading edge, or 3) it may load some cells more than others, changing the field distribution along the structure and with the risk of excitation of adjacent modes.

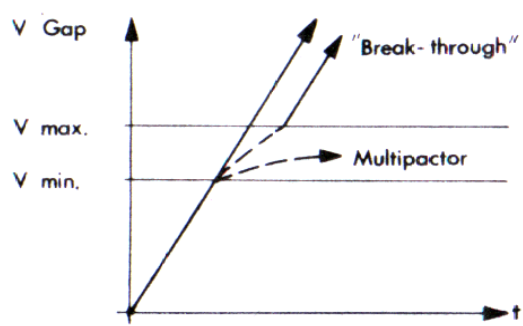

***Figure 7.16*** *Representation of gap voltage variation with time.*

At high fields the problem is the vacuum breakdown or sparking, which is a rapid ionisation by collision of gas molecules in the gaps, resulting in the collapse of the accelerating field. At high fields the problem of X-ray emission also arises. Figure 7.16 shows the schematic representation of gap voltage variation with time. In these mechanisms the quality surface plays a crucial role. It has been proven that surfaces without special cleaning may have values of multipactoring or breakdown four times higher than the cleaned and outgassed surfaces. The undesirable effects of these phenomena are that the growing electron current absorbs RF energy and may clamp the magnitude of electric fields at the multipactoring level. Considerable energy can be deposited in localised regions of the accelerating noses, resulting in outgassing or evaporation of material and, at high levels of electric fields, to a general cavity breakdown. Finally for very high field, "corrosion" of the surface copper material could also arise. With normal vacuum, in the range of $10^{-6}$ mbar, and undegassed surfaces, field emission can initiate discharges at electric fields lower than 10 MV/m.

*Table 7.6 First electrical breakdown field [C.20].*

| Material and production quality | First breakdown field  (MV/m) |
|---|---|
| Electro-formed | 10 (standard copper) – 40 (pure copper sulphate) |
| OFHC (lathe finished) | 20 |
| OFHC (electro-polished) | 16 |
| OFHC (Diamond finished) | 70 |

Table 7.6 shows typical results for the first breakdown field level for electro-formed, electro-polished, lathe and diamond finished copper in clean and good vacuum condition.

From this table it is evident that, in certain conditions and in a stable situation, strong electric fields greater than 10 MV/m can be sustained in RF accelerators. This results partly from the fact that there are no exposed insulators in regions of high electric field. In addition, RF accelerators run at high duty cycle, and it is possible to condition electrodes to remove surface whiskers. Moreover the accelerator operates for long periods of time at high vacuum, minimising problems of surface contamination on electrodes. In any case, whatever the mechanism of the discharge, they need a supply of gas molecules for ionisation, preferably it seems originating in or on the metal surface, since the introduction of clean gas into the cavity does not appear to increase sparking rate. The process of electrodes conditioning is attributed to the driving out of trapped gas through the surface of the electrode, and this process is normally aided by baking out at high temperature, by the sparking discharges themselves, or simply by the passage of time.

❑  *Details of prototype conditioning and RF tests*

The module has been connected to modulator 35 of the LIL tunnel and the klystron frequency has been set up at 2998.18 MHz, with the initial water cooling temperature fixed at 28.0 °C.

Remembering that the energy gain in MeV is $\Delta W = V \cdot T \cdot \cos(-19°)$, one can calculate the nominal voltage V of the prototype necessary to accelerate protons from 62 MeV to 73.8 MeV (table 5.5):

$$V = \frac{11.8}{0.9455 \cdot 0.846} = 14.75 \ MV$$

From measurements of the Q factor one obtains values of $Q_0 = 7900$ and $\hat{\beta}$ (matching factor) of 1.14. From measurements at low power of $r_{seff}/Q$ one can find a value of 6.34 k$\Omega$, so the total shunt impedance of the first module is 50 M$\Omega$ (measured) [D.51]. The power needed for the first prototype module is obtained by:

$$P = \frac{V^2}{Z} = \frac{217.6}{50} \cong 4.35 MW$$

Before to reach this power level one must perform the so-called conditioning of the cavity. Figure 7.17 shows the PC recording of the different physical parameters of the cavity during the power injection.

The RF parameters are the power level measured on the klystron control panel, the reflected power and the power levels in the tanks from the five pick-up loops.

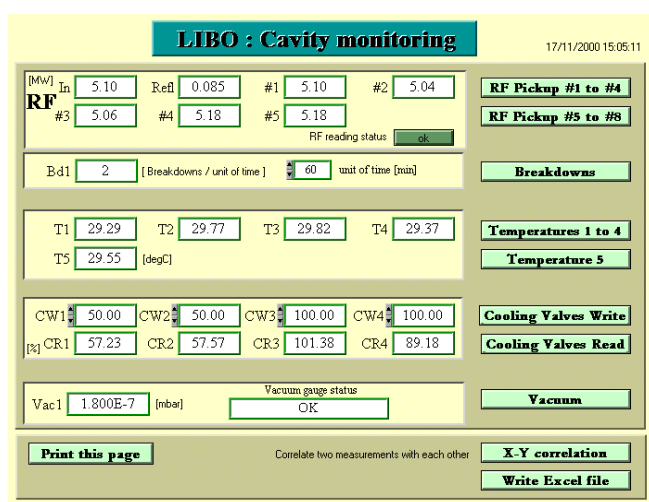

**_Figure 7.17_** _Data acquisition for RF high power tests of the prototype performed in LIL tunnel at CERN. The RF parameters are: injected power (in); reflected power (ref) and the power levels measured by the five pick-up loops (1, 2, 3, 4, 5). The thermal and vacuum parameters are: temperature of input cooling water ($T_1$), temperatures at the exit of the water cooling channels for the four tanks ($T_{2, 3, 4, 5}$), and vacuum level in the cavity (vac). Cri indicates the water valves conditions (%)._

At the second level the temperatures measured by the resistive probes. $T_1$ is the temperature of input cooling water, while $T_{2, 3, 4, 5}$ are the output water temperatures for the four tanks. Cri is the apertures of the water valves. The valves have been used for the thermal analysis of the powered prototype, while Vac1 indicates the vacuum level in the cavity. In a separate panel the tank temperatures measured by thermocouples are also shown. Figure 7.18 presents the above parameters as function of the injected RF power.

The conditioning has been performed in two different steps. The first RF power injection was performed on 8[th] November with a value of 27 kW (repetition rate 100 Hz, pulse length 2 μsec) and a cooling water temperature of 28 °C. At this stage the vacuum levels jumped from from 5 x $10^{-7}$ and 1 x $10^{-8}$ mbar for the cavity and manifold respectively to 1,1 x $10^{-6}$ and 5,3 x $10^{-8}$ mbar, with a consequent rapid stabilisation.

In this condition and after about 72 hours, the nominal accelerating field level of 15.8 MV/m (corresponding to a peak power of 4.7 MW) was exceeded in about forty increasing power levels, without multipactoring and with a very limited number of breakdowns. During all measurements the situation was remarkably steady, with a very reduced vacuum variation and fast return to the stable values.

Having extended the pulse length to 5 μsec, maintaining at the same time the same repetition rate (100 Hz), an accelerating field level of 27.5 MV/m (corresponding to a RF power level of 14.2 MW) was reached about 14 hours later. In this condition the corresponding computed maximum surface field is 2.6 times the Kilpatrick limit. The conditioning history of the prototype is shown in figure 7.19 and table 7.6, where the two different power conditions are visible. The field level in the module was determined by the input power and the previously measured shunt impedance.

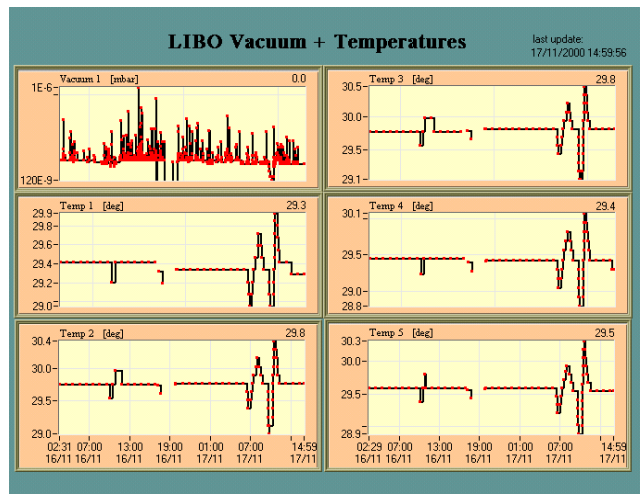

(a)

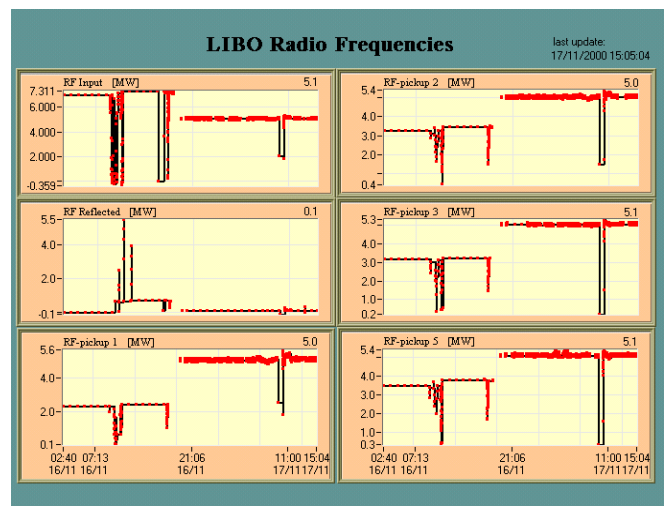

*(b)*

**Figure 7.18** *Prototype module parameters measured during conditioning and RF power tests performed at CERN.*

Changing the temperature of the LIBO structure, a fine-tuning of the frequency has been proved. This was achieved by adjusting the temperature of the water circulating in the cooling plates. The measured frequency dependence of the cavity with the temperature is about 60 kHz / °C, in accordance with the design value (see chapter 5).

It is important to note that the RF power flows from klystron gallery to the cavity in LIL tunnel, through about 20 m waveguide line. In order to calculate the real power injected in the prototype and to know precisely the power attenuation between the klystron and the module, measurements on reflected power and attenuation have been also performed. The signal coming from the coupler on the waveguide at the input of the module has been measured on the power meter upstairs in the gallery. At the nominal power the measurement gave an attenuation factor of 13.3%; adding ~1.6% of reflected power (see paragraph 6.3), the total attenuation between klystron and module was calculated as 15% [D.50].

**Table 7.6** *RF power test at LIL.*

| Main RF power steps for the conditioning period of the prototype |
|---|
| 100 Hz, 2 μsec, from power klystron (PK) 0 MW→ 4 MW (about 48 hours) |
| 100 Hz, 2 μsec, from power klystron (PK) 4 MW → 7 MW (about 72 hours) |
| 100 Hz, 5 μsec, from power klystron (PK) 1 MW → 6 MW (about 38 minutes) |
| 100 Hz, 5 μsec, from power klystron (PK) 7,2 MW → 13 MW (about 4 hours) |
| 100 Hz, 5 μsec, from power klystron (PK) 14,5 MW → 18 MW (about 5 hours) |
| 100 Hz, 5 μsec, from power klystron (PK) 18 MW → 20 MW (about 24 hours) |
| 100 Hz, 2 μsec, from power klystron (PK) 20 MW → 22 MW (about 24 hours) |

**Table 7.7** *RF power levels for the prototype tests taken with the peak power analyser.*

| Power at klystron (MW) | Reflected power (MW) | Attenuation (waveguide) | Reflected power | Attenuation + reflected |
|---|---|---|---|---|
| 5.94 | 0.079 | 0.1329 | 0.01534 | 0.1483 |
| 8 | 0.105 | 0.1425 | 0.01530 | 0.1578 |
| 13 | 0.166 | 0.1692 | 0.01537 | 0.1846 |

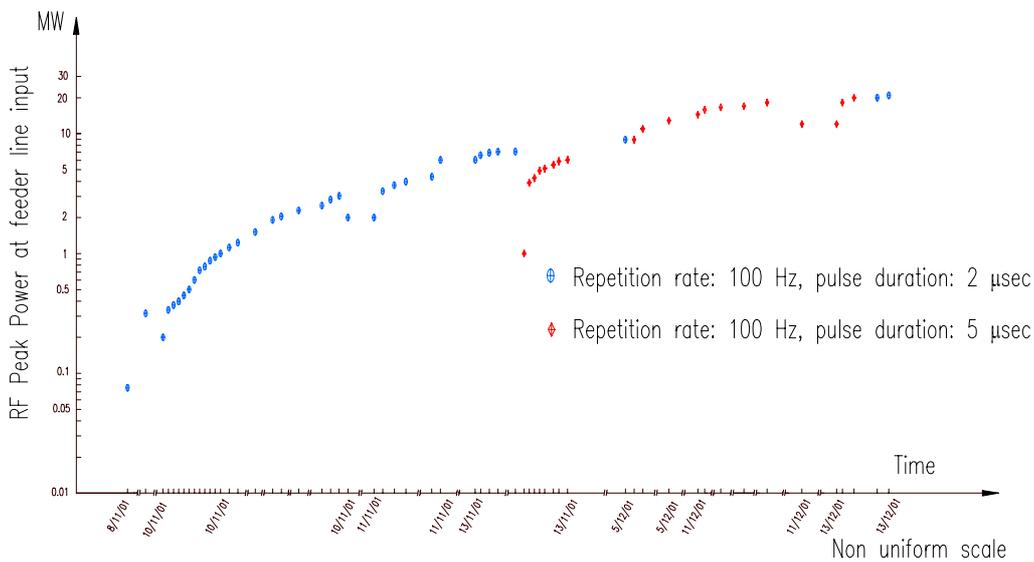

**Figure 7.19** *High power levels injected into LIBO during conditioning: the power levels are in logarithmic scale.*

Therefore in case of 6 MW provided by the klystron, 5.1 MW are really injected into the module. More measurements have been made for the other input RF power levels. As mentioned in table 7.7 the waveguide attenuation increases with RF power, while the reflected power stays constant in % (≈ 1.5%). For example for nominal 13 MW measured with the peak power meter at the klystron level, one then puts into the cavity $13 \times 0.815 = 10.6$ MW (where 0.815 takes into account the waveguide attenuation and the reflected power).

In this case the real maximum field is then calculated as:

$$\sqrt{\frac{10.6}{4.35}} \cdot 1.43 = 2.23 \ E_k \ \rightarrow \ E_s = 2.23 \cdot 46.77 \cong 104 \ \frac{MV}{m}$$

A further increase of the RF input power was unfortunately prohibited by the power limit of the installed circulator. There were, however, no indications that the limit of the field level into LIBO had been reached. After the tests, internal visual inspections on the structure by endoscope have been performed in order to control the RF surfaces and brazing alloy distribution. In this inspection no damages were evident.

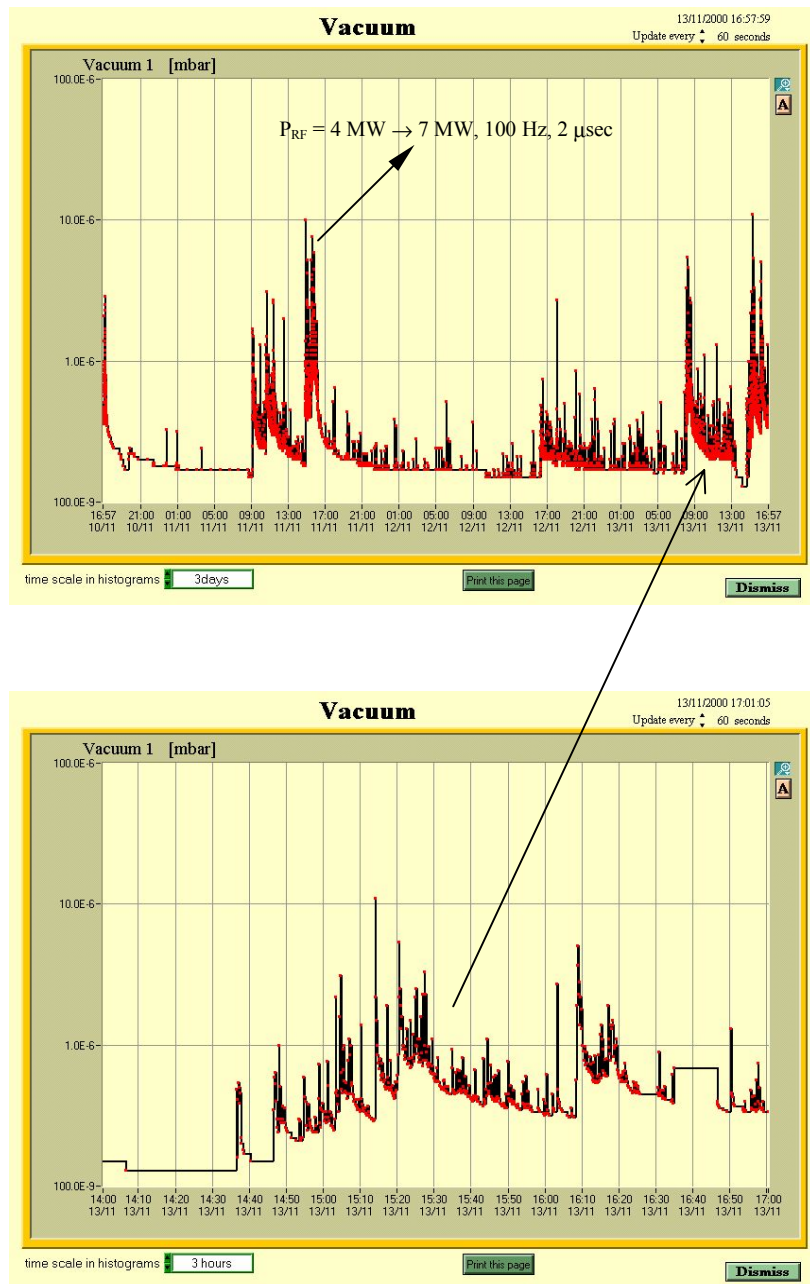

**_Figure 7.20.a_** _Examples of vacuum levels of LIBO prototype, during RF conditioning. A change of RF input power generates vacuum breakdowns with rapid stabilisation._

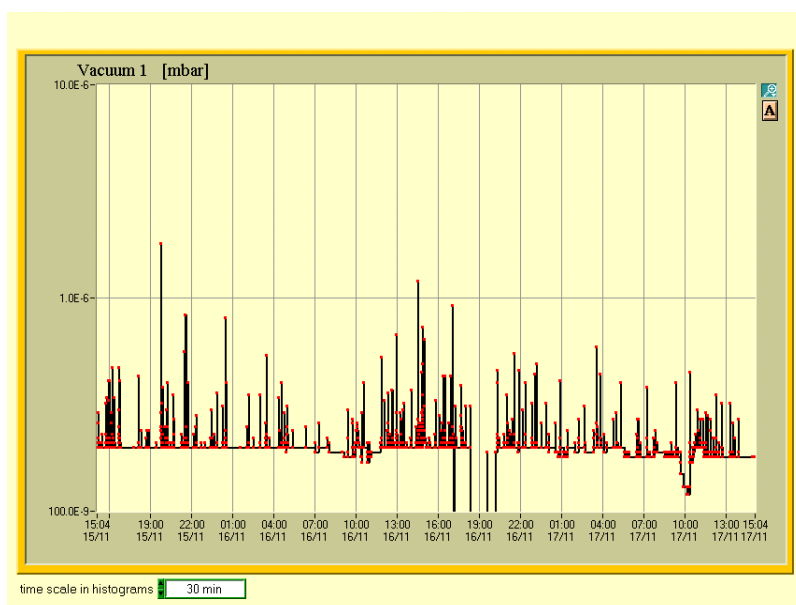

***Figure 7.20.b*** *Stable condition of vacuum level into LIBO during RF conditioning.*

❑   *RF power analysis via thermal measurements*

To get an independent confirmation of the high field level really present in the prototype and to understand the thermal behaviour of LIBO, systematic measurements have been compared with analytical and numerical calculations, as mentioned in chapter 5. A forced convection and conduction heat exchange between a tank, its cooling plates and water, and consecutive tanks have been investigated, following the figure A annex 5.3. More precisely the temperatures have been measured as a function of injected power and water flow. Thermocouples fixed on each tank, and platinum resistive probes (Pt100s) mounted on water tubes can provide a precise temperature measurement of the global system from the thermal point of view. Once the temperature set-up of tanks and cooling channels have been obtained with full water flux (all valves completely open) at 28.5 °C (table 7.8), transient and steady conditions have been investigated at different power levels and with different water fluxes [C.56, C.57, C.58].

In particular, regulating adequately the water flux in a tank (for example in tank 2, $Q_2$) with a valve and a flow meter, a steady state condition with reasonable differences in temperature between tanks has been obtained. This allowed us to appreciate the thermal behaviour of the global system, checking if the RF power injected from the klystron was really present in the prototype and correctly flowed through the bridge couplers.

Significant cases are illustrated from table 7.9 to table 7.13, where the measurements are compared with calculations, generated by the analytic model and corresponding to the applied peak power and to only a quarter of such power.

▪   Table 7.9 shows the changes of temperatures in tanks and cooling channels, moving from 0 MW to 6 MW. Measurements and calculations show that in this condition the 6 MW RF power is really injected into the module and it is uniformly distributed in four tanks ($\rightarrow$ we have RF and thermal continuity between consecutive tanks).

- Tables 7.10 and 7.11 show the different increases of temperature in the tanks ($\Delta T_{TK2-i} = T_{TK2}-T_{TKi}$) and in the water ($\Delta T_{w2-i} = T_{w2}-T_{wi}$) with different flow rates ($T_i$ indicates the average value over the three tanks 1, 3 and 4; i is the tank number), for RF power of 6 MW and 13 MW. Also here there is evidence that the real power injected in the module is the power read in the klystron control room. Moreover it is also evident that relatively small differences in temperature can be achieved between consecutive tanks.

- Tables 7.12 and 7.13 compare the absolute values of the temperature in tank 2 at different power levels (again measured and calculated), for a reduced water flow rate $Q_2$. The differences between the full applied power (for 6 MW and 13 MW respectively) and a quarter of this power are more visible.

*Table 7.8 Set-up apparatus for cooling tests.*

| Applied peak power: 0 MW | | | |
|---|---|---|---|
| Valves open $\rightarrow$ Water pump 200 $m^3/h$ max, $H=50$ m, $p_{in}=6$ bar, $p_{out}=0.5$ bar | | | |
| $Q_i$: 23 l/min with i = tank1,2,3,4 | | | |
| $T_{water\ control\ room\ (gallery)} = 28.5\ °C,\ T_{water\ input\ LIBO\ (tunnel)} = 27.67\ °C$ | | | |
| $T_{TK1}$ | 29.1 °C | $Pt_{100\ 1}$ | 27.85 °C |
| $T_{TK2}$ | 29.7 °C | $Pt_{100\ 2}$ | 27.90 °C |
| $T_{TK3}$ | 29.0 °C | $Pt_{100\ 3}$ | 27.66 °C |
| $T_{TK4}$ | 29.5 °C | $Pt_{100\ 4}$ | 27.67 °C |

*Table 7.9 Temperature gradients for copper tanks and water at the output of the cooling plates, when the prototype is heated by the RF power injection from 0 to 6 MW.*

| Applied peak power: 0 MW $\rightarrow$ 6 MW | | | |
|---|---|---|---|
| Duty cycle: 5 $10^{-4}$, power loss (att.+ refl.): 15% (table 7.7) $\rightarrow$ Average power into LIBO: 2550 W | | | |
| Valves open $\rightarrow Q_i$: 23 l/min with i = tank 1,2,3,4 | | | |
| $\Delta T_{TK1}$ | 1.3 °C | $\Delta T_w\ Pt_{100\ 1}$ | 0.54 °C |
| $\Delta T_{TK2}$ | 1.2 °C | $\Delta T_w\ Pt_{100\ 2}$ | 0.57 °C |
| $\Delta T_{TK3}$ | 1.6 °C | $\Delta T_w\ Pt_{100\ 3}$ | 0.54 °C |
| $\Delta T_{TK4}$ | 1.1 °C | $\Delta T_w\ Pt_{100\ 4}$ | 0.58 °C |
| Average $\Delta T_{TKi}$ (Measured) | 1.3 °C | Average $\Delta T_w\ Pt_{100}$ (Measured) | 0.56 °C |
| Average $\Delta T_{TKi}$ (Calculated @ 6 MW) | 1.14 °C | Average $\Delta T_w\ Pt_{100}$ (Calculated @ 6 MW) | 0.65 °C |
| Average $\Delta T_{TKi}$ (Calculated @ 1.5 MW) | 0.63 °C | Average $\Delta T_w\ Pt_{100}$ (Calculated @ 1.5 MW) | 0.5 °C |

Moreover figure 7.21 shows the transient thermal condition of a copper tank and its cooling water with RF power of 13 MW and a water flux in the cooling channel reduced from 23 l/min up to 2,8 l/min. Analogies of transients calculated with numerical codes have been found. Small discrepancies are due to the intrinsic time delays of the thermocouples, not present in the calculations.

**Table 7.10** _The different increases of temperature structure ($\Delta T_{TK2-i} = T_{TK2}-T_{TKi}$) and cooling water ($\Delta T_{w2-i} = T_{w2}-T_{wi}$) with different flow rates. The nomenclature is refereed to figure A annex 5.3._

| | Applied peak power: 6 MW | | |
| --- | --- | --- | --- |
| Duty cycle: 5 $10^{-4}$, power loss (att.+ refl.): 15% (table 7.7) → Average power into LIBO: 2550 W | | | |
| $Q_i$: 23 l/min with i = tank 1,3,4;  $Q_2$: 3 l/min | | | |
| | Measured @6 MW | Calculated @6 MW | Calculated @ 1.5 MW |
| $\Delta T_{w2-wi}$ | 0.75 °C | 0.74 °C | 0.18 °C |
| $\Delta T_{TK2-TKi}$ | 1.7 °C | 1.73 °C | 0.43 °C |

**Table 7.11** _The different increases of temperature structure ($\Delta T_{TK2-i} = T_{TK2}-T_{TKi}$) and cooling water ($\Delta T_{w2-i} = T_{w2}-T_{wi}$) with different flow rates. The nomenclature is refereed to figure A annex 5.3._

| | Applied peak power  P: 13 MW | | |
| --- | --- | --- | --- |
| Duty cycle: 5 $10^{-4}$, power loss (att.+ refl.) : 18% (table 7.7) → Average power into LIBO: 5330 W | | | |
| $Q_i$: 23 l/min with i = tank 1,3,4;  $Q_2$: 3 l/min | | | |
| | Measured @ 13 MW | Calculated @ 13 MW | Calculated @ 3.25 MW |
| $\Delta T_{w2-wi}$ | 1.61 °C | 1.54 °C | 0.38 °C |
| $\Delta T_{TK2-TKi}$ | 3.36 °C | 3.60 °C | 0.90 °C |

**Table 7.12** _Absolute values of the temperature in tank 2 at 6 MW for a reduced water flow rate $Q_2$._

| | Applied peak power  P: 6 MW | | |
| --- | --- | --- | --- |
| Duty cycle: 5 $10^{-4}$, power loss (att.+ refl.) : 15% (table 7.7) → Average power into LIBO: 2550 W | | | |
| $Q_i$: 23 l/min with i = tank 1,3,4;  $Q_2$: 3 l/min | | | |
| | Measured @ 6 MW | $\|\Delta T_{TK}\|_{measured-calculated}$ @ 6 MW | $\|\Delta T_{TK}\|_{measured-calculated}$ @ 1.5 MW |
| $T_{TK2}$ | 30.67 °C | 0.42 °C | 1.8 °C |

**Table 7.13** _Absolute values of the temperature in tank 2 at 13 MW for a reduced water flow rate $Q_2$._

| | Applied peak power  P: 13 MW | | |
| --- | --- | --- | --- |
| Duty cycle: 5 $10^{-4}$, power loss (att.+ refl.) : 18% (table 7.7) → Average power into LIBO: 5330 W | | | |
| $Q_i$: 23 l/min with i = tank 1,3,4;  $Q_2$: 3 l/min | | | |
| | Measured @ 13 MW | $\|\Delta T_{TK}\|_{measured-calculated}$ @ 13 MW | $\|\Delta T_{TK}\|_{measured-calculated}$ @ 3.25 MW |
| $T_{TK2}$ | 32.97 °C | 1.3 °C | 3.31 °C |

With the above considerations, one can conclude the following.

- By varying the RF power injected into the module and the water flow in one tank, _temperature gradients could be achieved with respect to the three others_. Then it is possible to control the thermal behaviour of the entire module and of each single tank with an accurate setting of the water valves. Thermal measurements then prove that there is RF continuity between consecutive tanks, so the injected power correctly flows through the bridge couplers and coupling slots.

Moreover it is possible to perform a fine frequency tuning of individual accelerating tank, comparing one to the others, by using cooling water.

▪ There is an evident correspondence between temperatures measured and calculated at the <u>applied peak power</u>. The results obtained for different RF power levels are compared with computations: the agreement for the whole set of measurements is within 5-10 %. One can reasonably conclude that *the power, read with the peak power analyser in the LIL gallery, is really present in the module and that <u>an accelerating field level of about 30 MV/m has been achieved during the RF tests</u>.*

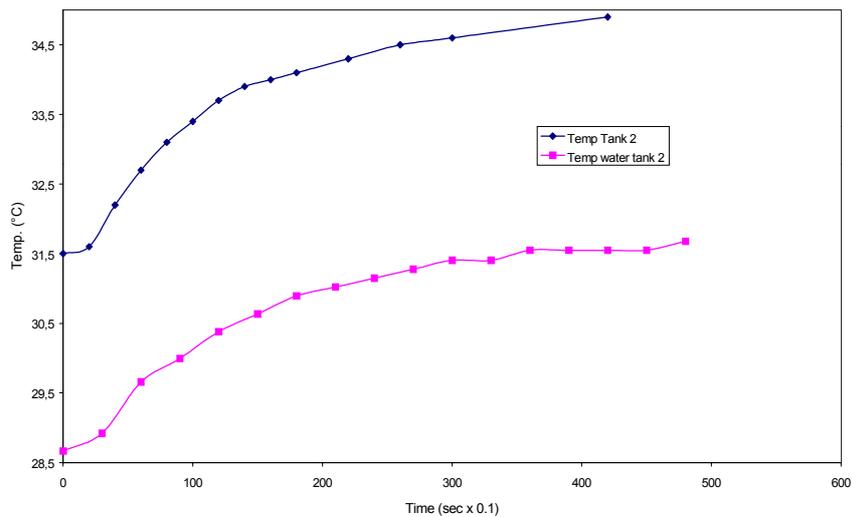

***Figure 7.21*** *Example of transient condition measured for copper structure (tank 2) and the cooling water.*

## 7.3     Acceleration tests at LNS in Catania of the LIBO-62 prototype[*]

❑   *General description*

In July 2001 the prototype module was transported and installed at the LNS in Catania (figure 7.22) [A.7]. A view of the accelerator placed in the proton beam line is shown in figures 7.23 and 7.24. For the installation, the collaboration has used the same apparatus of LIL tests, such as vacuum system, data acquisition and controls. Due to low repetition rate used for the acceleration tests (10 Hz), a simple closed loop water system has been used for cooling, enough to dissipate the average power of 200 W.

A set of three quadrupoles upstream of the prototype served to match the LNS Superconducting Cyclotron (SC) proton beam to the linac acceptance, which was calculated as being of the order of 12 π mm mrad (unnormalised). The proton beam delivered by the cyclotron had an energy of 62 ± 0.2 MeV and the operating value of the average current at the LIBO entrance was of the order of 1 nA.

Conventional diagnostic systems like Faraday cups and alumina screens were placed upstream and downstream of LIBO. Moreover, a set of movable thin scintillating fibers developed at LNS [C.61] was used to control the beam profiles. This system was extremely useful for centering the beam at very low intensities (fraction of pA). At the end of 2001 a compact solid-state modulator and a 3 GHz klystron, lent in the frame of a collaboration with the firm IBA/Scanditronix, was installed and coupled to LIBO.

The new conditioning at Catania LNS Laboratory has shown that the exposure of the structure to air for few days during the alignment procedures has not affected the already cleaned structure. First acceleration tests were performed during February and May of 2002 [A.2].

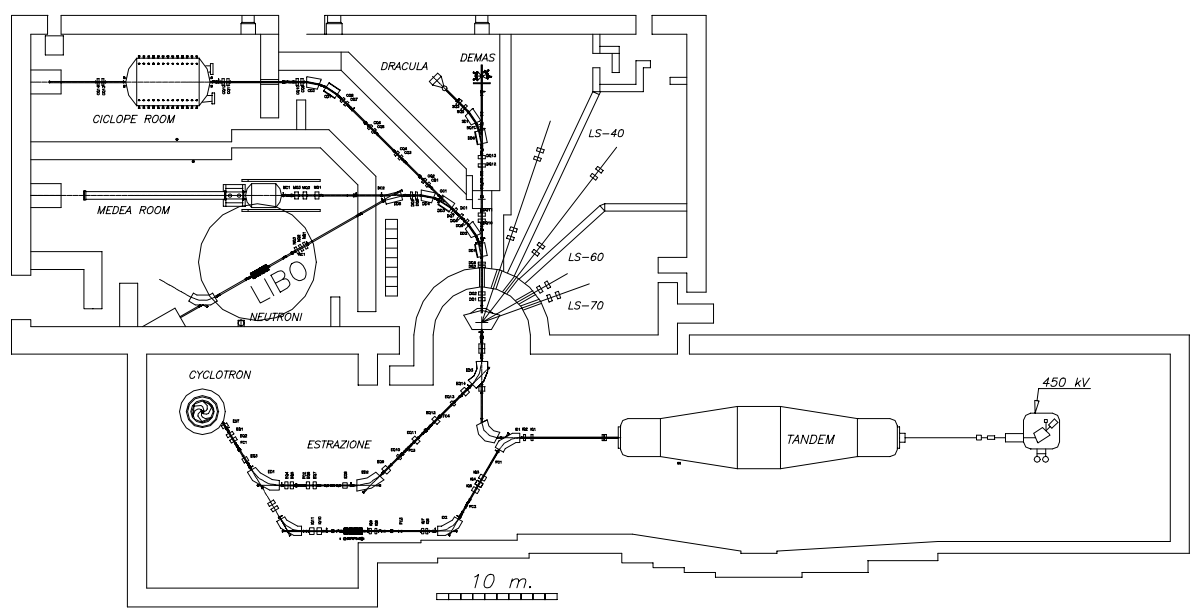

**Figure 7.22** *INFN-Laboratorio Nazionale del Sud (LNS) in Catania. The facility has been used for acceleration tests of LIBO prototype module, by using the existing 62 MeV proton Superconducting Cyclotron (SC).*

---

[*] The tests in Catania have been made in co-operation with Prof. Carlo De Martinis and the INFN group of Milan (D. Giove, C. Cicardi and L. Grilli), that were responsible for the acceleration measurements.

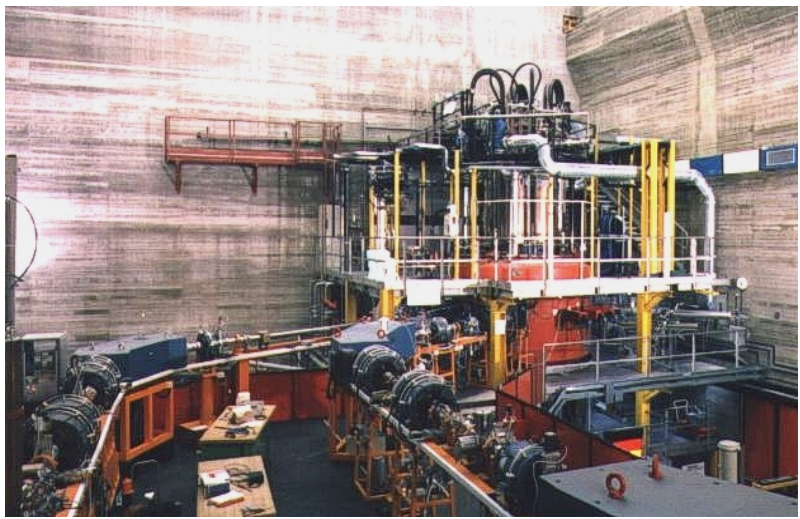

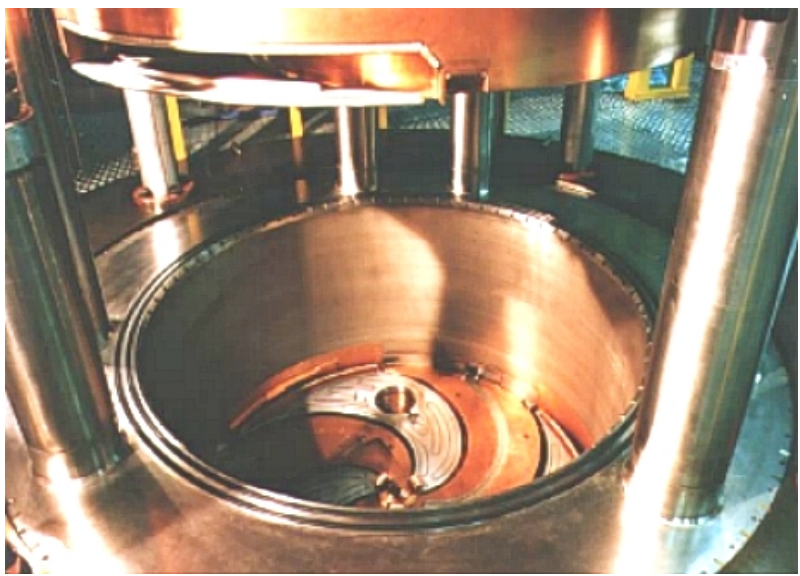

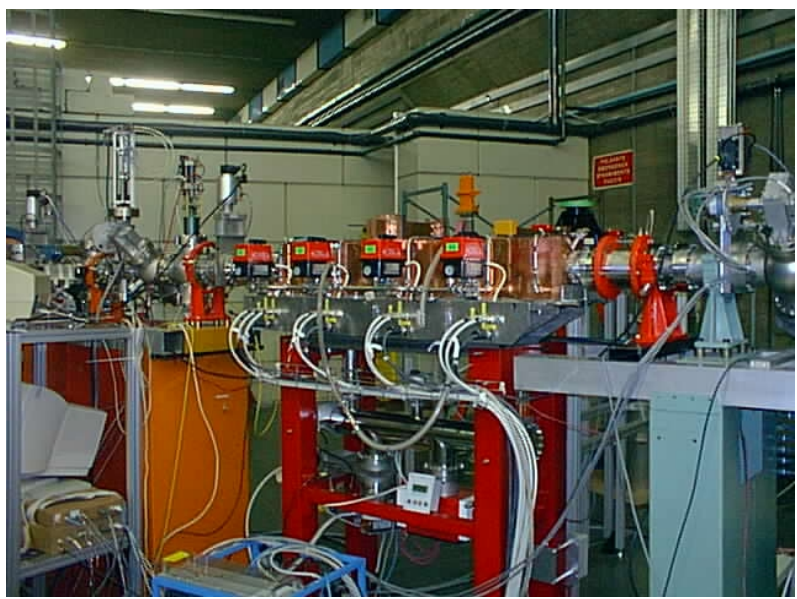

**_Figure 7.23_** _The SC cyclotron at LNS and the LIBO prototype installed in the test beam line (Catania)._

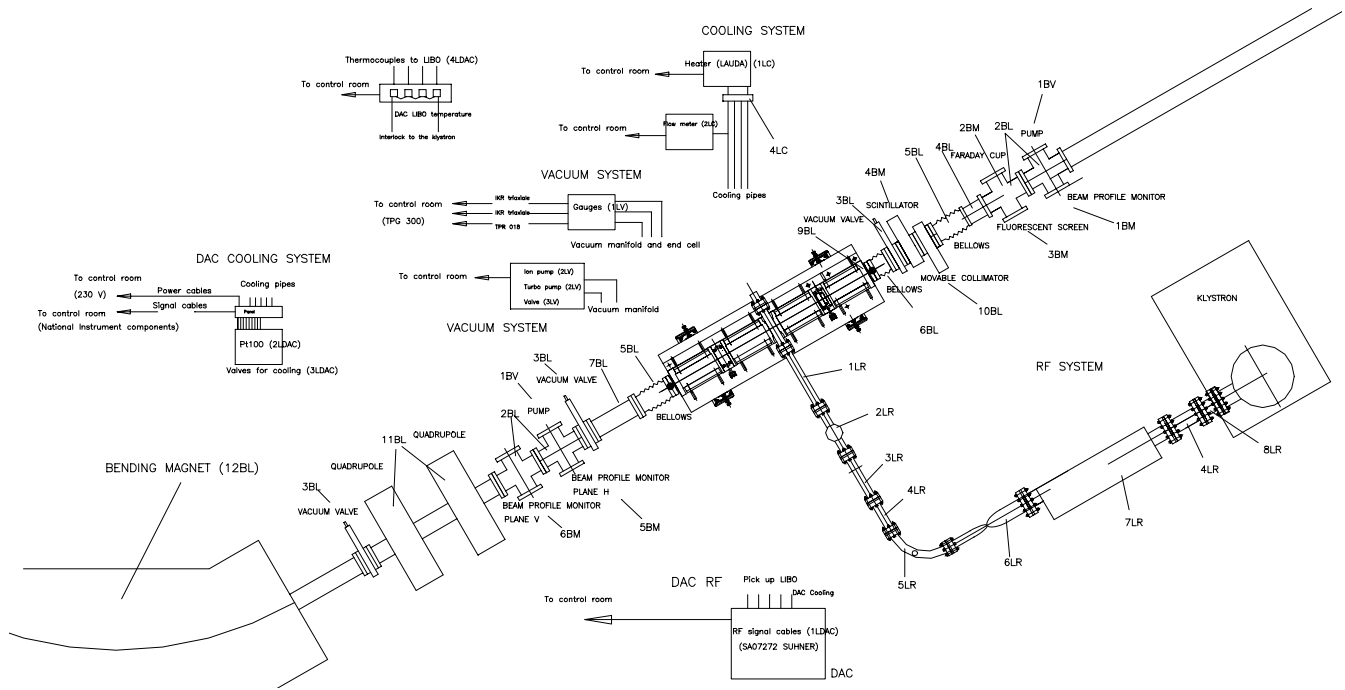

**_Figure 7.24_** _LIBO at LNS with a partial view of the beam line. The installation facility is visible with controls, DAQ apparatus, vacuum and cooling system, as well as power supply used for the acceleration tests. The injector is the 62 MeV INFN proton cyclotron (SC), while the klystron for RF power, needed to accelerate the particles up to 73 MeV, is produced by IBA/Scanditronix. Three quadrupoles (not visible in figure) are used to adjust correctly the cyclotron beam before injection into LIBO, and other two are placed downstream LIBO for beam focusing before the bending magnet, used for energy measurements. Beam profile monitors, faraday cup, fluorescent screen and scintillators are used for beam measurements before and after acceleration. Special collimators are also present to clean the particle divergences of the injected cyclotron beam._

_**Table 7.14** Main facilities used for acceleration tests of the prototype at LNS._

| ❑ Vacuum system LIBO | | ❑ Beam line elements | |
|---|---|---|---|
| 1.LV | Gauges | 1.BL | Vacuum tubes |
| 2.LV | Ion and turbo pumps | 2.BL | 4-cross-way vac. connections (3 units) |
| 3.LV | Valve | 3.BL | Vacuum valves (5 units) |
| ❑ Cooling system LIBO | | 4.BL | Vacuum tubes |
| 1.LC | Heater LAUDA | 5.BL | Bellows (CF100) |
| 2.LC | Flow meter | 6.BL | Bellows |
| 3.LC | Cooling valves | 7.BL | Special tubes (LIBO-beam line) |
| 4.LC | Manifold for cooling distribution | 8.BL | He bag for time of flight |
| ❑ RF system LIBO | | 9.BL | Fixed collimator (before LIBO) |
| 1.LR | Wave guide | 10.BL | Collimator in air (before LIBO) |
| 2.LR | WG-vacuum port | 11.BL | Quadrupoles (5 units) (from Catania) |
| 3.LR | WG-ceramic window | 12.BL | Bending magnet (from Catania) |
| 4.LR | WG-directional coupler (2 units) | 13.BL | Vacuum windows (50 μm of mylar foils) |
| 5.LR | WG-bend (SF6 port) | 14.BL | Alignment system LIBO-beam line |
| 6.LR | WG twister | ❑ Measurement hardware (beam line) | |
| 7.LR | Circulator | 1.BM | Beam profile monitor 1 |
| 8.LR | Flange adapter (IBA-Scanditronix) | 2.BM | Faraday Cup |
| ❑ Data acquisition and control (DAC) LIBO | | 3.BM | Fluorescent screen |
| 1.LDAC | RF signal cables | 4.BM | Scintillator |
| 2.LDAC | Pt100 with white signal cables | 5.BM | Beam profile monitor 2 (plane Horizontal) |
| 3.LDAC | Cooling valve (reading and steering) | 6.BM | Beam profile monitor 3 (plane Vertical) |
| 4.LDAC | Thermocouples and thermocontroller | 7.BM | Beam profile monitor 4 (plane Horizontal) |
| ❑ Vacuum line | | 8.BM | Beam profile monitor 5 (plane Vertical) |
| 1.BV | Pumps for beam line | 9.BM | Moveable target |
| | | 10.BM | Scintillator NaI(Tl) (2 units) |
| | | 11.BM | Plastic scintillator (2 units) |

❑ _Technical description of IBA compact klystron-modulator_

For the acceleration tests at LNS in Catania, a compact klystron modulator produced by IBA-Scanditronix has been used. This modulator is also used in the Scanditronix 50 MeV MM50 ARTS cancer treatment system, and it is now installed and operating at the Japan National Cancer Center (NCC) in Tokyo. All support subsystems, including the main charging power supply, the filament supply, DC reset supply and the computer interface electronics, are enclosed in the modulator cabinet, which has a total volume of less than 1 cubic metre. The klystron is mounted directly on the top cover of the modulator. Modulator components are reached through hatch-covers, visible next to the klystron (figure 7.25).

The basic specifications given for the klystron as printed on the tube manufacturers' label are as follows:

V=106 kV, I=66.4 A, Efficiency ~ 40%, RF drive 78 W peak. The pulse width of the modulator can be adjusted from few microseconds to about 10 μs, and it has the ability to operate from 0 up to 300 Hz.

However, this was not the case with the tests at LNS, where 10 Hz was sufficient. The total beam power in the klystron is given by 106 kV x 66.4 A = 7.04 MW. If the efficiency is 46%, then the maximum peak RF output power is about 7.04 x 0.46 = 3.24 MW. During accelerating tests at LNS in Catania, the klystron has operated above its normal operating level, so it has a much better efficiency than assumed here.

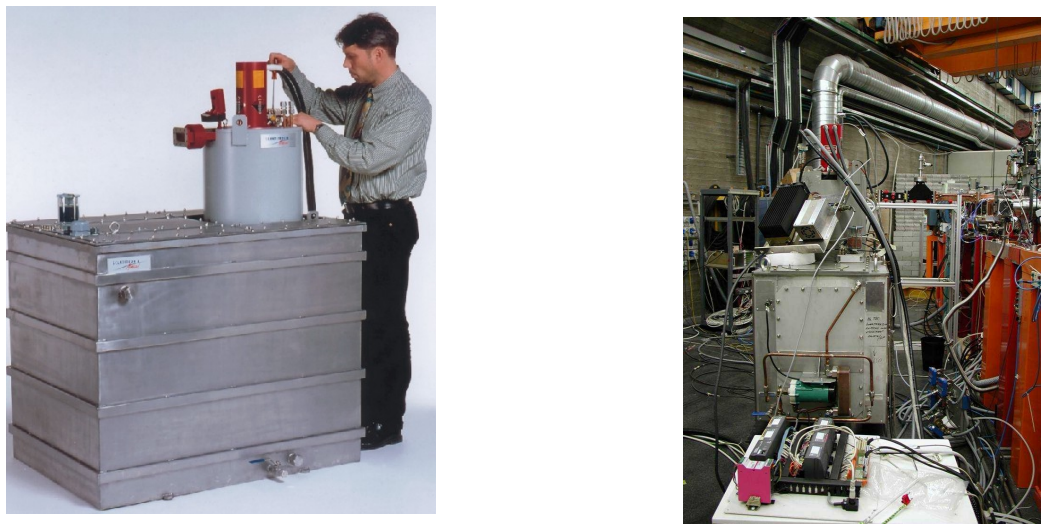

**Figure 7.25** *Commercial modulator produced by IBA, and used for LIBO acceleration tests. On the left the installation facility at LNS.*

❑   *Main results of acceleration tests at LNS*

The incoming proton beam was chopped in the injection line of the cyclotron, at the same repetition rate, and with a time window of about 30 µs. This value was much longer than the 3 GHz pulse, but it was necessary to compensate for the time jitter in the chopper. A drawback was that a large fraction of the 62 MeV was transmitted through LIBO without acceleration. Figure 7.27 shows this aspect as calculated by computer simulation.

To measure the energy of the accelerated protons, a 25 mm thick NaI(Tl)  crystal was used. The use of a nuclear detector was mandatory because of the very low beam intensity, the low duty cycle, and the computed overall acceleration efficiency. The NaI(Tl) was calibrated in energy resolution and linearity  by using as a reference the 62 MeV cyclotron beam. The results were quite good: the energy resolution at 62 MeV was better than 1% (figure 7.29) and the linearity down to 15 MeV better than 0.5%.

The detector was placed in air directly on the beam line, and protons came out from the prototype through a 50 µm mylar vacuum window. Measurements of the LIBO output beam were strongly affected by the not accelerated particles at 62 MeV. To avoid this effect, a 30 mm thick Perspex absorber was placed in front of the detector. The 62 MeV particles were then completely stopped in the Perspex, and only protons with energy above 65 MeV could reach the detector. The set-up experiment is shown in figure 7.24 and summarised in table 7.14. The typical spectrum of the accelerated proton beam is visible in figure 7.32, where the energy of the protons is corrected for the absorption in the beam path (Mylar windows, Perspex absorber, air and detector entrance windows). The peak energy of 73 MeV, corresponding to an energy gain of 11 MeV, was obtained by injecting into the cavity only 3.4 MW (error: ± 7%) of peak RF power, the maximum that could be delivered during the tests due to limitations in the driver amplifier. As mentioned before, the peak power for a nominal energy gain of 12 MeV and a synchronous phase of -19° is P= 4.4 MW. The formula linking the power P, the energy gain $\Delta W$ and the synchronous phase $\phi_s$ is:

$$P = \left( \frac{\Delta W}{q \cdot T \cdot \cos\varphi_s} \right)^2 \cdot \frac{1}{r_s}$$

where $T$ is the transit time factor, $q$ the proton charge and $r_s$ the total shunt impedance (50 MΩ).

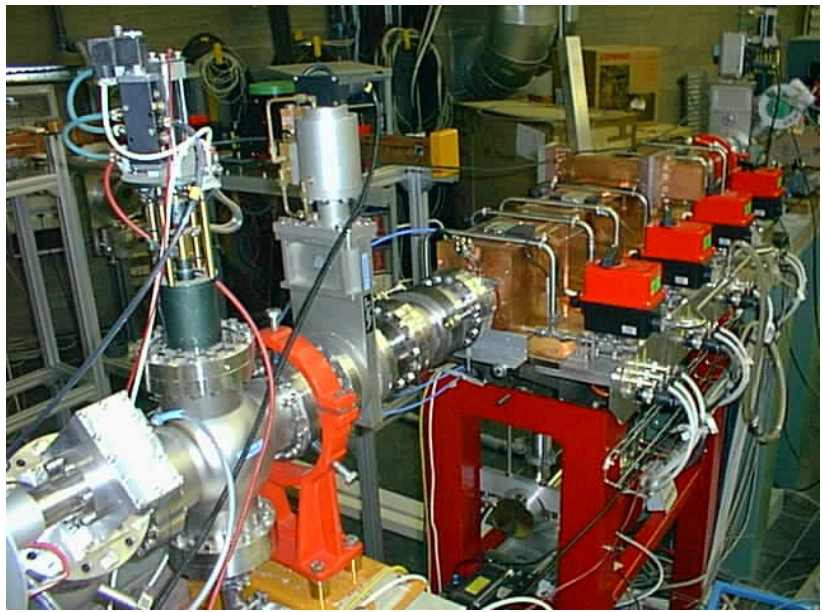

**Figure 7.26** *Partial view of the LIBO beam line for acceleration tests at LNS.*

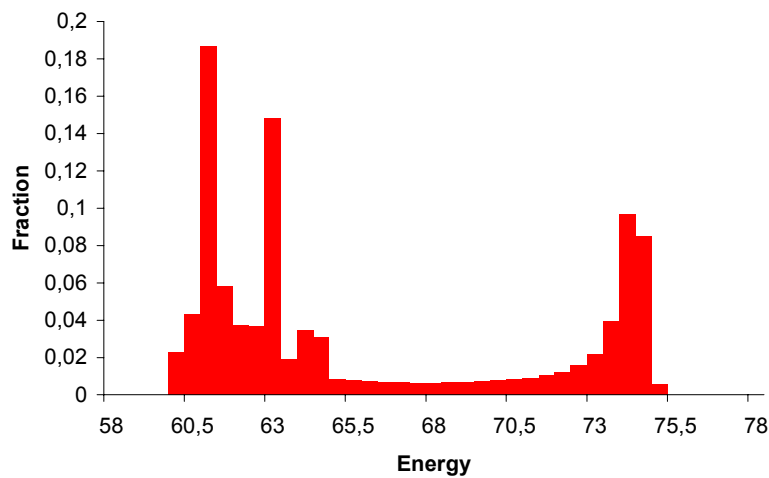

**Figure 7.27** *Energy distribution of proton beam calculated for the LIBO-62 prototype [D.51].*

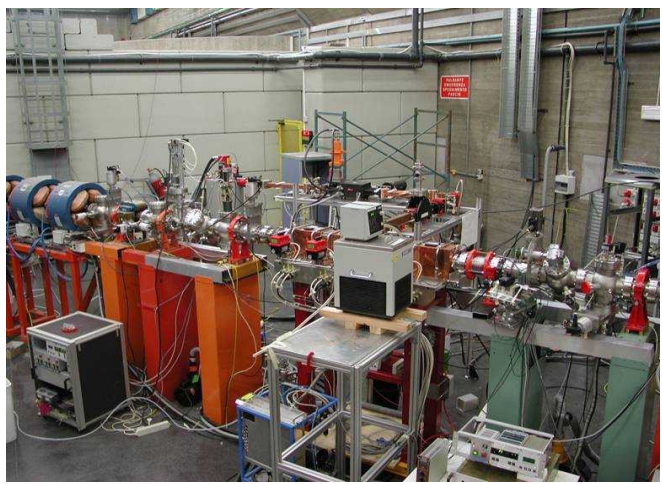

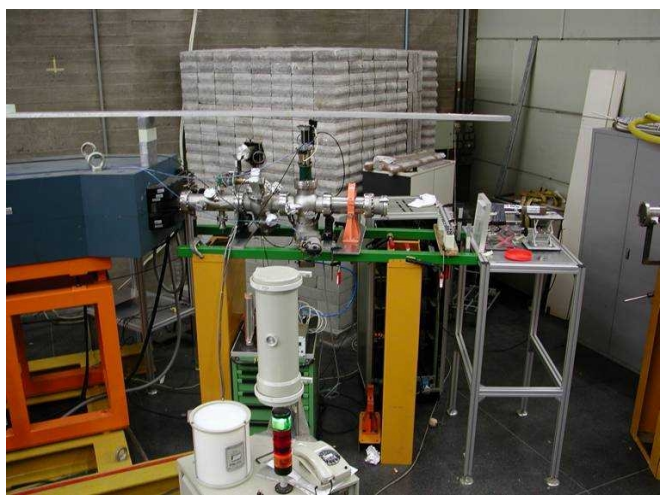

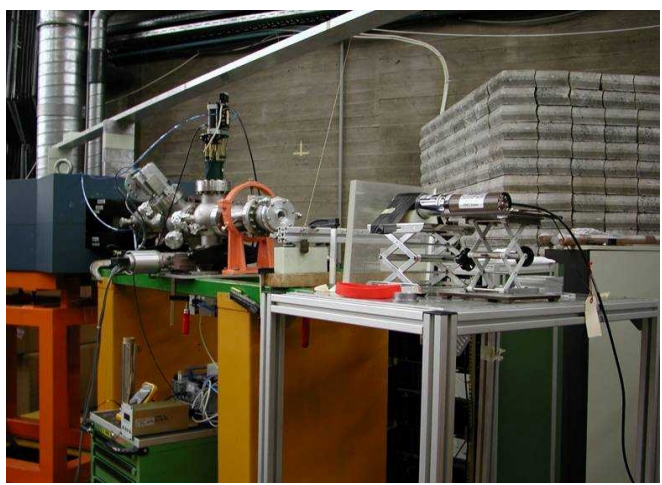

**_Figure 7.28_** _Installation of the prototype at LNS. Nuclear detector set up is shown, while the Perspex absorber is visible in front of the detector and after the bending magnet._

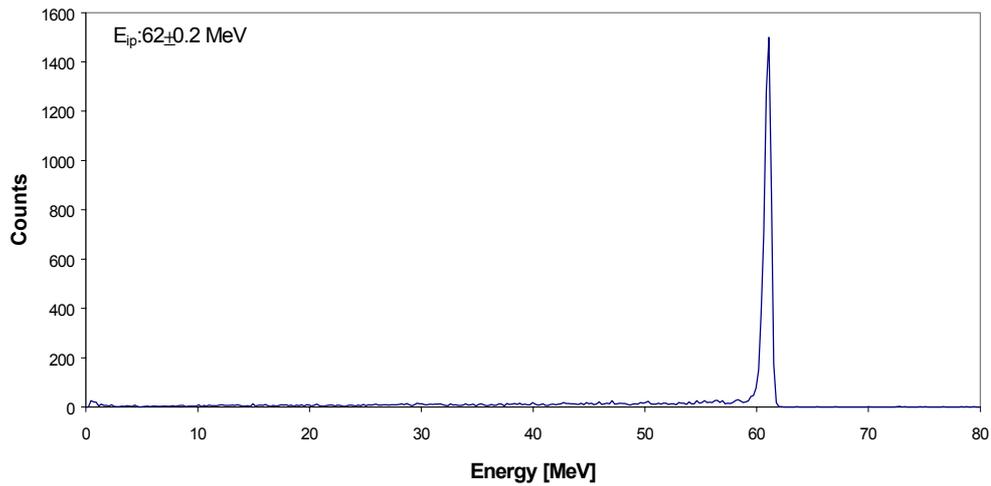

***Table 7.29*** *Measured spectrum of the 62 MeV proton beam from the Cyclotron and injected into LIBO. The energy resolution is about 1%.*

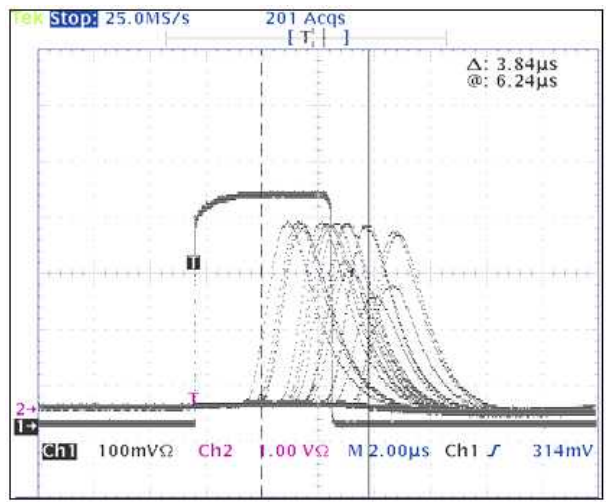

***Table 7.30*** *The 5 μsec RF pulse and accelerated proton pulse envelopes.*

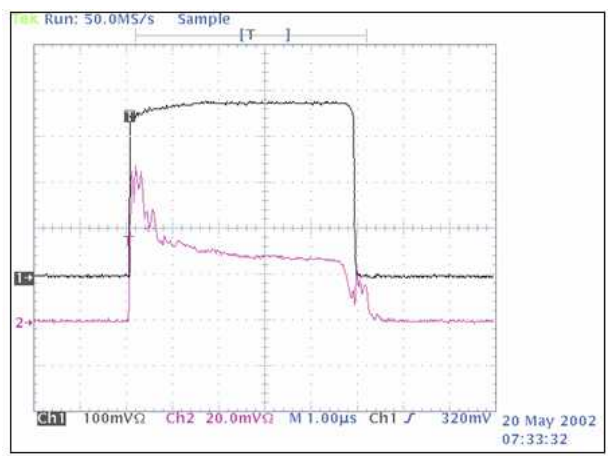

***Table 7.31*** *Transmitted and reflected RF power after injection into the module of 3.4 MW: the results are in agreement with the measurements done at CERN during conditioning (table 7.7).*

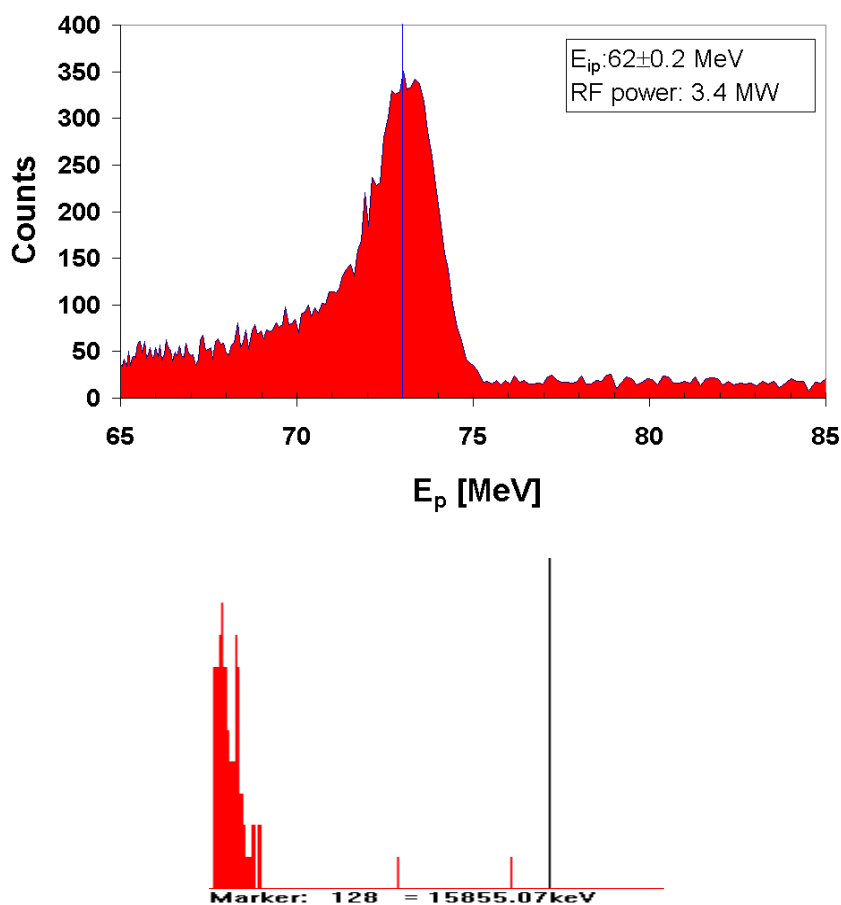

**Figure 7.32** _Energy spectrum obtained downstream LIBO prototype after passing through a 30 mm-thick Perspex absorber. The peak in the first figure (above) corresponds to an accelerated proton beam of 73 MeV. It was obtained injecting into the module only 3.4 MW of RF power (due the limitation in the driver amplifier), while the nominal RF power of the module is 4.3 MW. Second figure (below) was obtained with the same experimental conditions, but without RF power into LIBO: only the tail of the absorbed 62 MeV proton beam is visible. The black marker shows the 73 MeV peak energy._

With P = 3.4 MW and  ΔW = 11 MeV, the  spectrum of figure 7.32 corresponds to accelerated  protons  with |φ| < 5°. These results prove that the prototype works in accordance with the predictions. Further measurements using a deflecting magnet (figure 7.28), in order to separate the different energy components, confirmed the above results. Moreover the computer simulations of a 62 MeV beam accelerated by LIBO, with 3.4 MW of peak RF power and transmitted through the different absorbers, are in very good agreement with the measured values.

## 7.4    Conclusions

The successful tests on LIBO-62 prototype at CERN and LNS in Catania [A.2] prove, for the first time, that it is possible to inject and accelerate proton beams from a cyclotron into a 3 GHz SCL linac.

In particular the construction and tests of the module have shown the following items.

- The electric filed distribution as well as the acceleration rate and particles motion into LIBO, can be controlled inside of the design specifications. A final accelerating gradient of 29 MV/m has been reached, so even better than the design value (15.3 MV/m).
- The matching between klystron and the accelerating structure, as well as the transferring of RF power into particle energy have good efficiency and both are inside the design requirements.
- The accelerating structure, after a short conditioning, is free from electrical breakdown and the final vacuum levels are better than the design specifications.
- Construction procedures and mechanical tolerances of the accelerating structure allow covering the required physical performances. The prototype construction proves that standard technology at low costs can be adopted for fabrication, even for this unusual high frequency structure.

In conclusion, the short conditioning time to reach the nominal power (about 72 hours) and the very limited number of electrical breakdowns are due to the high surface cleanliness of the structure and the thorough and accurate machining and brazing operations. The careful design of the cells, with particular emphasis on the shape of the accelerating gap and the slot geometry, as well as the applied RF tuning procedures have contributed also to this success. These results are conclusive and they really open the way to a true technological transfer of the accumulated know-how to a consortium of companies, which could propose this technology to the medical market. In particular one could envisage having, with a 50-70 MeV cyclotron and a LIBO, a very advanced facility for deep protontherapy. Moreover due to the indications that the module can support even a higher accelerating gradient compared to the design value, the original configuration of 13-m length, composed of nine modules, can be modified.  Operating at higher accelerating fields, in the range of 18-20 MV/m, one can reduce the final linac booster length, and consequently the related costs. In this case also low energy protons extracted from a 30 MeV cyclotron can be accelerated to 200-250 MeV with a total length of about 15 m. In this framework TERA proposed the IDRA project in 2001, where a 30 MeV cyclotron, originally designed for isotopes production, can be used also as injector of LIBO. In this context the production of isotopes for PET and SPECT can be allowed for radioisotope therapy as well as the deep treatment of patients with protons, as widely mentioned in chapter 8. Moreover at the beginning of 2002 TERA started to look for companies interested in investing money, building together the first full LIBO and shearing the possible future advantages of such an industrial enterprise. It is foreseen to involve IBA as a possible supplier of 30 MeV cyclotrons, and two or three industries for construction of the accelerating structure (machining and brazing) and the RF klystrons. Another company could also be used for general diagnostics and controls.

*Chapter 8*

*New prospective generated by LIBO-62 prototype*

## 8.1    The IDRA project

As already mentioned in chapter 2 and in the conclusion of chapter 7, the successful tests at CERN and LNS in Catania have generated the definition of IDRA project of TERA Foundation, the Italian acronym that stands for "Institute for Advanced Diagnostics and Radiotherapy". This is a complete facility capable to produce isotopes for nuclear medicine as well as proton beams for eye melanoma and deep-seated tumours.

In particular in 2001 an agreement with the Ospedale Maggiore of Novara has been signed to study a preliminary design of this oncological centre with advanced techniques of diagnostics and radiotherapy [A.14]. Figure 8.1 shows the general layout of the accelerator facilities such as the 30 MeV cyclotron and the ten modules of LIBO needed to accelerate protons up to 200 MeV.

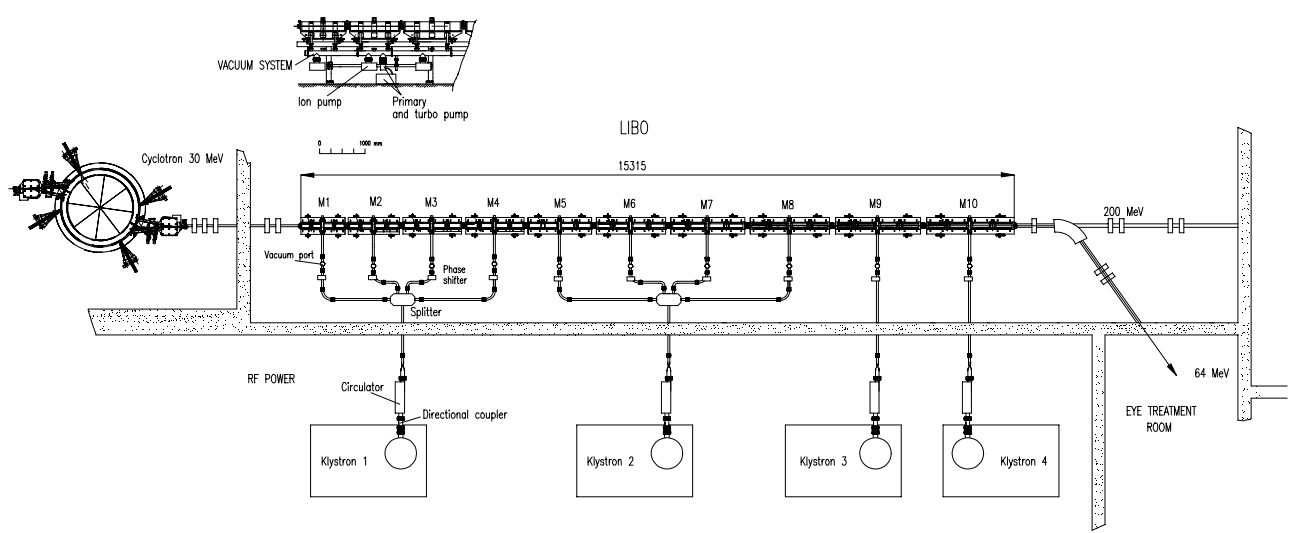

**_Figure 8.1_** _IDRA project foresees enough modules to accelerate the beam from 30 MeV up to 200 MeV. With this facility one could envisage to produce isotopes for nuclear medicine as well as protons for deep seated tumours._

Technically speaking one of the main problems of this facility is the big energy range to be covered by a single structure, especially if it is analysed in terms of efficiency. Normally several types of linacs, which are maximally efficient in a particular energy sub-range, have been developed [C.6]. So if a large range has to be covered, different linac structures, each optimally chosen in its frequency range, are serially disposed, with a consequent increased complexity and cost of the whole system (chapter 2). In this frame it has been accepted to used a low efficient (at low energy) LIBO structure for the entire energy range (30 MeV – 200 MeV), but with the advantage of cost saving by the study and production of only one type of accelerating structure.

The radioisotopes produced for medical application are used world-wide in nuclear medicine centres for imaging and diagnosis to provide dynamic information and studies of the organs function. For example the most commonly used commercial cyclotron isotope is Tl-201. It is absorbed within the muscle tissue of the heart, i.e. the myocardium, and it is used to demonstrate blood flow within the heart muscle by imaging a patient at rest and under stress. Ischaemic tissue and infractions can be also identified using 3D tomographic technique.

The labelled biomolecules and the radioisotopes involved for the purposes should have short half-time and a γ energy range, emitted from the radioisotope, between 100 and 300 keV. Moreover they must not be β-emitter. With these characteristics one can have then a maximum efficiency in the diagnosis and a minimum radiation dose to the patient[1]. After irradiation the produced isotopes are still in the solid target matrixes and they must be transferred from the irradiation station into a hot cell for further processing. This transfer must be safe, reliable and sometimes fast, so the possibility to generate own radioisotopes could be an important tool for dedicated centres of nuclear medicine[2].

The most important cyclotron produced isotopes for nuclear medicine are generated by (p,xn) nuclear reactions with proton energy ranging between 10 and 30 MeV. Concerning the commercial cyclotron, various firms sell 30 MeV cyclotrons. Currently the attention is concentrated on the cyclotron produced by IBA, the so-called Ciclone30. It is a fixed field, fixed frequency cyclotron with an extracted proton beam energy ranging between 18 and 30 MeV, with guaranteed current of 350 μA. It can also accelerate D- with an output energy between 9 and 15 MeV. It is equipped with up to 10 exit ports allowing the use of 10 targets and up to 2 beams extracted simultaneously. Stripping with thin carbon foils with 99.9% efficiency generates the extraction. It has to be pointed out that these machines are fully automated, with their control system and chemistry apparatus that allow the secure and easy production of the radioisotopes for most PET and SPECT applications.

The usual PET radioisotopes are: F-18, C-11, O-15 e N-13. They emit a positron that annihilates with an electron of the tissue and produces two opposing photons.

The reconstruction of the photon origin with a "PET camera" allows the quantitative three-dimensional analysis of the compound concentration. The tracers have a chemical structure close to the molecules naturally present in the tissues, so that it is possible to study the functions and metabolism of the various organs. The half-life of the PET radioisotopes is quite short and only the isotope F-18 (half-life = 110 minutes) can be produced and distributed to other hospitals that are not far from the production site.

---

[1] The physical properties show that a cyclotron, that for example produces I-123 (γ energy 159 keV, no β-, short half time), is more favourable for use in nuclear medicine than a reactor that produces I-131 (γ energy 364, 637 keV, β-emission, long half time).

[2] When irradiating a target material with charged beams from the cyclotron, the disintegration rate D of a produced radioisotope is (disintegration/sec):

$$D = I \cdot \frac{W \cdot K \cdot N_{Av}}{A} \cdot \sigma \cdot (1 - e^{-\lambda t})$$

where I is the intensity of the irradiating beams (number of particles /cm$^2$ sec), W is the weight of target material, K is the natural abundance of target element, A is the atomic weight, Nav is the Avogadro number, σ cross section (barn or $10^{-24}$ cm$^2$), λ is the decay constant, t is the duration of the irradiation. The amount of radioactivity produced depends on the intensity and energy (through the cross section) of the bombarding particles, the amount of target material, the half life of the produced radioisotopes and the duration of the irradiation. The term in parenthesis reaches unity when t is approximately 4-5 half-lives of the radionuclide. At that time the yield of the produced nuclide becomes maximum.

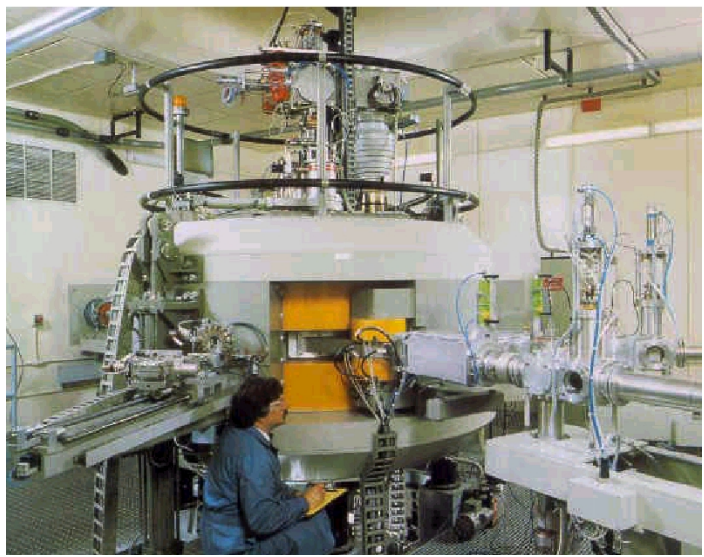

***Figure 8.2*** *30 MeV cyclotron produced by IBA (Ciclone 30).*

The cyclotron current should be enough to produce PET radioisotopes both for the internal uses of IDRA and of other facilities. Beside the production of PET radioisotopes, the 30 MeV cyclotron could be used for the production of single photon emitters (SPECT), like: Tl-201 used for myocardial scintigraphy, Ga-67 for the diagnosis of cancer and infections, Kr-81m for the diagnosis of ventilatory disorders of the lung, In-111 to label antibodies and to study the flow of cerebral fluid, I-123 for studies of thyroid and blood flow.

If wanted, the proton beam at 30 MeV can also be used to generate thermal or epithermal neutron beams for Boron Neutron Capture Therapy (BNCT) and Boron Neutron Capture Synovectomy (BNCS) applications.

These techniques are still experimental, but they appear to be quite good for the cure of radioresistant tumours (BNCT) and for the treatment of rheumatoid arthritis (BNCS).

BNCT is a dual treatment modality that consists in concentrating a boronated (B-10) compound in the tumour cells and the irradiation with thermal neutrons. The neutron irradiation breaks the boron nuclei and produces two short range and heavily ionising nuclei that locally destroy the tumour cells. This kind of treatment is promising for radioresistant tumours like glioblastoma and it has been studied by TERA for applications on explanted liver and skin melanoma.

***Table 8.1*** *Ciclone 30 (IBA)-Physical specification.*

| | |
|---|---|
| Energy of extracted protons [MeV] | 18-30 |
| Energy of the extracted deuterons [MeV] | 9-15 |
| Maximum proton current [expected] [μA] | 500 |
| Maximum proton current [guaranteed] [μA] | 350 |
| Number of exit ports | 10 |
| Number of simultanely extracted beams | 2 |
| Normalised emittance of the extracted beam: | |
| Horizontal [π mm mrad] | < 10 |
| Vertical [π mm mrad] | < 5 |
| Power consumption [kW] | 80-140 |
| Number of sectors | 4 |
| RF frequency [MHz] | 65.4 |

BNCS exploits the same modalities as BNCT for the treatment of rheumatoid arthritis [A.1]. In this case the local injection of the compound permits to obtain high gradients of boron concentration and a better spare of the healthy tissues included in the treatment field.

For all these applications about 0.3 mA of protons at 30 MeV are needed. The addition of LIBO to the cyclotron has no influence for radioisotopes and neutron production. Normally these cyclotrons need special designed shielding, due to the high current. For proton therapy the source will be pulsed at 400 Hz in order to obtain, with a peak current of 50 µA, an average current of about 0.1 µA. An image of the complete structure, cyclotron and LIBO, is presented in figure 8.4. Between two linear sections it should be possible to derive a beam line for the transport of protons at 62 MeV for the treatment of eye tumours. One of the characteristics of a linear structure as LIBO, is that it can be realised in modules that are independent and, eventually, the overall accelerator from 30 to 200 MeV can be realised in two parts: the first part from 30 to 62 MeV sufficient for treating ocular pathologies, and the second part from 62 to 200 MeV for deep seated tumours. The protontherapy part foresees treatment rooms with horizontal beams as well as rotating gantry.

A novel protontherapy facility represents an expensive investment for a hospital (in general of the order of 30-50 MEuro and depending on the number of treatment rooms used) respect to the 3-5 MEuro needed for the corresponding units in a conventional radiotherapy department equipped with electron linacs. This is more important if one considers the particular attitude of all health system that do not want to pay for extra-costs. Very rough estimates indicate that the difference in costs between proton and conformal photon therapy is very modest in relation to the general medical costs of handling cancer, independently of the results (see chapter 1). If the treatment with protons avoids some treatment complication, they could be justified purely on the basis of economical arguments.

In this frame an analysis has been successfully performed by TERA, proving the feasibility of the IDRA facility in terms of costs and revenues. Here the 30 MeV cyclotron and the equipment for the radioisotope production can be predicted rather precisely being commercial products, but the costs of a novel structure like LIBO are difficult to quantify, so the experience with the prototype construction at CERN has been useful at this stage. In particular from the indications of the prototype presented in annex 6.5, one could extrapolate for example the technical manpower needed for the construction of a full linear accelerator with the help of external industries (see also next paragraph).

***Table 8.2** Typical Production Yield – PET isotopes.*

| Isotope | Target Reaction | Beam Energy Of Target (MeV) | Irradiation Time (minutes) | Synthesis Time (minutes) | Recovered Activity (mCi) |
|---|---|---|---|---|---|
| $^{11}$C | $^{14}$N(p,$\alpha$) $^{11}$C | 15 | 30 | 10-25 | 2500-700 |
| $^{13}$N | $^{16}$O(p,$\alpha$) $^{13}$N | 16 | 20 | (--) | 400 |
| $^{15}$O | $^{16}$O(p,pn) $^{15}$O | 30 | On line | (--) | 500 mCi/min (flow) |
| $^{18}$F | $^{18}$O(p,n) $^{18}$F | 18 | 120 | (--) | 5000 |

**Table 8.3** *Typical Production Yield – SPECT isotopes.*

| Isotope | Target Reaction | Beam Energy of Target (MeV) | Typical Beam Intensity (μA) | Typical Radiochemical Yield (mCi/μA) |
|---|---|---|---|---|
| $^{111}$In | $^{112}$Cd(p,2n) $^{111}$In | 27 | 120 | 1.5 |
| $^{123}$I | $^{124}$Xe(p,2n) $^{123}$I | 28 | 100 | 10 |
| $^{201}$Tl | $^{203}$Tl(p,2n) $^{201}$Pb $\rightarrow$ $^{201}$Tl | 27.5 | 200 | 0.3 |
| $^{67}$Ga | $^{68}$Zn(p,2n) $^{67}$Ga | 25 | 120 | 1.2 |

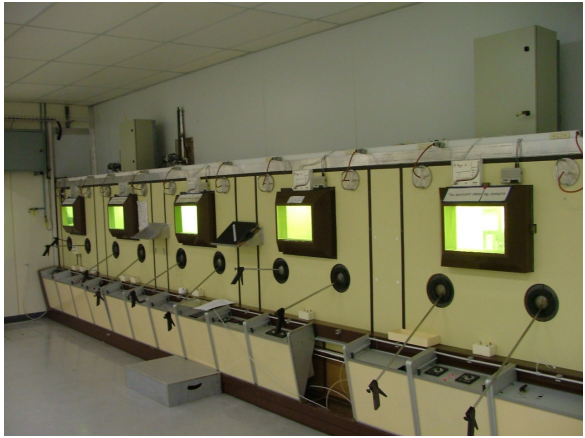 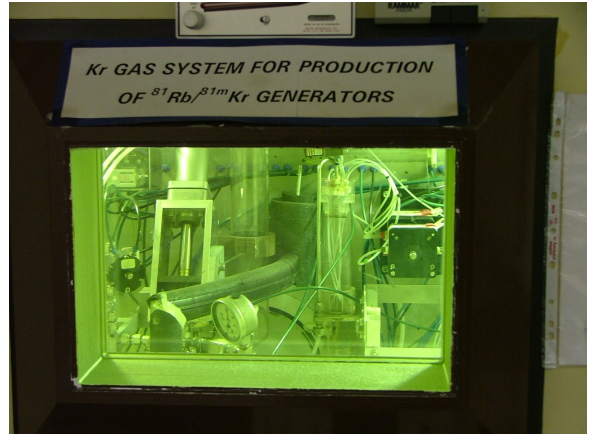

**Figure 8.3** *Example of hot cells for isotope production at NAC (Cape Town, South Africa) [B.39].*

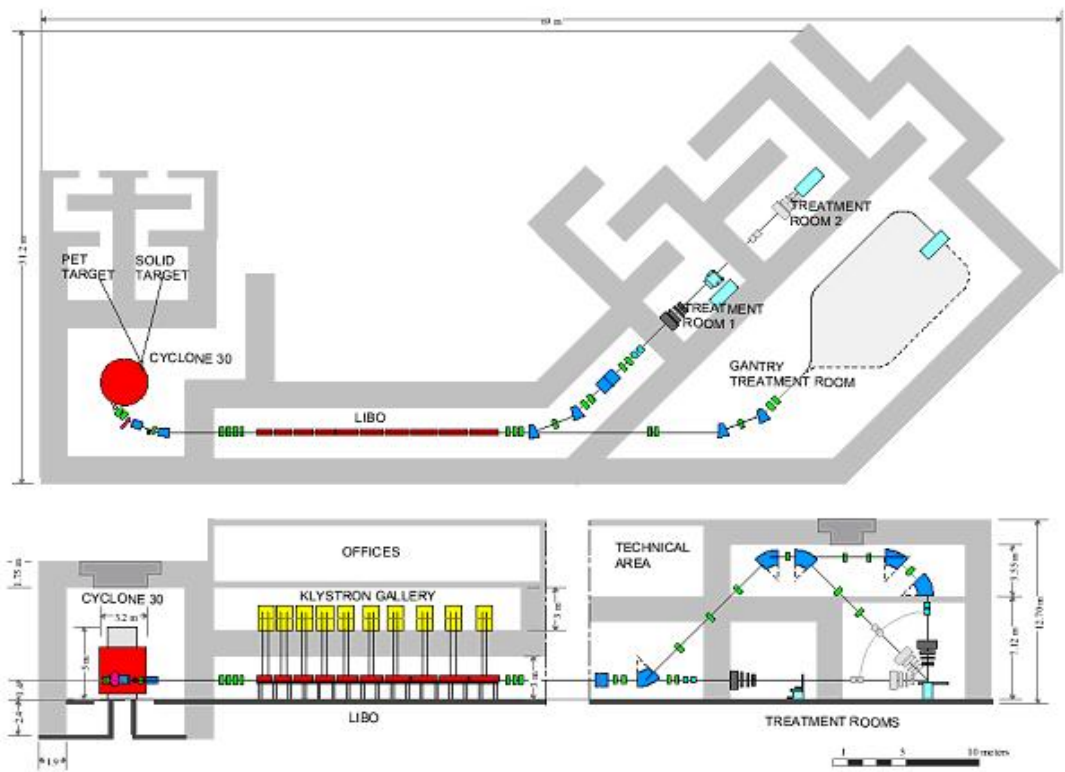

**Figure 8.4** *IDRA project foresees the use of the SCL technology developed with the first LIBO prototype [A10, A.14].*

| | | | 1Y | | | | 2Y | | | | 3Y | | | | 4Y | | | | Menyears | |
|---|---|---|---|---|---|---|---|---|---|---|---|---|---|---|---|---|---|---|---|---|
| | | | 1Q | 2Q | 3Q | 4Q | 1Q | 2Q | 3Q | 4Q | 1Q | 2Q | 3Q | 4Q | 1Q | 2Q | 3Q | 4Q | | |
| ENGINEERING | Mech. | Mech. Eng. A | | | | | | | | | | | | | | | | | 4 | Design |
| | | Drafts. A | | | | | | | | | | | | | | | | | 4 | Production |
| | | Drafts. B | | | | | | | | | | | | | | | | | 4 | Installation Design |
| | | Mech. Eng. B | | | | | | | | | | | | | | | | | 4 | Tuning |
| | | Tech. Mech. | | | | | | | | | | | | | | | | | 0,5 | Physics Design |
| | Vac. | Vac. Eng. | | | | | | | | | | | | | | | | | 1 | Beam Instrumentation Design |
| | | Vac. Tech. | | | | | | | | | | | | | | | | | 2 | Construction instrum. and test |
| | Control | Control Eng. | | | | | | | | | | | | | | | | | 3,5 | Control Design |
| | | | | | | | | | | | | | | | | | | | | Cont. Inst. Des. |
| Physics | RF | RF Eng. A | | | | | | | | | | | | | | | | | 2,5 | Installation |
| | | RF Eng. B | | | | | | | | | | | | | | | | | 2,5 | |
| | | RF Eng. C | | | | | | | | | | | | | | | | | 1 | |
| | | RF Tech. (Electr.) | | | | | | | | | | | | | | | | | 2 | |
| | | RF Tech. (Operator) | | | | | | | | | | | | | | | | | 1,5 | |
| | BEAM | Beam Eng. A | | | | | | | | | | | | | | | | | 1,5 | |
| | | Beam Eng. B (Instrum.) | | | | | | | | | | | | | | | | | 4 | |
| | | | | | | | | | | | | | | | | | Tot. men years | 38 | |

_**Figure 8.5** Proposal for technical manpower for the full LIBO construction._

To give some indications of the technical complexity, one could envisage four people (engineers and draftsmen) for 1,5 years, needed to produce the full set of drawings (about 500). For the construction of ten modules in industry the main figure should range between 25000 and 30000 hours, performed for full machining and metrology, while cleaning, heat treatments and brazing will take approximately 5000 hours (considering a good safety factor). Outsourcing with external industries could perform these steps. RF tuning and measurements could take about 8000-10000 hours for the ten modules, so four people (RF engineers and technicians) for 1,5 years should be enough. Finally the global figure of internal technical manpower for a full LIBO construction should be about 38 men-years.

.

## 8.2    What's next: technology transfer to industry for a LIBO facility construction*

The LIBO-62 prototype tests have opened the way to a true technological transfer for a possible creation of a company consortium, which will propose the full-scale facility to the medical market.

In this framework, during the year 2002-2003, TERA has looked to qualify companies capable of building a complete LIBO-30 in the near future, checking also the real industrial cost of this facility. An Italian company for machining has been found and tested by producing twelve half-cell plates in three different steps. The first one was devoted to reproducing the construction sequences used at CERN for the prototype, while the second and third steps have been used to machine new half-cells for the LIBO-30. The last two steps have been significant to prove the machining feasibility of the thin 30 MeV half-cells, especially in terms of mechanical stability (see paragraph 2.1). For these reasons, cells with a thickness of about 6 mm, corresponding to the first tank sizes of LIBO-30, have been built.

Metrology and RF measurements performed on both cell types show very promising results for the future production in industry. Figure 8.6 shows the high accuracy of the accelerating cell, with a geometrical average value in the range of 10 µm. Metrology has then checked the different dimensions, all evaluated inside of the mechanical tolerances.

Moreover the RF results show very low spreads in frequency. These range from 3035.2 to 3037.7 MHz for the accelerating cells (spread of 0.09 %), while the coupling cells frequencies range from 3023.2 to 3028.1 MHz (spread of 0.16 %). The quality factor Q ranges then from 1000 to 2100.

These measurements prove the high accuracy of the machining, confirming at the same time the possibility to use the same procedures developed for the prototype. Due to the low frequency spread, significant and useful data have been finally obtained even without RF tuning by re-machining of the ring.

Following these tests, a full production of 30 MeV module was launched in 2005, by using the same design developed for the 62 MeV prototype. By the end of 2005 TERA will complete this important step, in collaboration with two industrial partners, which are in charge of the production of the mechanical components and the brazing [A.10].

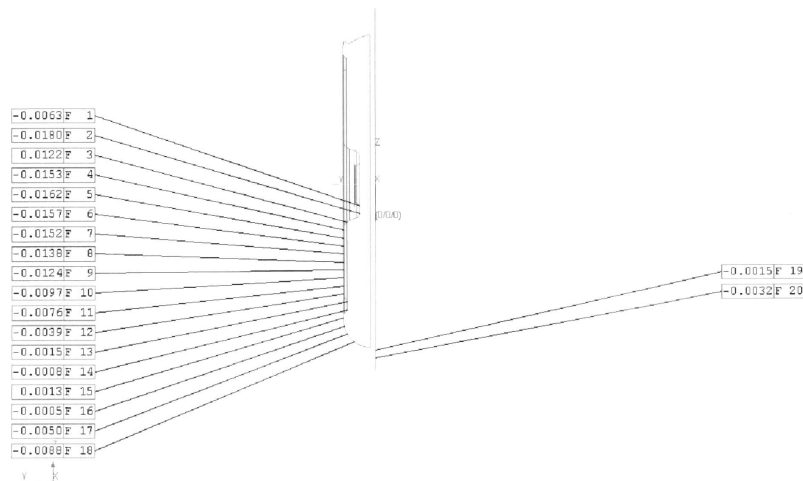

(a)

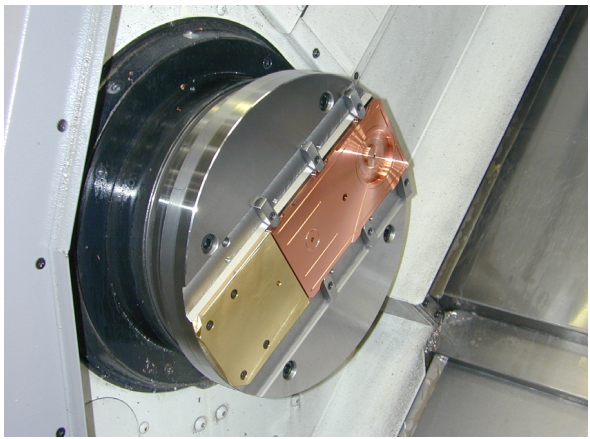 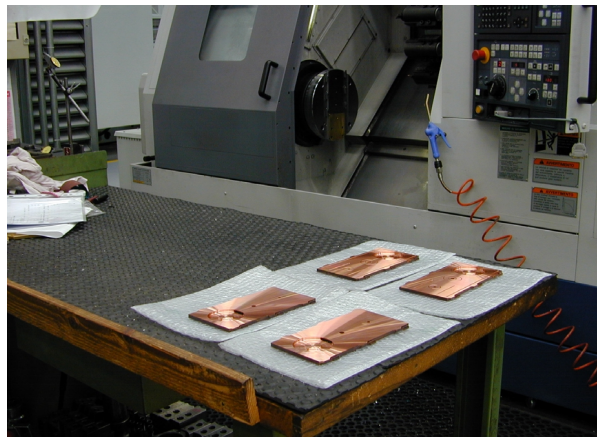

(b)

**Figure 8.6** *Results for the technology transfer in the period 2003-2004. Half-cells have been produced in a high-tech Italian company, by using the procedures developed at CERN for the prototype. The figures show the high accuracy of the accelerating cell (metrology results) (a), as well as the first 30 MeV half-cell plates for the new LIBO-30 (b). The tests have proven the mechanical and RF feasibility for the construction of LIBO-30 in industry. As a consequence, the first modules of LIBO-30 have been launched for a full production in 2005.*

**8.3     A long-term solution: the TERA project of a novel compact linac for ions***

As for protons, also for ions the energy range covered by linacs is of several tens of kilo-electron-volts per nucleon (keV/u) to hundreds of million-electron-volts per nucleon (MeV/u) (i.e. a β range from about 0.05 to about 0.8). It is also known that deep cancer therapy with light ion beams requires β ≤ 0.6, which is in the range of standing wave linacs. For this reason after the successful tests at CERN and Catania, TERA has started the design of a new linear accelerator for ions [U. Amaldi, M. Crescenti, R. Zennaro, A.11, A.20], by using similar technology developed for the LIBO prototype.

This project foresees the basic concept of a drift tube linear accelerator to be used for the acceleration of low energy ion beams. The particles enter into the linac at low energy and are accelerated and focused along a straight line in a series of resonant accelerating cavities interposed by coupling structures up to the desired energy needed for deep therapy. In these accelerating structures the drift tubes are supported by stems, alternatively horizontal and vertical. A basic module is composed of two accelerating structures and an interposed coupling structure (figure 8.7), or a modified coupling structure connected to a RF power generator and vacuum system. Such basic structure can be repeated to get modules that present alternating coupling and accelerating cavities. The proposed linac, composed by different modules aligned in series allows obtaining a large accelerating gradient and a very compact structure.

This linac is called Coupled-cavity Linac USing Transverse Electric Radial fields (CLUSTER). The accelerating resonant structures are excited on H-mode standing wave electromagnetic field pattern, with high working frequency [C.6]. Output beam energy can be modulated by varying the incoming RF power, whereas the output beam intensity can be modulated by adjusting the ion beam injection parameters and dynamics. It should be pointed out that conventional H-type cavities are currently used for the acceleration of low velocity, high intensity and high mass-over-charge ion beams.

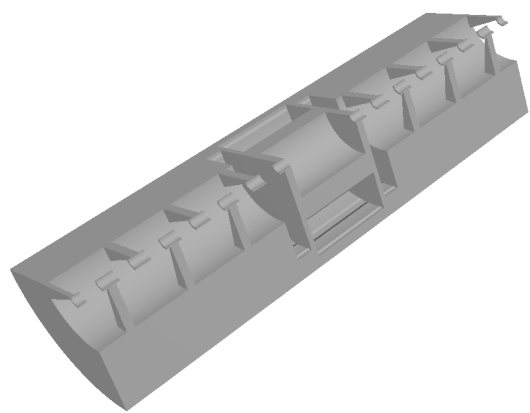

**Figure 8.7** *General layout of the new ion linac for hadrontherapy proposed by TERA Foundation in the period 2002-2004. A drift tube linac can be used for the acceleration of low energy ion beams. The particles are accelerated and focused along a straight line in a sequence of resonant accelerating cavities interposed by coupling cavities up to the desired energy. The accelerating structures are excited by an H-type resonant electromagnetic field. Each basic module is composed of two accelerating structures and an interposed coupling structure, equipped with permanent magnet quadrupoles.*

In such applications, the beam transverse dimensions are rather high, and therefore the beam hole must also be correspondingly large. As a consequence, the cavities built and working under known concepts are bound to work on a low frequency range, i.e. from about a few MHz (cavities with diameters of about 1 m) up to a few hundreds MHz (cavities with diameters of the order of about 0.3 m). In medical applications, since low intensity beams are required, a beam hole of the order of few mm is large enough. In order to simplify the installation in hospitals, the length of such structures should be as short as possible. Instead of using mid or low working frequencies, CLUSTER can use high working frequencies of about 0.5 GHz to several GHz. Today, the progress in mechanical technologies allows the production of such small structures with the required precision. It should be also pointed out that the field stability decreases with the increase in frequency and length. This severely limits the development of long conventional accelerating structures. The present linac can solve the problem by creating a sequence of accelerating cavities of moderate length coupled together. The use of H-type structures at much higher frequencies is new. In fact, it is well known that the higher the frequency, the higher the allowable field, with consequent increase of the energy gain per meter and reduction of the overall accelerator length. In the proposed CLUSTER solution the ions are accelerated and longitudinally focused at the same time by RF electric fields in the accelerating gaps up to the design energy for cancer therapy. Transverse focusing is given separately by magnetic fields. In the CLUSTER the focusing quadrupoles can be located directly inside the coupling structures.

For medical applications it is possible to accelerate the ion beam up to about 4000 MeV (330 MeV/u), which is the present optimal maximum beam energy considered for deep cancer therapy. The cavities in the modules tuned at the same working frequency, are coupled in order to resonate in the mode $\pi/2$, with the coupling cavities nominally not excited or partly excited, where such configuration greatly contributes to the stability of the system. The resonant working mode of the accelerating cavities can then be classified as an $H_{m10}$ mode. Effective acceleration is possible at each accelerating gap because the distance between said accelerating gaps is $\beta\lambda/2$. Different solutions of CLUSTER modules, working at different frequencies (1.5, 3 and 6 GHz) have been studied by TERA, where $^{12}C^{6+}$ (Q = 6, A = 12) is the accelerated particle. More information on this new carbon ion linac for hadrontherapy can be found on [A.20].

*Annexes*

**Annex 1.1      Clinical experience with proton therapy and main general results\***

At the beginning of the proton therapy history the majority of patients have been treated in non-hospital based centres, using machines not ideally designed for clinical use and with several patient access restrictions due to fact that such centres were built and devoted for physic research and not for medicine.

This led to the use of unconventional fractionations not dictated from clinical evidences. In spite of these limitations, results were very positive for several types of tumours, with improved local control. The main clinical indications for proton therapy are summarised as follows: 1) proximity of the target volume to critical organs, where high selectivity of the dose distribution is fundamental; 2) low tumour radiosensitivity necessitating higher doses; 3) high benefits, due to the improved local control on the long term survival. Several types of tumours meet these criteria, and in these cases proton therapy may be superior to photon beam therapy. Table A annex 1.1 shows a brief summary of the main current indications for the clinical use of protons [B.21, B.22]. Based on the experience for a limited set of pathologies, where it is possible to prove advantages in the use of protons, indications have been divided in two categories. Category A includes all tumours where the advantage of protons has clearly demonstrated and where this kind of therapy is the only way to give a curative dose, minimising the side effects. In table B annex 1.1 are mentioned some indications for proton therapy and potential patients to be treated yearly in Italy in this category. Category B includes tumours characterised by a local evolution, with a limited likelihood of distant spread and therefore cured by the loco regional control. In these cases protons could represent the only possible treatment. Table C annex 1.1 shows the possible indications, as mentioned in table B annex 1.1, but for category B. In this category are also included some pediatric tumours. Childhood cancer shows about 110 new cases per million children under 16 years of age and leads to death from disease. Moreover it is important to control this aspect for other reasons. First the sick child certainly affects the entire family, with a relevant social and economic impact. Second for a child the long term survival is measured in terms of fifty or more years, so it is important to avoid possible cronical illness that could affect quality of life of the adult of tomorrow. Children are very sensitive and delayed complications of radiation therapy can be extremely pronounced.

*Table A annex 1.1 Clinical indication for proton therapy [B.21].*

| Ocular diseases | Head and neck tumours |
|---|---|
| Uveal melanomas | Paranasal sinuses carcinomas |
| Angiomas and hemangiomas | Pharynx |
| Intraocular metastases | Salivary glands tumours |
| Retinoblastoma | |
| Macular degeneration | |
| **Brain tumours and non malignant lesions** | **Others tumours** |
| Meningiomas | Chordomas and chondrosarcomas of the skull base and spinal cord |
| Gliomas | Rectum |
| Acoustic neurinomas | Lung |
| Craniopharyngiomas | Cervix and endometrium |
| Hypophysis adenomas | Oesophagus |
| MAVs | Pancreas, biliary tract, liver |
| | Soft tissue sarcomas |

For this reason reduction of the treated volume and local protection of radiosensitive organs by using proton beams can minimise the side effects after a successful cancer treatment, such as, for example, reduction of intellectual capacity or fertility. For these types of tumours the percentage of patients to be treated with protons has been estimated at 13% with a wide range varying from 50% for thyroid tumours to 5% for oesophageal cancer. 830 potential patients have been estimated in Italy for category A and 15023 for category B, with a total of more than 15800 new cases to be treated every year, representing about 13% of the 120000 patients currently irradiated in Italy with conventional radiation therapy each year. Of all patients treated with protons at 13 facilities world-wide, 46% of the patients had small benign lesions within the cranium, 32% eye tumour, and remaining 22% miscellaneous lesions (mainly malignant tumours).

**Table B annex 1.1** *Indications for proton therapy and number of potential patients to be treated in Italy (Category A).*

| Category A | Patients/year | Suitable for protons | |
|---|---|---|---|
| | | Number | Percentage (%) |
| **Uveal melanomas** | 310 | 310 | 100 |
| **Chordomas of the base of the skull** | 45 | 45 | 100 |
| **Chondrosarcomas of the base of the skull** | 90 | 90 | 100 |
| **Meningiomas of the base of the skull** | 250 | 125 | 50 |
| **Paraspinal tumours** | 140 | 140 | 100 |
| **Schwannomas** | 300 | 45 | 15 |
| **Hypophysis adenomas** | 750 | 75 | 10 |
| **Total** | 1885 | 830 | 44 |

**Table C annex 1.1** *Indications for proton therapy and number of potential patients to be treated in Italy (Category B).*

| Category B | Patients/year | Suitable for protons | |
|---|---|---|---|
| | | Number | Percentage (%) |
| **Neuroepithelial brain tumours** | 2600 | 520 | 20 |
| **Brain metastases** | 1000 | 100 | 10 |
| **Head and neck cancers** | 6780 | 1017 | 15 |
| **Undifferentiated cancer of the thyroid** | 100 | 50 | 50 |
| **NSCLL** | 31000 | 1550 | 5 |
| **Thymomas** | 110 | 11 | 10 |
| **Oesophageal cancers** | 2840 | 142 | 5 |
| **Billiary tract tumors** | 4300 | 430 | 10 |
| **Liver tumours** | 13340 | 1334 | 10 |
| **Pancreatic cancers** | 9050 | 1810 | 20 |
| **Cancer of the uterus** | 2990 | 598 | 20 |
| **Bladder cancers** | 16950 | 1695 | 10 |
| **Prostate cancers** | 22330 | 5582 | 25 |
| **Pelvic recurrences after surgery** | > 500 | > 250 | 50 |
| **Pediatrics tumours** | 970 | 144 | 15 |
| **Non-neoplastic lesions** | | | |
| MAVs | 130 | 40 | 30 |
| Macular degeneration | NA | NA | NA |
| **Total** | > 114490 | > 15023 | 13 |

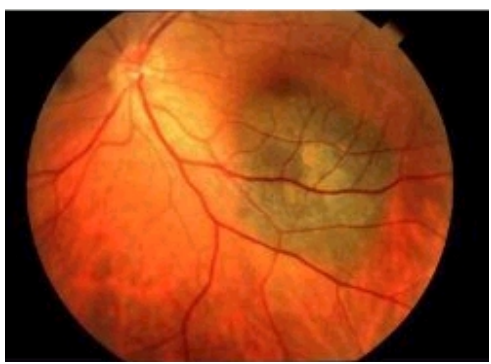

**Figure A annex 1.1** *Choroidal melanomas are the most frequent primary intraocular tumours: the estimated incidence rate is about 6-7 new cases per million inhabitants, corresponding to some 500-600 new cases per year in Germany.*

The following is a brief review of the currently available literature [B.2, B.13, B.14, B.15, B.19, B.22, B.23] that compares the toxicity of conventional photon and proton beams and presents main results on protontherapy to be used as a clinical reference (University of Pennsylvania).

❑ *Eye cancer*

Today protontherapy, if available, is the best choice for about one third of the intraocular tumours. In many cases protontherapy permits successful treatments of ocular tumours, preserving the disease eye and at least a minimum useful visual acuity. The homogeneous irradiation field encloses even large tumours, which is a feature that cannot be satisfied by conventional radiotherapy. Here, proton beam irradiation might be the only eye-saving therapy. As it accounts for 70% of ocular tumours in Caucasians, several proton facilities dedicated for treatments of these tumours have been established world-wide.

❑ *Prostate cancer*

A large number of patients treated in radiation oncology centres has prostate cancer. Side effects of treatment generally include gastrointestinal (GI) and genitourinary (GU) damage. Significant number of patients has urinary frequency and diarrhea during treatment, and long-term, may suffer impotence, incontinence, rectal fibrosis and bleeding, and extensive bowel fibrosis. These side effects may cause a reduction in the quality of life and result in delays of a typical radiation therapy treatment course. Tables D and E compare the acute and long-term complications of localised prostate cancer treated with protons, conventional X-rays, and radical prostatectomy respectively.

❑ *Lung cancer*

Lung-cancer is the most common malignancy seen in men and women in the developed countries, and a very substantial source of cancer mortality. A significant percentage of lung cancer patients is treated with radiation therapy. Since many of these patients have poor lung function due to years of smoking tobacco, preservation of functioning lung tissue is paramount. Tables F and G compare the acute and long-term complications of lung cancer patients treated with protons versus conventional X-rays.

**Table D annex 1.1** *Acute complications associated with the treatment of prostate cancer.*

| Acute Toxicity | Protons | Conventional Radiotherapy | Prostatectomy |
|---|---|---|---|
| ≥ Grade 2 GU toxicity (frequency, nocturia, dysuria) | 0% | 28% | N/A |
| ≥ Grade 2 GI toxicity (diarrhea, rectal/abd pain) | 0% | 35% | N/A |
| Either GU or GI morbidity | 0% | 53% | N/A |
| Hospitalisation | None | None | 5-7 days |
| Absence from work | None | None | 4-6 weeks |
| Death | 0% | 0% | 0.3% |
| Pulmonary embolism/ DVT | 0% | 0% | 2.6% |
| Myocardial infarction or arrhythmia | 0% | 0% | 1.4% |
| Wound Complications | None | None | 1.3% |
| Lymphocele | None | None | 0.6% |
| Surgical Rectal Injury | N/A | N/A | 1.5% |

**Table E annex 1.1** *Long-term complications associated with the treatment of prostate cancer.*

| Chronic Toxicity | Protons | Conventional Radiotherapy | Prostatectomy |
|---|---|---|---|
| Impotence | 30% | 60% | 60% |
| Incontinence requiring a pad | < 1% | 1.5% | 32% |
| Bladder Neck contracture | 0% | 3% | 8% |
| Chronic Cystitis | 0.4% | 5% | N/A |
| Grade 3 GU toxicity Severe frequency q 1 hr dysuria | 0.3% | 2% | 36% |
| Grade 3 GI toxicity rectal bleeding requiring transfusion severe pain (>70 Gy) | 0% | 7% | N/A |
| Rectal stricture | 0% | 0.5% | N/A |

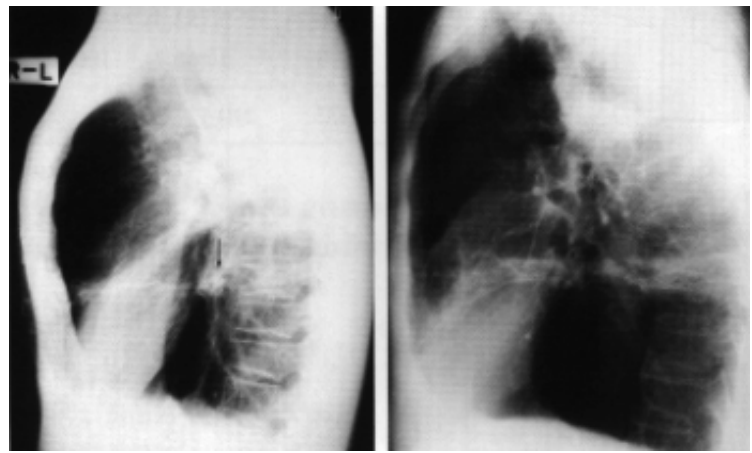

**Figure B annex 1.1** *X-ray photograph of lung cancer (left) treated with the proton beam through the anterior portal. Along the proton path, a horizontal linear shadow is produced and it ends at the tumor (arrow head). Treated with Co60 gamma rays (right). Fibrosis is more conspicuous than that of the proton beams, and its shadow is a horizontal belt traversing the lung (Tsukuba, Japan).*

*Table F annex 1.1  Acute complications associated with the treatment of lung cancer.*

| Acute Side Effects | Protons | Conventional Radiotherapy |
|---|---|---|
| Nausea/Vomiting | 0% | 30% |
| Dyspnea | 0% | 16% |
| Esophagitis | <5% | 31% |
| Fatigue | <5% | 23% |
| > 5 lb. weight loss | 0% | 34% |

❑ *Head and neck cancer*

The morbidity associated with the treatment of head and neck cancer with protons and conventional photons has been reviewed. Specifically, cancers of the paranasal sinuses, tonsillar region, and nasopharynx have been evaluated. In each of these cancers, protontherapy should result in an improvement of local control with a reduction in the morbidity associated with conventional photon treatment. There has been a significant reduction in the rates of blindness seen in the treatment of paranasal sinus tumours as shown in Table H. Comparative plans for the treatment of tonsillar and nasopharyngeal cancer revealed also that proton beam therapy can deliver higher doses to the tumour volumes, with significantly reduced radiation to the salivary glands and mandible, than can photon beam irradiation.

*Table G annex 1.1 Long-term complications associated with the treatment of lung cancer.*

| Chronic Side Effects | Protons | Conventional Radiotherapy |
|---|---|---|
| Lung Fibrosis by CT scan | 33% | 85% |
| Normal Lung Destroyed | 8% | 29% |
| Lung injury ≥ Score 2 | 0% | 62% |
| Decreased pulmonary function testing (VC, $FEV_1$, diffusion capacity) | 0% | 20% |
| Dyspnea | 0% | 32% |
| ≥ Grade 2 Esophagitis/Stricture | 0% | 10% |
| ≥ Grade 2 Pneumonitis | 5% | 15% |
| Cardiac Complications | 0% | 7% |

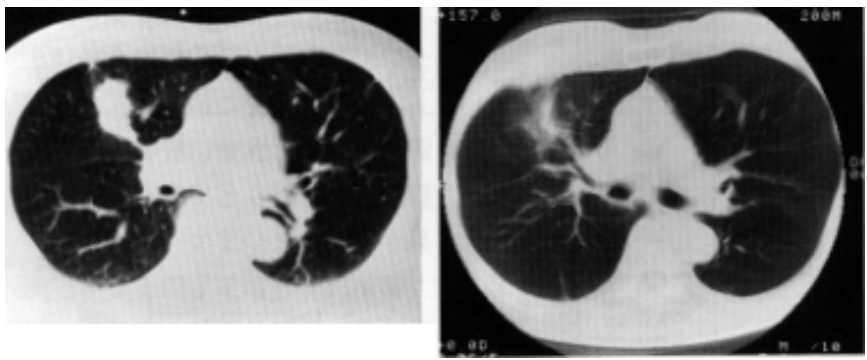

*Figure C annex 1.1 A tumour of 4 cm diameter in the right lung before treatment (left). After proton beam irradiation the tumour disappeared with minimal change in normal tissue (right) (Tsukuba, Japan).*

It should be noted that essentially 100% of all patients treated for head and neck cancer with X-rays will experience severe xerostomia (dry mouth), which although it may not be life-threatening severely impairs quality of life. This complication is totally unavoidable with X-rays because of their through penetrating nature requiring to treat both parotid glands even for well lateralized lesions, and which can be totally avoided with protons.

**Table H annex 1.1** *Major side effects associated with treatment of head and neck cancer.*

| Side Effects | Protons | Conventional Radiotherapy |
|---|---|---|
| Blindness (maxillary sinus tumours) | 2% | 15% |
| Xerostomia (Dry mouth) | < 5% (with protons alone) | 100% |
| Dysphagia | 12 % | 100% - 80% require liquid nutrition |
| Require PEG for nutrition | 0% | 30% |

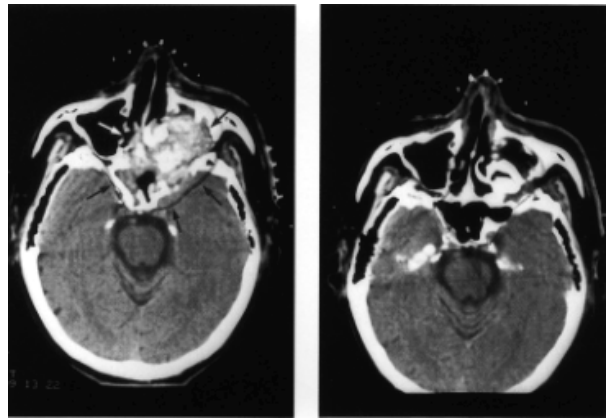

**Figure D annex 1.1** *Maxillary sinus tumour with skull base invasion before (left) and after proton irradiation (Tsukuba, Japan).*

❑ *Pediatric tumours*

The treatment of pediatric tumours with protontherapy also provides a unique opportunity to significantly reduce the acute and long-term complications associated with conventional radiation therapy. The pediatric population is sensitive to the effects of radiation therapy. Long-term complications including growth abnormalities, second malignancies, neurologic complications, cardiac and pulmonary toxicities, and infertility may all be reduced with the use of protons. X-ray therapy causes effects on the hearts and lungs of pediatric patients. Proton beams should be able to entirely avoid these complications. Well-recognised side effects of conventional photon irradiation of the brains of young children include neuropsychologic and intellectual deficits. The side effects vary directly with the volume of brain tissue irradiated and the dose of radiation delivered. By decreasing both the volume and dose of radiation to normal brain tissue through the use of protons, these side effects should be reduced. Table I outlines the reduced toxicity associated with protontherapy compared to conventional radiotherapy for this category of patients.

**Table I annex 1.1** *Complications associated with cranial spinal irradiation in pediatrics.*

| Side Effects | Protons | Conventional Radiotherapy |
|---|---|---|
| Restrictive Lung Disease | 0% | 60% |
| Reduced exercise capacity | 0% | 75% |
| Abnormal EKG | 0% | 31% |
| Growth abnormality-Vertebral body receiving significant dose | 20% | 100% |
| IQ drop of 10 points at 6 yrs | 1.6% | 28.5% |
| Risk of IQ score < 90 | 15% | 25% |

❑ *Pancreatic cancer*

Also for treatment of pancreatic cancer one has significant reductions in dose to normal structures. The tolerance of normal tissues has prevented effective dose escalation for this malignancy. Considering the experience to date, protontherapy offers important advantages over X-rays. First of all protontherapy allows a significant reduction in treatment related morbidity respected to X-rays. Because of this reduction in normal tissue toxicity, dose escalation studies are currently under investigation. This should further increase the local control, and ultimately survival, while minimising treatment induced complications.

**Table L annex 1.1** *Comparison between X-ray and proton doses for pancreatic cancer.*

| Structure | X-ray Dose (Gy) | Proton Dose (Gy) | Dose Reduction |
|---|---|---|---|
| Spinal Cord | 27 | 6 | 78% |
| Liver | 22 | 10 | 55% |
| Right Kidney | 14 | 8 | 43% |
| Left Kidney | 11 | 3 | 73% |

❑ *Clinical conclusions about tumour control with proton therapy*

For many sites, increasing the dose of radiation therapy to the tumour may increase the ultimate cure rates. The following data are from sites already evaluated with protontherapy.

- One of the most difficult areas to treat in the human body is a tumour that arises in the base of skull region. Damage to normal structures such as the brainstem, brain, cranial nerves, and optic chiasm can cause significant morbidity, thus limiting standard treatments. Surgical resection of this area is typically incomplete. Postoperative X-ray therapy achieves local control in only 35-40% of patients.

It has been shown substantially higher doses of radiation therapy can be delivered with protontherapy. By delivering a median dose of 68.5 Gy with protons, significant improvements have been made in both local control and survival with these tumours. The 5-year local control rates for proton therapy are 91% for chondrosarcomas and 65% for chordomas. The 5 year overall survival rates range from 62%-88%.

*Finally proton therapy has become the standard of care for tumours of the skull base.*

- Uveal melanomas have historically resulted in loss of vision from the tumour or from the treatment, which consists of surgical removal of the eye. Over 2500 patients have been treated with proton therapy for uveal melanoma. The typical dose is 70 Gy over 5 treatments. The 5-year local control with protons is reported at 96%. The eye retention rate is 90% while the metastases free survival is 80%.

- Loma Linda University Medical Center has treated over 1000 patients with prostate cancer using proton therapy. Using doses comparable to standard X-ray treatments they have shown significant reductions in side effects as noted above. They have currently devised dose escalation studies to find the maximum dose that can be safely delivered with protons to the prostate gland. Until the maximum dose is reached, final improvements in survival will not be known. However, the initial results reported based on PSA level with a very modest elevation of dose to 75 Gy are encouraging.

Unfortunately, some patients experience a local recurrence of their cancer after treatment with radiation therapy. Only a minority of patients is curable after a recurrence because the normal tissues can not tolerate significant doses of additional radiation.

*Table M annex 1.1* *Tumour control based on PSA at time of diagnosis.*

| PSA Level | Proton Therapy | Conformal X-ray therapy | Radical Prostatectomy |
|-----------|----------------|-------------------------|-----------------------|
| < 4       | 100%           | 91%                     | 92%                   |
| 4-10      | 89%            | 69%                     | 83%                   |
| 10-20     | 72%            | 62%                     | 56%                   |
| >20       | 57%            | 38%                     | 45%                   |

Because protons can spare normal tissues, many patients that were not previously considered treatable again with X-rays may be treated with protons. This may further increase the cure rates in some specific tumours. Currently there is also significant interest in the treatment of a number of benign diseases. This includes functionally abnormal areas that can be safely ablated by protons for diseases such as seizures, Parkinson disease, arteriovenous malformations, macular degeneration, and severe rheumatologic conditions. There is even interest in evaluating protons for the prevention of coronary artery restenosis after angioplasty and prevention of stenosis of peripheral vascular shunts that are created in-patients requiring dialysis.

There are some preliminary data available on the treatment of macular degeneration. This is the leading cause of adult onset blindness in the developed countries. Unfortunately current treatments (laser ablation, X-ray therapy, etc.) have not been extraordinarily successful for most patients. There are very encouraging preliminary studies from Loma Linda University Medical Center where over 200 patients have been treated with protons with a single fraction of 14 Gy. The lesion control is 95% with either improvement in vision or no worsening of vision.

**Annex 1.2      Specifications for a novel proton medical facility: the experience of Harvard Medical School and the Massachusetts General Hospital (MGH) in Boston\***

Today there is a real need to improve the quality and the efficiency of hadron facilities in terms of technical and economical aspects. Specifications for a proton medical facility are extremely different from those used for high-energy physics. Below follows a list of goals that a novel proton accelerator should reach to be competitive, by using the accumulated experience of Harvard Medical School and the Massachusetts General Hospital (MGH) for the new Northeast Proton Therapy Centre (NPTC) in Boston [B.31].

The most important specifications are related to: safety, reliability and availability, ease of operation and ease of maintenance. Moreover all these aspects are application driven and the equipment requirements are inferred from these specifications. The design must also feature a thoughtful allocation between the capital equipment costs and the running (or operating) costs.  In the long term, the running costs are likely to be dominant. Moreover the interaction between the equipment and the building is much more complicated than in usual constructions, so that the equipment interface needs particular attention. All these aspects will be analysed below.

▪ *Safety*. A fatal failure must not occur more often than 1 in $10^8$ treatment field applications.[1] To achieve it, a systematic analysis must be made and safe design strategies must be considered. The analysis should include failure mode effects analysis, while design strategies (such as the requirements for redundancy) should be developed and documented [C.56].

▪ *Availability/Reliability*. Conventional linac-based radiation therapy machines operate with an availability (% of time machine is available during the period of scheduled operation) of $\geq 98\%$. A lower level of reliability is widely considered to be clinically unacceptable.  This leads to the requirement that a proton medical facility should have each treatment room available $\geq 98\%$ of the time. This is usually taken to imply that a machine with 3 treatment rooms should have an overall availability (all rooms available) at least 95% of the time. Moreover scheduled down time for routine maintenance should not intrude upon a 2-shift, 5 days/week, 52 weeks/year operation. For safety a machine should shut down on the slightest hint of malfunction: this clearly impacts negatively upon availability. Strategies to keep availability high while keeping the system safe need to be adopted.

▪ *Operability*. Ideally, a proton medical facility should feature push-button operation. Ease of operation is important because it reduces operating costs and it may increase reliability by reducing or eliminating operator error.

▪ *Maintainability*. Availability is high when failures are infrequent and, when they occur, can be quickly repaired.  This means long time between failures and short time to repair. For a medical facility these aspects are necessary to maintain availability at the required level.  A maintainability and reliability analysis should be performed on the proton medical facility design to increase the chances that its operation will be satisfactory. Some points which support a high level of maintainability are: modular components, easily swappable sub-assemblies, good diagnostics, a good supply of spare parts, components available from commercial sources wherever possible, a strong preventative maintenance program.

---

[1] No fatality over 30 years for 50 machines, each delivering some 50,000 fields per year.

- *Extensibility.* Technology changes rapidly. The proton medical facility should be designed to allow the addition of new features and the modification of existing ones.

- *Intensity Modulated Proton Therapy (IMPT).* It is widely accepted that any proton medical facility in the future will have to provide an intensity modulated proton therapy capability.

- *Field size and geometry:* There is no reason to think that, for general cancer therapy, proton field sizes should be much smaller than for X-ray therapy. Moreover the development of a parallel beam of protons, in which the virtual particle source is at infinity, provides some quite advantageous fields with minimal junction problems.

- *Minimise overall cost.* Many factors contribute to the cost of proton beam therapy. First of all the cost of equipment. Here value engineering can reduce the production cost of the facility, together with competition and efficiencies of scale connected with a greater demand and industrialisation. The size of the equipment, such as the number of treatment rooms, has an important impact on costs. Up to a certain point, the more rooms in a facility, the higher the capacity and the lower the cost/fraction. Unfortunately, this observation has led to probably unrealistic ideas of how big a proton medical facility should be. One should not minimise the risk of increasing the cost per fraction in the event that a large facility does not operate at capacity. A safer approach is to improve the throughput of a smaller number of rooms. Another aspect of costs is connected to the building. Here an integration of the proton facility with an existing therapy facility, sharing for example the service infrastructure, could save considerable costs. Reliability, operability, availability, maintainability and extensibility should also be taken seriously into consideration as important cost reduction possibilities in connection with a robust and redundant design of the facility, as mentioned above. Moreover important should be a detailed analysis of costs of pre-treatment work-up and treatment planning as well as the cost of treatment delivery.

**Annex 2.1      An active scanning solution for a LIBO facility: the PSI spot scanning**

We have seen that LIBO can perfectly fit the physical requirements of an active scanning technique. For this reason the apparatus developed at Paul Scherrer Institute in Villingen (Switzerland), has been used as a reference [B.40, B.42]. The compact gantry at PSI has been designed exclusively for the spot scanning in order to achieve a very compact rotating structure. The diameter is about 4 meters, which is determined by the space occupied by the 90 degrees bending magnet, by the nozzle (with monitor and range shifter system) and by the patient table. No drift space is used between bending magnet and patient. The proton beam is deflected by two 35 degrees bending magnets to be parallel (at 1.27 m distance) to the rotation axis of the gantry. It is then deflected on the patient by a large 90 degrees bending magnet. The length of the gantry, measured from the point where the beam leaves the gantry axis to the isocenter plane, is 7.4 m. The beam used at PSI for hadrontherapy comes by degrading the 590 MeV beam down to energies ranging between 80 and 270 MeV, so the phase space of the beam is large and magnets and quadrupoles with big apertures are used, with a final weight of about 110 tons.

Due to the strong degradation of the beam, the optics has been designed to transport a large momentum band (± 1%) with complete achromaticity of the beam at the isocenter. Without these constraints the gantry could be smaller in size and weight. The reduction in size of about a factor of 3 compared to other designs is achieved by combining the beam delivery within the beam optics and by mounting the patient table eccentrically (by 1.3 m) on the front wheel of the gantry. The price to pay for the last is a second axis of rotation $\phi$ needed to maintain the patient table horizontal during gantry rotation. Gantry rotation $\alpha$ is ± 185 degrees. Stereotactic head treatments, which are usually performed in horizontal beam lines, are executed at PSI directly on the gantry. For this purpose a special support can be mounted on the gantry table. This support provides an additional computer controlled rotation of the patient couch in the horizontal plane.

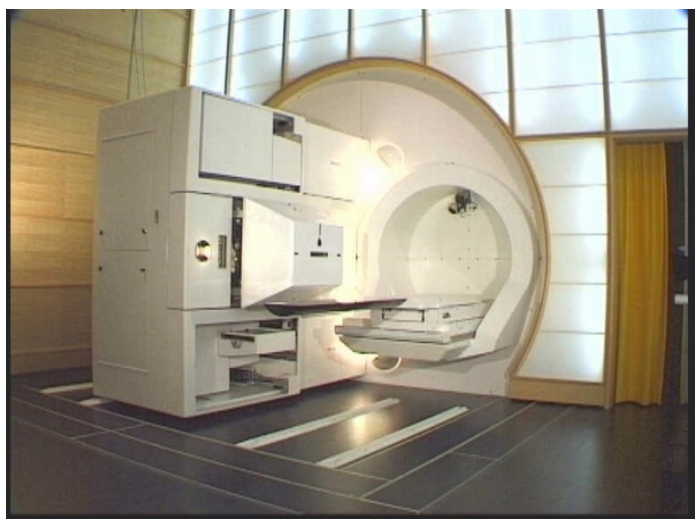

***Figure A annex 2.1*** *Compact proton gantry at PSI (Switzerland) designed for the spot scanning technique [B.42]. This system can be also implemented for a LIBO facility.*

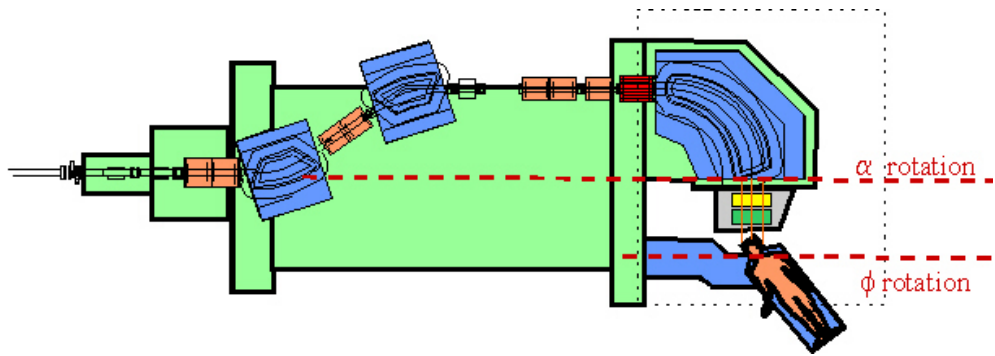

**Figure B annex 2.1** *Longitudinal section of the compact proton gantry design at PSI.*

The rotation angle is selectable between 0 and 120 degrees. By combining together the horizontal rotation with the usual axial rotation of the gantry, an increased flexibility to direct the beam on the skull of the supine patient from almost any direction is obtained.

The measured precision of the beam profiles is better than 1 mm as well as the precision at the isocenter lies within a sphere of 1 mm radius.

The dose delivery is performed by <u>spot scanning</u> already ahead of the 90 degrees bending magnet. The principle of this method is the static application of a sequence of individual beam spots. When all parameters of the beam for a given spot are set and stable, the beam is switched with a kicker magnet. A monitor system measures the integrated beam flux and when the required pre-set value has been reached, the beam is switched off automatically using the fast kicker magnet. The system has been designed for the superposition of typically 10000 spots in 2 minutes for a 1 l target volume. The reaction time of monitor system and of the kicker magnet together is designed to be about 0.1 ms, which corresponds to a dose error of 1% for a mean dose spot time of 10 ms. At the end of each spot deposition the parameters of the beam are changed to the next setting defined in a computer steering file and the cycle is repeated until all spots have been deposited. In this way the target is irradiated completely under computer control, thus providing an automated three-dimensional conformation of the dose on the target volume without collimators and compensators. The scanning of the beam in space is performed in one direction by the action of a sweeper magnet placed in front of the last 90° bending magnet. The magnetic scanning motion is the most rapid one and is therefore the most often used in the scan sequence. A range shifter system is used for moving the Bragg peak in depth. The third scanning motion is the one that is less often used and is performed by moving the patient table.

The computer in the treatment planning system chooses the position and dose of each spot. In this way dose distributions of complex shape can be constructed at will.

The advantages of spot scanning over a continuous scanning method can be summarised as follows.

❑ This method is very insensitive to beam intensity fluctuations of the split beam delivered by the accelerator (better dose control). The PSI scanning system is expected to work with a large variety of accelerators, provided that some form of a slow extraction of the beam is available.

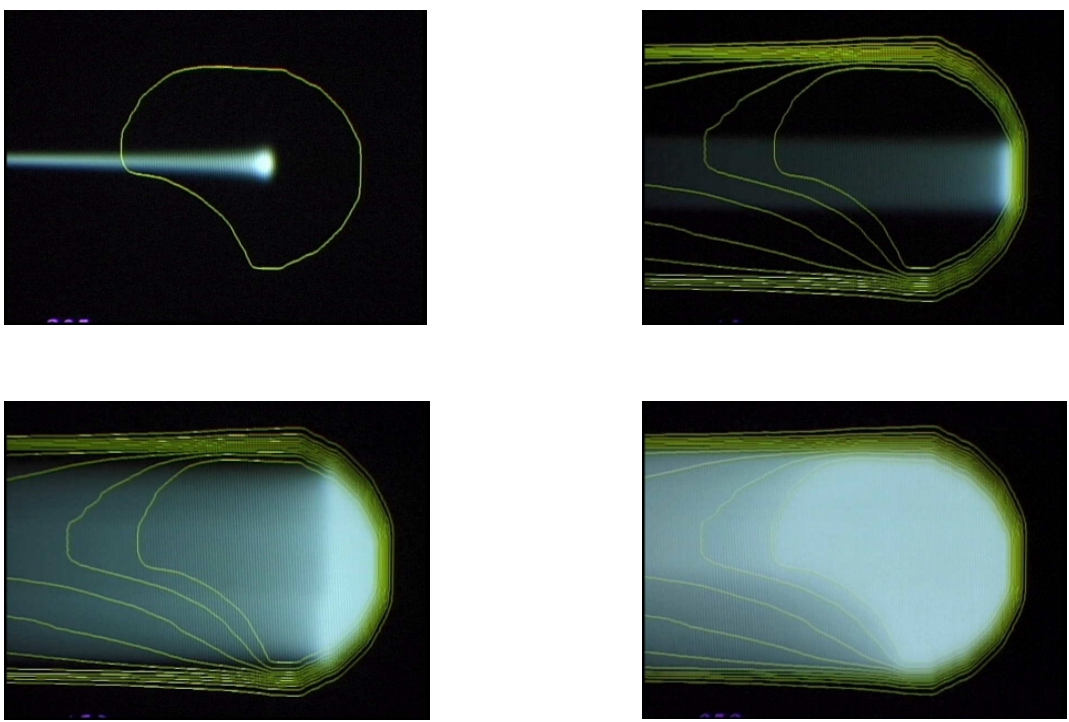

**Figure C annex 2.1** *First figure shows the dose distribution of a single static proton pencil beam (a "dose spot"). Through the superposition of a very large number of these individual dose applications, a total conformation of the dose to the target volume can be achieved (PSI, Switzerland).*

❑ The steering system and the safety analysis are simpler. This method of dynamic scanning is nothing more than a repeated series of single static irradiations.

❑ The dose calculation for the treatment planning system is easier (discrete summation of static spots).

❑ This method allows the combination of slow and fast steering devices in the same scanning system without affecting the dose distribution.

❑ The synchronisation of the treatment with external conditions, such as for example the phase of breathing of the patient, is easier to implement in a "discrete" rather than in a continuous scanning method.

The main advantage of the spot scanning method is given by its flexibility in shaping the dose distribution. However, a major concern is the necessity of controlling the beam delivery quickly and reliably, since the dose rates at the spot peak are remarkably high (in the order of few Grays per sec). An extensive safe system based on redundancy and diversity, which applies and checks the dose to the patient, eliminates completely the risk of uncontrolled dose delivery to the patient.

The realisation of the spot scanning technique required the development of new software for treatment planning. The dose calculation is performed as the superposition of the dose distributions of single proton pencil beams, calculated individually in three dimensions for each spot of the individual treatment plan. The dose is calculated in the patient on the basis of the information delivered by sequential CT slices, which defines in three dimensions the anatomy of the patient. The physician defines then a three-dimensionally shaped target volume by entering individual contours on each CT slice displayed on a workstation. The computer chooses automatically from the list of all possible dose "spot" positions, those which are within some safety margin inside the target volume and defines the dose intensity of each spot according to a predefined Spread-Out Bragg Peak scan law. At the same time a dose distribution "template" is generated on the base of the target contour information using only geometrical criteria. This fictitious dose distribution is chosen to be the goal function for an optimisation procedure, in which the dosage of each spot is allowed to vary individually.

Figure D shows an example of axial, sagital and coronal image of the CT data with an overlaid dose distribution represented as a colour wash. This picture presents the excellent shaping capability of the PSI spot scanning technique. An adequate knowledge of the possible errors occurring during therapy is a point of crucial importance: a dedicated Monte Carlo program has been developed and provides a precise dose calculations in the presence of complex anatomical inhomogeneities, such as bony structures and air cavities.

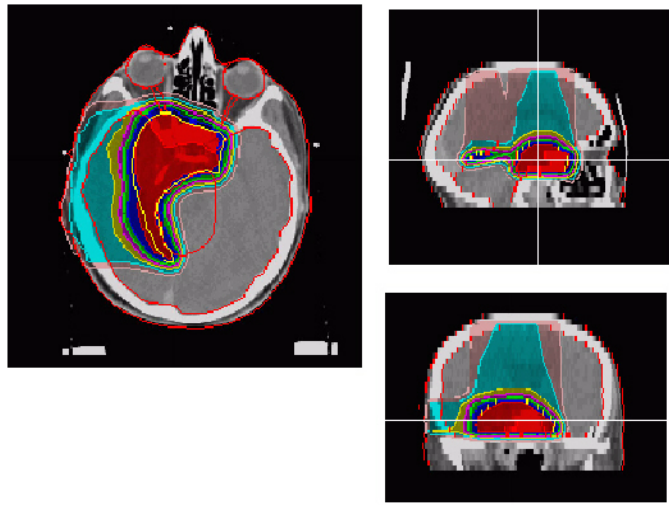

**Figure D annex 2.1** *Example of dose distribution obtained with the spot scanning technique (PSI, Switzerland).*

**Annex 3.1       Coupled oscillators analysis for a biperiodic structure**

To describe the behaviour of the cavities in standing wave regime, one can identify a chain of resonating cells considering N coupled electrical oscillators (figure A annex 3.1) [C.6, C.28, C.32]. The coupling between these oscillators is assumed to be magnetic, through the mutual inductance. Without coupling there are N independent oscillators. Moreover we consider that each cell resonates in the same cavity mode, such as a $TM_{010}$ mode.

The excitation of this system can be described in terms of normal modes of oscillation of the system (see chapter 3). When we consider the cavity modes, it is to these normal modes that we refer, and not to individual $TM_{010}$ cell electromagnetic mode. Each of these individual electromagnetic modes has its own set of normal modes of oscillation in the cavity. Generally speaking, for a chain of N+1 cells, each mode begets a family of N+1 normal mode that occupies a frequency band near the resonant frequency of the isolated oscillator. The properties of these N+1 normal modes can be determined by solving an eigenvalue problem. Applying the Kirchoff'' law to the N circuits, the resulting N equations are solved for the eigenfrequencies and the corresponding eigenvectors. The eigenvector components give the current distribution in the oscillators for each normal mode. Normally a tank is described in terms of N+1 coupled oscillators: N-1 are identical oscillators in the middle and half oscillator on each end.

Under these assumptions, let consider now a biperiodic chain, as shown in figure B annex 3.1.b, composed of accelerating (even) and coupling cells (odd). k is defined as the nearest neighbour coupling and $k_1$ and $k_2$ are the next or second nearest neighbour coupling. For this case, the resonator equations are:

$$I_{2n} = X_{2n} \cdot \left(1 - \frac{f_1^2}{f^2} + \frac{1}{jQ_1} \cdot \frac{f_1}{f}\right) + \frac{1}{2} \cdot k \cdot \left(X_{2n-1} + X_{2n+1}\right) + \frac{1}{2} \cdot k_1 \cdot \left(X_{2n-2} + X_{2n+2}\right)$$

$$I_{2n+1} = X_{2n+1} \cdot \left(1 - \frac{f_2^2}{f^2} + \frac{1}{jQ_2} \cdot \frac{f_2}{f}\right) - \frac{1}{2} \cdot k \cdot \left(X_{2n} + X_{2n+2}\right) + \frac{1}{2} \cdot k_2 \cdot \left(X_{2n-1} + X_{2n+3}\right)$$

and the appropriate end equations are:

$$X_{-1} = X_1, \quad X_{-2} = X_2$$

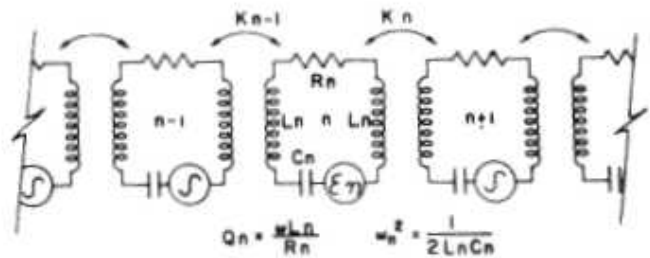

**_Figure A annex 3.1_** _Generalised equivalent magnetically coupled resonant circuits. Each cell is defined by the capacitance (C), inductance (L) and resistance (R), and connected to the adjacent cell by a magnetic coupling._

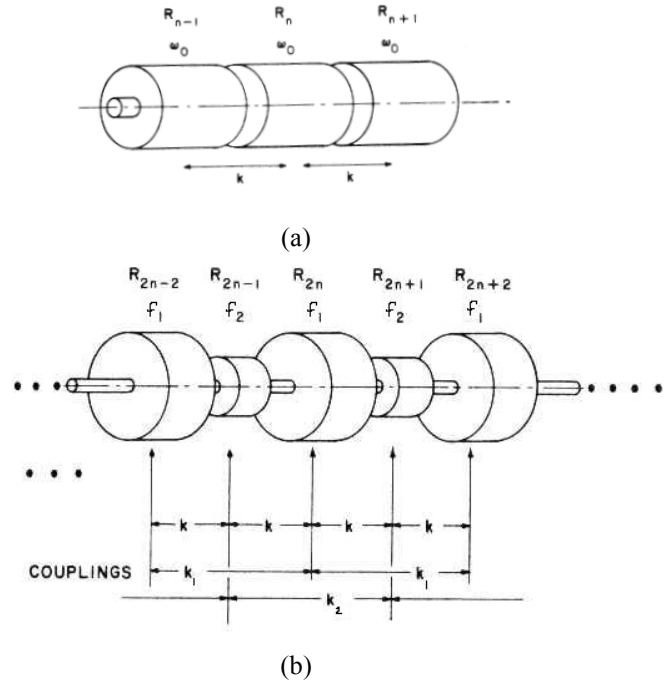

(a)

(b)

**Figure B annex 3.1** *Figure (a) shows coupled resonator model for a singly periodic cavity chain with coupling k, while figure (b) represents a biperiodic cavity chain with nearest neighbour coupling k and second nearest neighbour coupling $k_1$ and $k_2$. This is the typical schema used for the study of the side coupled structure, where the accelerating cells are the even cells ($f_1$, $k_1 \rightarrow f_a$, $k_a$) and the coupling cells are the odd cells ($f_2$, $k_2 \rightarrow f_c$, $k_c$).*

where $I_n$ is connected with voltage drive in the nth circuit equations while $X_n$ to the circulating current and it may be identified with the field level in the cavity. Finally $\frac{1}{2} \cdot \left| X_n^2 \right| \cdot W(n)$ may be identified with the stored energy in the cell n (with W(n) = 1, n = 1…N-1; W(n) = ½, n = 0, N) [C.32]. The solution to the homogeneous equation (representing the electromagnetic field values in the odd and even cells of a biperiodic cavity chain), with the nomenclature used in [C.33] and in the losses case is:

$$X_{2n} = A \cdot \cos(2 \cdot n \cdot \phi), \quad X_{2n+1} = B \cdot \cos(2n+1)\phi,$$

$$\frac{A}{B} = k \cdot \frac{\cos\phi}{\frac{f_1^2}{f^2} - 1 - k_1 \cdot \cos 2\phi} = \frac{\frac{f_2^2}{f^2} - 1 - k_2 \cdot \cos 2\phi}{k \cdot \cos\phi}$$

where $\phi = \frac{1}{2} \cdot \frac{\pi \cdot q}{N}$, q is the normal mode number (q = 0, 1, …..2N), N is the number of cells.

The dispersion relation is [C.28]:

$$k^2 \cdot \cos^2 \phi = \left( 1 - \frac{f_1^2}{f^2} + k_1 \cdot \cos 2\phi \right) \cdot \left( 1 - \frac{f_2^2}{f^2} + k_2 \cdot \cos 2\phi \right)$$

The dispersion relation has several important features. In general there is a stopband (SB) between two passbands corresponding to the difference in frequency between odd and even cells. The second nearest neighbour couplings tend to alter the shape of the dispersion curve, especially near the 0 and π modes. An important consideration in choosing an operating mode for a linac tank is the tolerance of the cavity field distributions in that mode to errors in cavity manufacture, that is to errors in individual cavity frequencies and couplings. The coupled equations given by the coupled resonator model are suitable for analysis by standard perturbation techniques, giving important results with regard to choice of operating mode for linac tanks. With N+1 cells numbered 0, 1, ….N, each mode has a N+1 dimensional eigenfunctions $X^q$ with components $X_n^q$, and a dispersion relation connecting the natural frequency or eigenvalue of the mode $f_q$ to q. If the frequencies of individual cells ($f_0$) are allowed to have random errors $f_{0n}^2 = f_0^2 + \delta f_{0n}^2$, then the change in the eigenvalue is given by:

$$\frac{1}{2} \cdot \frac{N}{W(q)} \cdot \frac{\delta f_q^2}{f_q^2} = \sum_n W(n) \cdot X_n^q \cdot \frac{\delta f_{0n}^2}{f_0^2} \cdot X_n^q$$

The change in the eigenfunctions of the various modes is calculated as an expansion in the unperturbed eigenfunctions:

$$\delta X^q = \sum_{q' \neq q} \alpha_{qq'} \cdot X^{q'}, \text{ where } \alpha_{qr} = \frac{\sum_n W(n) \cdot X_n^r \cdot \dfrac{\delta f_{0n}^2}{f_0^2} \cdot X_n^q}{\left( \dfrac{f_q^2}{f_r^2} - 1 \right)} \cdot \frac{2W(r)}{N}$$

The above equations indicate that the effect of a cell frequency error on the overall frequency of the mode depends on the stored energy in the cell in question in that mode. Moreover the last equation shows that near π and 0 modes the denominator gets very small (close mode spacing) and sensitivity to errors gets quite high. Important results can be derived from the π/2 mode (q = N/2). In this case an explicit solution yields (here shown for simplicity for a periodic structure) [C.1, C.32]:

$$\delta X_n^{\frac{\pi}{2}} = \frac{1}{k} \cdot \sum_{p=1}^{N} \varepsilon_p \cdot \frac{\sin(\dfrac{\pi \cdot p \cdot n}{N}) \cdot \sin(\dfrac{\pi}{2} \cdot n)}{\sin(\dfrac{p \cdot \pi}{N})}$$

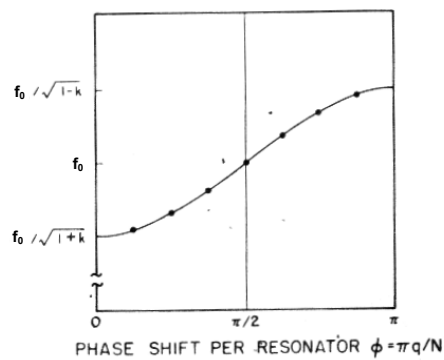

**_Figure C annex 3.1_** _Dispersion curve for a periodic structure of simple coupled cavities and for a specific coupling k. φ is the average cell-to-cell phase shift, q is the mode number. For N+1 cells there are N+1 normal modes spaced as indicated._

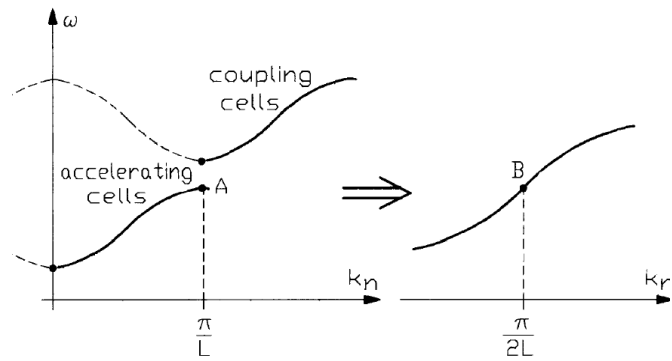

**Figure D annex 3.1** *Dispersion curve for a biperiodic structure with a difference of frequency between accelerating and coupling cells: the stopband (SB). The stopband can be closed with appropriate RF tuning. In this case the resonant frequencies of both cavities are theoretically the same. Continuity between RF frequencies is connected to RF stability of the overall structure.*

with

$$X_n^{\frac{\pi}{2}} = \cos(\frac{\pi}{2} \cdot n) \quad \text{and} \quad \varepsilon_p = \frac{2}{N} \cdot \sum_{r=0}^{N} W(r) \cdot \frac{\delta f_{0r}^2}{f_0^2} \cdot \cos(\frac{\pi \cdot p \cdot r}{N}) \,.$$

Examination of the equations shows that *for the π/2 mode the field levels in even cavities (n = 0, 2, ...N), the only excited in this tuned mode, are independent of frequency errors* in first order. *So a cavity chain operated in the resonant π/2 mode should be less sensitive from problems associated with cavity mistuning* with respect to the other resonant modes.

## Annex 4.1 SUPERFISH output summary*

❑ *The accelerating cell of the first LIBO module [D.50]*

```
SUPERFISH output summary for problem description:
AC Cell for LIBO 62 MeV
Cell length = 17.5 mm (half length 8.75 mm)
Bore radius = 0.4 cm
Diameter=7.0cm
Ez=15.4 MV/m
Tune on gap length
Septum Thickness=0.4 cm
Equator flat 1 mm; Routcor=5.75 mm
Nose flat=1.0 mm
Adjusting gap, currently =  0.6122581,  g/bl = 0.1749309
Problem file: D:\TERA\LIBO\MODEL\FINMOD\TMODAC31.REG 10-06-98  17:35
---------------------------------------------------------------------------
All calculated values below refer to the mesh geometry only.
Field normalization (NORM = 0):    EZERO  =    15.40000 MV/m
Frequency (starting value = 3050.000)  =   3049.99969 MHz
Particle rest mass energy              =    938.279640 MeV
Beta =  0.3560796        Kinetic energy =      65.813 MeV
Normalization factor for E0 = 15.400 MV/m =  115473.418
Transit-time factor                    =     0.8506006
Stored energy                          =     0.0140071 Joules
Using standard room-temperature copper.
Surface resistance                     =    14.82684 milliOhm
Normal-conductor resistivity           =     1.82574 microOhm-cm
Operating temperature                  =    35.0000 C
Reference temperature                  =    20.0000 C
Reference resistivity                  =     1.72410 microOhm-cm
Temperature coefficient of resistance  = 3.93000E-03 per deg C
Power dissipation                      =    32.4213 kW
Q   =   8279.41       Shunt impedance  =    64.006 MOhm/m
Rs*Q =   122.757 Ohm                Z*T*T =    46.310 MOhm/m
r/Q  =    48.942 Ohm  Wake loss parameter =   0.23448 V/pC
Average magnetic field on the outer wall =  27208.54 A/m,  548.819 W/cm^2
Maximum magnetic field on boundary       =  35765.23 A/m,  948.289 W/cm^2
Maximum electric field on boundary       =     70.757 MV/m,  1.5007 Kilp.

Wall segments:
Segment   Zend      Rend      Emax     Power      P/A       dF/dZ      dF/dR
          (cm)      (cm)     (MV/m)    (kW)     (W/cm2)    (MHz/mm)   (MHz/mm)
---------------------------------------------------------------------------
          0.00000   3.50000
    9     0.10000   3.50000   0.1263   1.2040   547.4752   0.000E+00 -1.111E+01
   10     0.20613   3.49012   0.2856   1.2841   547.6134  -1.096E+00 -1.178E+01
   11     0.42193   3.40143   0.7326   2.8406   557.2651  -9.877E+00 -2.402E+01
   12     0.54340   3.29108   1.1005   2.0190   582.7007  -1.240E+01 -1.364E+01
   13     0.67500   2.92500   2.9784   5.0020   644.0767  -4.137E+01 -1.490E+01
   14     0.67500   1.06245  14.5794  17.3218   742.4049  -5.342E+01  0.000E+00
   15     0.67500   0.94432  13.6385   0.5439   730.3810   2.674E+01  0.000E+00
   16     0.65747   0.86245  10.3447   0.3849   805.0836  -1.642E+00 -3.802E-01
   17     0.60430   0.79174   9.9033   0.4128   892.8579  -1.602E+00 -1.194E+00
   18     0.54340   0.75638  14.0241   0.3240   943.5797  -3.858E-01 -5.774E-01
   19     0.42193   0.71217  29.6882   0.5325   892.7825   2.891E+00  7.944E+00
   20     0.37193   0.69397  45.1159   0.1789   761.0489   4.478E+00  1.230E+01
   21     0.30613   0.60000  70.7571   0.2572   512.5808   7.908E+00  4.958E+01
   22     0.30613   0.50000  63.2623   0.0838   242.4037   5.154E+01  0.000E+00
   23     0.40613   0.40000  63.7263   0.0295   68.6181   4.350E+01  2.591E+01
   24     0.42193   0.40000  21.9167   0.0004   10.8788   0.000E+00  6.811E-01
   25     0.54340   0.40000  16.8736   0.0014   4.5219   0.000E+00  1.559E+00
   26     0.87500   0.40000   6.1073   0.0003   0.3778   0.000E+00  3.523E-01
---------------------------------------------------------------------------
              Total           32.4213

Coupling-slot effects for power of   32.421 kW on boundary segments   9
through  26 for 3.000% increase in outer wall power per percent coupling:

Coupling  Slot Power  Total Power    Q          Z          ZTT
  (%)       (kW)         (kW)                 (MΩ/m)     (MΩ/m)
0.0000     0.00       32.421     8279.41    64.006     46.310
1.0000     0.973      33.394     8038.26    62.142     44.961
2.0000     1.945      34.367     7810.76    60.383     43.688
3.0000     2.918      35.339     7595.79    58.721     42.486
4.0000     3.891      36.312     7392.33    57.148     41.348
5.0000     4.863      37.284     7199.49    55.657     40.269
6.0000     5.836      38.257     7016.45    54.242     39.245
```

❑ *The coupling cell of the first LIBO module [D.50]*

```
SUPERFISH output summary for problem description:
CC Cell for LIBO 62 MeV; "cylindrical" cell type
Beta=.35
Diameter=50, 60, 70 mm
Tune on gap length
Cell length 2*8.75 =17.5 mm
Web thickness = 4 mm
All radii of curvature  = 1 mm
Nose diameter 20 (for D=50 and 60), 25 (for D=50, 60 and 70), 30 (for D=70)
Nose flat=8.8, 11.3, 13.8 mm
Adjusting gap, currently =   0.6882319,  g/bl = 0.1966377
Problem file: D:\TERA\LIBO\MODEL\FINMOD\TMODCC44.REG 10-05-98  14:27
---------------------------------------------------------------------------
All calculated values below refer to the mesh geometry only.
Field normalization (NORM = 0):    EZERO =    1.00000 MV/m
Frequency (starting value = 3120.000)     =  3120.00032 MHz
Particle rest mass energy                 =  938.279640 MeV
Beta =  0.3642520         Kinetic energy =   69.214 MeV
Normalization factor for E0 = 1.000 MV/m =   8089.466
Transit-time factor                       =  0.9365142
Stored energy                             =  6.66821E-05 Joules
Using standard room-temperature copper.
Surface resistance                        =  14.99602 milliOhm
Normal-conductor resistivity              =  1.82574 microOhm-cm
Operating temperature                     =  35.0000 C
Reference temperature                     =  20.0000 C
Reference resistivity                     =  1.72410 microOhm-cm
Temperature coefficient of resistance = 3.93000E-03 per deg C
Power dissipation                         =  184.2263 W
Q    =  7095.65        Shunt impedance =   47.496 MOhm/m
Rs*Q =  106.407 Ohm               Z*T*T =   41.657 MOhm/m
r/Q  =   51.369 Ohm  Wake loss parameter =  0.25175 V/pC
Average magnetic field on the outer wall  =  2148.68 A/m,    3.462 W/cm^2
Maximum magnetic field on boundary        =  3260.24 A/m,    7.970 W/cm^2
Maximum electric field on boundary        =    6.456 MV/m,  0.1355 Kilp.

Wall segments:
Segment    Zend     Rend     Emax     Power     P/A      dF/dZ      dF/dR
           (cm)     (cm)    (MV/m)    (W)     (W/cm2)   (MHz/mm)   (MHz/mm)
-------------------------------------------------------------------------------
        0.00000   3.00000
   5    0.29412   3.00000   0.0010   19.1910   3.4616   0.000E+00 -3.762E+01
   6    0.48774   3.00000   0.0031   12.6297   3.4604   0.000E+00 -2.476E+01
   7    0.57500   3.00000   0.0063    5.6880   3.4583   0.000E+00 -1.115E+01
   8    0.67500   2.90000   0.0458   10.3065   3.5378  -1.295E+01 -1.286E+01
   9    0.67500   1.72678   0.6686   79.8923   4.6849  -1.279E+02  0.000E+00
  10    0.67500   1.52678   0.6751   12.0291   5.8843  -1.434E+01  0.000E+00
  11    0.67500   1.38536   0.6204    8.3612   6.4624  -1.260E+01  0.000E+00
  12    0.67500   1.35000   0.4017    2.1202   6.9783  -3.771E+00  0.000E+00
  13    0.57500   1.25000   0.4137    9.7370   7.6733  -1.158E+01 -1.160E+01
  14    0.48774   1.25000   0.9776    5.2915   7.7214   0.000E+00 -6.967E+00
  15    0.44412   1.25000   1.6188    2.4810   7.2405   0.000E+00  7.951E-01
  16    0.34412   1.15000   3.4862    6.9787   5.8281   7.299E+01  5.234E+01
  17    0.34412   0.02000   3.1005    9.5201   2.2921   2.471E+02  0.000E+00
  18    0.35412   0.01000   2.6137    0.0000   0.0003   5.045E-02  2.312E-02
  19    0.48774   0.01000   6.4564    0.0000   0.0000   0.000E+00  2.265E-04
  20    0.87500   0.01000   0.0000    0.0000   0.0000   0.000E+00  1.669E-24
-------------------------------------------------------------------------------
                    Total        184.2263

Coupling-slot effects for power of   184.226 W on boundary segments    5
through  20 for 3.000% increase in outer wall power per percent coupling:

Coupling  Slot Power  Total Power     Q          Z        ZTT
  (%)       (W)          (W)                   (MΩ/m)    (MΩ/m)
0.0000      0.000     184.226     7095.65    47.496    41.657
1.0000      5.527     189.753     6888.98    46.113    40.443
2.0000     11.054     195.280     6694.01    44.807    39.299
3.0000     16.580     200.807     6509.77    43.574    38.217
4.0000     22.107     206.333     6335.40    42.407    37.194
5.0000     27.634     211.860     6170.13    41.301    36.223
6.0000     33.161     217.387     6013.26    40.251    35.302
```

## Annex 5.1    Basic concepts on the elastic problem for structural analysis*

Let us consider the figure A annex 5.1, where $x^i$ is the spatial reference system for the configuration B, while $\overline{x^i}$ is the spatial reference system for the elastically deformed configuration $\overline{B}$.

We define the metric of $x^i$ as:

$$dS^2 = a_{ik} \cdot dx^i \cdot dx^k, \quad \text{with} \quad a_i \times a_k = a_{ik},$$

the metric of $\overline{x^i}$ as:

$$d\overline{S}^2 = \left(a_{ik} + S_{i/k} + S_{k/i} + S_{n/i} \cdot S^n /_k\right) \cdot d\overline{x}^i \cdot d\overline{x}^k,$$

and the vector displacement as:

$$S = S_i \cdot a^i$$

For i = 1, 2, 3 = x, y, z, S is represented by a vector explicated as:

$$S = S(u,v,w) = S_1 \cdot a_1 + S_2 \cdot a_2 + S_3 \cdot a_3 = u \cdot a_x + v \cdot a_y + w \cdot a_z$$

We define the Ricci tensor as:

$$\in_{ikr} = a_i \wedge a_k \times a_r$$

We consider now an infinitesimal element of material where the applied stress is indicated with the tensor $\sigma_{ik}$, while the strains are defined with the tensor $\varepsilon_{ik}$. We define the stress and strain problem with the well-known equations written with a tensor formulation [C.54, C.55]:

$$\begin{cases} \sigma_{ik/k} = -F \\ \sigma_{ik} \cdot n^k = f_i \\ \varepsilon_{ik} = \dfrac{1}{2}\left(S_{i/k} + S_{k/i}\right) \\ S_{i(c)} = \overline{S_i} \\ \sigma_{ik} = c_{ikrs}\varepsilon^{rs} \end{cases}, \quad \text{with} \quad c_{ikrs} = \lambda \cdot a_{ik} \cdot a_{rs} + \mu \cdot \left(a_{ir} \cdot a_{ks} + a_{is} \cdot a_{kr}\right)$$

where λ and μ are the Lame' constants.

The first two equations represent the relation between force and stress for internal (F) and boundary (f) conditions respectively, as illustrated in figure A annex 5.1. The third and fourth fix the relation between strains and displacements, while the last equation defines the relation between stress and strain. To complete the problem definition one can also validate the equation [C.55]:

$$\in_{ilr} \cdot \in_{kms} \cdot \varepsilon^{lm/rs} = 0$$

The following relation calculates the principal stresses [C.35, C.55]:

$$\left(\sigma_i - \lambda \cdot a_{ik}\right) \cdot n^k = 0 \quad \Rightarrow \quad \sigma_{princ}\left(\sigma_I, \sigma_{II}, \sigma_{III}\right) \Leftrightarrow \quad \det(\sigma_{ik} - \lambda \cdot a_{ik}) = 0$$

The Von Mises criteria for material resistance is then shown as a reference:

$$\sigma_{Von-Mises} = \sqrt{\sigma_I^2 + \sigma_{II}^2 + \sigma_{III}^2 - \sigma_I \cdot \sigma_{II} - \sigma_{II} \cdot \sigma_{III} - \sigma_I \cdot \sigma_{III}} \leq \sigma_{amm} = \frac{\sigma_{elast\_Material}}{\eta}$$

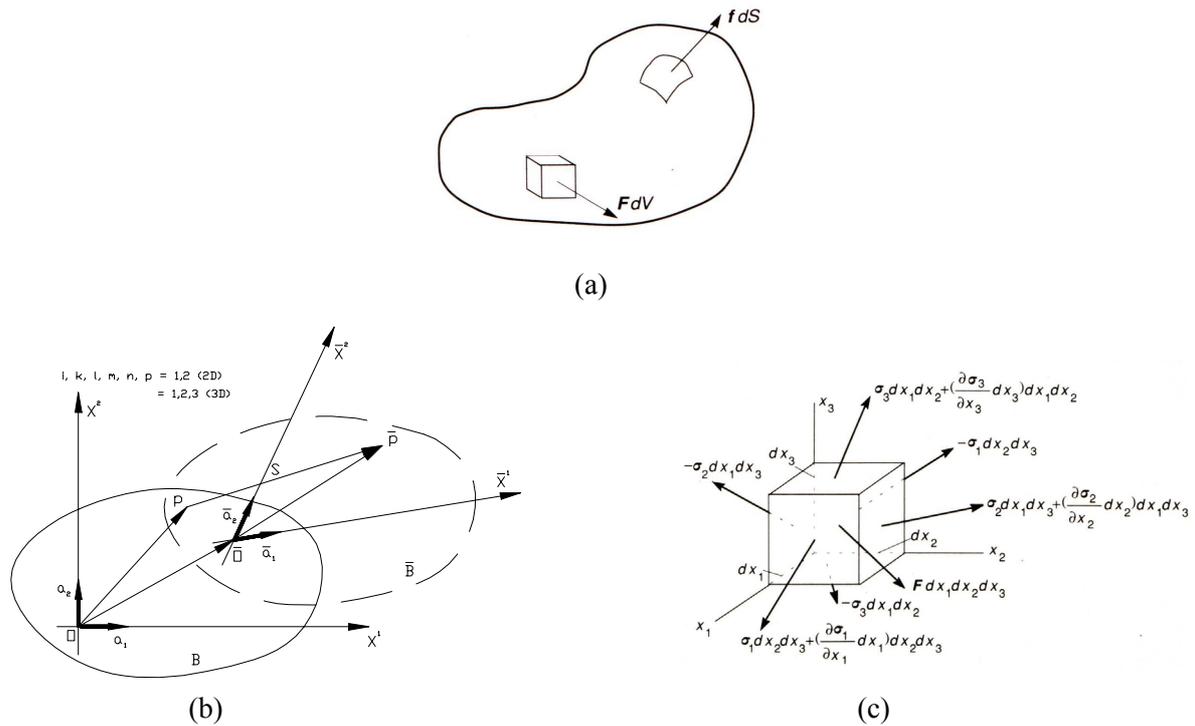

(a)

(b)                                                           (c)

**_Figure A annex 5.1_** _Stress and strain problem definition [C.55]._

$\sigma_{elast\_Material}$ is the elastic limit of the material and $\eta$ is a safety margin coefficient. More about the resistance criteria of materials can be found in literature [C.54]. The stress and strain relation in a compact format is:

$$\varepsilon_{ik} = \frac{1}{2\mu} \cdot \sigma_{ik} - \frac{\lambda}{2\mu \cdot (3\lambda + 2\mu)} \sigma_l^l \cdot a_{ik}$$

or with the explicit 3D engineering form [C.35]:

$$\begin{cases}
\varepsilon_{11} = \varepsilon_x = \dfrac{\sigma_{11}}{E} - \dfrac{\sigma_{22} + \sigma_{33}}{\nu \cdot E} = \dfrac{\sigma_x}{E} - \dfrac{\sigma_y + \sigma_z}{\nu \cdot E} \\[2mm]
\varepsilon_{22} = \varepsilon_y = \dfrac{\sigma_{22}}{E} - \dfrac{\sigma_{11} + \sigma_{33}}{\nu \cdot E} = \dfrac{\sigma_y}{E} - \dfrac{\sigma_x + \sigma_z}{\nu \cdot E} \\[2mm]
\varepsilon_{33} = \varepsilon_z = \dfrac{\sigma_{33}}{E} - \dfrac{\sigma_{11} + \sigma_{22}}{\nu \cdot E} = \dfrac{\sigma_z}{E} - \dfrac{\sigma_x + \sigma_y}{\nu \cdot E} \\[2mm]
2 \cdot \varepsilon_{12} = 2 \cdot \varepsilon_{xy} = \gamma_{12} = \dfrac{\tau_{12}}{G} = \dfrac{\sigma_{xy}}{G} \\[2mm]
2 \cdot \varepsilon_{13} = 2 \cdot \varepsilon_{xz} = \gamma_{13} = \dfrac{\tau_{13}}{G} = \dfrac{\sigma_{xz}}{G} \\[2mm]
2 \cdot \varepsilon_{23} = 2 \cdot \varepsilon_{yz} = \gamma_{23} = \dfrac{\tau_{23}}{G} = \dfrac{\sigma_{yz}}{G}
\end{cases}$$

E is the Young module, $\nu$ is the Poisson module, while G is the shear module. $\sigma_{ik}$ are the stress components while $\varepsilon_{ik}$ are the equivalent strains. The nomenclature reefers at the figure in note 4 of paragraph 5.7.1.

Let us consider also an anelastic effect, such as deformations produced by temperature gradients ($\Delta T$). The stress and strain relation can be written as:

$$\sigma_{ik} = c_{ikrs} \cdot \left( \varepsilon^{rs} - \varepsilon_{th}^{rs} \right) \quad \text{and} \quad \varepsilon_{th\ ik} = \alpha \cdot \Delta T \cdot \delta_{ik}$$

where $\alpha$ is the thermal expansion coefficient and $\delta_{ik}$ is the well-known Kronecker term.

**Annex 5.2**     **Analytic base of a finite element code for thermal and structural analysis***

The finite element method is a numerical approach to calculate the thermal and stress distribution in all the cases that do not have an exact mathematical solution. Starting from the analytical description of the physical phenomena, this method is extended to all problems that can be analysed in terms of a variational formulation, such as the temperature distribution analysis or the structural analysis. For practical reasons the problem is developed with a matrix formulation [C.35].

❑   *Thermal analysis*

The first law of thermodynamics states that thermal energy is conserved. Specialising this to a differential control volume:

$$\rho \cdot c \cdot \left( \frac{\partial T}{\partial t} + \{v\}^T \cdot \{L\} \cdot T \right) + \{L\}^T \cdot \{q\} = \ddot{q}$$

where $\rho$ is the density, c is the specific heat, T (x, y, z) is the temperature distribution, t is the time.

$\{L\} = \left\{ \begin{array}{c} \frac{\partial}{\partial x} \\ \frac{\partial}{\partial y} \\ \frac{\partial}{\partial z} \end{array} \right\}$ is the vector operator, $\{v\} = \left\{ \begin{array}{c} v_x \\ v_y \\ v_z \end{array} \right\}$ is the velocity vector for mass transport of heat, $\{q\}$ is the heat flux vector, $\ddot{q}$ is the heat generation rate per unit volume. One can define also $K_{xx}, K_{yy}, K_{zz}$ as the conductivity values of the elements in x, y and z direction. Next, Fourier's law is used to relate the heat flux vector to the thermal gradients:

$$\{q\} = -[D] \cdot \{L\} \cdot T$$

where

$$[D] = \begin{bmatrix} K_{xx} & 0 & 0 \\ 0 & K_{yy} & 0 \\ 0 & 0 & K_{zz} \end{bmatrix} = \text{Conductivity matrix.}$$

Let consider the physical parameters of the heat transfer phenomena mentioned above and defined in a continuos domain. We mesh then this continuum with ideal surfaces and lines, the so-called finite elements. One can write the following assumption:

$$T(x, y, z) = \{N(x, y, z)\}^T \cdot \{T_e\}$$

where T(x, y, z) is the temperature distribution over the entire domain, N(x, y, z) are the element shape functions, and $\{T_e\}$ is the nodal temperature vector.

The combination $\{L\} \cdot T$ is written as:

$$\{L\} \cdot T = [B] \cdot \{T_e\}$$

where

$$[B] = \{L\} \cdot \{N\}^T .$$

The static heat transfer equation can be written as:

$$([K_e^{tm}] + [K_e^{tb}] + [K_e^{tc}]) \cdot \{T_e\} = \{Q_e^f\} + \{Q_e^c\} + \{Q_e^g\}$$

where:

$[K_e^{tm}] = \rho \int\limits_V c \cdot \{N\} \cdot \{v\}^T \cdot [B] \cdot dV$ = element mass transport conductivity matrix;

$[K_e^{tb}] = \int\limits_V [B]^T \cdot [D] \cdot [B] \cdot dV$ = element diffusion conductivity matrix;

$[K_e^{tc}] = \int\limits_S h_f \cdot \{N\} \cdot \{N\}^T \cdot dS$ = element convection surface conductivity matrix;

$\{Q_e^f\} = \int\limits_S \{N\} \cdot q^* \cdot dS$ = element mass flux vector;

$\{Q_e^c\} = \int\limits_S T_B \cdot h_f \cdot \{N\} \cdot dS$ = element convection surface heat flow vector;

$\{Q_e^g\} = \int\limits_V \ddot{q} \cdot \{N\} \cdot dV$ = element heat generation load.

Inverting the above equation, one can find:

$$\{T_e\} = [K_e^{tot}]^{-1} \cdot \{Q_e^{tot}\}$$

This equation describes the thermal distribution at the nodal points over the entire domain. Solving the linear system, one can obtain the temperature values. The dynamic system is represented adding to the equation the terms $\dfrac{\partial T}{\partial t} = \{N\}^T \cdot \{\dot{T}_e\}$ and integrating the linear equations step by step in time.

❑ _Structural analysis_

Once the thermal distribution is calculated, one can also proceed to the structural analysis of the system. Starting from the stress and strain equations under thermal and structural loads, one can write the following:

$$\{\sigma\} = [D] \cdot \{\varepsilon^{el}\}$$

Let us consider the vector displacement, as presented in annex 5.1.

We write the complete relations between the strains and the physical displacements (S(u, v, w)) of the material as:

$$\begin{cases} \varepsilon_x = \dfrac{\partial u}{\partial x} \quad & \varepsilon_{xy} = \dfrac{\partial u}{\partial y} + \dfrac{\partial v}{\partial x} \\[2mm] \varepsilon_y = \dfrac{\partial v}{\partial y} \quad & \varepsilon_{xz} = \dfrac{\partial u}{\partial z} + \dfrac{\partial w}{\partial x} \\[2mm] \varepsilon_z = \dfrac{\partial w}{\partial z} \quad & \varepsilon_{yz} = \dfrac{\partial v}{\partial z} + \dfrac{\partial w}{\partial y} \end{cases}$$

The equations connecting the material displacements and the thermal (T (x, y, z)) and structural (X, Y, Z) loads are then:

$$\begin{cases} \dfrac{E}{2(1+\mu)\cdot(1-2\mu)}\cdot\left(\dfrac{\partial^2 u}{\partial x^2}+\dfrac{\partial^2 v}{\partial x\partial y}+\dfrac{\partial^2 w}{\partial x\partial z}\right)+\dfrac{E}{2(1+\mu)}\cdot\left(\dfrac{\partial^2 u}{\partial x^2}+\dfrac{\partial^2 u}{\partial y^2}+\dfrac{\partial^2 u}{\partial z^2}\right)-\dfrac{E\cdot\alpha}{1-2\mu}\cdot\dfrac{\partial T}{\partial x}+X=0 \\[3mm] \dfrac{E}{2(1+\mu)\cdot(1-2\mu)}\cdot\left(\dfrac{\partial^2 u}{\partial x\partial y}+\dfrac{\partial^2 v}{\partial y^2}+\dfrac{\partial^2 w}{\partial y\partial z}\right)+\dfrac{E}{2(1+\mu)}\cdot\left(\dfrac{\partial^2 v}{\partial x^2}+\dfrac{\partial^2 v}{\partial y^2}+\dfrac{\partial^2 v}{\partial z^2}\right)-\dfrac{E\cdot\alpha}{1-2\mu}\cdot\dfrac{\partial T}{\partial y}+Y=0 \\[3mm] \dfrac{E}{2(1+\mu)\cdot(1-2\mu)}\cdot\left(\dfrac{\partial^2 u}{\partial x\partial z}+\dfrac{\partial^2 v}{\partial y\partial z}+\dfrac{\partial^2 w}{\partial z^2}\right)+\dfrac{E}{2(1+\mu)}\cdot\left(\dfrac{\partial^2 w}{\partial x^2}+\dfrac{\partial^2 w}{\partial y^2}+\dfrac{\partial^2 w}{\partial z^2}\right)-\dfrac{E\cdot\alpha}{1-2\mu}\cdot\dfrac{\partial T}{\partial z}+Z=0 \end{cases}$$

The above equations can be solved numerically. One can then obtain the stress and strain distribution under the thermal and structural loads in the finite element nodal positions.

The procedure for the structural analysis is similar to the method used for the thermal analysis, and can be summarised as follows.

a) The continuum is meshed with ideal surfaces and lines: the finite elements.

b) The elements are connected through a discrete number of nodal points on the boundary. The displacements in the nodal points are the problem solutions.

c) The relation between the displacements in a single element and the displacements at the nodal points is a typical function of the problem. One can define the displacements over the domain as:

$$\{f(x,y,z)\}=\begin{Bmatrix} u\ (x,y,z) \\ v\ (x,y,z) \\ w\ (x,y,z) \end{Bmatrix}$$

and its relation with the nodal displacements as:

$$\{f(x,y,z)\}=[N(x,y,z)]\cdot\{\delta_e\}$$

where N(x, y, z) are the element shape functions and { $\delta_e$ } is the nodal displacements vector.

d) With the elastic properties one can calculate the stress distribution in all the elements, included on the boundary. The relation between strains and displacements reads as:

$$\{\varepsilon\}=[B]\cdot\{\delta_e\}$$

where $[B]$ is the matrix with the terms $\dfrac{\partial N_i}{\partial x},\dfrac{\partial N_j}{\partial y}$, etc.

The relation between stress and strain is:

$$\{\sigma\}=[D]\cdot[\{\varepsilon\}-\{\varepsilon_{th}\}]$$

where $[D]$ is the matrix of elasticity, and $\{\varepsilon_{th}\}$ is the anelastic strain vector connected to the thermal deformations.

e) It is then possible to define a load configuration concentrated on the nodal points and equivalent to the global loads.

For a single element we write:

$$\{F_e\} = \begin{Bmatrix} F_i \\ F_j \\ F_m \end{Bmatrix}$$

f)  Through the equilibrium equations to the nodes and the virtual works principle, one can obtain, for each single element, a linear system, where the displacements are the solutions:

$$\{F_e\} = \{\delta_e\} \cdot \int_V [B]^T \cdot [D] \cdot [B] \cdot dV - \int_V [B]^T \cdot [D] \cdot \{\varepsilon_0\} \cdot dV - \int_V [N]^T \cdot \{p\} \cdot dV$$

The stiffens matrix of the element is defined as:

$$[k_e] = \int_V [B]^T \cdot [D] \cdot [B] \cdot dV$$

One should repeat the process for each element, imposing the equilibrium in each node. At the end we have a 3n linear equation system with 3n variables, where n is the number of nodes with respect to the 3 possible displacements (*u, v, w*).

Due to the fact that the stiffness matrix is symmetric and positive, the above equation can be inverted:

$$\{\delta_e\} = \left[K_e^{tot}\right]^{-1} \cdot \left\{F_e^{tot}\right\}$$

This is the final equation that describes the physical phenomena at the nodal points. Solving the linear system, one obtains the displacement distribution in each nodal point and then the deformation and stress. If the boundary conditions are defined in terms of displacements, one can apply them directly. Otherwise with conditions expressed in terms of loads, it is necessary to transform them as forces in equivalent nodal points. The most important calculations are two: the stiffness matrix definition for each element and the solution of the global linear system. Both can be solved numerically using several optimised methods, such as Gauss, Gauss-Jordan or Choleski methods [C.35].

**Annex 5.3        Transient thermal analysis of the prototype***

To understand the dynamic behaviour of the system, one could solve the problem by anti-transforming directly the Laplace equations of chapter 5 or by using a numerical solution. With respect to the analytical model mentioned in paragraph 5.7.3, in order to have a more realistic scenario of the physical phenomena, one should take into consideration a more complex system, including also the heat transmission between consecutive tanks. Forced convection and conduction heat exchanges between a tank, its cooling plates and water, and two consecutive tanks have been then investigated, as shown in figure A annex 5.3. This figure can be described with the following system (the nomenclature refers to tables 5.7 and 5.8 in chapter 5, the indexes 1 and 2 indicate the tank number):

$$\begin{cases} c_w \cdot M_{w1} \cdot \dfrac{\partial}{\partial t} T_{w1} = -2 \cdot c_w \cdot \Gamma_1 \cdot (T_{w1} - T_{in1}) + h_1 \cdot S_{ext} \cdot (T_{c1} - T_{w1}) \\[2mm] c_c \cdot M_c \cdot \dfrac{\partial}{\partial t} T_{c1} = -h_1 \cdot S_{ext} \cdot (T_{c1} - T_{w1}) + P - \dfrac{K \cdot S_{BC}}{d} \cdot (T_{c1} - T_{c2}) \\[2mm] c_w \cdot M_{w2} \cdot \dfrac{\partial}{\partial t} T_{w2} = -2 \cdot c_w \cdot \Gamma_2 \cdot (T_{w2} - T_{in1}) + h_2 \cdot S_{ext} \cdot (T_{c2} - T_{w2}) \\[2mm] c_c \cdot M_c \cdot \dfrac{\partial}{\partial t} T_{c2} = -h_2 \cdot S_{ext} \cdot (T_{c2} - T_{w2}) + P + \dfrac{K \cdot S_{BC}}{d} \cdot (T_{c1} - T_{c2}) \end{cases}$$

This system can be solved with a conventional numerical approach:

$$\begin{cases} \dfrac{T_{w1}^{i+1} - T_{w1}^{i}}{\Delta t} = \dfrac{v_1}{\Gamma_1 \cdot L_{cs} \cdot c_w} \cdot \left( h_1 \cdot S_{ext} \cdot T_{c1}^{i} - h_1 \cdot S_{ext} \cdot T_{w1}^{i} - 2 \cdot c_w \cdot \Gamma_1 \cdot T_{w1}^{i} + 2 \cdot c_w \cdot \Gamma_1 \cdot T_{in1} \right) \\[2mm] \dfrac{T_{c1}^{i+1} - T_{c1}^{i}}{\Delta t} = \dfrac{1}{c_c \cdot M_c} \cdot \left( P - \dfrac{K \cdot S_{BC}}{d} \cdot T_{c1}^{i} + \dfrac{K \cdot S_{BC}}{d} \cdot T_{c2}^{i} - h_1 \cdot S_{ext} \cdot T_{c1}^{i} + h_1 \cdot S_{ext} \cdot T_{w1}^{i} \right) \\[2mm] \dfrac{T_{w2}^{i+1} - T_{w2}^{i}}{\Delta t} = \dfrac{v_2}{\Gamma_2 \cdot L_{cs} \cdot c_w} \cdot \left( h_2 \cdot S_{ext} \cdot T_{c2}^{i} - h_2 \cdot S_{ext} \cdot T_{w2}^{i} - 2 \cdot c_w \cdot \Gamma_2 \cdot T_{w2}^{i} + 2 \cdot c_w \cdot \Gamma_2 \cdot T_{in1} \right) \\[2mm] \dfrac{T_{c2}^{i+1} - T_{c2}^{i}}{\Delta t} = \dfrac{1}{c_c \cdot M_c} \cdot \left( P + \dfrac{K \cdot S_{BC}}{d} \cdot T_{c1}^{i} - \dfrac{K \cdot S_{BC}}{d} \cdot T_{c2}^{i} - h_2 \cdot S_{ext} \cdot T_{c2}^{i} + h_2 \cdot S_{ext} \cdot T_{w2}^{i} \right) \end{cases}$$

The system, the base of our numerical code, is then written in terms of matrix formulation as:

$$\begin{bmatrix} T_{w1} \\ T_{c1} \\ T_{w2} \\ T_{c2} \end{bmatrix}^{i+1} = \left[ Matr\_System \right] \cdot \begin{bmatrix} T_{w1} \\ T_{c1} \\ T_{w2} \\ T_{c2} \end{bmatrix}^{i} + \left[ Vect\_System \right] \cdot \Delta t$$

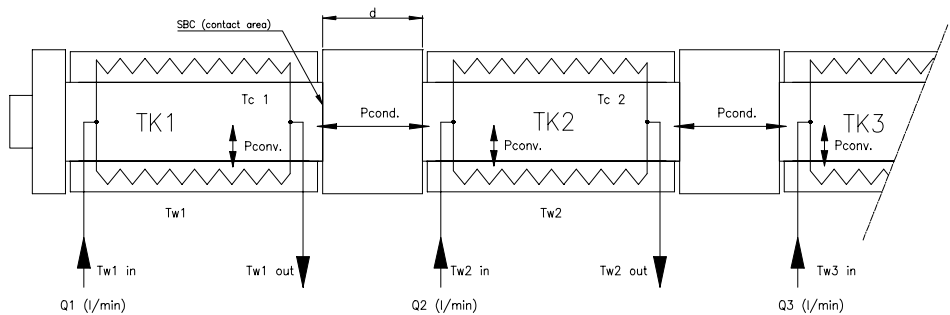

***Figure A annex 5.3***   *Prototype module with tanks and cooling channels. Each tank can exchange heat with its lateral cooling plate through convection and with the neighbour tank through conduction with the bridge coupler.*

where the matrix and vector of the system are ($T_{in1} = T_{in2} = T_{in}$ (input water temperature)):

$$[Matr\_System] = \begin{bmatrix} -\dfrac{h_1 \cdot S_{est} + 2 \cdot c_w \cdot \Gamma_1}{\Gamma_1 \cdot L_{cs} \cdot c_w / v_1} \cdot \Delta t + 1 & \dfrac{h_1 \cdot S_{est}}{\Gamma_1 \cdot L_{cs} \cdot c_w / v_1} \cdot \Delta t & 0 & 0 \\[3mm] \dfrac{h_1 \cdot S_{est}}{c_c \cdot M_c} \cdot \Delta t & -\dfrac{h_1 \cdot S_{est} + K \cdot \dfrac{S_{BC}}{d}}{c_c \cdot M_c} \cdot \Delta t + 1 & 0 & \dfrac{K \cdot \dfrac{S_{BC}}{d}}{c_c \cdot M_c} \cdot \Delta t \\[3mm] 0 & 0 & -\dfrac{h_2 \cdot S_{est} + 2 \cdot c_w \cdot \Gamma_2}{\Gamma_2 \cdot L_{cs} \cdot c_w / v_2} \cdot \Delta t + 1 & \dfrac{h_2 \cdot S_{est}}{\Gamma_2 \cdot L_{cs} \cdot c_w / v_2} \cdot \Delta t \\[3mm] 0 & \dfrac{K \cdot \dfrac{S_{BC}}{d}}{c_c \cdot M_c} \cdot \Delta t & \dfrac{h_2 \cdot S_{est}}{c_c \cdot M_c} \cdot \Delta t & -\dfrac{h_2 \cdot S_{est} + K \cdot \dfrac{S_{BC}}{d}}{c_c \cdot M_c} \cdot \Delta t + 1 \end{bmatrix}$$

$$[Vect\_System] = \begin{bmatrix} \dfrac{2 \cdot c_w \cdot \Gamma_1 \cdot T_{in}}{\Gamma_1 / v_1 \cdot L_{cs} \cdot c_w} \\[3mm] \dfrac{P}{c_c \cdot M_c} \\[3mm] \dfrac{2 \cdot c_w \cdot \Gamma_2 \cdot T_{in}}{\Gamma_2 / v_2 \cdot L_{cs} \cdot c_w} \\[3mm] \dfrac{P}{c_c \cdot M_c} \end{bmatrix}$$

A simple MATHCAD program solves the problem easily. Figure B annex 5.3 shows the case for tank 1 and 2 with an applied RF power of 13 MW. A deep study of transients and the time delays gives a clear scenario of the thermal behaviour of the prototype under full RF power. Final stable conditions have then been obtained, confirming also the analysis of chapter 5.

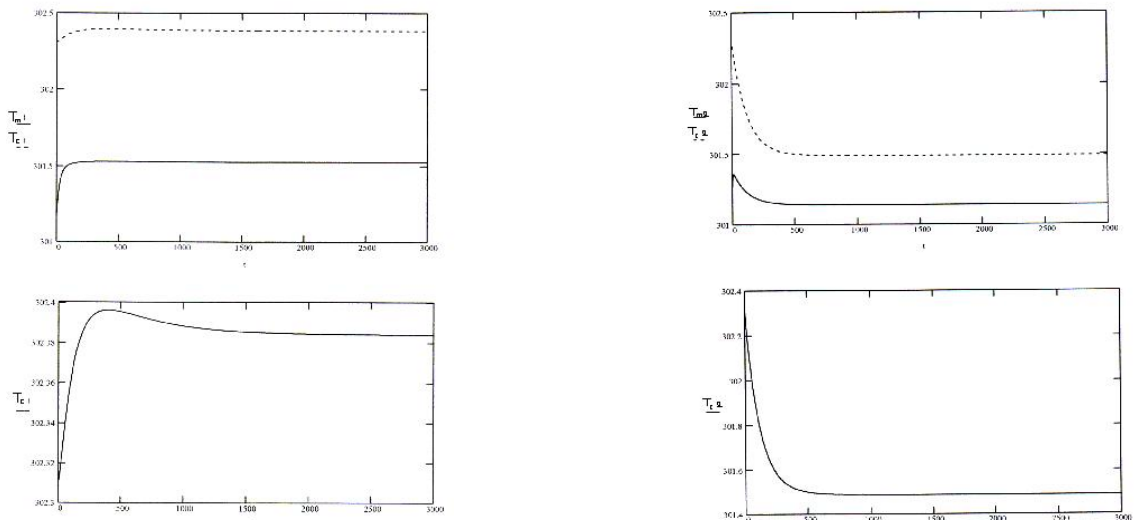

**_Figure B annex 5.3_** _Transient analysis of tank 1 and 2 under full RF power as function of time (applied peak power  P 13 MW, duty cycle 5 $10^{-4}$, power loss 18% → Average power into LIBO (tank 1 to 4) 5330 W). The water valve in tank 1 is practically closed ($\Gamma_1$ 0.025 l/sec), while it is open in tank 2 ($\Gamma_2$ 0.19 l/sec.): the low cooling increases the temperature of tank 1 up to a stable condition._

## Annex 6.1      Tools for RF measurements of single LIBO half-cell plate*

One of the main parameters of a cavity resonator is its Q value. It measures the width Δf of the resonance defined as the bandwidth between the frequencies where the stored energy has fallen to half its resonant value while being driven from a constant RF power source (paragraph 6.3). Normally high Q value implies low spreads in frequency measurements.

In order to perform systematic RF measurements on single half-cell plates, new tools have been designed for the construction of the prototype. The tools are shown in figures A and B annex 6.1.

Here a plate with two RF antennas can squeeze (with a constant load P) the material on a 2 mm-width-circular-area around the cell. In this area the copper is compressed up to 10 N/mm² (the elastic limit of heat treated material has been estimated to 50 N/mm²), creating a well-defined area with high electric contact. This aspect implies high Q values, and consequently low spreads in frequency measurements. Figure C annex 6.1 presents the Q values measured on the half-cell plates of the first tank. Q of about 2000-3000 can generate low frequency spreads (about 1-1.5 MHz), small enough to be corrected by lateral tuning rods. Measurement of a non-brazed full tank is still an open problem. In this case Q values of about 1000 generate high frequency error (about 3 MHz): anyway this can be also corrected by the large tuning range during the lateral rod insertion. Finally measurements on tanks already brazed (Q ranges between 2000 and 6000) have shown very low frequency spreads.

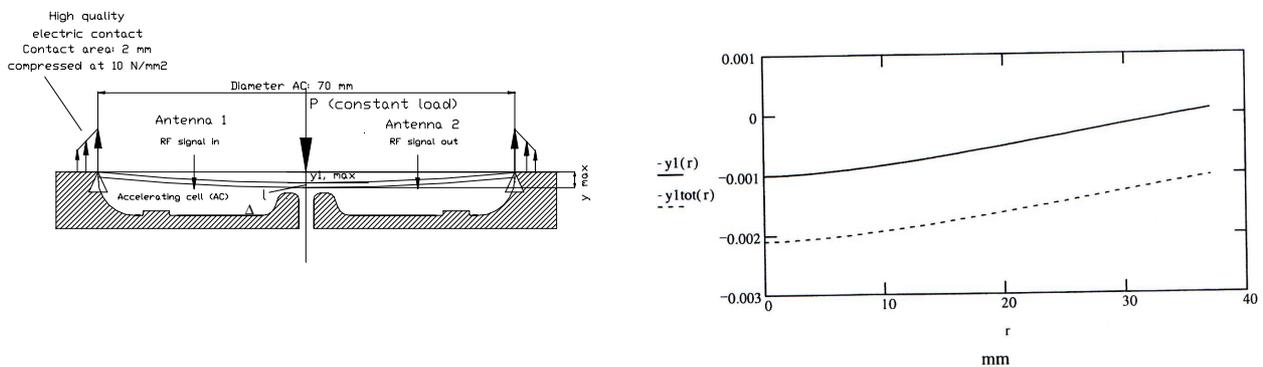

**Figure A annex 6.1** *Schematic picture of tools for frequency measurements of single half-cells and the cover deflection under the load P² [C.54].*

---

[2] The displacement of the cover in copper for frequency measurements of single half-cell is generated by a load P, calculated by:

$$y_1(r) = \frac{12 \cdot P \cdot (1 - v^2)}{16 \cdot \pi \cdot E \cdot th^3} \cdot \left[ \frac{3 + v}{1 + v} \cdot (R_{in}^2 - r^2) - 2 \cdot r^2 \cdot \ln\left(\frac{R_{in}}{r}\right) \right]$$

where v is the Poisson module (0,35), E is the module of elasticity (1.2 10⁵ N/mm²), th is the cover thickness, $R_{in}$ is the inner radius of the compressed circular area, r is the radial position. To have a realistic scenario one should take into account also the displacement of the compressed copper of the cavity (calculated by the simple relation Δl / l = σ / E).
In this case one can estimate a maximum displacement of the cover, respect to the original shape of the cell, as the sum of two terms: $y_{max} = y_{1,max} + Δl$. Changes of RF volumes produce changes in frequencies. Final maximum cover deflection is about 2 μm (figure A annex 6.1), with accuracy in frequency for the measurement much smaller than the required tolerances.

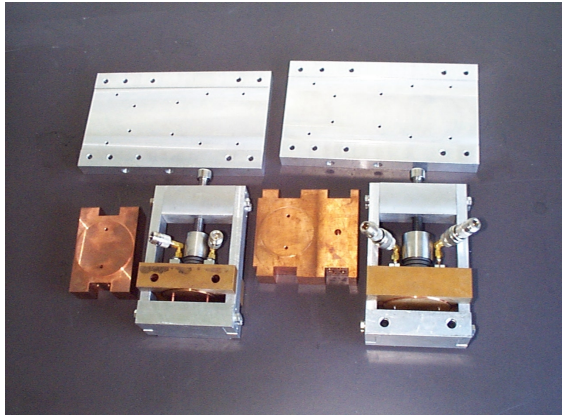 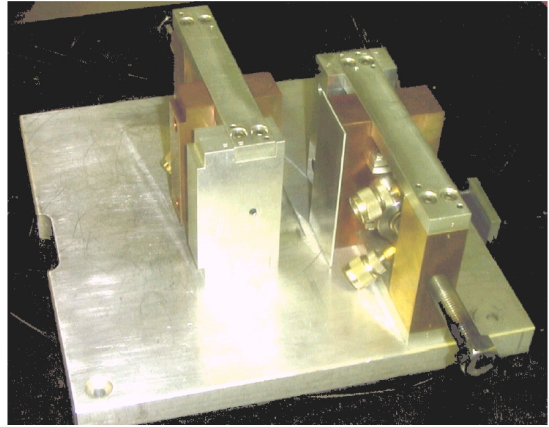

**Figure B annex 6.1** *Tools for frequency measurements of half-cells. On the left the full set of tools. On the right the two main tools for measurements of the accelerating and coupling cells. The covers in contact with the cells are in copper.*

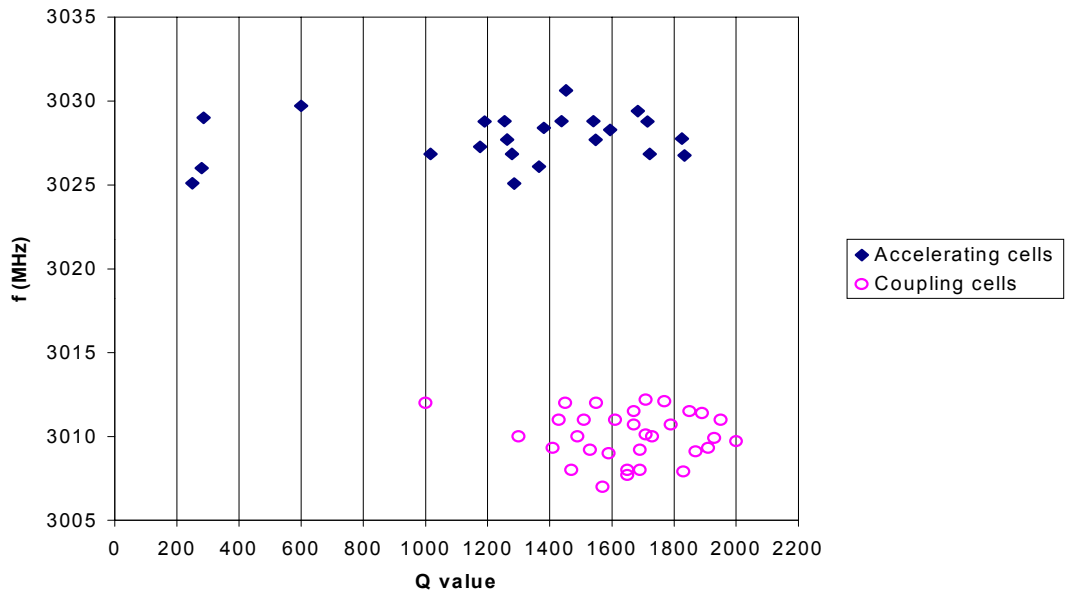

**Figure C annex 6.1** *Example of Q values measured for accelerating and coupling cells (refereed to tank 1). These measurements are repetitive and within a reasonable spread.*

**Annex 6.2        Slater perturbation theorem and the bead pulling technique for field measurements\***

The perturbation of a single oscillator resulting in a change of the stored energy will generally result in a resonant frequency shift. For a cavity on resonance, the electric and magnetic stored energies are equal. If a small perturbation is made on the cavity wall, this will generally produce an imbalance of the electric and magnetic energies, and the resonant frequency will shift to restore the balance. The Slater perturbation theorem describes the shift of the resonant frequency, when a small volume $\Delta V$ is removed from the cavity of volume V. The general result is [C.25, C.26]:

$$\frac{\Delta f}{f_0} = \frac{\int_{\Delta V}(\mu_0 H^2 - \varepsilon_0 E^2)dV}{\int_V (\mu_0 H^2 + \varepsilon_0 E^2)dV} = \frac{\Delta U_m - \Delta U_e}{U}$$

where U is the total stored energy, given in terms of the unperturbed field amplitudes E and H by:

$$U = \frac{1}{4}\int_V (\mu_0 H^2 + \varepsilon_0 E^2)dV$$

One can also write:

$$\Delta U_m = \int_{\Delta V}\mu_0 H^2 dV / 4 , \ \ \Delta U_e = \int_{\Delta V}\mu_0 E^2 dV / 4$$

where the first term is the time average of the stored magnetic energy removed and the second is the time average of the stored electric energy removed as a result of the reduced volume. The frequency increases if the magnetic field is large where the walls are pushed in, and it decreases if the electric field is large there. This result is easier to remember if one identifies a decrease in the effective inductance where the magnetic fields is large, and an increase in the effective capacitance where the electric field is large. There can be cases where the electric and magnetic effects cancel. For example, this would occur if the end walls of a pillbox cavity, excited in a $TM_{010}$ mode, were pushed in uniformly. The contribution from the electric fields near the axis is exactly cancelled by the dominant magnetic field effect at larger radius. This is why the frequency is independent of the cavity length for this case.

Anyway the Slater theorem provides the basis for field measurements in cavities: if a small bead is inserted into the cavity, the perturbation shifts the resonant frequency, and the electric field can be measured. This method is called bead pulling technique (paragraph 6.3).

**Annex 6.3    RF tuning solutions by "dinging" technique: the experience of Los Alamos National Laboratory and the R&D program for LIBO performed at CERN for future applications\***

One solution for RF tuning of the accelerating and coupling cells could include, <u>for future applications,</u> the so-called "dinging" technique. Here the frequency is raised by cavity volume deformation, dimpling a copper septum as shown in figure A annex 6.3. Wide experience has been developed for the 800 MHz SNS side coupled cavity linac [C.40] at Los Alamos National Laboratory.

Each accelerating and coupling cell has blind holes and a thin membrane is plastically deformed with appropriate tooling to tune the cavity in a predominantly magnetic field region. Slater-perturbation is then used to calculate the shift in frequency based on displaced volume. In this context an accurate stress and strain curve (figure B) is needed for the analysis of plasticity and the elastic behaviour of the material, once the applied load is released. It is important also to check the grain size and orientation, inclusions, voids and material composition of the thin septum under stress (figure C).

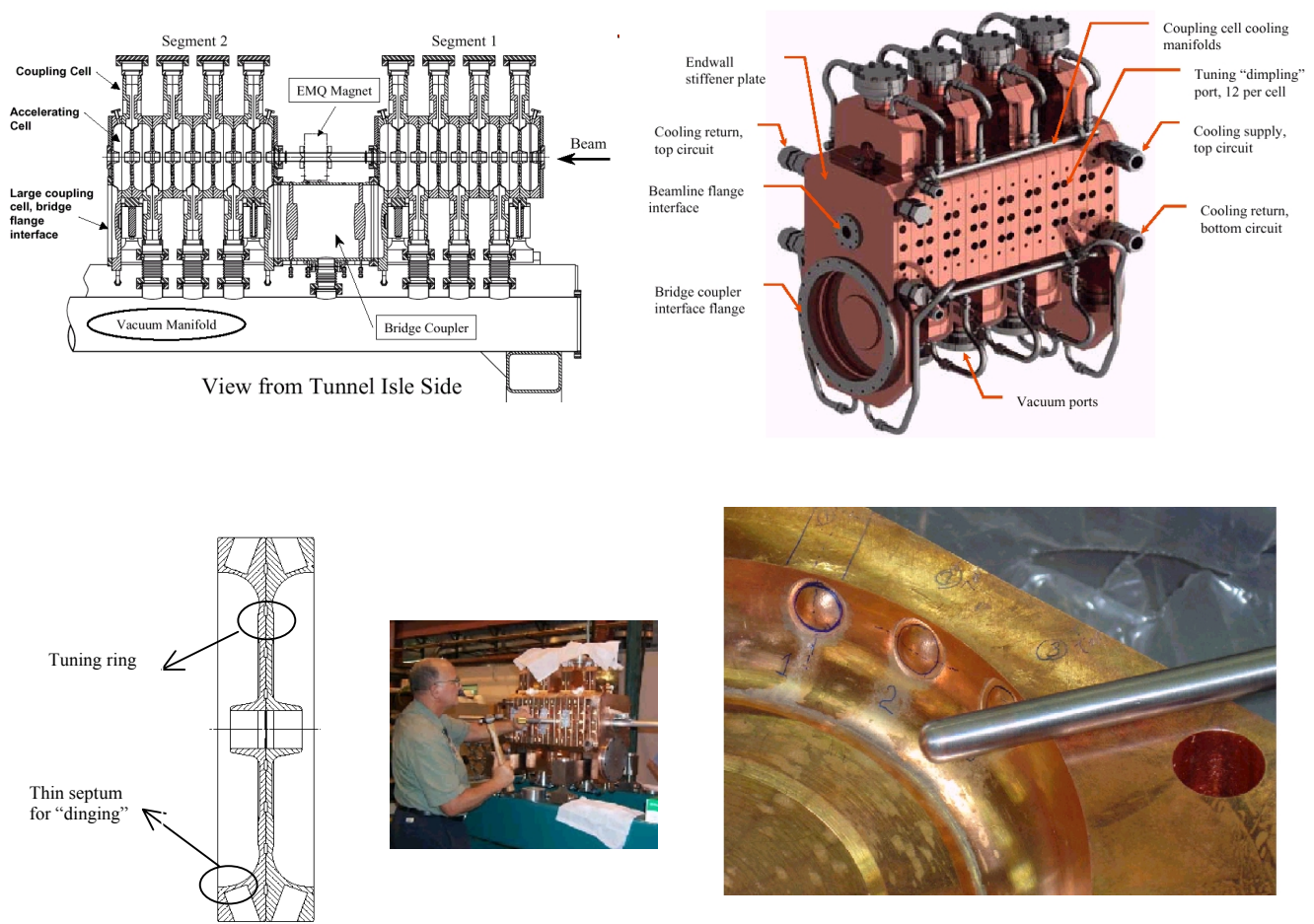

*Figure A annex 6.3* Experience with RF tuning by "dinging" technique, developed for the SNS project. Here two tuning facilities are used for the accelerating cells: machining of the tuning ring (with a range of 2.5 MHz) does the first while the second one includes the "dinging" holes (with a range of 400 kHz). The figures show a view of the side coupled linac: tuning holes are on the lateral side of the accelerating tank (Los Alamos National Laboratory).

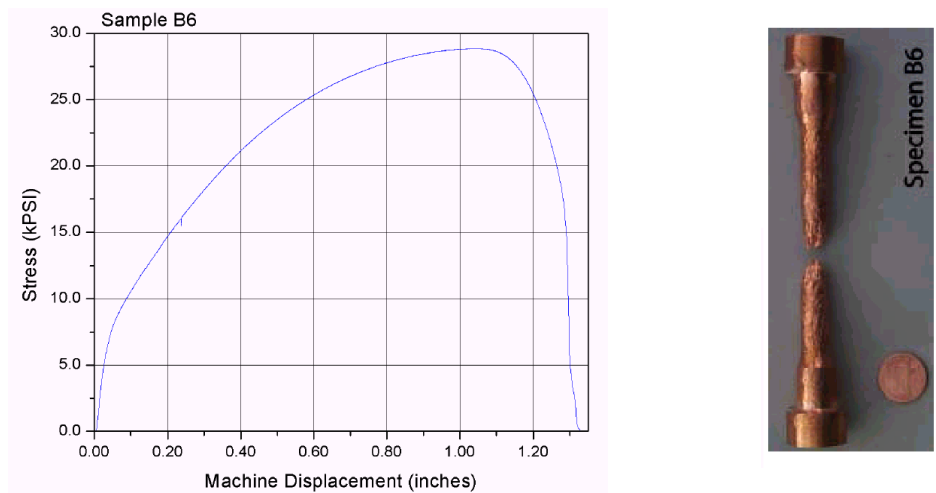

**Figure B annex 6.3** *Stress and strain behaviour of OFHC copper is shown: "dinging" technique foresees a deep knowledge of the mechanical properties of the materials [C.40] (Los Alamos National Laboratory).*

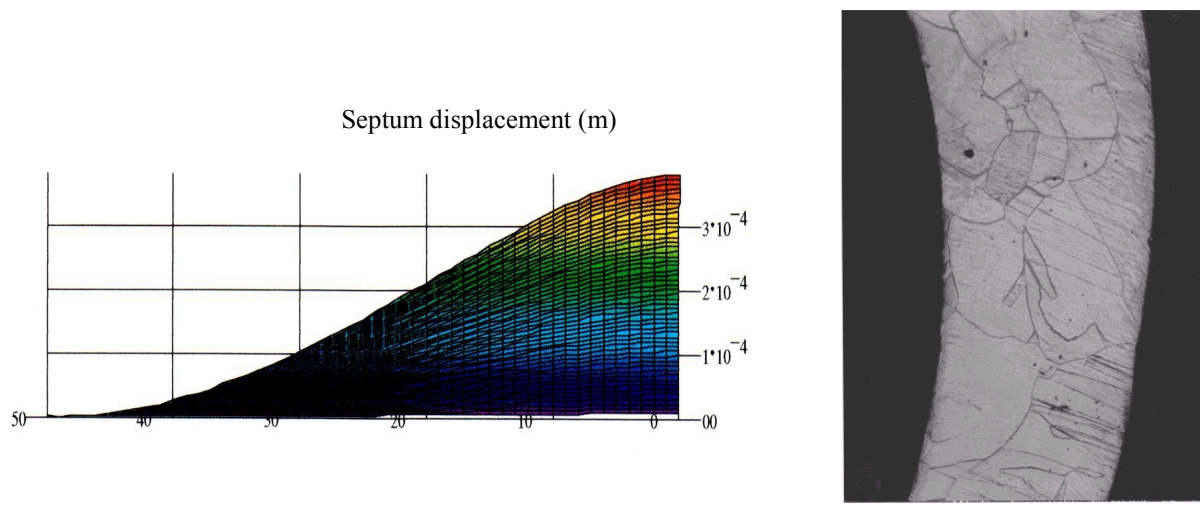

**Figure C annex 6.3** *The study of the "dinging" technique as an RF tuning option. On the left a computer simulation of the mechanical behaviour for the thin septum: figure shows the case of 0.4 mm of maximum plastic deformation (CERN). On the right, the crystallographic structures of the copper thin wall for the "dinging" hole (Los Alamos National Laboratory). The septum connects the internal side of the cavity (under vacuum) with the external atmosphere, and because high temperature brazing produces grain growth, the risk of possible disomogenities must be seriously considered and avoided.*

Detailed calculations (figure C) and tests (figure D) have been carried out at CERN to determine all these aspects. In particular they have clarified the mechanical performance, the frequency sensitivity and the design margin of this RF tuning system. Finally, absences of vacuum leaks have been also proved once the septum is deformed at the maximum displacement.

The mechanical tests have been performed on simple pieces in forged and laminated copper after three heat treatments, ranging between 760 °C and 860 °C (brazing simulation). The geometry of the "dinging" hole is shown in figure D annex 6.3. The maximum acceptable displacement of the septum has been fixed between 0.6 and 0.8 mm, corresponding to a tuning volume of about 6 mm³.

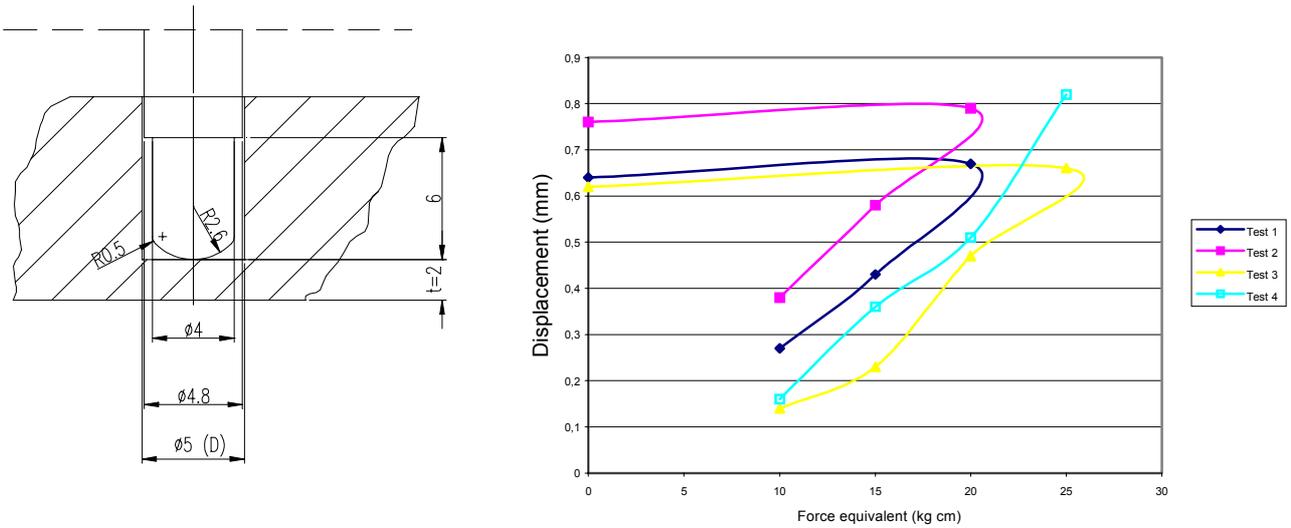

**Figure D annex 6.3** *To have a clear understanding of the "dinging" technique as a RF tuning facility, several pieces have been deformed, checking then the maximum acceptable displacement before septum braking. On the right the stress and strain relation of thin septum is shown. The copper has been heat-treated before the tests. The elastic return (less then 0.1 mm) is also visible when the load is released (CERN).*

In particular the connection between the orthogonal force applied on the septum and the maximum displacement under plastic and irreversible deformation is shown. The elastic return[3] of copper material has been also measured with repetitive results. In all cases no evidence of leakage has been detected.

---

[3] In analogy to the analysis of annex 5.1, we briefly define the basic relation in the plastic domain. We define the stress and strain problem for an incremental process ($\Gamma$ is the normal domain, while $\delta\Gamma$ is the incremental situation when the stress is applied). The tensor formulation can be summarised by the following equations:

$$\begin{cases} \delta\sigma_{ik/k} = -\delta F \\ \delta\sigma_{ik} \cdot n^k = \delta f_i \\ \delta\varepsilon_{ik} = \dfrac{1}{2}\left(\delta S_{i/k} + \delta S_{k/i}\right) \\ \delta S_{i(c)} = \overline{\delta S_i} \\ \delta\sigma_{ik} = c_{ikrs}\left(\underbrace{\delta\varepsilon^{rs}}_{\substack{ELASTIC \\ COMPONENT}} - \underbrace{\delta\vartheta^{rs}}_{\substack{THERMAL \\ COMPONENT}} - \underbrace{\delta\varepsilon p^{rs}}_{\substack{PLASTIC \\ COMPONENT}}\right) \\ \delta\varepsilon p_{ik} = \dfrac{\partial\varphi}{\partial\sigma_{ik}} \cdot \delta\lambda \end{cases}$$

where $\delta\varepsilon p_{ik}$ is the plastic deformation, while $\delta\varepsilon_{ik}$ and $\delta\theta_{ik}$ are the elastic and thermal anelastic terms respectively. The final deformation is linearly composed by these three components.
A simplified expression for a one-dimension case (simple traction → $\sigma_{11} = \sigma_x$ and $\sigma_{ik} = 0$ for i = 2,3 and i≠k) is shown above as a reference. The different domains are indicated, including the permanent plastic deformation ($\varepsilon p_x$) when the stress ($\sigma_x$) is realised. As usual E is the elastic coefficient of the material.

Moreover, there is no evident worsening in terms of crystallographic structure between one and three thermal cycles as well as there is no evident difference between laminated and forged material with the same brazing simulations. It has been also proved that heat-treated soft material generates less repetitive measurements. Figure E annex 6.3 shows a final geometry of the plastically deformed septum measured by optical techniques at CERN Metrology Laboratory. The deformed volume, analysed on fourteen test pieces, is about 6 mm$^3$ (average value), corresponding to a shift in frequency of about 0.3 MHz and 1.8 MHz for the accelerating and coupling cells respectively. To have then a tuning range of 1 MHz for the cavity, four "dinging" holes for each full accelerating cell and two "dinging" holes for each full coupling cell should be enough.

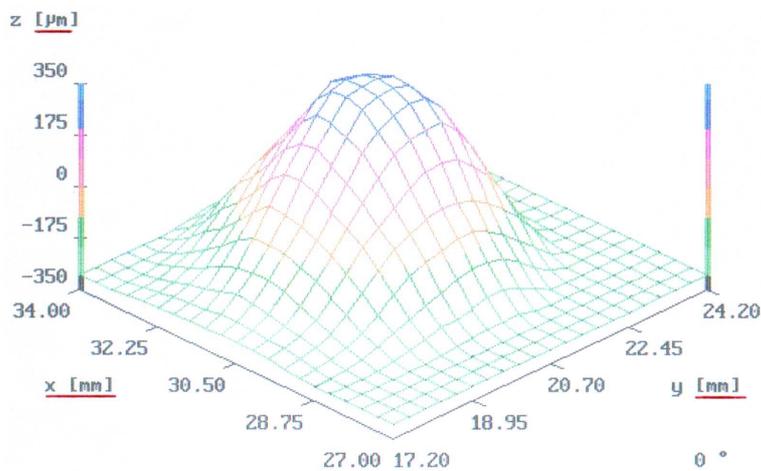

**_Figure E annex 6.3_** _Metrology result by optics technique of the plastically deformed septum. A maximum real displacement of 0.7 mm has been reached without vacuum leaks or crystallographic damage (Metrology Laboratory, CERN)._

**Annex 6.4    Brief summary of RF measurements on LIBO-62 prototype***

❑  *Measurements on test pieces for the definition of the physical parameters (see chapter 5 and 6)*

▪  Coupling coefficient (k) ⇒ determine the size of the coupling slot.

▪  Cell frequencies ⇒ to be compared with simulation codes (SUPERFISH, MAFIA).

▪  Second or next neighbour couplings ($k_a$ and $k_c$) ⇒ indication of difference cell / tank frequencies.

▪  Range and sensitivity of tuning facilities ⇒ eventually, change number and/or position.

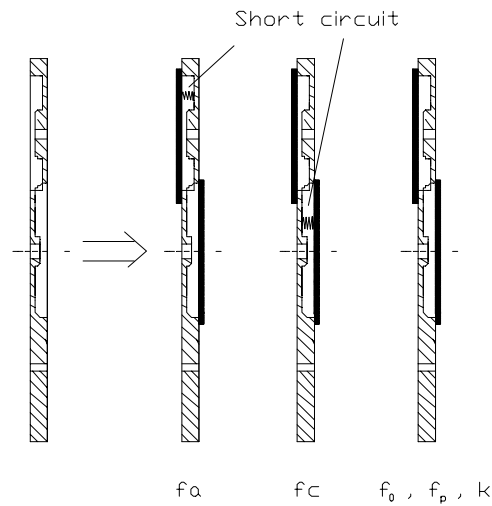

***Figure A annex 6.4*** *Set up for RF measurements of singles half-cells performed with the network analyser. For each configuration the relative frequencies are measured ($f_a$, $f_c$, $f_0$, $f_\pi$ and coupling coefficient k between accelerating and coupling cells). With the next nearest neighbour couplings between accelerating ($k_a$) and coupling cells ($k_c$), $f_a$ and $f_c$ ≠ $f_{\pi/2}$. $k_a$ is about –0,74%, while $k_c$ is about 0. With these considerations $f_c ≈ f_{\pi/2}$, while $f_a > f_{\pi/2}$ [D.50, D.51].*

❑  *Measurement sequences on test pieces*

▪  Measurement of frequency and coupling on all cells.

▪  Open progressively the slot of one half-cell  →  define the slot size.

▪  Machining of all the slots at the correct size.

▪  Measurement of the cells in all possible configurations.

▪  Brazing of 2 (or 4) cells and RF measurement after brazing → brazing effect on RF frequency.

▪  Change the frequencies by tuning facilities (machining of rings and rods insertion).

❑  *Measurements and tuning of tank 1*

▪  Machining of all the cells for tank 1.

▪  Machining of the entire cell slots (same dimension).

▪  Measurement in the RF laboratory of all individual cell frequencies ($f_a$ ,$f_c$).

▪  Assembly of each tank in RF lab and measurement of π/2 frequency and electric field distribution on axis with:  - covers (pressure on both sides to have a <u>good electric contact</u> *)*,

        - bead-pulling measurement system – (bead through cell centres).

- Tuning of each cell on the lathe (by machining the tuning ring) ⟹ minimise the spread between cells.

- Correction of the average cell frequencies accordingly to what measured above.

- Assembly of the tank in the lab and measurement of π/2 frequency and electric field distribution on axis.

- Brazing of the tank (25 half-cell plates).

- Measurement of π/2 frequency and electric field distribution on axis (in the RF lab).

- Preliminary tuning evaluation of the tank by lateral rods insertion (no brazing) to compensate:

  - the errors in the determination of the final tank frequency (← uncertain electric contact during the assembly before brazing),

  - errors/differences in the determination of the brazing effect on RF frequency,

  - differences between individual cells appearing during brazing (deformations, different thickness, …)

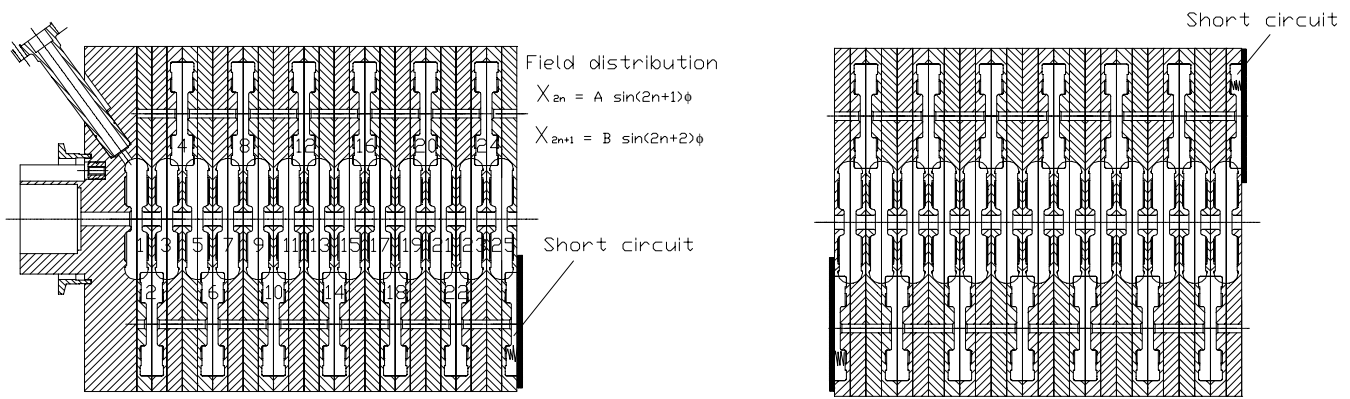

***Figure B annex 6.4*** *Set up for RF measurements of pre-brazed tanks 1 and 4 (left) and tanks 2 and 3 (right). For each configuration the $f_{\pi/2}$ frequency has been measured. The electric field distributions on axis are also indicated. The $f_{\pi/2}$ drops with increased number of accelerating cells and is correctly determinated only with the total number of cells [D.50, D.51].*

❑  *Tuning of tanks 2, 3 and 4*

Follow the same procedures mentioned for tank 1.

❑  *Tuning of the module*

- Tuning and brazing of the bridge couplers with the three piston tuners (see figure 5.10).

- Assembly of the bridge couplers with the four tanks already brazed (4 tanks + 3 bridge couplers) before brazing (good pressure on 1.3 m length!).

- Measurement of the overall frequency and electric field distribution on axis (bead-pulling).

- Cutting at the final length the lateral tuning rods (rods for AC for electric field flatness <u>inside of each tank,</u> rods for CC to close the stop band (SB)).

- Brazing of the module (tanks with end cells + bridge couplers + lateral rods (piston tuners in the cells)).

- Final tuning with tuners in the bridge couplers for final frequency and electric field adjustment <u>between consecutive tanks.</u>

**Annex 6.5     Engineering management for LIBO-62 prototype construction\***

The experience developed at CERN with the prototype could be useful to extrapolate, due to its modularity, important considerations for time and costs construction of a full scale LIBO. Figures A and B annex 6.5 show the cash flow and the budget distribution of the LIBO-62 prototype during the construction period (1999-2001) [C.62, C.63, C.65]. The cost analysis includes: working hours of labour (machining, brazing, cleaning, metrology, etc.), raw materials and special instrumentation (PMQs, water valves, cables, DAQ and controls for cooling used for RF tests). Salaries are not included at this stage, as well as the complete vacuum and RF system (klystron, DAQ and controls), borrowed from CERN. Figure C presents the repartition of the total working hours needed for the prototype construction. In this frame one should divide the total amount into two main tasks: the R&D for final design and the real construction of the prototype. Figure D shows a good balance of tuning procedures for the different sub-components. Finally, figure E presents details of brazing in terms of working hours. This figure should be important due to the high industrial cost of a brazing process under vacuum. From this analysis, one can envisage the importance of the assembly procedure for brazing in terms of working hours, and an industrialisation of the assembly procedures should be investigated in depth with the help of industrial partners.

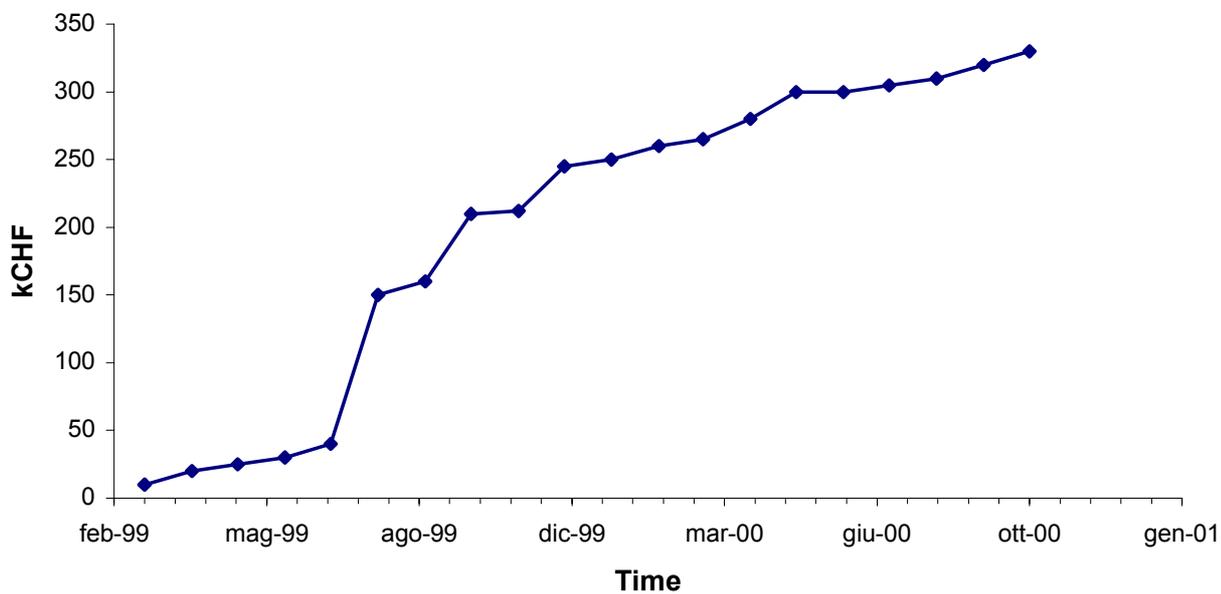

**_Figure A annex 6.5_** _Cash flow for LIBO prototype built at CERN in the period 1999-2001 (kCHF). The analysis includes: 1) working hours of labours at CERN Central Workshop, Surface Laboratory (for brazing and cleaning) as well as all external experts and companies involved in the prototype development (salary of LIBO collaboration scientists are not included at this stage), 2) raw materials and special instrumentation (PMQs, water valves, cables, DAQ and controls)._

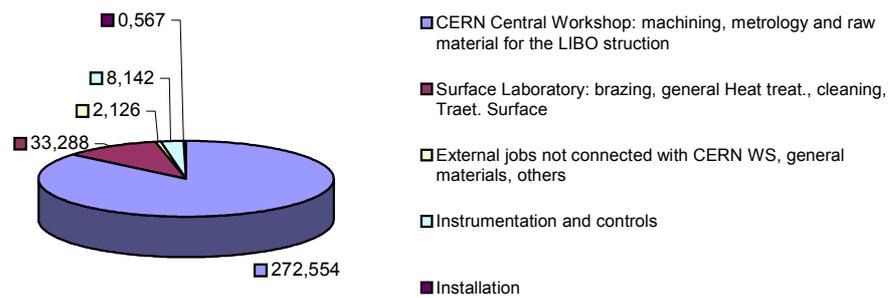

**Figure B annex 6.5** *Budget repartition for LIBO prototype built at CERN (kCHF). The budget is divided into five main groups: 1) CERN Central Workshop, 2) Surface Laboratory, 3) Raw material and external jobs, 4) Commercial instrumentation and controls, 5) Installation.*

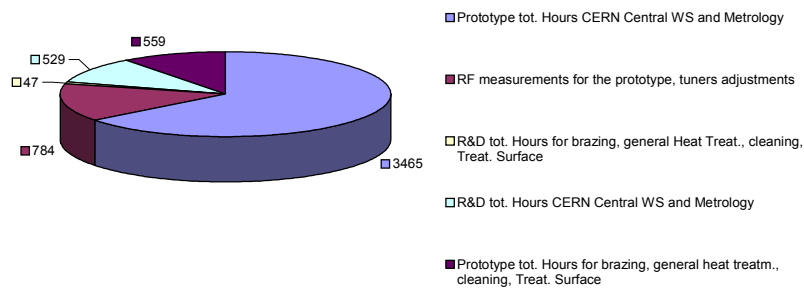

**Figure C annex 6.5** *Repartition of the working hours for LIBO prototype construction at CERN. The construction is divided into five categories, grouped into two main tasks: the R&D for final design and the real construction of the module. The numbers refer to the real hours effectively worked.*

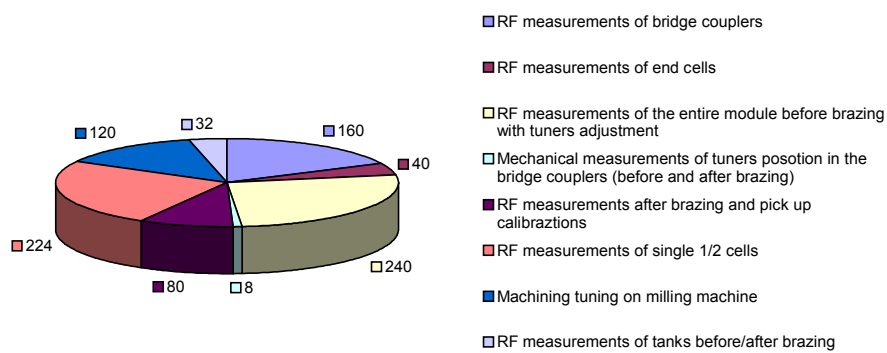

**Figure D annex 6.5** *Repartition of working hours for RF measurements and tuning on the prototype. Experts of LIBO Collaboration have performed the measurements, so the relative costs are not included in the overall analysis.*

*RF measurements and mechanical tuning procedures take about 900 hours. These figures are important for the estimation of a full LIBO construction in industry, where external contractors should perform systematic and repetitive RF measurements in situ during production following precise RF protocols.*

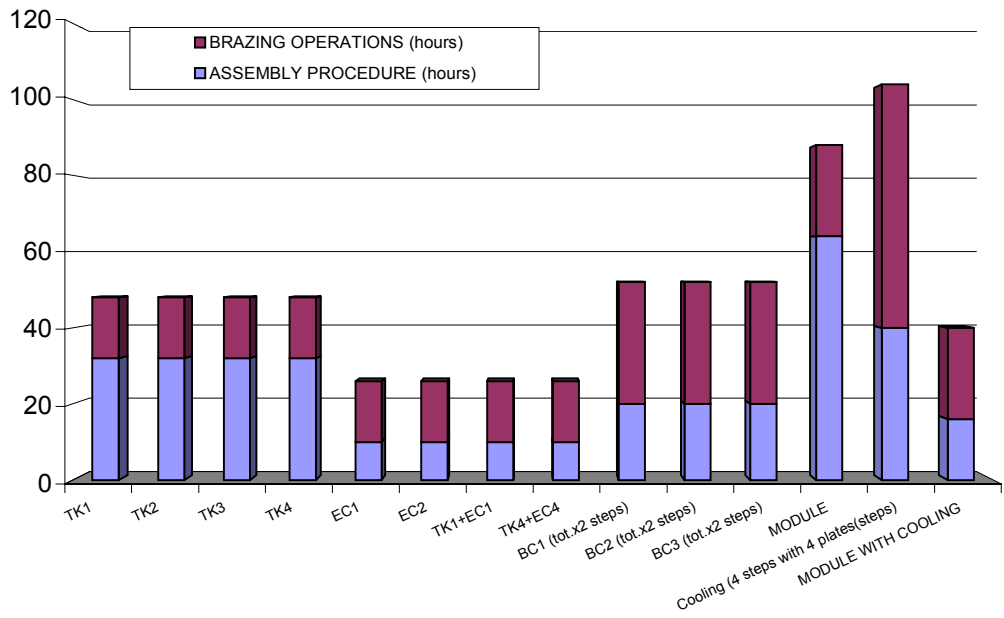

**Figure E annex 6.5** *Time analysis of brazing for the prototype. The figure shows the importance of the assembly procedure. An industrialisation of the assembly for brazing should be investigated with the help of industrial partners.*

## Annex 7.1      Vacuum system*

The function of the vacuum system is to minimise interactions between the beam and residual gases within the accelerating structure. In a linear accelerator for cancer therapy, the choice of geometry, materials and pumping is determined by the compromise between vacuum and mechanical stability on the one side and the necessity to cover all the physical constraints dictated by the clinical requirements. For the vacuum stability of a warm structure, one must maximise the gas pumping and reduce ion induced desorption. Moreover to suppress electron multipactoring one must select surface materials with low secondary electron emission yield. For economical reasons linear accelerator cavities are normally designed for maximum, operationally maintainable, axial fields. Therefore, in order to sustain the high field between the noses of accelerating cells, the vacuum system parameters are determined by the requirement that the pressure in the linac cavity should be of the order of $10^{-6}$ mbar or lower [C.44]. Another consideration here is the possibility of the beam losses due to interaction with the residual gas in the cavities. Due to the very low LIBO beam current needed for therapy, this aspect is of minor importance. With standard pumps it is today possible to avoid "backstreaming" of the pumping vapour, since this will condense on the cavity surfaces, with a resultant deteriorating effect on field holding capability. Moreover contamination of the cavity surface might result in a lowering of the threshold level for multipactoring type discharges (see chapter 7). Particular care is then needed for the design of the vacuum system [C.44]. Moreover the pumps have different pumping speeds for various gases and vapours and especially for the noble gases, such as argon, the pumping speed is relatively low. For example in an evapor-ion pump, with a pumping speed for air of 1500 l/sec at $10^{-6}$ Torr, the pumping speed for argon is less than 5 l/sec. Notwithstanding the relatively low abundance of argon in air, with an imperfect vacuum envelope this could lead to a build-up of argon in the vacuum system. Another practical problem has been found on the difficulty of starting the discharge type pump at pressures above $10^{-4}$ Torr, especially if water or organic vapours are present.

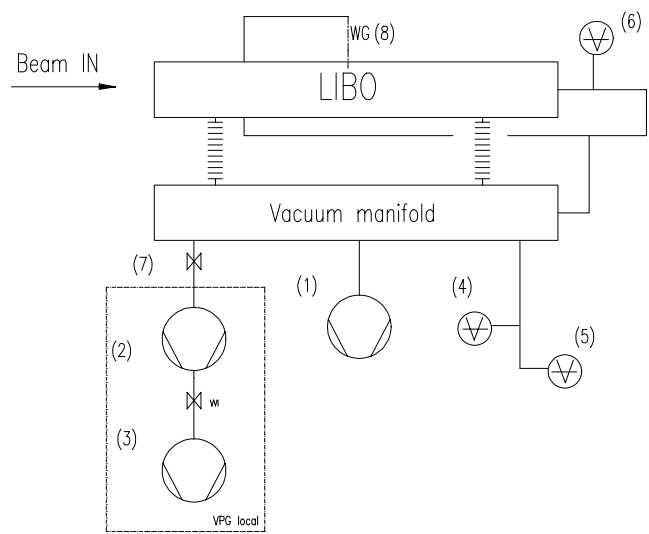

**Figure A annex 7.1** _Vacuum system configuration for LIBO-62 prototype. In figure are visible: 1) ion pump, 2) turbo pump, 3) primary pump, 4) Pirani gauge, 5-6) Penning gauges, 7) vacuum valve, 8) waveguide [C.52]._

For these reasons it has been found desirable to complement a system of ion pumps with turbo pumps, which would be used for start-up procedures. For pre-evacuation of the linac structure, standard rotating vane type pumps (primary group) are used. It is worthwhile to mention that the outgassing rate for copper surfaces in the linac cavities is taken to be $5 \times 10^{-9}$ Torr l/sec/cm$^2$. This number is based on the outgassing rate for copper and other materials used in the linac assembly, such as gasket, etc. After long periods of operation, with additional clean-up produced by the usage of high RF fields in the cavity, the outgassing rate is considerably reduced. Typically this outgassing rate can be reduced by a factor 100 since initial operation [C.10, C.52].

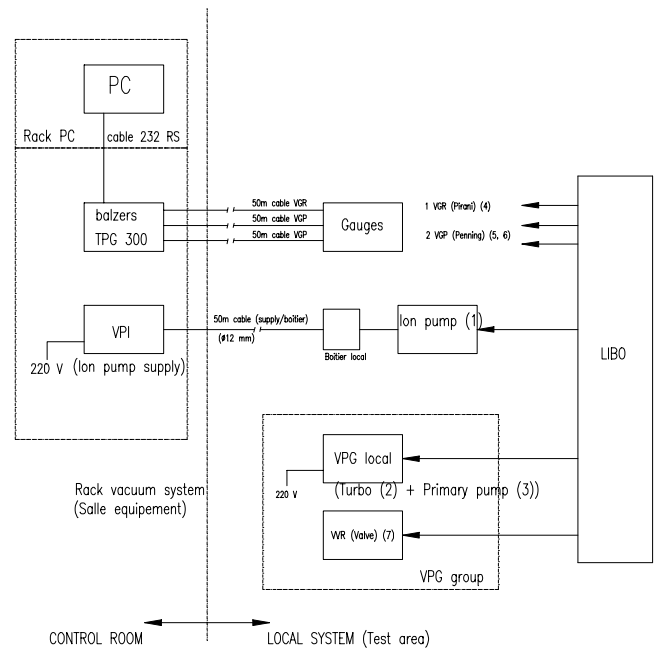

**_Figure B annex 7.1_** _Controls for vacuum system of LIBO-62 prototype._

The final vacuum configuration for the prototype is shown in figure A annex 7.1. The primary requirement for this system is to provide sufficient pumping to overcome the surface outgassing of vacuum facing components and maintain a beam tube pressure that is below the values required for the beam current operation. The prototype vacuum system is composed by an ion pump (1) to maintain the vacuum at the proper operating level and a primary system (2 + 3) with the relative valve (7). Moreover the module has a valve (not indicated in figure) and connection for back-filling with inert gas in the event that maintenance involves breaking the vacuum. Instrumentation is provided for monitoring valve status and vacuum pressure as shown in figure B annex 7.1. An ion pump is needed near the RF window of the central bridge coupler (iris) to supply the additional pumping requirements needed to compensate for potentially high outgassing rates from the window during conditioning.

The structure design foresees also particular attention for the vacuum aspects (see chapter 5). For example the apertures of the pumping ports in the bridge couplers and in the active RF zones are studied to attenuate the RF power flow and the slot orientations are in the same direction as the RF currents. The total vacuum pumping conductance per port is high enough to allow an adequate pumping at the operational base pressure at least of $10^{-6}$ mbar.

## *Acknowledgements*

This thesis is a summary of the scientific journey performed for the design, construction and tests of LIBO prototype in the period 1998-2002 at CERN, in co-operation of many people.

First of all I would like to thank the founder of TERA Foundation, Professor Ugo Amaldi, who made possible everything that is described in these pages. For these reasons I have enormously appreciated the opportunity he offered me more than seven years ago when, after an interview in Geneva, he gave me the chance to be part of this challenging and outstanding adventure. Special thanks are for the founder and the Scientific Director of the ETOILE Project, Professor J. Remillieux as well as Professor A. Demeyer of the Université Claude Bernard, Lyon 1, for the patience, kindness and support.
To all three go my thanks for the assumed responsibility, helping me to realise what I had had in my mind for a long time, even in the unusual condition forced by my unexpected current responsibilities.

A special thanks is for the project leader of LIBO, Dr. Mario Weiss for his patience and constant instructive support during the construction of the prototype. I appreciated very much his thorough competence and I still remember when, more then ten years ago, I studied particle accelerator physics on his papers at graduate level. Moreover I'm deeply grateful for his several suggestions and help in the reading of the manuscript, especially during our long phone conversations, sometimes even till midnight.

At this point I MUST say thanks to Dr. Balazs Szeless, who introduced me to the field of engineering design for particle accelerators, and constantly taught me the importance of the systematic approach during a scientific project in all these years. Moreover I would like to thank him once again for his courage. More than seven years ago and as a chairman of the committee, he allowed a young and inexperienced engineer to be part of the Production Readiness Review for ATLAS experiment, normally performed only at senior level. I will be always grateful for all this.

I would like also to thank my CERN fellowship supervisor Dr. Ettore Rosso, for his friendship, shearing with me different subjects with stimulating discussions. Moreover I appreciated very much his constant help, supporting me in all the initiatives and the never-ending requests we proposed to the ETT Division.

To Prof. Ugo Amaldi, Mario, Balazs and Ettore go my sincere thanks. In all these years they were always ready to help and protect me whatever the problem and it has really been an honour to learn and grow with them for what, I'm sure, will be part of my memory for all my life.

I want to thank now Riccardo Zennaro for his friendship: seven years ago we started together a fascinating and mysterious project that has become a unique experience over the years.

Thanks also to Professor C. De Martinis, D. Giove, L. Grilli and C. Cicardi of the INFN and Milan University group and Professor V. Vaccaro, Dr. D. Davino and M. R. Masullo from University Federico II of Naples. It is also a pleasure to acknowledge all the other old and new colleagues of LIBO Collaboration, such as S. Allegretti, S. Braccini, K. Crandall, M. Crescenti, M. Di Rosa, G. Magrin, P. Pearce, D. Toet, M. Vretenar. To all these people go my thanks for their ideas, discussion, support, efforts, and goodwill.

Moreover the construction and tests of the LIBO prototype is the result of the collaboration of many people which I worked with. First of all the CERN Directors have given a continuous and concrete support to the project. In particular I would like to mention the Director of the CERN Technology Transfer Dr. H. F. Hoffmann, that first allowed me to work in the technical Co-ordinator group of ATLAS Collaboration in parallel with LIBO project, and later pushed me for a CERN fellowship position created for LIBO.

Mechanical construction and vacuum brazing of the module were done at CERN in co-operation with dedicated teams led by J. C. Gervais and Dr. S. Mathot, whom I appreciated very much to work with.

I would like to mention the Chief of Surface Laboratory and Material Dr. Bacher as well as the collaboration of the vacuum section of the LHC Division, in particular D. Allard.

Thanks also to Dr. P. Bourquin for the consulting in technology of particle accelerators, giving the permission to have access to the archive of PS and CLIC drawing offices: the past and the present historical memory of CERN.

Most of the work for the prototype tests has been done in the CERN-PS Division and I acknowledge the help of the Division leaders and all the PS staff. In particular I must mention Dr. Rinolfi, G. Rossat and G. Yvon for their friendship and professional competence during the installation at LIL.

Thanks also to A. Catenaccio of CERN and the Los Alamos colleagues Dr. J. Stovall and N. Bultman.

I must mention now the spokesmen of ATLAS Collaboration Dr. P. Jenni and all the technical Co-ordinators (Drs H. F. Hoffmann, M. Price and M. Nessi) for the hospitality in building 40 at CERN during the last seven years as well as all people of ATLAS that supported me, even during our animated weekly LIBO meetings.

I am then indebted to all the members of the jury for accepting to devote their valuable time to me.

Finally, my special thanks go to the person who supported me for all these years, even during difficult moments, and shearing with intelligence and patience different and uncommon subjects: my wife.

❑ *Main references on physics and engineering of particle accelerators*

▪ *Particle accelerators theory*

▪ *Radio Frequency for particle accelerators*